\documentclass[%
aip,
amsmath,amssymb,nofootinbib,
reprint,%
]{revtex4-1}

\usepackage{graphicx}% Include figure files
\usepackage{dcolumn}% Align table columns on decimal point
\usepackage{bm}% bold math
\usepackage[utf8]{inputenc}
\usepackage[T1]{fontenc}
\usepackage{mathptmx}
\usepackage{amssymb,bm}
\usepackage[usenames]{color}
\usepackage{url}
\usepackage{enumerate}
\usepackage{natbib}
\usepackage{color}
\usepackage{xcolor}

\def \aap {A\&A}

\def \apj {ApJ}
\def \apjl {ApJL}
\def \apjs {ApJS}

\def \mnras {MNRAS}

\def \prd {Phys. Rev. D}

\begin{document}
\definecolor{color1}{RGB}{191, 0, 255}
\newcommand{\DV}[1]{{\textcolor{orange}{DV: #1}}}
\newcommand{\CP}[1]{\textcolor{red}{CP: #1}}
\newcommand{\RA}[1]{\textcolor{color1}{RA: #1}}
\newcommand{\sign}{{\rm sgn}}
\newcommand{\vort}{\vec{\omega}}

\title[SGS model for compressible MHD instabilities]{Extension of the sub-grid-scale gradient model for compressible magnetohydrodynamics turbulent instabilities}

\author{Daniele Vigan\`o}\email{daniele.vigano@uib.eu}
\author{Ricard Aguilera-Miret}%
\author{Carlos Palenzuela}%
\affiliation{Departament  de  F\'{\i}sica, Universitat  de  les  Illes  Balears, Institut d'Estudis Espacials  de  Catalunya and Institut d'Aplicacions Computacionals de Codi Comunitari (IAC3),  Palma  de  Mallorca,  Baleares  E-07122,  Spain}

\date{\today}

\begin{abstract}
Performing accurate large eddy simulations in compressible, turbulent magnetohydrodynamics is more challenging than in non-magnetized fluids due to the complex interplay between kinetic, magnetic and internal energy at different scales. Here we extend the sub-grid-scale gradient model, so far used in the momentum and induction equations, to account also for the unresolved scales in the energy evolution equation of a compressible ideal MHD fluid with a generic equation of state. We assess the model by considering box simulations of the turbulence triggered across a shear layer by the Kelvin-Helmholtz instability, testing cases where the small-scale dynamics cannot be fully captured by the resolution considered, such that the efficiency of the simulated dynamo effect depends on the resolution employed. This lack of numerical convergence is actually a currently common issue in several astrophysical problems, where the integral and fastest-growing-instability scales are too far apart to be fully covered numerically. We perform a-priori and a-posteriori tests of the extended gradient model. In the former, we find that, for many different initial conditions and resolutions, the gradient model outperforms other commonly used models in terms of correlation with the residuals coming from the filtering of a high-resolution run. In the second test, we show how a low-resolution run with the gradient model is able to quantitatively reproduce the evolution of the magnetic energy (the integrated value and the spectral distribution) coming from higher-resolution runs. This extension is the first step towards the implementation in relativistic magnetohydrodynamics.

\end{abstract}

\maketitle

\section{Introduction}

In the context of computational fluid dynamics,  direct numerical simulations (DNS) are employed to resolve the equations in all the dynamically relevant scales, including the dissipative ones. This means that the numerical grid size $\Delta$ has to be smaller than the dissipative scale. Since the computational cost of DNSs scales as the cube of the Reynolds number\citep{sagautbook}, in many cases it is unfeasible to have a small enough grid size and a large enough domain at the same time, thus representing all the scales of interest. This is the case for a variety of problems within different disciplines, ranging from industrial, laboratory-testable applications (such as gas turbines, steady turbulent flows in channels...), to astrophysical scenarios (such as convection zones of stars, planetary atmospheres, accretion disks, relativistic jets...). The increasing computational resources and the gradual sophistication of numerical techniques are constantly improving the achievable resolution and widening the number of scenarios where DNS will be feasible; however, currently there are many cases with very large Reynolds number, where simulations are still very far from capturing all the scales, and an alternative to the brute force of increasing resolution is then needed.

The two most widely used numerical approaches to bypass the resolution limitations in turbulent flows are the Reynolds Average Numerical Simulations (designed for steady-state flows and used especially in industrial applications) and the Large Eddy Simulations (LES). For both cases, the idea is to directly evolve the quantities on the {\em resolved scales}, i.e. from the integral scales (the largest ones, corresponding to dominion size or to the typical scale where energy is injected), down to the grid scales, and to describe the effect of the smaller ones through a suitable sub-grid-scale (SGS) model. LES have been developed and applied to many different computational fluid dynamics applications, with a plethora of variants, applicable to different numerical schemes. (see\cite{piomelli14} for an historical and prospective point of view of LES development).
While LES were originally designed for simple, steady flows, applications to decaying\cite{yang04} transitional flows have become more popular (see recent reviews\cite{yang15} and references within). Due to the variety of turbulent scenarios, LES with SGS models represent an active field of inter-disciplinary research, including engineering, plasma physics and astrophysics. For the latter, which is also our ultimate aim of implementation, the use of LES is still not very common\cite{miesch15,schmidt15}, compared to terrestrial scenarios.

Generally speaking, the purely hydrodynamic turbulent dynamics is relatively well understood due to the universality of the spectral properties, and an improvement of the proposed SGS models have allowed an impressive accuracy in reproducing experimental data or results coming from the DNS\cite{yang15}. On the other hand, the dynamics becomes more problem-dependent when magnetic fields are included, and prevent one from defining universal laws, valid for different scenarios. Magneto-hydrodynamic (MHD) turbulence consists of a non-linear interplay, with transfer from the kinetic to the magnetic (dynamo mechanism) and internal (dissipation) energies, and, at the same time, between different scales.\cite{beresnyak15} In general, at the smallest scales the dynamo mechanism is more effective. A further non-trivial effect of the magnetic fields is the breaking of isotropy and self-similarity of the flow. Despite the huge theoretical and numerical progress in half a century, there are still many open questions. The main reason is that experimental data for high magnetic Reynolds number are limited, despite recent interesting results\cite{tzeferacos18}. A living review\cite{beresnyak19} extensively summarizes the current state-of-the-art of MHD turbulence, focusing especially in astrophysical scenarios, with a focus on numerical methods, and linear and non-linear regimes of the dynamo effects.

In these scenarios, the smallest relevant dynamical scales cannot be captured by the available resolution, so that numerical simulations are still far from reaching the numerical convergence. In particular, LES have been very rarely applied to study the development of MHD instabilities. This is the case, for instance, of the Kelvin-Helmholtz instability (KHI), which growth rate is inversely proportional to the scale of the perturbation. The fastest-growing modes are characterized by a scale inversely proportional to the Reynolds number, so that if the latter is large enough, the KHI cannot be fully captured numerically. Among different astrophysical cases where KHI takes place, we mention the shear layer developed in a binary neutron star merger, source of the recently detected gravitational waves: the smallest relevant scales are likely a fraction of a meter, and the integral scales are thousands of kilometers, so that the state-of-the-art general relativistic MHD simulations\cite{kiuchi15,kiuchi18} reach at its very best only 12.5 meters (at a price of about 60 million CPU-hours), thus not showing numerical convergence yet.

The arguably most used SGS model in LES is the Eddy-dissipative one, introduced by Smagorinsky half a century ago in the context of purely hydrodynamic turbulence\cite{smagorinsky63}. It allows to mimic the fluid's viscosity, and, in its magnetic extension \cite{theobald94}, the turbulent magnetic diffusivity. The multiple variants of the Eddy-dissipative model have allowed to set-up optimal coefficients in order to fit with an increasing accuracy the results coming from experiments or DNSs, especially for non-magnetic problems\cite{schumann75,kraichnan76,chollet81,germano91,vreman94,yang15}. Other SGS models have also been proposed relying mainly on dimensional analysis (cross-helicity model, for instance) or physical assumptions, like the dynamo mechanism. Comparisons between them have been studied in a variety of scenarios.\cite{muller02a,chernyshov07,grete17phd}

In this paper, we focus on the gradient model\cite{clark79,yeo87,liu94,muller02a,balarac13,grete17phd}, based on a mathematical formulation which relates the finite resolution to an effective filtering of the equations. By expanding the filtered non-linear terms up to the first order of the Taylor series, one can approximate the residuals appearing in a given evolution equation with a functional form which includes the derivatives of the resolved fields, thus effectively extrapolating to the SGS the trends seen at the smallest resolved scales.

Recently, new advances in SGS modeling have been focused mostly in proposing advance methods for the dynamical evaluation (inspired by the pioneering work of Germano\cite{germano91}) of the free coefficient in different SGS formulations\cite{hendra19,zhou18}, including the gradient model for scenarios including, e.g., combustion flame \cite{lar18}, channel flow \cite{ghaisas16,jiang16,jiang18,dupuy19,lu19}, or a set of these standard computational fluid dynamics problems \cite{vollant16}.

In this paper, we propose an extension of the gradient model to include the residuals appearing in the energy evolution equation, for a generic equation of state, with a particular focus on the ideal gas.

In order to test the extended gradient model, we run box simulations of a non-forced magnetic KHI evolving according to the compressible MHD equations for an ideal gas. This problem, commonly used in several studies, allows us to test the different phases of a highly turbulent, magnetized flow: the development of the instability at small scales (the higher the resolution, the smaller such scales and the faster is the growth of the instability), and its decay after reaching isotropy and homogeneity. We test different initial conditions. We use high-order-accurate finite-difference methods, that are in general less suitable than spectral methods for turbulent flows (due to their higher dissipation), but are adequate in astrophysical simulations to deal with the development of strong shocks.

We will perform a-priori and a-posteriori tests. The former relies on the filtering of high-resolution runs, in order to fit the filtering residuals with an SGS model. We compare the performance of the gradient model with a few other models for the different non-linear residuals. The second kind of test is more challenging and consists in the effective implementation of the SGS model in the simulation, thus including its feedback in the evolution. One can compare the results of a low-resolution case with SGS model to a high-resolution one, without it.

Very detailed works about LES with SGS models have been performed in the context of either incompressible\cite{muller02a,muller02b,kessar16,vollant16} or compressible \cite{grete15,grete16,vlaykov16,grete17,grete17phd,balarac13} MHD, but not, to our knowledge, including an ideal gas equation of state. While the momentum and induction equations have been explored in many different works, the energy evolution has usually been neglected, or only partially considered, since they impose either incompressibility or a polytropic/isothermal equation of state \cite{grete17b}.

In \S~\ref{sec:mhd} we summarize the MHD equations and their filtered versions implicitly evolved in a LES. In \S~\ref{sec:sgs} we introduce the extended gradient model, and a few more SGS models, for comparison. In \S~\ref{sec:simulations} we describe the three-dimensional (3D) KHI simulations, together with some details about the platform and the numerical methods used. In \S~\ref{sec:apriori} we present the main highlights of the a-priori results obtained by filtering snapshots at different times of simulations with a variety of initial conditions, equations of state, and resolutions. In \S~\ref{sec:les} we show the results coming from the implementation of the proposed model in low-resolution LES. Conclusions are drawn in \S~\ref{sec:conclusions}. The Appendices provide  some details about the spectral and fitting calculations, and the numerical validation of our code with well-known, numerically-converging two-dimensional (2D) benchmark tests.

%%%%%%%%%%%%%%%%%%%%%%%%%%%%%%%%%%%%%%%%%%%%

%%%%%%%%%%%%%%%%%%%%%%%%%%%%%%%%%%%%%%%%%%%%
\section{LES and sub-filter-scale residuals in compressible ideal MHD}\label{sec:mhd}

\subsection{Evolution equations}

The conservative formulation of the compressible ideal (non-relativistic) MHD equations consists of the continuity, momentum, induction and energy evolution equations, respectively:

\begin{eqnarray}
&& \partial_t\rho + \partial_k \left[ \rho v^k \right] = 0~, \label{eq:mhd} \\
&& \partial_t S^i + \partial_k \left[ \rho v^k v^i - B^k B^i + \delta^{ki}\left(p + \frac{B^2}{2}\right) \right] = 0~, \nonumber \\
&& \partial_t B^i + \partial_k \left[ v^k B^i - v^i B^k  \right] = 0~, \nonumber \\
&& \partial_t U + \partial_k \left[\left(U  + p + \frac{B^2}{2}\right) v^k -(v_j B^j) B^k \right] = 0~, \nonumber
\end{eqnarray}
where $i$ and $k$ represent the spatial components, 
$\delta^{ki}$ is the standard Kronecker delta, $S^i = \rho v^i$ is the lineal momentum. The magnetized fluid is described by the mass density $\rho$, the fluid pressure $p$, the internal energy density $e$, the velocity field $v^i$, and the magnetic field $B^i$. The total energy density is defined as
\begin{equation}
U = e + \frac{\rho v^2}{2} + \frac{B^2}{2}~,\label{eq:energy_definition}
\end{equation}
where for simplicity we have absorbed the $1/4\pi$ factor in $B^2$, and set the speed of light and the magnetic permeability $c = \mu_0 = 1$.

The closure of the set of equations is defined by an equation of state, which relates the pressure to the other physical variables, $p=p(\rho,e,B^i)$. At each step of a simulation, the set of the discretized conserved quantities, ${\cal C}^a = \{ \rho, S^i, B^i, U\}$, is numerically evolved, and the internal energy is recovered by inverting eq.~(\ref{eq:energy_definition}). The physical variables are recovered accordingly: the velocity is simply given by dividing $S^k$ by $\rho$, while the fluid pressure is a equation-of-state-dependent function of the conserved variables, $p({\cal C}^a)$.

\subsection{Filtering and inversion of variables}\label{sec:inversion}

In a LES, the effect of finite resolution can be thought as equivalent to a low-pass spatial filter applied to the evolution equations. The spatial scale cut-off, $\Delta_f$, is of the order of the grid cell, $\Delta$. Indicating the filter kernel with $G$, the filtering operator over a field $f$ can be written as
\begin{equation}
{\overline f}({\bf x},t) :=  \int_{-\infty}^{\infty} G({\bf x}-{\bf x'}) f({\bf x'},t) d^3 x'~.
\end{equation}
An extended literature is dedicated to the mathematical representation of the filter.\cite{yeo87,pope00,grete17phd} 
The functional form of the filter kernel is intrinsically given by the numerical method, and can be evaluated by specific studies of mode propagation, which can show both a damping and dispersive character for high wavenumbers $k \gtrsim 2\pi/\Delta$.\cite{zhao14} Generally speaking, finite-difference methods, here used, are much more dissipative (i.e., the effective filter is wider) than spectral methods, even when they are built for high-order accuracy, like the ones used here.

When we evolve any discretized set of equations with a conservative approach, we effectively solve for the filtered conserved fields, indicated hereafter by $\overline{{\cal C}}^a$. Any other primitive or auxiliary field $f$ (e.g., velocity, pressure or internal energy) can be recovered by expressing it as a function of the conserved fields, $f(\overline{{\cal C}}^a)$. These implicit dependences will be indicated by $\widetilde{f}$. The most common example of this approach consists in expressing the velocities as a function of the filtered conserved density and momenta: the well-known mass-weighted or Favre-filtered velocities
\begin{equation}\label{eq:vel_filter}
\widetilde{v}^i := \frac{\overline{S}^i}{\overline{\rho}}~~.
\end{equation}
Hereafter we will write the relevant expressions as functions of $\widetilde{v}^k$, but all equations could be easily rewritten in terms of the conserved filtered momentum, $\overline{S}^k$, simply using eq.~(\ref{eq:vel_filter}) and employing the identity $\overline{\rho v}^k = \overline{\rho}~\widetilde{v}^k$.

Besides the velocity, we will make use of the same notation for other two non-conserved fields: the pressure, $\widetilde{p}:=p(\overline{\rho},\overline{S}^i,\overline{B}^i,\overline{E})$, for which the functional form depends on the equation of state, and a quantity closely related to it.

\subsection{Filtered equations}

The filtered continuity equation is simply
\begin{equation}\label{eq:continuity_filter}
\partial_t\overline{\rho} + \partial_k \left[ \overline{\rho}~\widetilde{v}^k \right] = 0~.
\end{equation}
The filtered momentum and induction equations are well known to be
\begin{eqnarray}
&& \partial_t (\overline{\rho}\widetilde{v}^i) + \partial_k\left[ \overline{\rho}~ \widetilde{v}^k\widetilde{v}^i - \overline{B}^k\overline{B}^i + \delta^{ki}\left(\widetilde{p} + \frac{\overline{B}^2}{2} \right) \right] = \partial_k\overline{\tau}_{\rm mom}^{ki}~, \label{eq:momentum_filtered} \\
&& \partial_t \overline{B}^i + \partial_k \left[ \widetilde{v}^k~\overline{B}^i - \widetilde{v}^i~\overline{B}^k \right] = \partial_k \overline{\tau}_{\rm ind}^{ki}~,  \label{eq:ind_filtered}
\end{eqnarray}
where
\begin{eqnarray}
\overline{\tau}_{\rm mom}^{ki} := && 
\overline{\tau}_{\rm kin}^{ki} - \overline{\tau}_{\rm mag}^{ki} + \overline{\tau}_{\rm pres}\delta^{ki}~.
\label{eq:ns_mom}
\end{eqnarray}
The sub-filter scale (SFS) terms account for the non-commutativity of the filtering operator with the non-linear fluxes. There are then three SFS tensors related to the products $v^kv^i$, $B^kB^i$ and $v^kB^i$ in the fluxes,\footnote{In literature, different sign conventions are used. Hereafter we will always define the SFS terms as the product of the filter minus the filter of the product.}
and a scalar related to the pressure:

\begin{eqnarray}
\overline{\tau}_{\rm kin}^{ki} = && \overline{\rho}~\widetilde{v}^k ~\widetilde{v}^i - \overline{\rho v^k v^i} ~,
\label{eq:tau_kin}\\
\overline{\tau}_{\rm mag}^{ki} = && \overline{B}^k ~\overline{B}^i - \overline{B^k B^i}~,
\label{eq:tau_mag} \\
\overline{\tau}_{\rm ind}^{ki} = && (\widetilde{v}^k~ \overline{B}^i - \widetilde{v}^i~ \overline{B}^k) - (\overline{v^k B^i} - \overline{v^i B^k})~, \label{eq:tau_ind} \\
\overline{\tau}_{\rm pres} = && \widetilde{p} - \overline{p} + \frac{1}{2}\overline{\tau}_{\rm mag}^{jm}\delta_{jm}~.\label{eq:tau_pres} 
\end{eqnarray}
Notice that, by construction, $\overline{\tau}_{\rm kin}$ and $\overline{\tau}_{\rm mag}$ are symmetric tensors, while $\overline{\tau}_{\rm ind}$ is anti-symmetric and its inclusion is equivalent to having a SFS electric field with components given by $\epsilon_{ijk}\overline{\tau}_{\rm ind}^{ki}$.

By following a similar approach, we can write the energy evolution equation as

\begin{equation}
 \partial_t \overline{U} + \partial_k \left[\widetilde{\Theta}~ \widetilde{v}^k - (\widetilde{v}_j\overline{B}^j)\overline{B}^k \right] = \partial_k [\overline{\tau}^k_{\rm adv} - \overline{\tau}^k_{\rm hel} ]~, \label{eq:energy_filtered}
\end{equation}
where we have introduced the advected quantity
\begin{equation}
\widetilde{\Theta} = \overline{U} + \widetilde{p} + \frac{\overline{B}^2}{2}~, \label{eq:def_theta}
\end{equation}
and the additional vectorial SFS terms on the right-hand side of eq.~(\ref{eq:energy_filtered}) arise from the non-linear terms in the flux of the filtered equation. The first SFS term arises from the product of two non-conserved fields, $\widetilde{v}^k$ and $\widetilde{\Theta}$:

\begin{eqnarray}
 \overline{\tau}^k_{\rm adv} & = & \widetilde{\Theta}~\widetilde{v}^k - \overline{\Theta v^k}  = (\widetilde{\Theta} - \overline{\Theta})~\widetilde{v}^k + (\overline{\Theta}~\widetilde{v}^k - \overline{\Theta v^k}) = \nonumber \\
 & = & \overline{\tau}_{\rm pres}\widetilde{v}^k + (\overline{\Theta}~\widetilde{v}^k - \overline{\Theta v^k})~,
 \label{eq:tau_adv}
\end{eqnarray}
and is related to $\overline{\tau}_{\rm pres}$ by construction, since $\widetilde\Theta$ contains the same factor $\widetilde p + \overline{B}^2/2$, while the additional term $\overline{U}$ does not imply any SFS residual since it is conserved. The second SFS term in the right hand side of eq.~(\ref{eq:energy_filtered}) is the cross-helicity term, which is partially related to the induction equation tensor, $\overline{\tau}_{\rm ind}$, but cannot be written only as a function of it (or a simple contraction with another filtered quantity):
\begin{equation}
\overline{\tau}^k_{\rm hel} =  (\widetilde{v}_j~\overline{B}^j)~\overline{B}^k - \overline{(v_jB^j) B^k}~. \label{eq:tau_hel}
\end{equation}
We stress that the functional forms of the terms related to the pressure ($\overline\tau_{\rm pres}$, $\overline\tau_{\rm adv}^k$) depend on the equation of state and, like $\tau_{\rm adv}^k$, had been neglected or only partially considered before. The pressure term $\overline{\tau}_{\rm pres}$ has been studied for the non-magnetized fluid in a few scenarios\cite{borghesi14,borghesi15,taskinoglu10,taskinoglu11}, for instance with a Peng-Robinson equation of state\cite{ma15}, considering the performances of the gradient model\cite{selle07}. Here we consistently include all the terms at play.

To summarize, we need to give the prescriptions for 22 independent components of: three tensors (two symmetric and one anti-symmetric), eqs.~(\ref{eq:tau_kin})-(\ref{eq:tau_ind}), the scalar term $(\widetilde{p}-\overline{p})$ appearing in eqs.~(\ref{eq:tau_pres}) and (\ref{eq:tau_adv}), and two vectors, eqs.~(\ref{eq:tau_adv})-(\ref{eq:tau_hel}).

\subsection{A specific equation of state: ideal gas}

So far, our formalism is general and can extend to any equation of state, with the only difference arising from the dependence $p({\cal C}^a)$. In the next sections, we will present numerical implementation and results for an ideal gas equation of state, namely $p_{\rm ideal} = (\gamma-1) e$, where $\gamma>1$ is the ideal gas adiabatic index. In this case, considering eqs.~(\ref{eq:energy_definition}) and (\ref{eq:def_theta}), we have:
\begin{eqnarray} 
&& \widetilde{p}_{\rm ideal} = (\gamma - 1)\left( \overline{U} - \frac{\overline{\rho}~\widetilde{v}^2}{2} - \frac{\overline{B}^2}{2}\right)~,\label{eq:p_tilde} \\
&& \widetilde{\Theta}_{\rm ideal} = \gamma~\overline{U} + \frac{1-\gamma}{2}\overline{\rho}~\widetilde{v}^2 + \frac{2-\gamma}{2}\overline{B}^2~. \label{eq:Theta_ideal}
\end{eqnarray}
One of the 22 SFS independent functional forms (see previous sub-section), the pressure term, is automatically set by the traces of the tensors $\overline{\tau}_{\rm kin}$ and $\overline{\tau}_{\rm mag}$, since it reads:

\begin{equation}
\overline{\tau}_{\rm pres,ideal} = \left(
\frac{1 - \gamma}{2} \overline{\tau}_{\rm kin}^{lm}  + \frac{2 - \gamma}{2} \overline{\tau}_{\rm mag}^{lm}\right)\delta_{lm}~.\label{eq:tau_pres_ideal} \end{equation}
In the pressure residuals, besides the well-known magnetic pressure contribution, there is an additional (negative) term related to the internal energy (i.e., to the fluid pressure $p$), proportional to $(1 - \gamma)/2$ times the SFS kinetic and magnetic energy. In the limit of isothermal gas ($\gamma \rightarrow 1^+$), such extra contributions are nullified: in that case, or if fluid incompressibility is assumed, the fluid pressure does not depend on the filtered conserved fields, so that $\widetilde{p}=\overline{p}$, and the only surviving term is the magnetic pressure, in agreement with the commonly used formalism\cite{grete16}.

%%%%%%%%%%%%%%%%%%%%%%%%%%%%%%%%%%%%%%%%%%%%
%%%%%%%%%%%%%%%%%%%%%%%%%%%%%%%%%%%%%%%%%%%%%%
%%%%%%%%%%%%%%%%%%%%%%%%%%%%%%%%%%%%%%%%%%%%%%
\section{Sub-grid-scales modeling}\label{sec:sgs}
%%%%%%%%%%%%%%%%%%%%%%%%%%%%%%%%%%%%%%%%%%%%%%
%%%%%%%%%%%%%%%%%%%%%%%%%%%%%%%%%%%%%%%%%%%%%%
%%%%%%%%%%%%%%%%%%%%%%%%%%%%%%%%%%%%%%%%%%%%%%
The explicit expressions of the entire set of SFS terms above (which by definition cannot be computed or known a-priori) represent the closure of the system of equations in a LES.
The closures, or SGS models, considered hereafter are local, i.e. depend only on the derivatives of the fields at the smallest resolved scales: we do not consider the multi-scale models\cite{sondak12,sondak15,sagautbook}, where two different resolved scales are compared in order to extrapolate, by self-similarity, the SGS model at the unresolved scales. There are virtually infinite possible combinations of mutual contractions and functional forms involving the first-derivatives of the resolved fields, which can be considered.\citep{speziale91,kosovic97,lu16}
Far from being exhaustive, below we list the ones we explicitly use in this work, as a comparison with the extended gradient model. Note that the comparison is possible only for some of the SGS terms appearing in the momentum and induction equations: $\tau_{\rm kin}$, $\tau_{\rm mag}$, (or their combination $\tau_{\rm mom}$), and $\tau_{\rm ind}$. Since the terms $\tau_{\rm hel}$ and $\tau_{\rm adv}$ were never formalized before to our knowledge, there is no available SGS models alternative to the one proposed in this paper to compare with. The physical meaning of these new terms is less intuitive, so it is not clear how to model them on a physically-based grounds.

Hereafter we will indicate with an overline the SFS terms $\overline{\tau}$ resulting from either the formal analysis (like in the previous section) or the post-process filtering of a numerical solution (see \S~\ref{sec:apriori}), and with a simple $\tau$ the analytical SGS models, which are implicitly defined proportional to a constant parameter $C$, that are meant to approximate the SFS terms:
\begin{equation}\label{eq:precoefficient}
  \overline{\tau} \approx C ~\tau~.
\end{equation}

\subsection{Extended gradient model}

The {\em gradient} (or {\em non-linear}) {\em model}~\cite{leonard75,clark79,yeo87,liu94,muller02a,balarac13,vollant16,grete17phd} arises from the analytical Taylor expansion of the SFS terms appearing in the MHD equations, under the hypothesis of having a filter with a Gaussian kernel. In the space domain this kernel can be written as 
\begin{equation}
G_i(|x_i-x'_i|) =  \left( 4 \pi \xi \right)^{-1/2} \exp\left(-|x_i-x'_i|^2/4 \xi\right)~,
\end{equation} 
where $\xi=\Delta_f^2/24$ provides a filtering width  equivalent (up to the second moment) to a box filter with size $\Delta_f$. When we apply such filter to a product of fields, $fg$, the Taylor expansion of its Fourier transform and its inverse in terms of the filter scale allow to write the following rules\cite{yeo87,grete17phd}:
\begin{eqnarray}
\overline{fg} &\simeq&  {\overline f} ~{\overline g} + 2\, \xi\, \partial^i {\overline f} \, \partial_i {\overline g}~, \label{eq:expansion_rule} \\
\overline{vg} &\simeq& \widetilde{v}~\overline{g} + 2\, \xi\, \left( \partial^i \widetilde{v} \, \partial_i {\overline g} - \frac{\overline{g}}{\overline{\rho}}~\partial_i\overline{\rho} \, \partial^i \widetilde{v} \right)~, \label{eq:expansion_rule_v}
\end{eqnarray}
which are accurate up to ${\cal O}(\xi^2)$. The second equation is obtained by recurrence relations and is applicable to the product involving the mass-weighted velocity $\widetilde{v}$, eq.~(\ref{eq:vel_filter}). Analogous (although possibly more elaborated) relations can be obtained for other non-conserved fields.

Applying these simple rules to the non-linear products of the fields appearing in the SFS tensors, we can obtain the expression for each tensor. The well-known expansions of the terms~(\ref{eq:tau_kin}), (\ref{eq:tau_mag}), (\ref{eq:tau_ind}), at leading order in $\xi$, read:

\begin{eqnarray}
\tau_{\rm kin}^{ki} &=& - 2\, \xi\, \overline{\rho}\,  \partial_j \widetilde{v}^k \, \partial^j \widetilde{v}^i ~,
\label{eq:taukin_grad}\\
\tau_{\rm mag}^{ki} &=&  - 2\, \xi\, \partial_j \overline{B}^k \, \partial^j \overline{B}^i ~,
\label{eq:taumag_grad}\\
\tau_{\rm ind}^{ki} &=& - 2\, \xi\, \left[\partial_j \widetilde{v}^k  
\left( \partial^j \overline{B}^i - \frac{\overline{B}^i}{\overline{\rho}}\partial^j \overline{\rho}\right)\right. 
\nonumber \\
&& ~ \left. -  \partial_j \widetilde{v}^i  
\left( \partial^j \overline{B}^k - \frac{\overline{B}^k}{\overline{\rho}}\partial^j \overline{\rho} \right) \right] ~.
\label{eq:tauind_grad_compres}
\end{eqnarray}
These expressions have been commonly derived and used, both in a-priori and a-posteriori studies cited above.

Here we propose the general extension to the SFS terms related to the energy evolution equation, eqs.~(\ref{eq:tau_adv})-(\ref{eq:tau_hel}), for a general equation of state in compressible ideal MHD:

\begin{eqnarray}
\tau^k_{\rm adv} & = & \tau_{\rm pres}\widetilde{v}^k - 2\xi\left(  \partial_j  \widetilde{\Theta}\partial^j \widetilde{v}^k  - \frac{\widetilde{\Theta}}{\overline{\rho}}\partial_j\overline{\rho}\partial^j \widetilde{v}^k  \right) ~, \label{eq:tau_adv_grad} \\
\tau^{k}_{\rm hel} &=& - 2 \, \xi \, \left[ \partial_j (\widetilde{v}_m \overline{B}^m) \partial^j \overline{B}^k + \right. \nonumber \\
&& + \left. \overline{B}^k\partial^j \widetilde{v}^m\left(  \partial_j \overline{B}_m - \frac{\overline{B}_m}{\overline{\rho}} \partial_j \overline{\rho}\right) \right]~, \label{eq:tau_hel_grad}
\end{eqnarray}
where we have used eq.~(\ref{eq:expansion_rule_v}) and noted that $\xi \partial_j  \overline{\Theta} = \xi \partial_j  \widetilde{\Theta} + {\cal O}(\xi^2)$.

Last, we have to define the functional form of the pressure term $\widetilde{p} - \overline{p}$, which enters in eqs.~(\ref{eq:tau_pres}) and (\ref{eq:tau_adv_grad}). In general, one has to consider the functional form $p({\cal C}^a)$, and apply the rules above to the non-linear products of conserved variables contained in it.

For an ideal gas equation of state, $\tau_{\rm pres}$ is given by construction by the traces of the kinetic and magnetic SFS tensors, according to eq.~(\ref{eq:tau_pres_ideal}), so that:
\begin{equation}
\tau_{\rm pres,ideal}^{ki} = \xi[(\gamma - 1)\overline{\rho}\partial_j\widetilde{v}^m \partial^j\widetilde{v}_m  + (\gamma - 2)\partial_j\overline{B}^m \partial^j\overline{B}_m]~. \label{eq:tau_pres_grad_ideal}
\end{equation}

\subsection{Eddy-dissipative models}

As a comparison, we consider a simple version of the Eddy-dissipative SGS model (also known as Smagorinsky model).\cite{smagorinsky63} It accounts for the fluid dissipation at small scales, introducing an effective viscosity tensor proportional to the strain tensor $\widetilde{S}^{ki}$. A subsequent work\cite{theobald94} extended the Eddy-dissipative model to the MHD, introducing the analogous turbulent magnetic diffusivity in the induction equation, proportional to the current density tensor. In this paper, we will test the following tensors:
\begin{eqnarray}
\tau^{ki}_{\rm kin} = \Delta_f^2~\overline{\rho}~|\widetilde{S}|~\widetilde{S}^{ki} &\qquad& \widetilde{S}_{kj}:=\frac{1}{2}(\partial_k \widetilde{v}_j + \partial_j \widetilde{v}_k) ~, \label{eq:tau_kin_smag} \\
\tau^{ki}_{\rm mag} =  \Delta_f^2~ |\overline{M}|~\overline{M}^{ki}  &\qquad& \overline{M}_{kj}:= \frac{1}{2}(\partial_k \overline{B}_j + \partial_j \overline{B}_k)~, \label{eq:tau_mag_smag} \\
\tau_{\rm ind}^{ki} =  \Delta_f^2~ \frac{|\overline{J}|}{\overline{\rho}^{1/2}} &\qquad& \overline{J}_{kj} := \frac{1}{2}(\partial_k \overline{B}_j - \partial_j \overline{B}_k) ~, \label{eq:taui_lin}
\end{eqnarray}
where we have introduced the standard resolved fluid strain tensor $\widetilde{S}_{kj}$, the resolved magnetic strain tensor $\overline{M}_{kj}$, and the resolved current tensor $\overline{J}_{kj}$. The kinetic and induction SGS models are the ones commonly used in the literature, with the addition of the factor $1/\overline{\rho}^{1/2}$ in the last term, introduced on dimensional basis; we add a further term to account for the magnetic non-linear term.

By construction, the contractions of eq.~(\ref{eq:tau_kin_smag}) with the fluid strain tensor and of eq.~(\ref{eq:taui_lin}) with the current tensors provide definite positive values of large-to-small-scale transfer energy rate: this shows their dissipative nature. In its original version, $\tau_{\rm kin}^{ki} = (c_s\Delta_f)^2 \sqrt{2}|\widetilde{S}|\widetilde{S}^{ki}$, the pre-factor is usually found to be $c_s \sim 0.1-0.2$\citep{lu16} for incompressible steady flows. In more elaborated versions, $c_s$ can also be determined with a dynamical procedure, instead of being fixed, as proposed originally for a channel flow problem.\cite{germano91,oberai05} Besides the normalization, different recipes are available to treat differently the diagonal and off-diagonal components of $\tau_{\rm mom}$ (for instance, to see how the energy transfer is affected \cite{grete17b}). A plethora of detailed studies\cite{bardina80,rogallo84,moin84,germano91,moin91,lilly92,brun06,leveque07} are dedicated to fine-tune the dissipation rate for specific problems (especially hydro-dynamical, but see a recent work\cite{grete17} and the references within).

Since the fine-tuning of this model is not this purpose and we aim at the comparison of its performances with the gradient model in a turbulent MHD instability case, we will consider a constant parameter $C$, as in eq.~(\ref{eq:precoefficient}).

\subsection{Cross-helicity model}

This model\citep{muller02a} is a variant of the Eddy-dissipative one, with a different normalization, which include products of derivatives of the velocity and magnetic field:
\begin{eqnarray}
&& \tau_{\rm mom}^{ki} = \Delta_f^2~ |\overline{\rho}~\widetilde{S}:\overline{M}|^{1/2}\widetilde{S}^{ki}~, \nonumber \\
&& \tau_{\rm ind}^{ki} = \Delta_f^2 ~\sign(\overline{J}:\widetilde{\Omega})~\frac{|\overline{J}:\widetilde{\Omega}|^{1/2}}{\overline{\rho}^{1/4}} \overline{J}^{ki}~,\\
&& \widetilde{\Omega}_{ki}:= \frac{1}{2}(\partial_k \widetilde{v}_i - \partial_i \widetilde{v}_k)~, \nonumber
\end{eqnarray}
where we have corrected dimensionally the original version (by introducing $\overline{\rho}$ factors), and we have introduced the resolved vorticity $\widetilde{\Omega}_{ki}$. As for the previous one, this model also transfers kinetic energy from resolved scales to SGS. On the other hand, in the induction equation the signum function allows for a back-scatter of magnetic helicity and magnetic energy (inverse cascade) if $\sign(\overline{J}:\widetilde{\Omega})<0$.

\subsection{Vorticity and Alfv\'en velocity models (for induction only)}

Finally, we propose two new further models for the induction equation only. The first one is the vorticity model, defined by:

\begin{equation}
\tau_{\rm ind}^{ki} = \Delta_f^2~{\overline{\rho}^{1/2}}~|\widetilde{\Omega}|~\widetilde{\Omega}^{ki}~.
\end{equation}
Qualitatively, similar models (in their co-variant version) have been proposed in general relativity for neutron star mergers, as extra-terms in the induction equation. They are based on the vorticity, assumed to be a physical tracer of the dynamo mechanism (see its use in a relativistic MHD scenario\cite{palenzuela15,giacomazzo15}), but they consist in an arbitrary functional form.

Finally, we test a new model based on the current, proportional to the resolved Alfv\'en velocity:
\begin{eqnarray}
\tau_{\rm ind}^{ki} = \Delta_f~ \frac{|\overline{B}|}{\overline{\rho}^{1/2}}~\overline{J}^{ki}~. \label{eq:taui_dynamo}
\end{eqnarray}
Based on dimensional grounds only, this tensor is proportional to $\Delta_f$, instead of $\Delta_f^2$.

\section{Simulations}\label{sec:simulations}

\subsection{Platform and numerical schemes}

The code presented here has been generated by using {\it Simflowny}~\cite{arbona13,arbona18} to run under the {\it SAMRAI} infrastructure,\cite{hornung02,gunney16} which provides efficient parallelization and adaptive mesh refinement (not used in this work). {\it Simflowny} is an open-source and user-friendly platform developed by the IAC3 group since 2008 to facilitate the use of HPC infrastructures to non-specialist scientists. It allows to easily implement scientific dynamical models, by means of a domain specific language, based on {\tt MathML} and {\tt SimML}, and a browser-based integrated development environment, which automatically generates efficient parallel code for simulation frameworks. {\it Simflowny} splits the physical models and problems from the numerical techniques. The automatic generation of the simulating code allows to properly include the parallelization features, which in this case rely on the SAMRAI infrastructure. The combination of these two platforms provides a final code with a good balance of speed, accuracy, scalability, ability to switch physical models (flexibility), and the capacity to run in different infrastructures (portability). 

The MHD system of equations, given by eq.~(\ref{eq:mhd}), can be written formally in conservation law form, namely 
\begin{eqnarray}\label{PDEequationdecomposed}
\partial_t {\cal C}^a + \partial_k {\cal F}^k({\cal C}^a) = 0~,
\end{eqnarray}
where ${\cal C}^a=\{ \rho, \rho~v^i, B^i, U, \Phi\}$ is the list of evolved conserved fields and ${\cal F}^k({\cal C}^a)$ their corresponding fluxes which might be non-linear but depend only on the fields and not on their derivatives. Here $\Phi$ is a scalar field introduced to dynamically control the divergence of the magnetic field by means of a divergence-cleaning scheme\cite{dedner02}. For instance, in the simulations presented below, the dimensionless ratio between the L2-norm of $\Delta^2(\vec{\nabla}\cdot\vec{B})$ and the volume-integrated magnetic energy rise at the beginning of the simulation, reach a maximum of less than $10^{-4}$ and then stably decrease to $\sim 10^{-5}$, thus ensuring a negligible violation of the constrain. For further details about the divergence-cleaning formalism and numerical results, see our previous works \cite{palenzuela18,vigano19}.

The time integration of the discretized equations is performed by using a fourth-order Runge-Kutta scheme, which ensures the stability and convergence of the solution for a small enough time step $\Delta t \leq 0.4 ~\Delta$. For the spatial reconstruction, we employ High-Resolution Shock-Capturing (HRSC) methods\citep{toro97} based on  high-order accurate finite-difference schemes. They are commonly used in astrophysical scenarios and allow to deal with the appearance of shocks and to take advantage of the existence of weak solutions in the equations. The fluxes at the cell interfaces are calculated by combining the Lax-Friedrichs splitting with high-order non-oscillatory reconstruction schemes.\cite{shu98}

In Appendix~\ref{app:2D_validation}, we show in detail the validation of the code through benchmark tests in 2D, comparing different reconstruction schemes, of third and fifth order: FDOC3, WENO3YC, FDOC5, MP5, and WENO5Z. The latter is our final choice for all the results here presented. We defer the reader to previous works\cite{palenzuela18,vigano19} for further details on these numerical schemes and an extensive analysis of the performance with different discretization schemes for different problems, including MHD. Here we only stress that these schemes, although very accurate, are more dissipative than spectral methods (more commonly used in turbulence, see below for the implications).

\subsection{Initial conditions}\label{sec:setup}

The KHI has been studied by a plethora of works with many different numerical schemes, also in the compressible MHD case\cite{keppens99}. We employ a setup similar to what found in literature, \cite{beckwith11} consisting in a periodic cubic 3D box of domain $[-L/2,L/2]^3$  (described by Cartesian $\{x,y,z\}$ coordinates), with shear layers at $y = \pm y_l$ planes, and initial conditions given by:

\begin{eqnarray}
&& \rho=\rho_0 + \rho_1~\sign(y)\tanh\left(\frac{|y| - y_l}{a_l}\right)~, \\
&& v_x = v_{x0} ~\sign(y)\tanh\left(\frac{|y| - y_l}{a_l}\right) + \delta v_x~, \label{eq:vx} \\
&& v_y = \delta v_y~, \\
&& v_z = v_{z0}~\sign(y)\exp\left[-\frac{(|y| - y_l)^2}{\sigma_z^2}\right] + \delta v_z~, \label{eq:vz}
\end{eqnarray}
where $a_l$ is the shear layer thickness and $\sigma_z$ is the extension scale the profile of $v_z$, respectively. The pressure and the magnetic fields are assumed constant, $p=p_0$, $\vec{B}=\vec{B}_0$.

\begin{table}[t]
	\begin{center}
		\begin{tabular}{l c c c c c}
			\hline
			\hline
			Model 	  & $N^3$ & $\gamma$       &$v_{x0}$&$v_{z0}$& $\rho_1$ \\
			\hline
			\hline
			{\tt KH3D250}   & $ 250^3$ & $4/3$& 0.50 & 0 & 0.5 \\
			\hline
			{\tt KH3D500}   & $ 500^3$ & $4/3$ & 0.50 & 0 & 0.5 \\
			{\tt KH3D500h}  & $ 500^3$ & $4/3$ & 0.50 & {\it 1} & 0.5 \\
			{\tt KH3D500x}  & $ 500^3$ & $4/3$ & {\it 0.25} & {\it 1} & 0.5 \\
			{\tt KH3D500d}  & $ 500^3$ & $4/3$ & 0.50 & 0 & {\it 1.0} \\
			{\tt KH3D500i}  & $ 500^3$ & {\it 1.0001} & 0.50 & 0 & 0.5 \\
			\hline
			{\tt KH3D1000}  & $1000^3$ & $4/3$ & 0.50 & 0 & 0.5 \\
			{\tt KH3D1000h} & $1000^3$ & $4/3$ & 0.50 & {\it 1} & 0.5 \\
			\hline
			{\tt KH3D2000}  & $2000^3$ & $4/3$ & 0.50 & 0 & 0.5 \\
			\hline
			\hline
		\end{tabular}
	\end{center}
	\caption{3D KHI models: initial setup parameters (marked in italic the variations over the fiducial ones). See text for the other parameters, common to all simulations.}
	\label{tab:models}
\end{table}

\begin{figure*}[t] 
	\centering
	\includegraphics[width=0.31\linewidth]{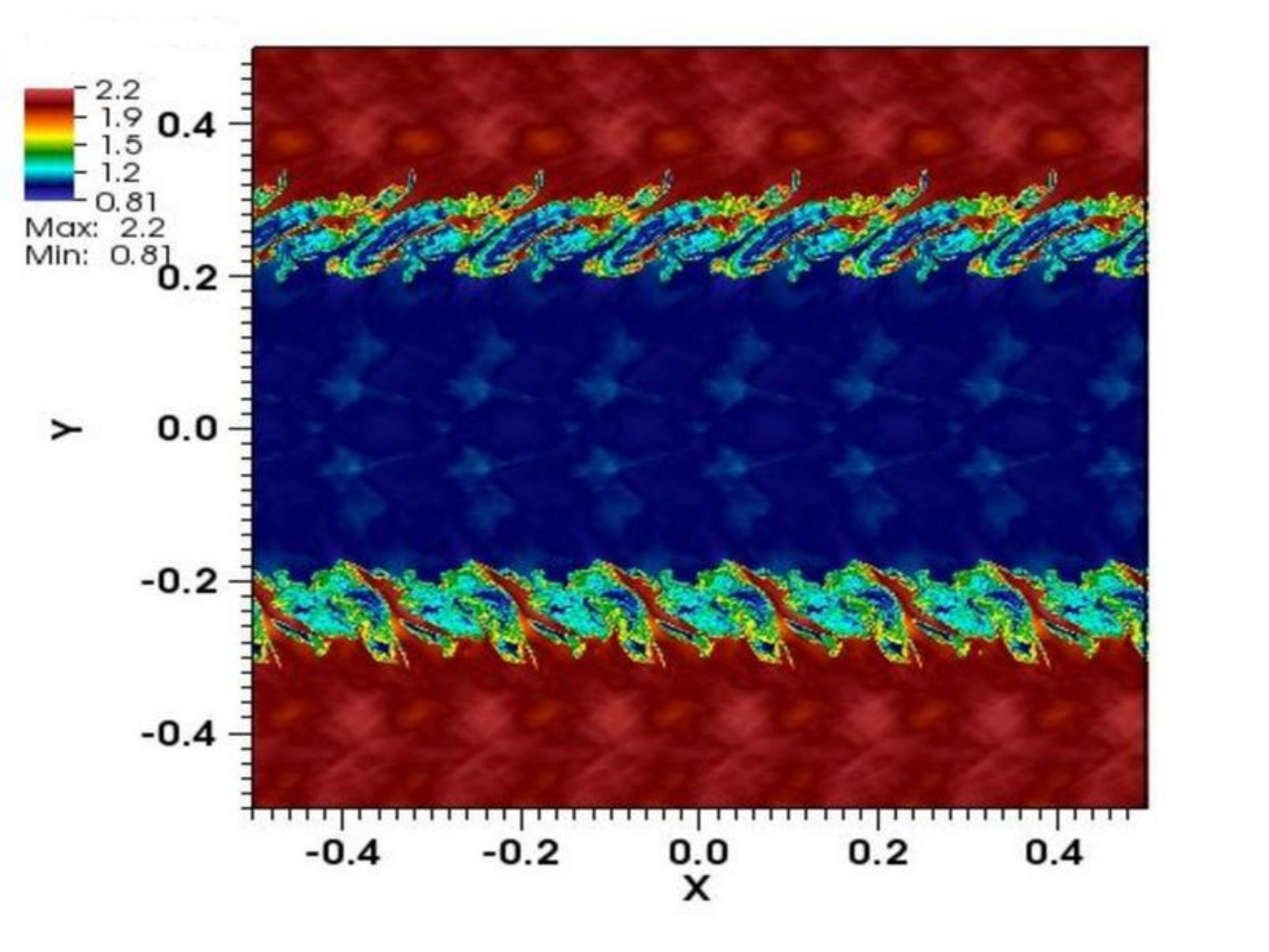}
	\includegraphics[width=0.31\linewidth]{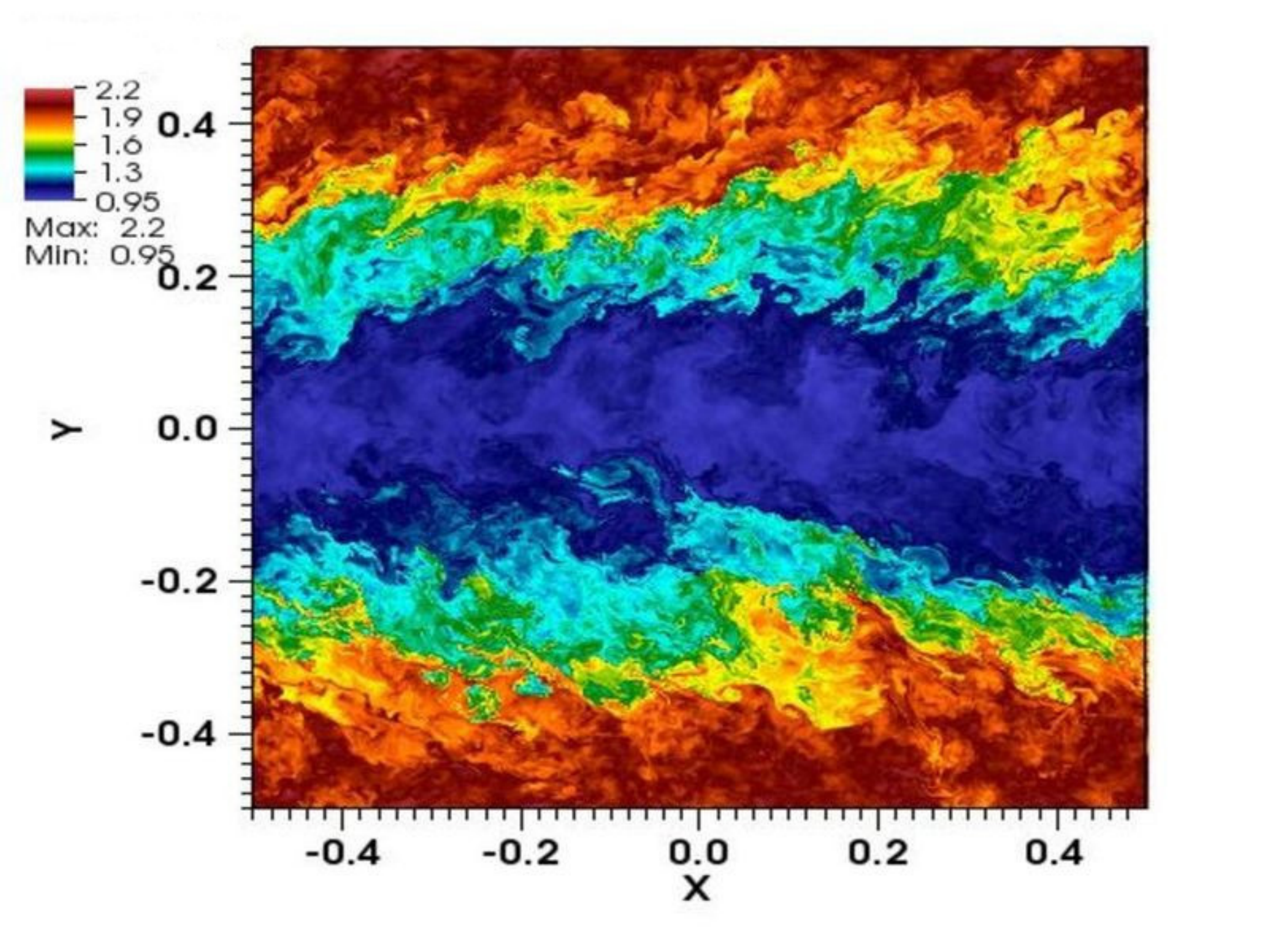}
	\includegraphics[width=0.31\linewidth]{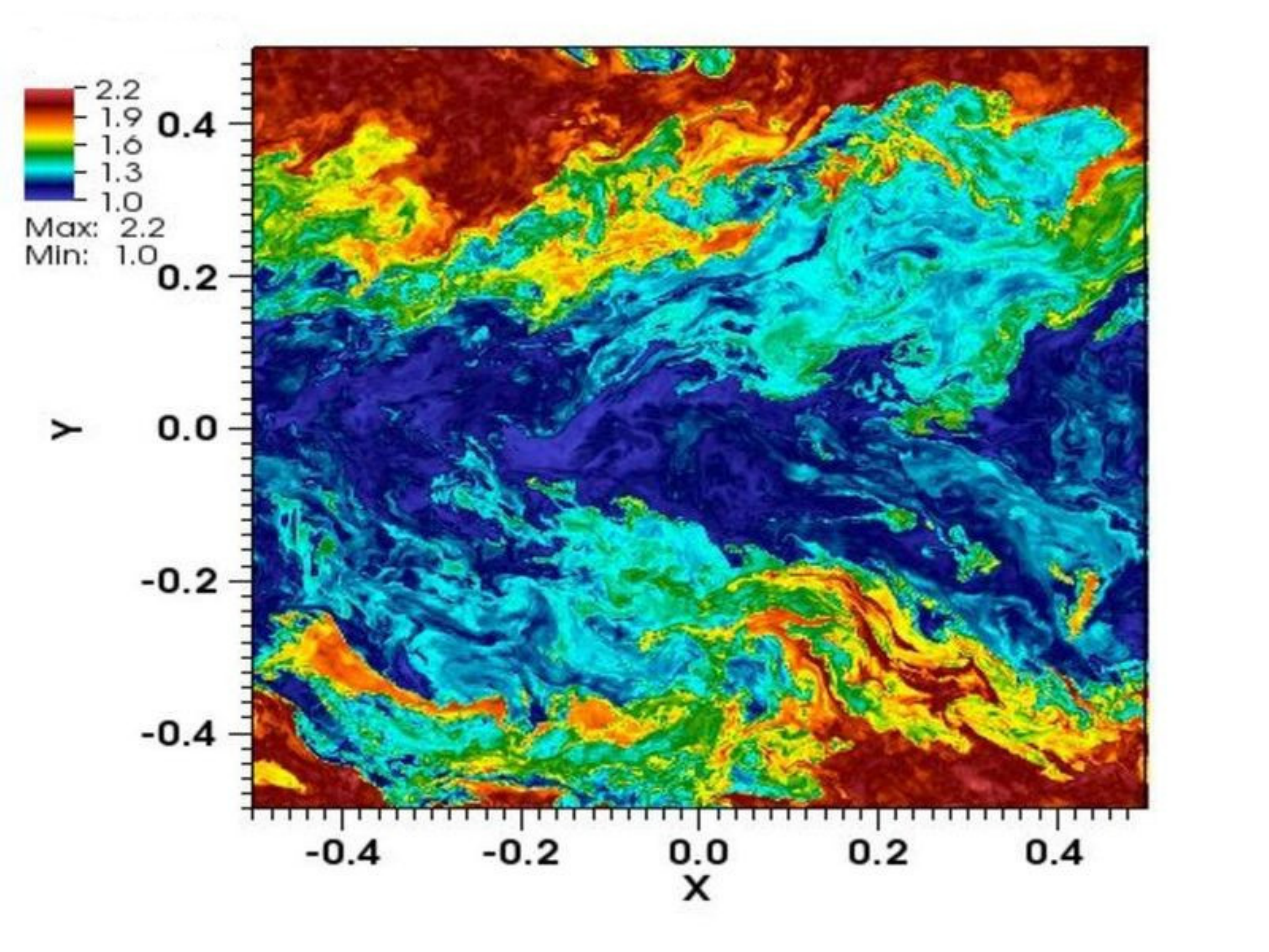}
	\caption{Evolution of the density distribution in the $z=0$ plane, for the highest resolution run, {\tt KH3D2000}, at (from left to right) $t=1,5,9$.}
	\label{fig:kh3d_evo_rho} 
\end{figure*}

The main flow is then initially given by $v_{x0}$. For our initial conditions, the vorticity, $\vec{\omega}=\vec{\nabla}\times\vec{v}$, is concentrated around the shear layers $y \pm y_l $, with maximum values of the order $|\omega_z| \sim v_{x0}/a_l$ and $|\omega_x| \sim v_{z0}/\sigma_z $. The integrated values in the volume are $\int_V ~\vec{v}~ dV = {\cal O}(\delta v)$ and $\int_V \vort ~dV = {\cal O}(k~\delta v)$ due to the symmetry of the main flow with respect to each layer, where $k$ is the inverse of the typical scale of the perturbation.
Therefore, $v_{z0}$ is used to primarily control the initial net kinetic helicity ${\cal H}_k = \int_V~\vec{\omega}\cdot\vec{v}~dV \sim 2~v_{z0}v_{x0}a_l$.

We consider fiducial values of $L=1$, $\rho_0=1$, $\rho_1=0.5$, $a_l=0.01$, $y_l=0.25$, $v_{x0}=0.5$, $\sigma_z^2=0.01$, $\vec{B}_{0}=0.001~\hat{x}$, $p_0=1$, assuming in general an ideal gas equation of state with $\gamma=4/3$. We consider models with number of points between $250^3$ and $2000^3$, and a possible variation of the fiducial parameters as described in Table~\ref{tab:models}. The setup is initially dominated by the large-scale kinetic energy, similarly to what happens in astrophysical scenarios, where the instability is actually the trigger for the growth and amplification of magnetic fields. The initial value of the internal energy depends on the choice of the pressure, and, in our case, we have initially $E_{\rm int} \gg E_{\rm  kin} \gg E_{\rm mag}$. Note that in the model {\tt KH3D500i}, we consider an isothermal fluid, approximated by $\gamma=1.0001$, in order to study the effect of the evolved pressure-related contributions. 

We evolve the system up to $t=20$, except for the {\tt KH3D2000} model, which required over two million CPU-hours to reach $t=10$. Similar 2D setups and related results are described and analyzed in the Appendix~\ref{app:kh2d_periodic}.

\begin{figure*}[t] 
	\centering
	\includegraphics[width=0.32\linewidth]{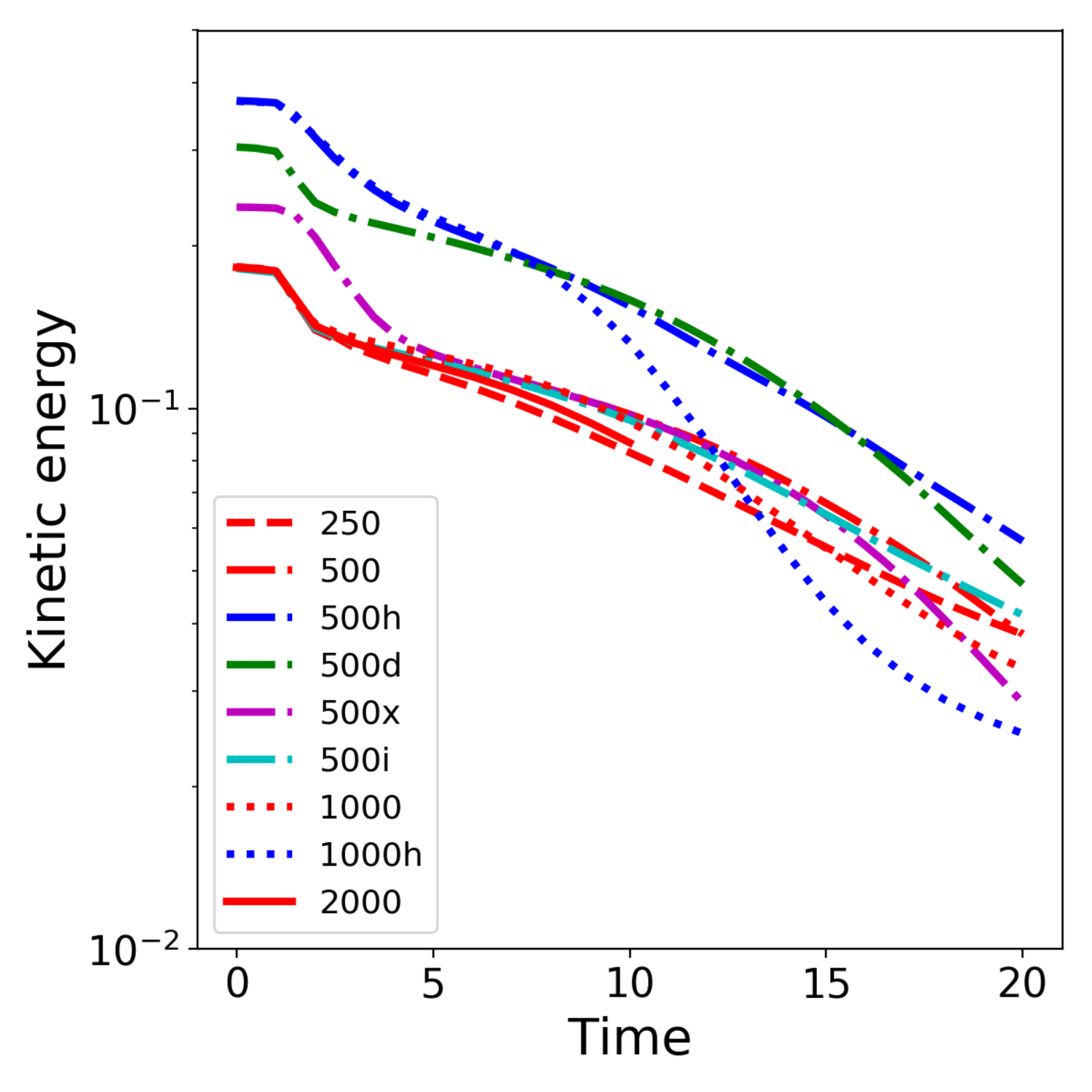}
	\includegraphics[width=0.32\linewidth]{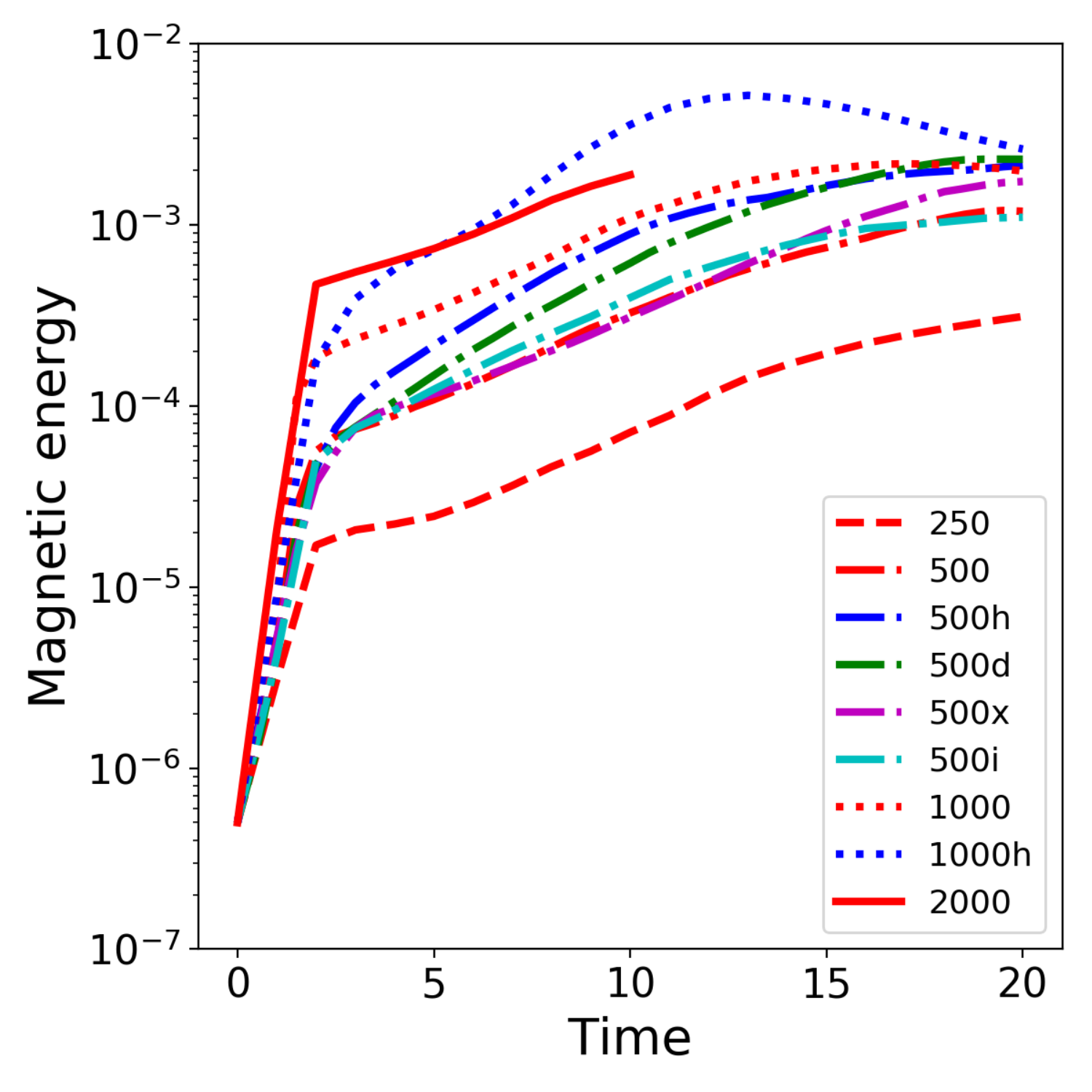}
	\includegraphics[width=0.32\linewidth]{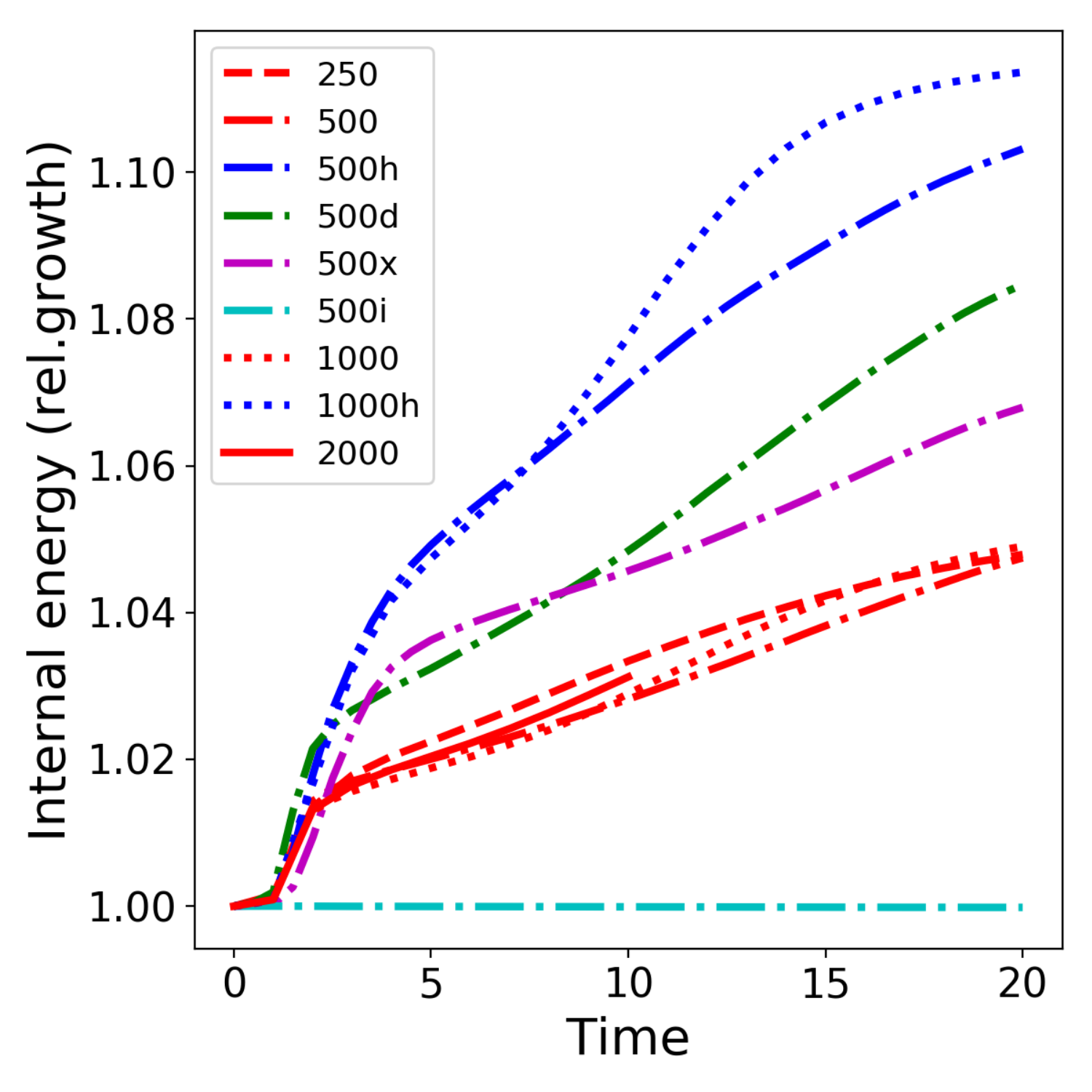}
	\caption{Evolution of the kinetic (left panel), magnetic (central) and internal energy (right, normalized to its initial value), comparing all models. The linestyles identify the resolution, while the color correspond to the different initial conditions (in red, the fiducial ones).}
	\label{fig:kh3d_evo_overall} 
\end{figure*}

\begin{figure*}[t] 
	\includegraphics[width=0.31\linewidth]{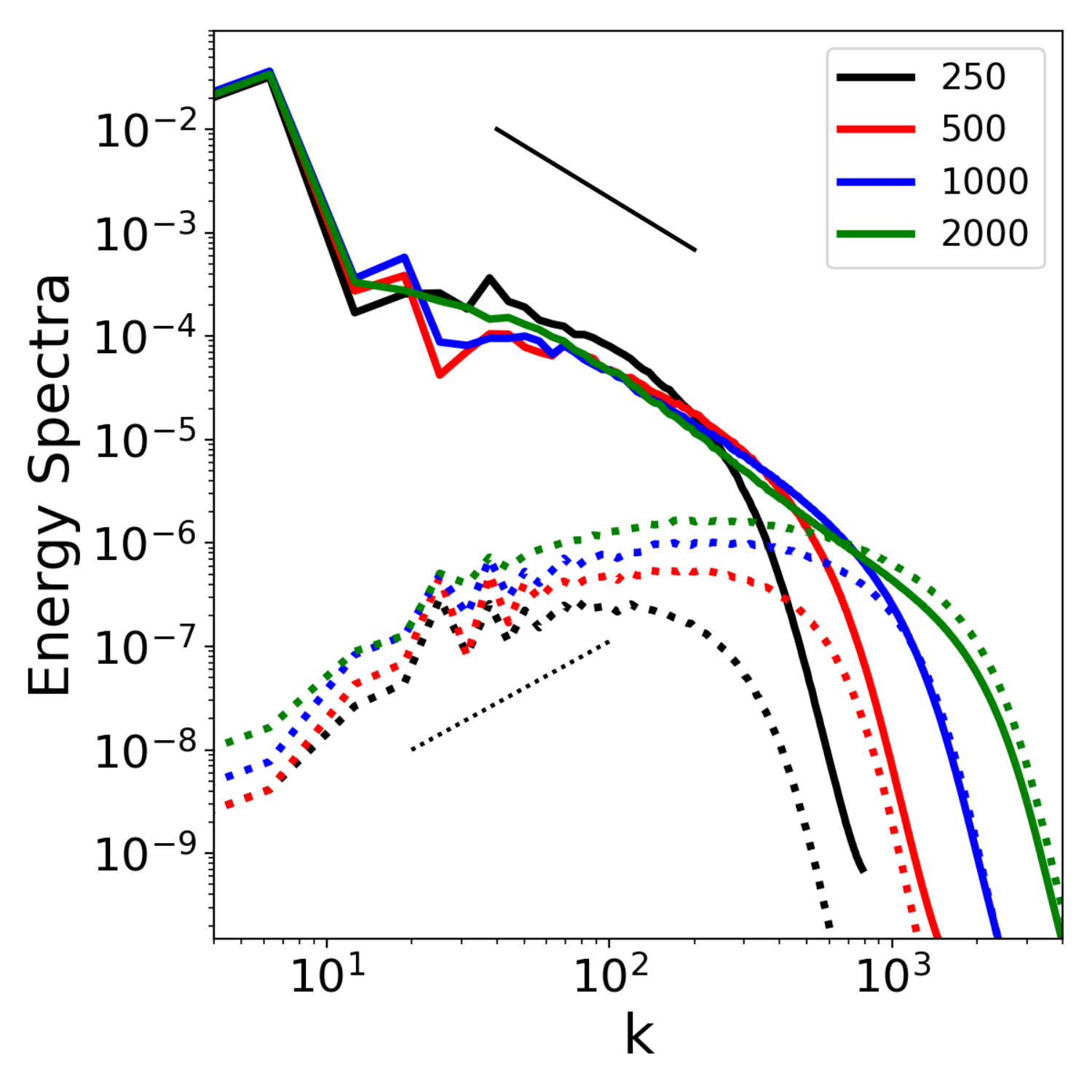}
	\includegraphics[width=0.31\linewidth]{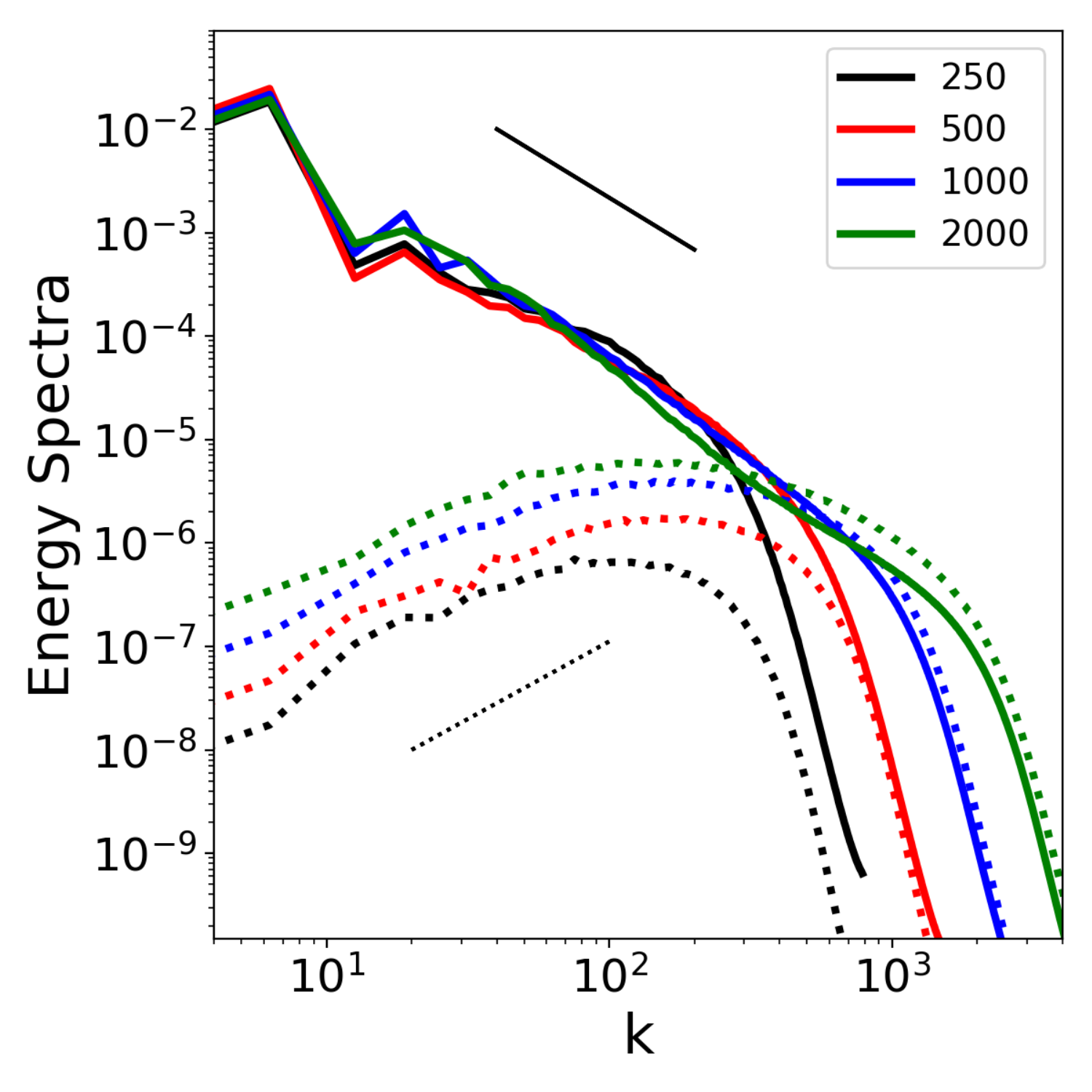}
	\includegraphics[width=0.31\linewidth]{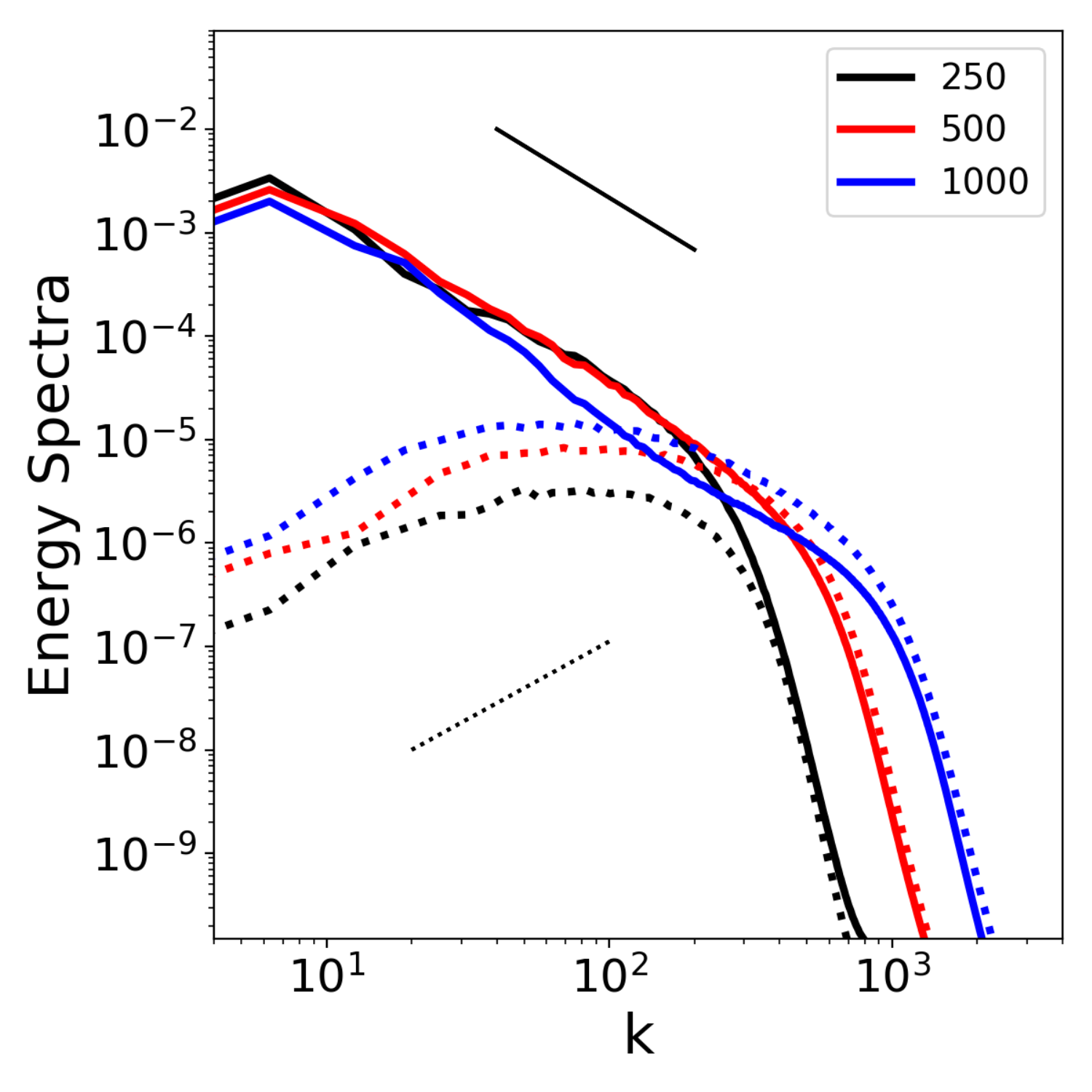}
	\caption{Kinetic (solid lines) and magnetic spectra (dots) at $t=5$ (first panel), $t=10$ (second), $t=20$ (third), respectively, for the same fiducial initial conditions and different resolutions. The wavenumber $k$ includes the factor $\pi$. The solid and dotted slopes indicate as references the Kolmogorov and Kazanstev spectral laws, $\propto k^{-5/3}$ and $\propto k^{3/2}$, respectively.} 
	\label{fig:kh3d_spectra} 
\end{figure*}

\begin{figure}[t] 
	\includegraphics[width=0.62\linewidth]{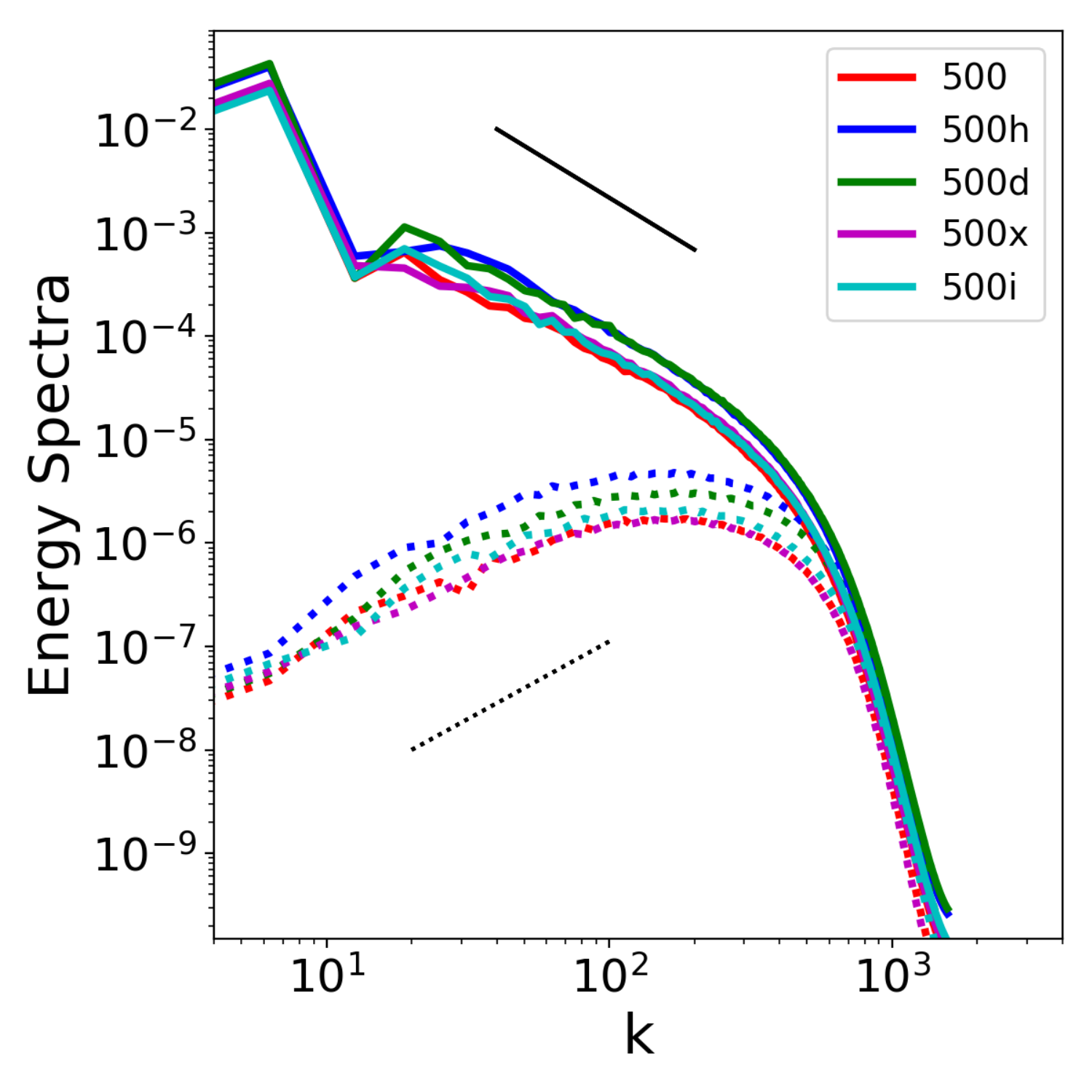}
	\caption{Kinetic (solid lines) and magnetic spectra (dots) at $t=10$ for the different initial conditions and $N=500^3$.} 
	\label{fig:kh3d_spectra_500} 
\end{figure}

\subsection{The initial perturbation and its implications}\label{sec:initial_perturbation}

The components of the three initial velocity perturbations appearing in eqs.~(\ref{eq:vx})-(\ref{eq:vz}), are periodic, single modes with a wavenumber related to an integer $n_i \in [1,N/2]$, and a different propagation direction: 

\begin{eqnarray}
&& \delta v_x=\delta v_{x0} \sin(2\pi n_x~z/L)~, \\
&& \delta v_y=\delta v_{y0} \sin(2\pi n_y~x/L)\sign(y)  \exp\left[-\frac{(|y| - y_l)^2}{\sigma_y^2}\right]~, \\
&& \delta v_z=\delta v_{z0} \sin(2\pi n_z~y/L)~.
\end{eqnarray}
Hereafter, we use $n_x=11$, $n_y=7$, $n_z=5$, $\delta v_{x0} = \delta v_{z0} = 0.01$, $\delta v_{y0} = 0.1$, $\sigma_y^2=0.1$. It is important to note that this multi-modal perturbation in different directions, with different prime numbers for the wavenumber values, allows us to excite virtually all the modes spectrum. The non-linear interaction of the three initial different modes is effectively similar to imposing a random perturbation. The specific forms (values of $\delta v_{i0}$, $\sigma_y^2$ and $n_i$) of the initial perturbations have no influence on the asymptotic turbulent behavior, as long as the spectrum is entirely excited, which is guaranteed if the wavenumbers are prime number between them.

This choice gives a qualitatively difference compared to the same case with a partial excitation of the modes (e.g., $\delta v_{i0}\neq 0$ only for one component, or $n_x=n_y=n_z$): in that case, the turbulence will be restrained to a single mode and its harmonics, and, if the mode wavelength is much larger than the grid resolution, numerical convergence can be easily reached. In this sense, see the 2D, single-mode perturbed benchmark tests, taken from \cite{obergaulinger10}, successfully replicated in Appendix~\ref{app:2D_tests}, where one can also get an insight on the growth rates of the KHI. 

The excitement of the entire spectrum, together with the ideal flow assumption (absence of a viscous scale), has also important implications for the general behaviour and the numerical convergence, as explained in the next section.

\begin{figure*}[ht] 
	\centering
	\includegraphics[width=0.31\linewidth]{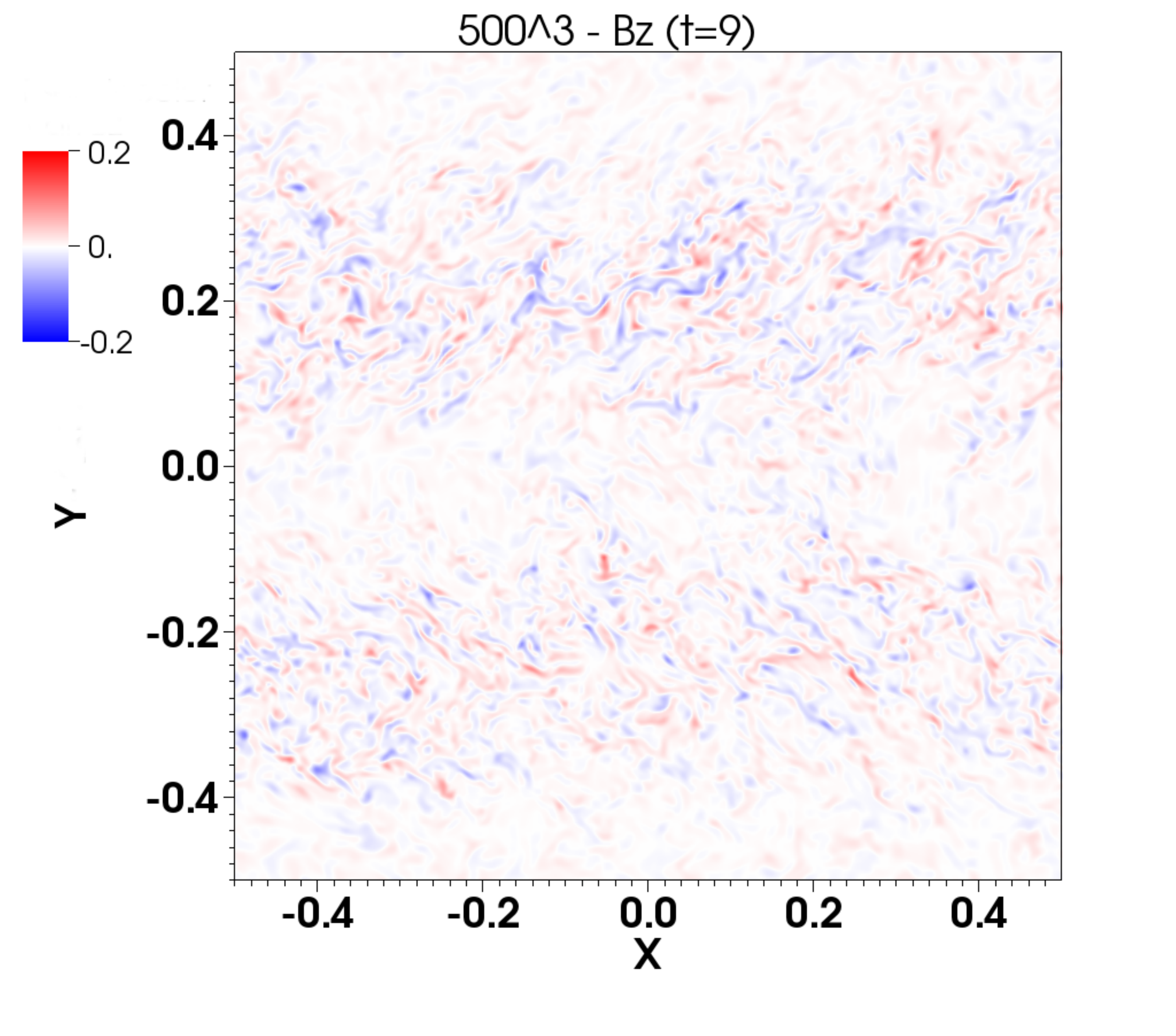}
	\includegraphics[width=0.31\linewidth]{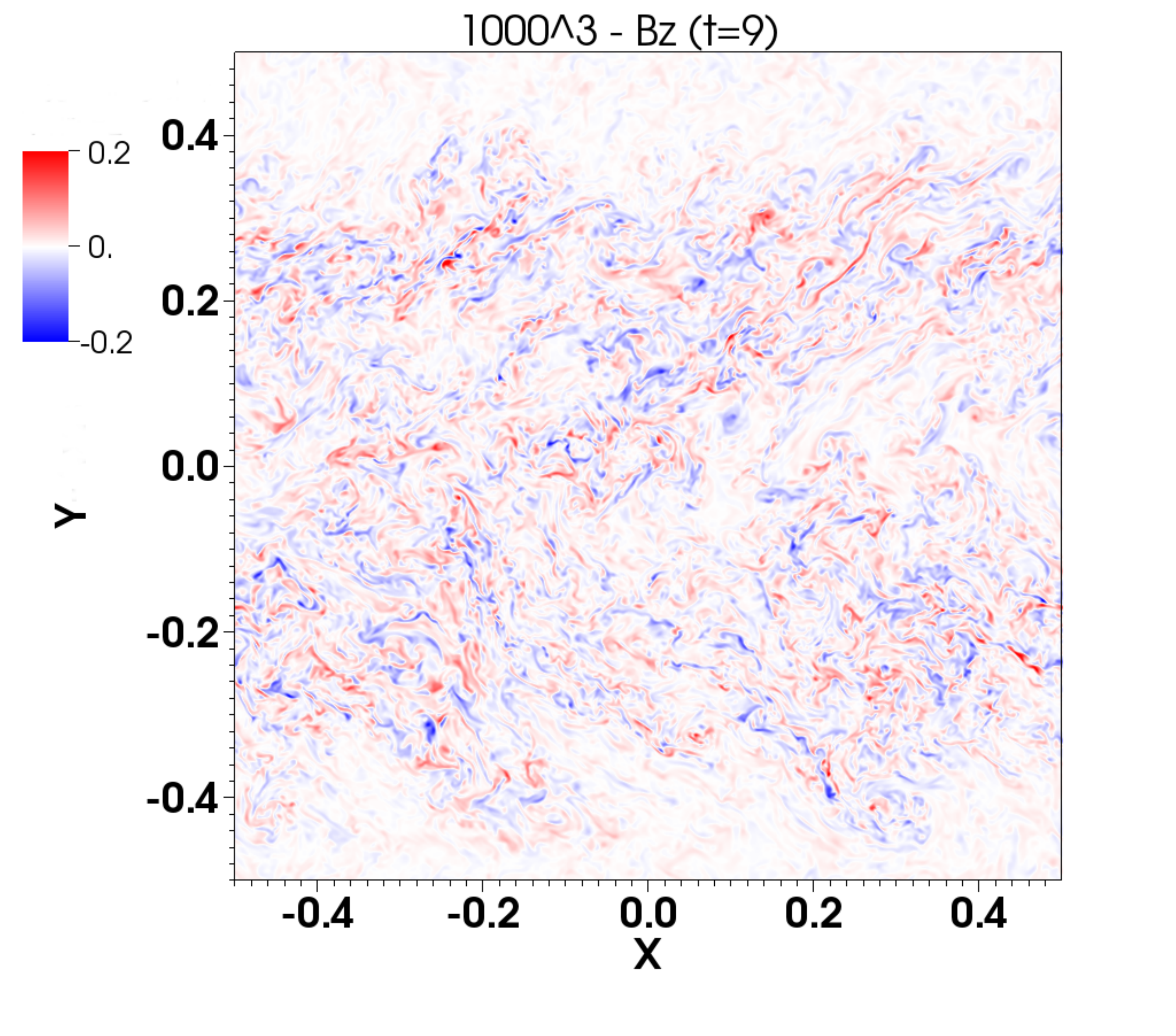}
	\includegraphics[width=0.31\linewidth]{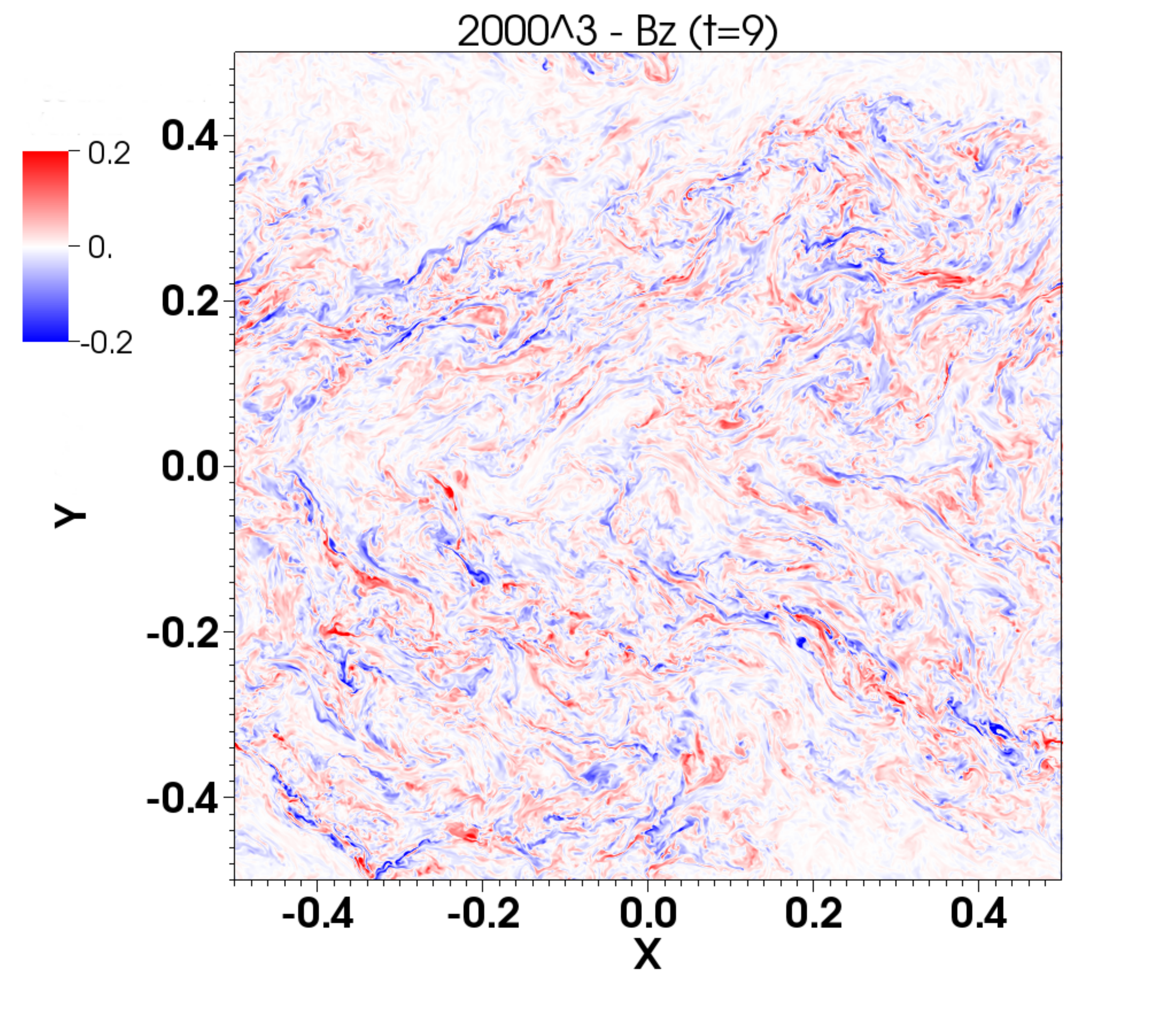}
	\caption{Values of the $B_z$ component in the $xy$-plane at $t=9$, for different resolutions (from left to right), for the models {\tt KH3D500}, {\tt KH3D1000}, {\tt KH3D2000} (from left to right).}
	\label{fig:kh3d_img} 
\end{figure*}

\begin{figure*}[ht] 
	\centering
	\includegraphics[width=0.24\linewidth]{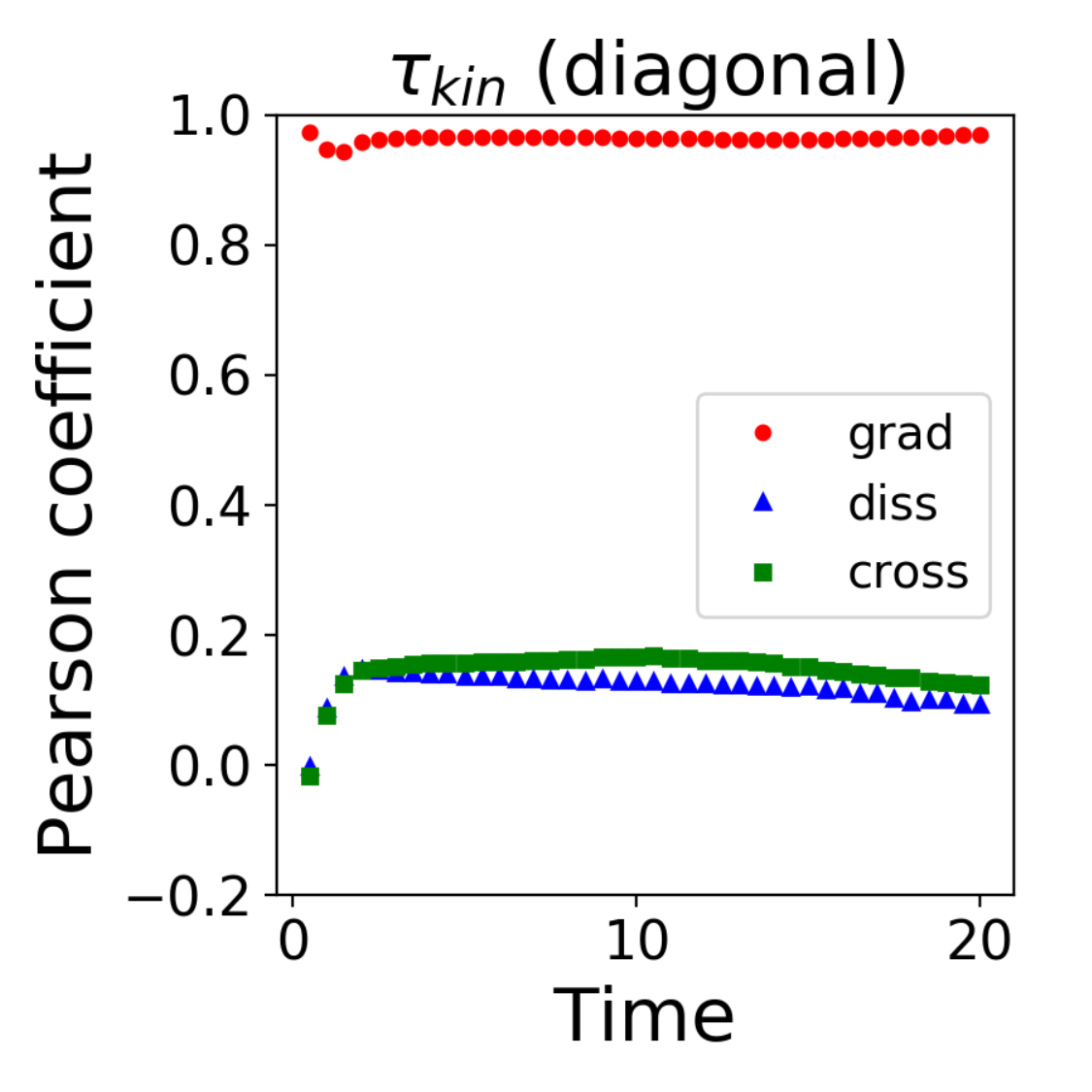}
	\includegraphics[width=0.24\linewidth]{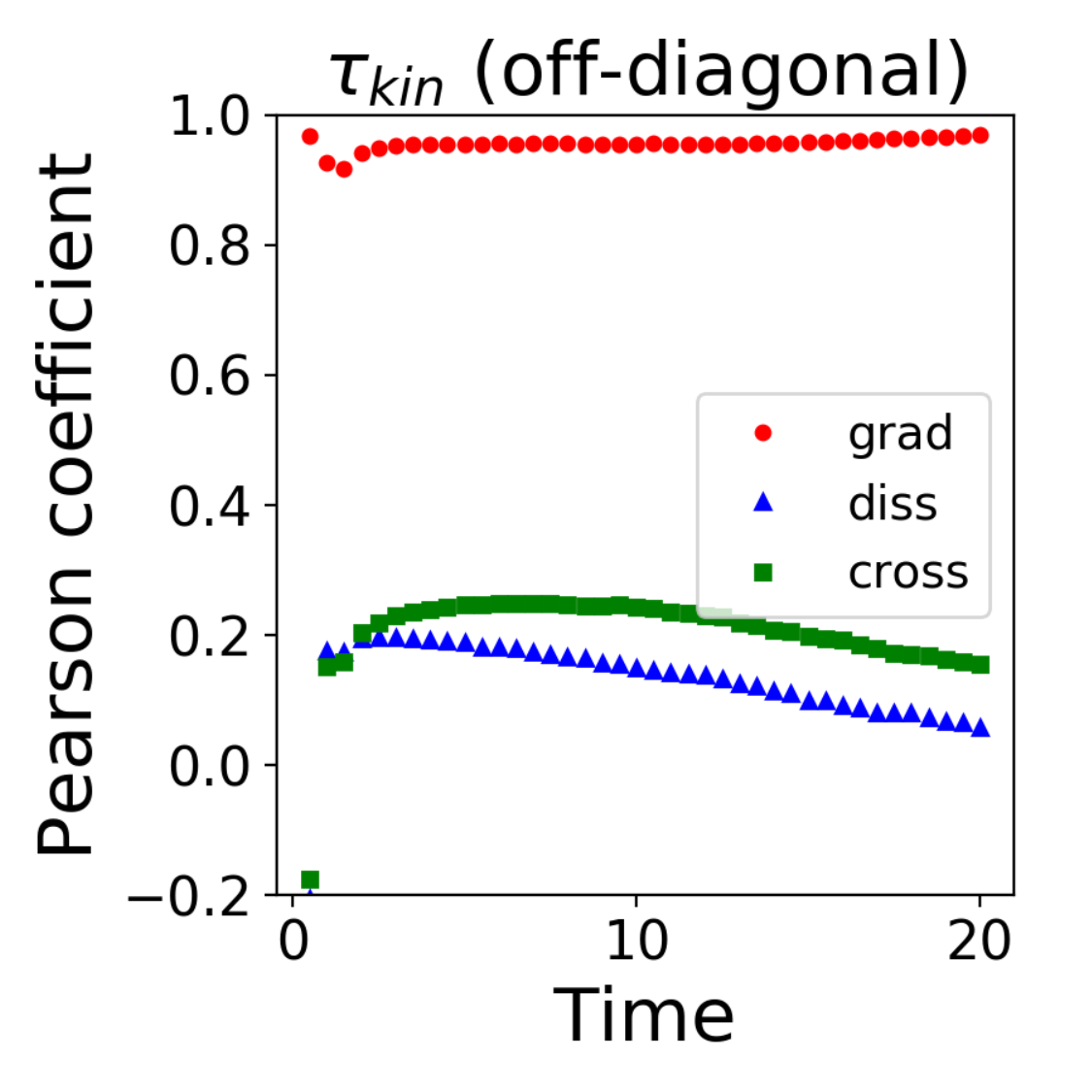}
	\includegraphics[width=0.24\linewidth]{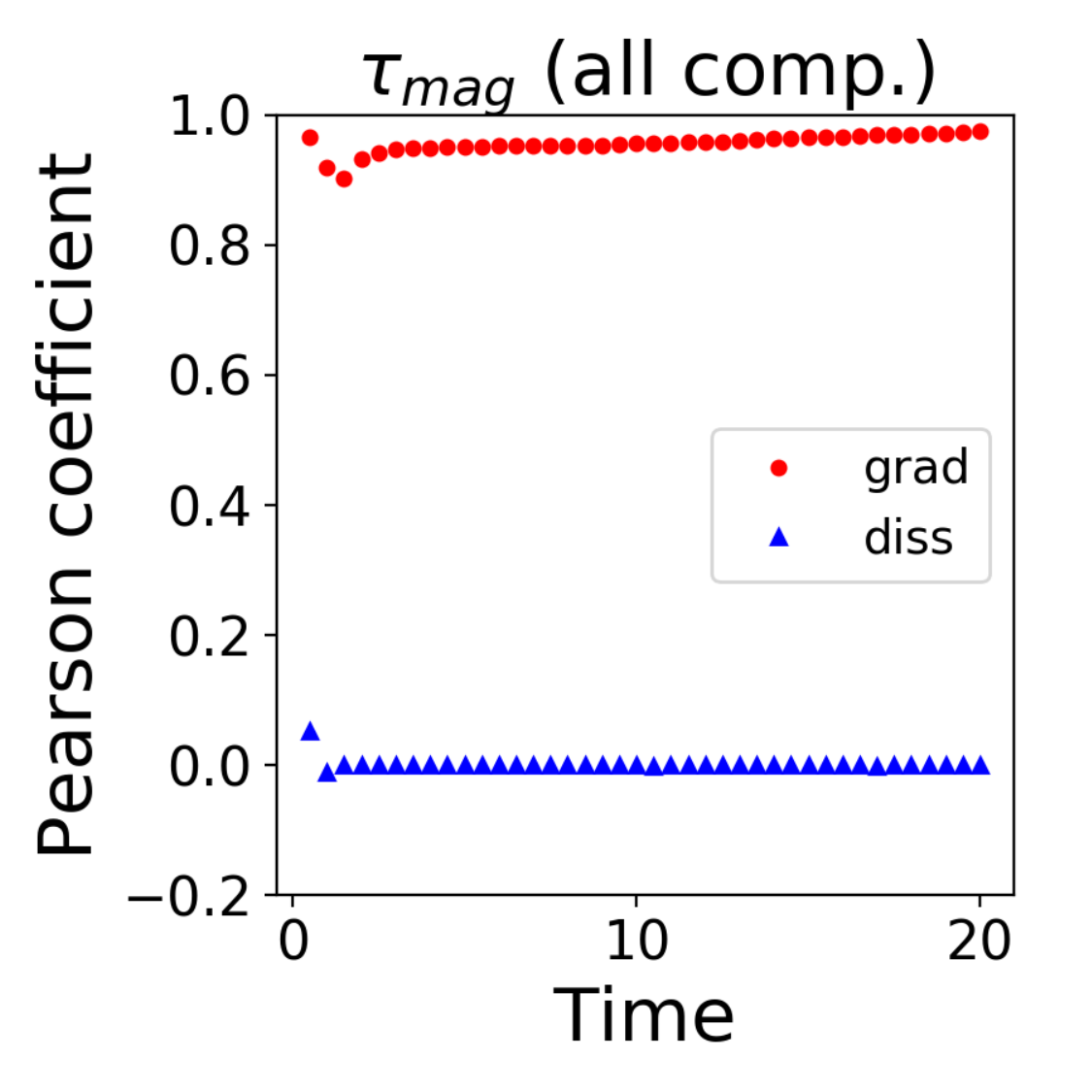}
	\includegraphics[width=0.24\linewidth]{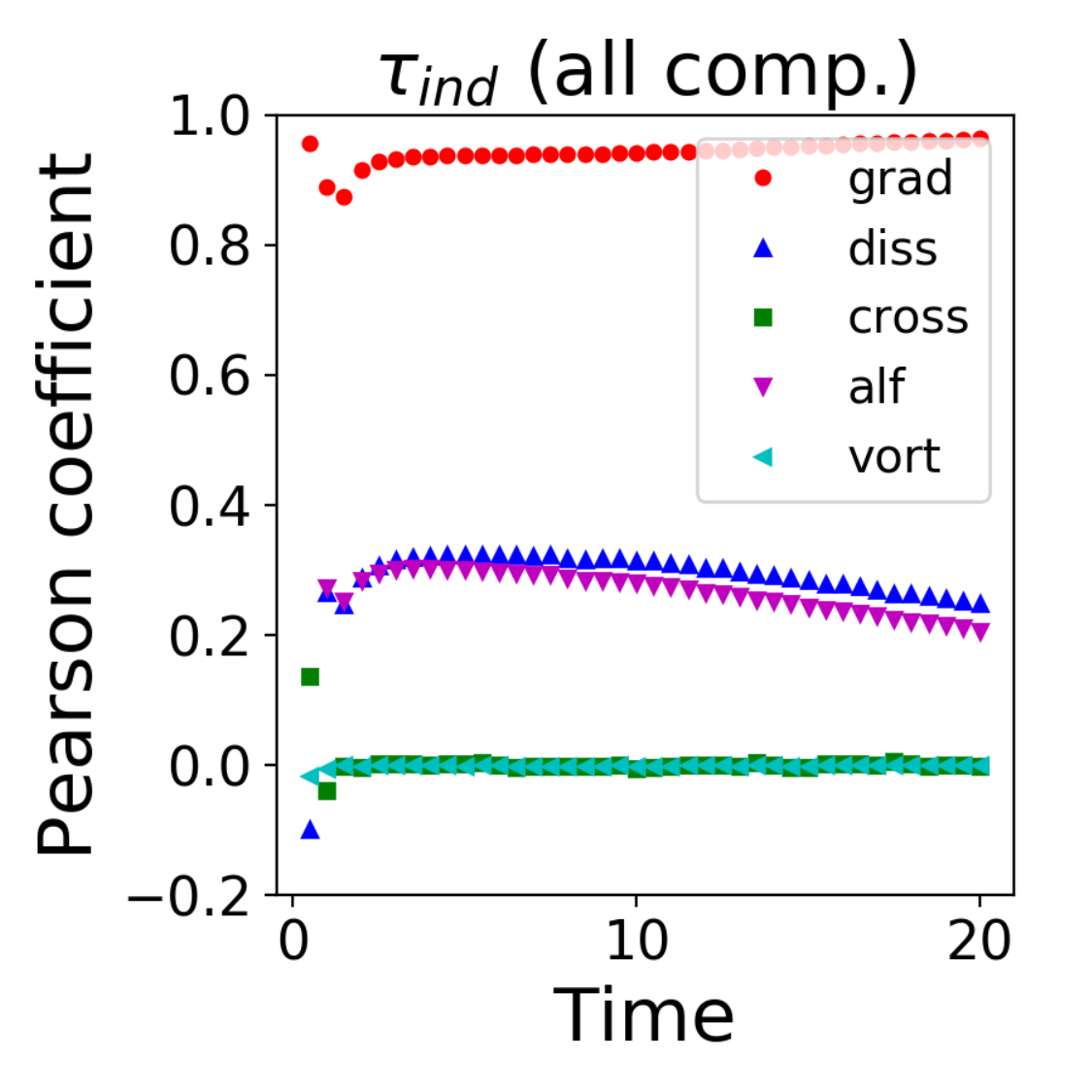}\\
	\includegraphics[width=0.24\linewidth]{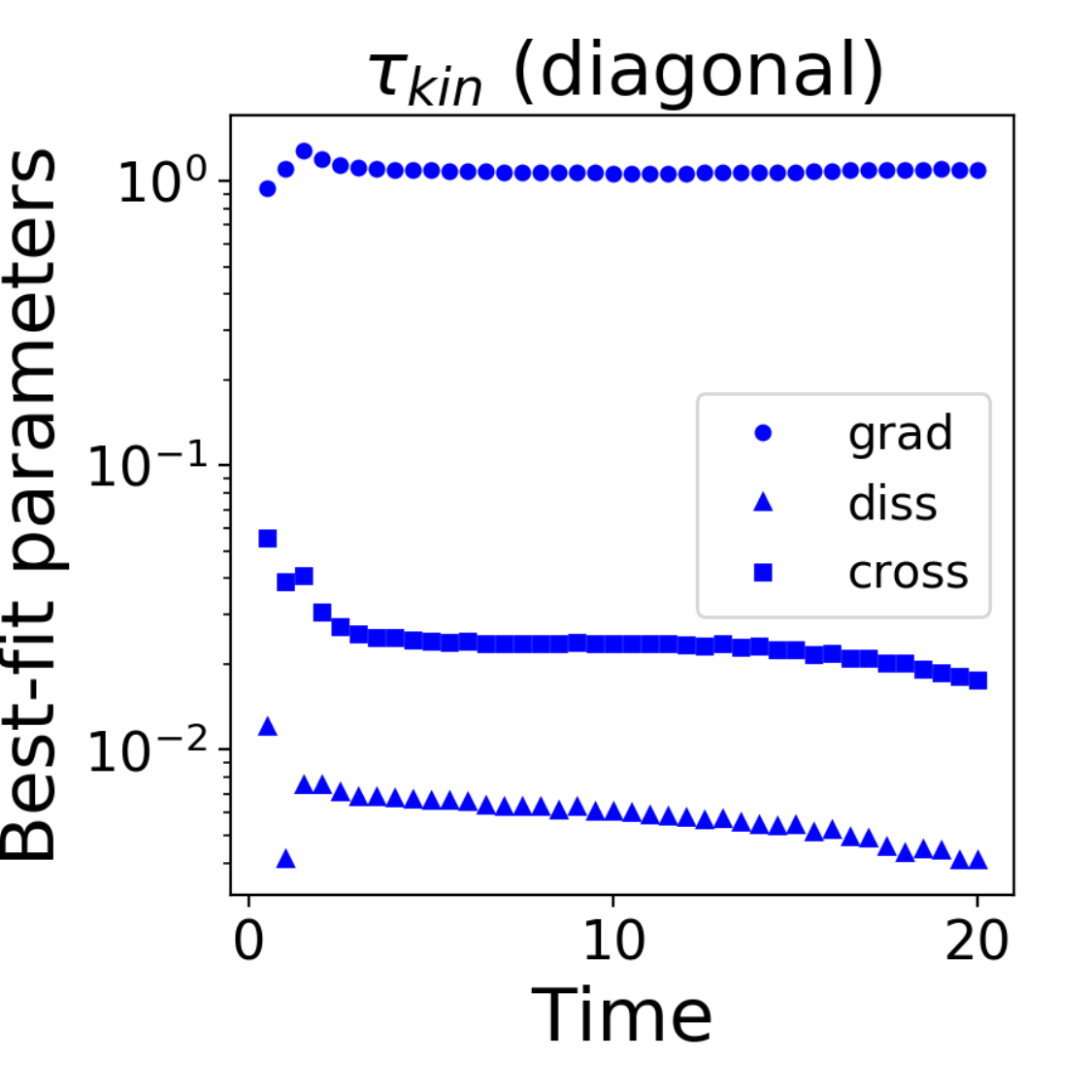}
	\includegraphics[width=0.24\linewidth]{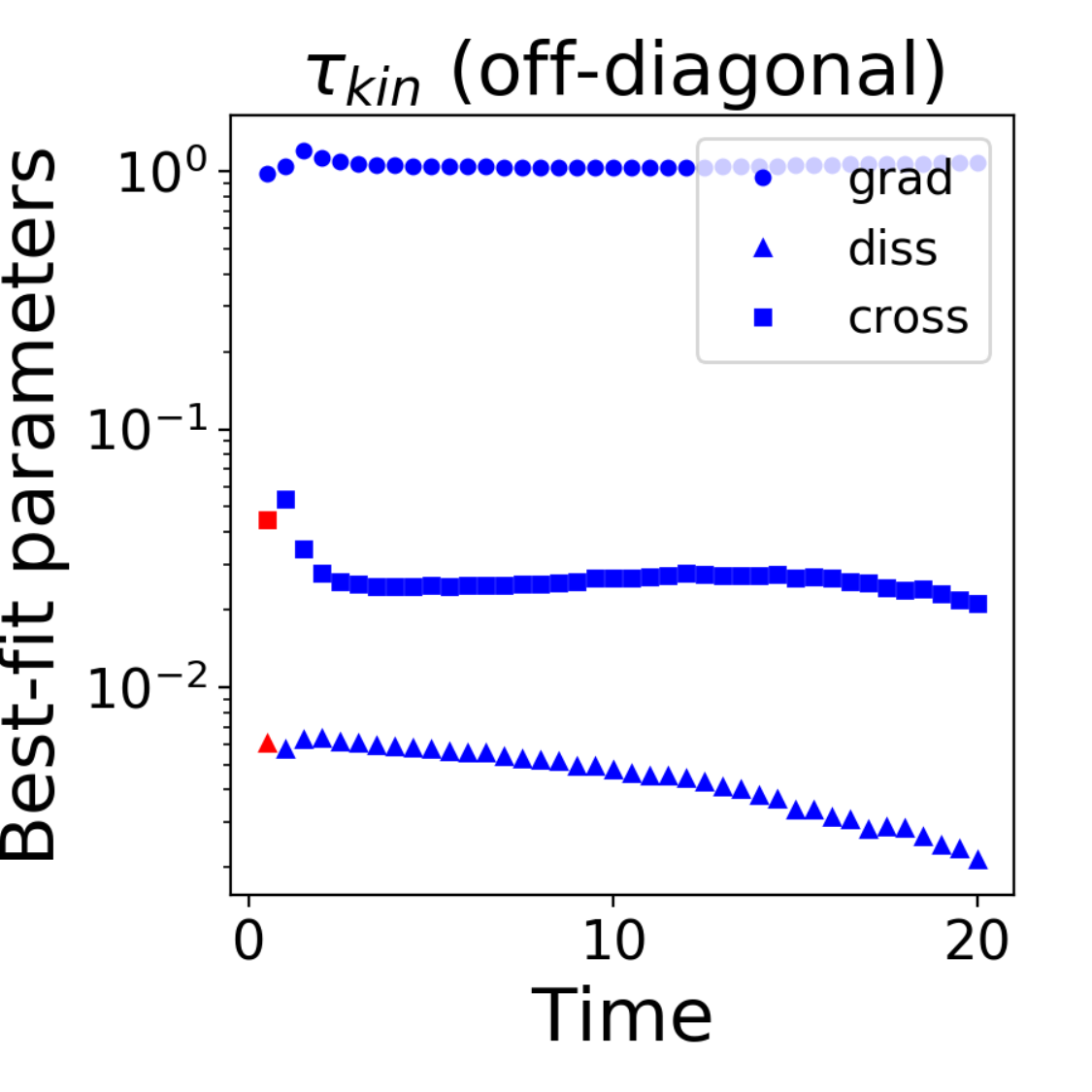}
	\includegraphics[width=0.24\linewidth]{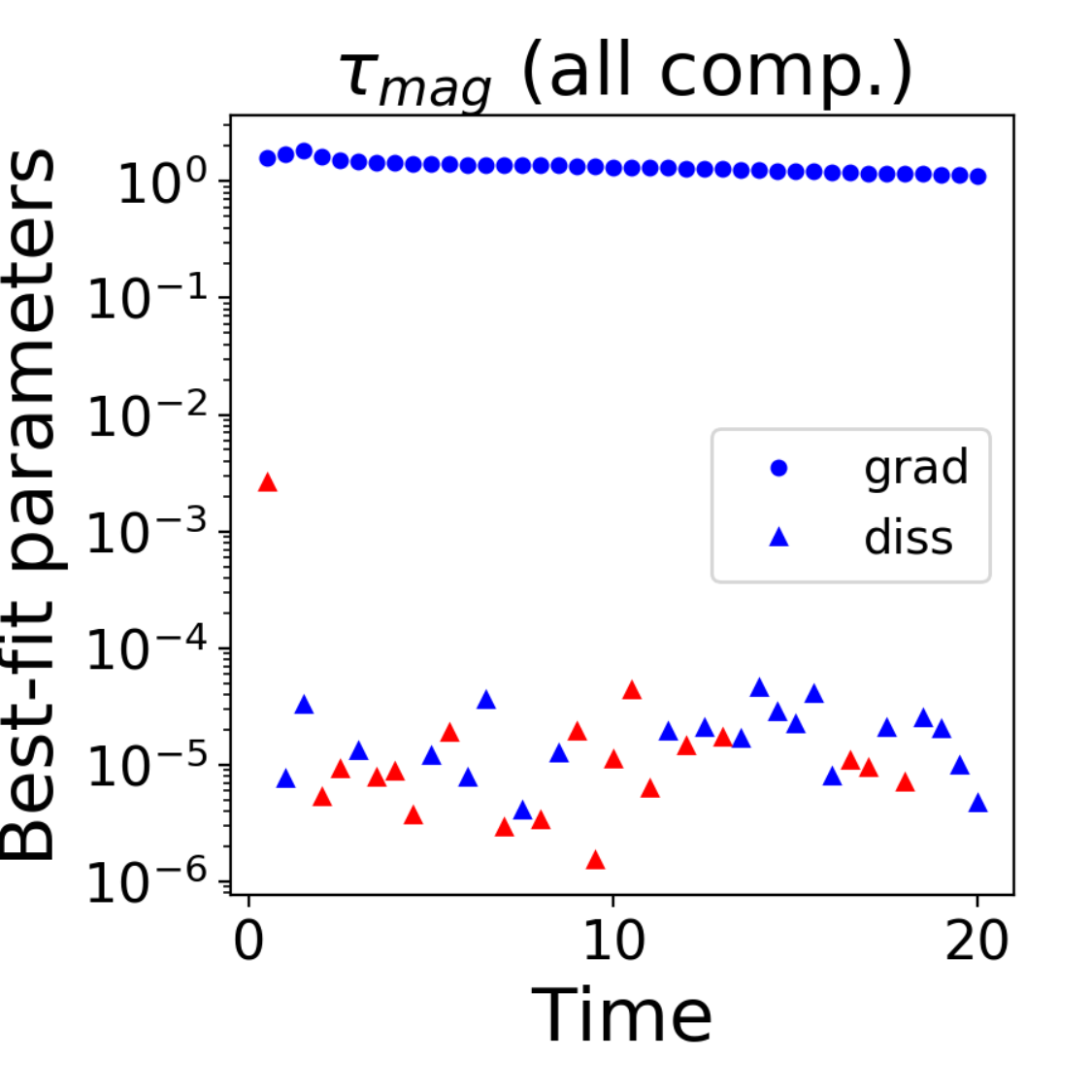}
	\includegraphics[width=0.24\linewidth]{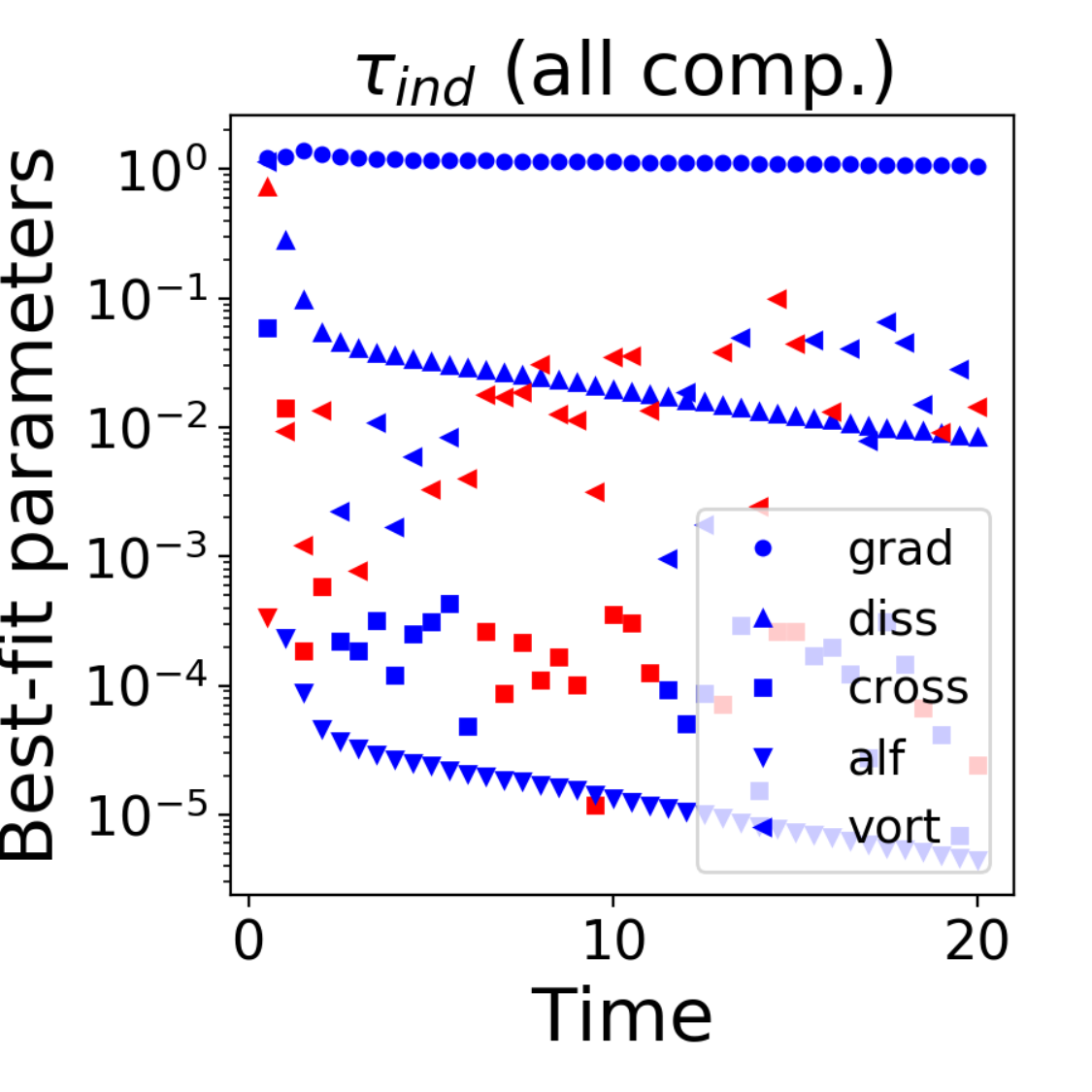}
	\caption{Evolution of the Pearson correlation coefficients ${\cal P}$ (top panels) and the best-fit parameters $C_{\rm best}$ (bottom) for the {\tt KH3D1000} model with $S_f=2$, comparing different SGS models, indicated by symbols as in legends. {\em First and second columns}: diagonal and off-diagonals components of $\tau_{\rm kin}$ ($\tau_{\rm mom}$) for the Eddy-dissipative and gradient models (cross-helicity model). {\em Third column}: $\tau_{\rm mag}$ for the Eddy-dissipative and gradient models. {\em Fourth column}: $\tau_{\rm ind}$ for the gradient, Eddy-dissipative, cross-helicity, vorticity and Alfv\'en model. All values here represented are the average over all the independent, non-zero tensor components (3, 3, 6, and 3, respectively). For the sake of visibility, we employ a semi-log plot for the best-fit parameters, where colors indicate positive (blue) or negative values (red).}
	\label{fig:pearson_evo_1000sf2} 
\end{figure*}

\begin{figure*}[ht] 
	\centering
	\includegraphics[width=0.31\linewidth]{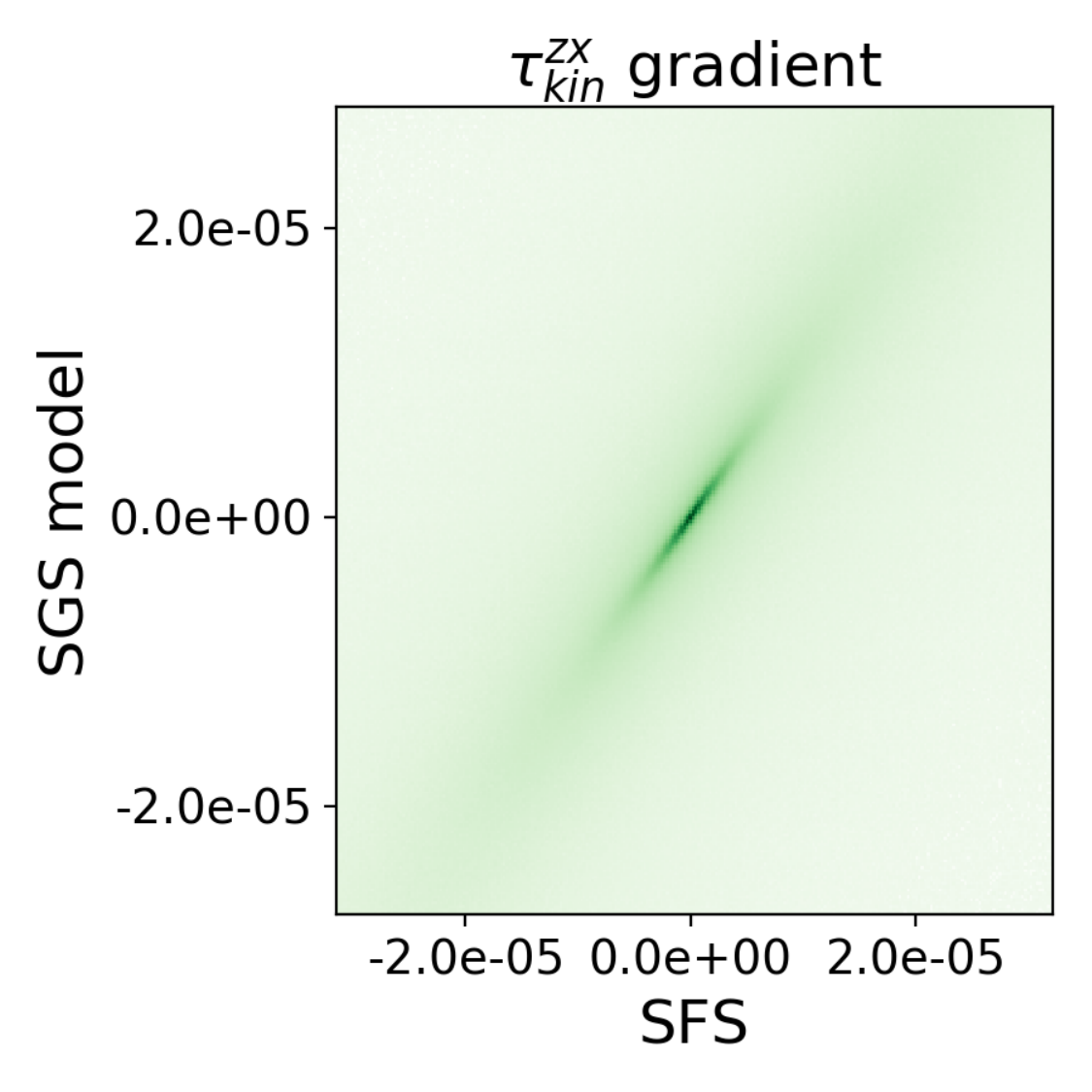}
	\includegraphics[width=0.31\linewidth]{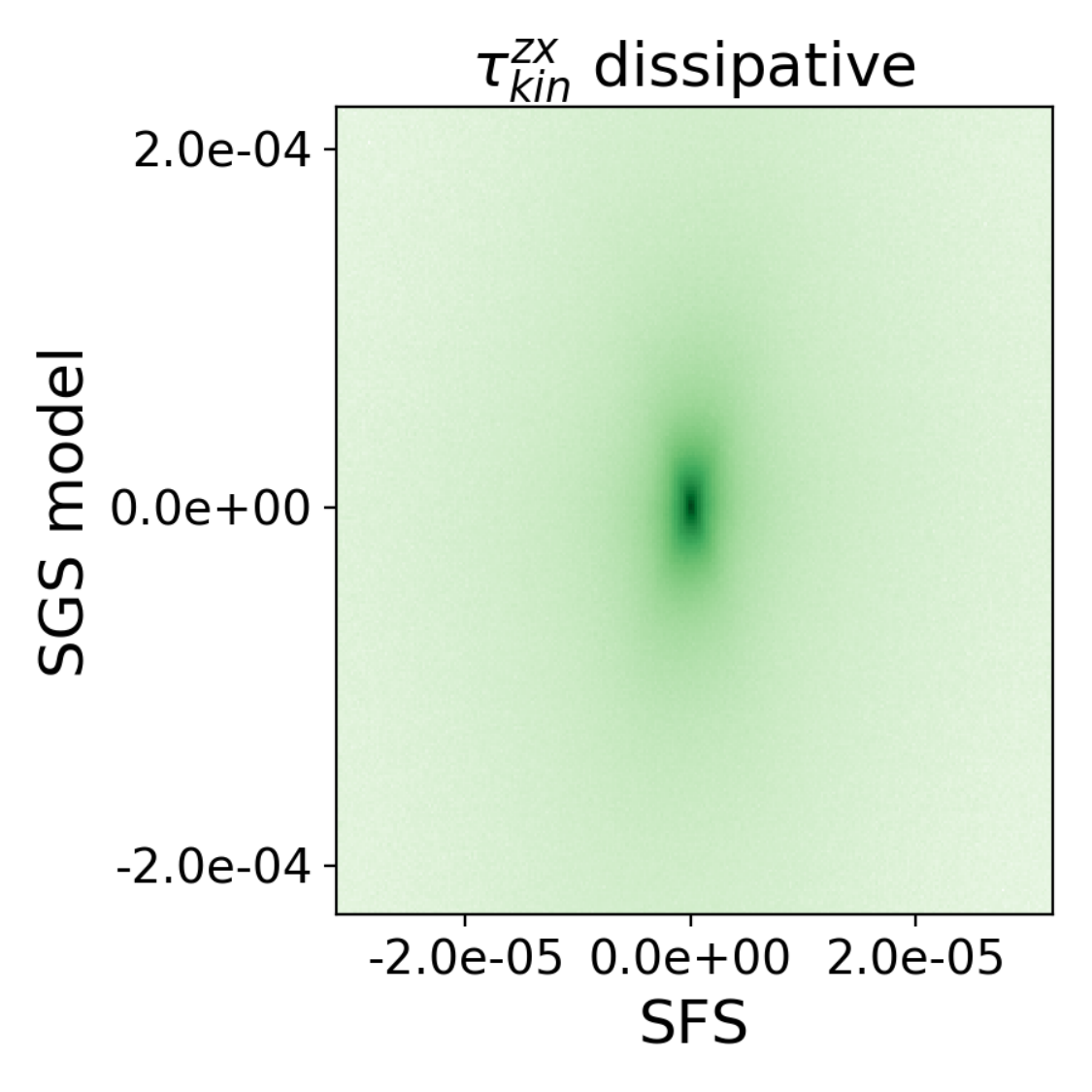}
	\includegraphics[width=0.31\linewidth]{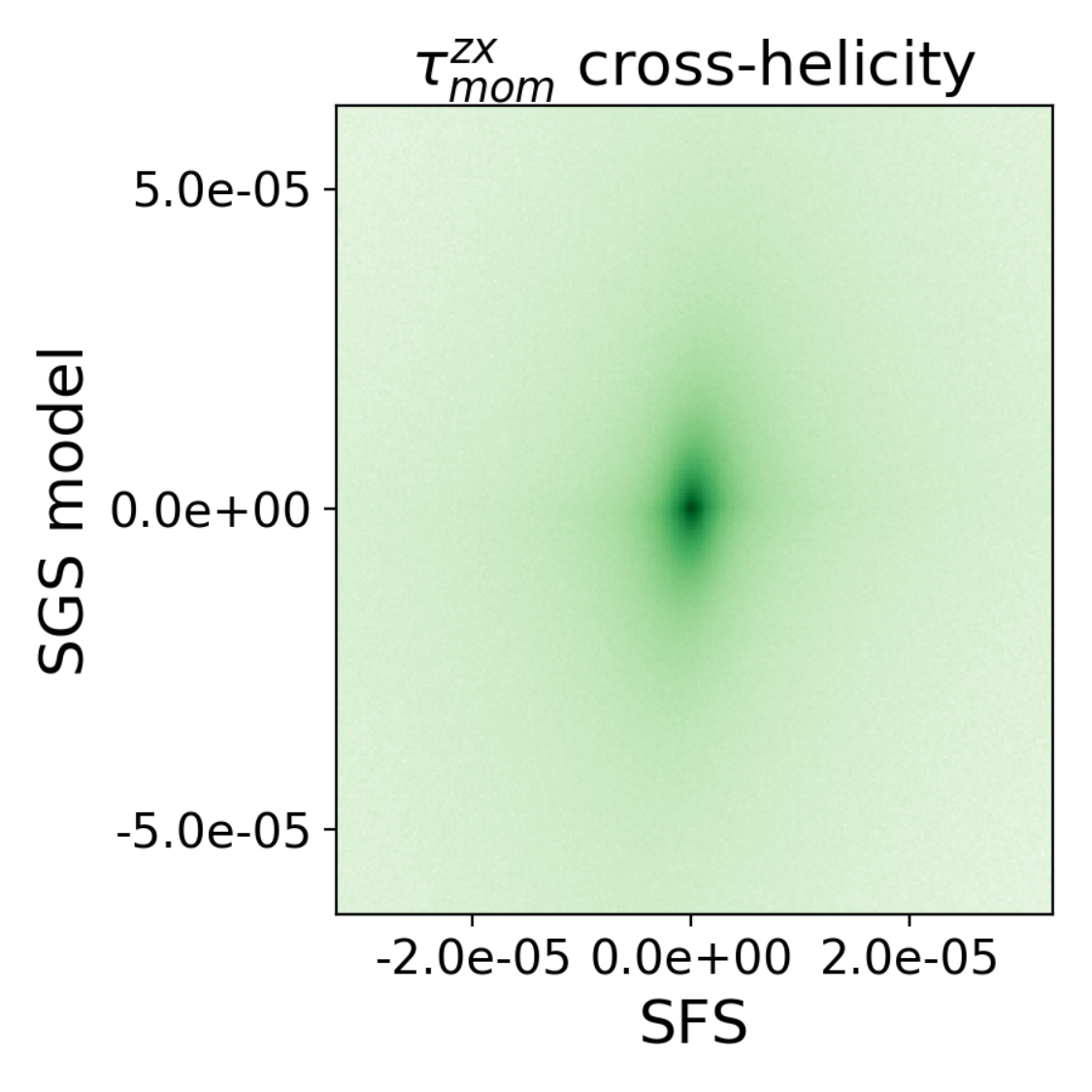}\\
	\includegraphics[width=0.31\linewidth]{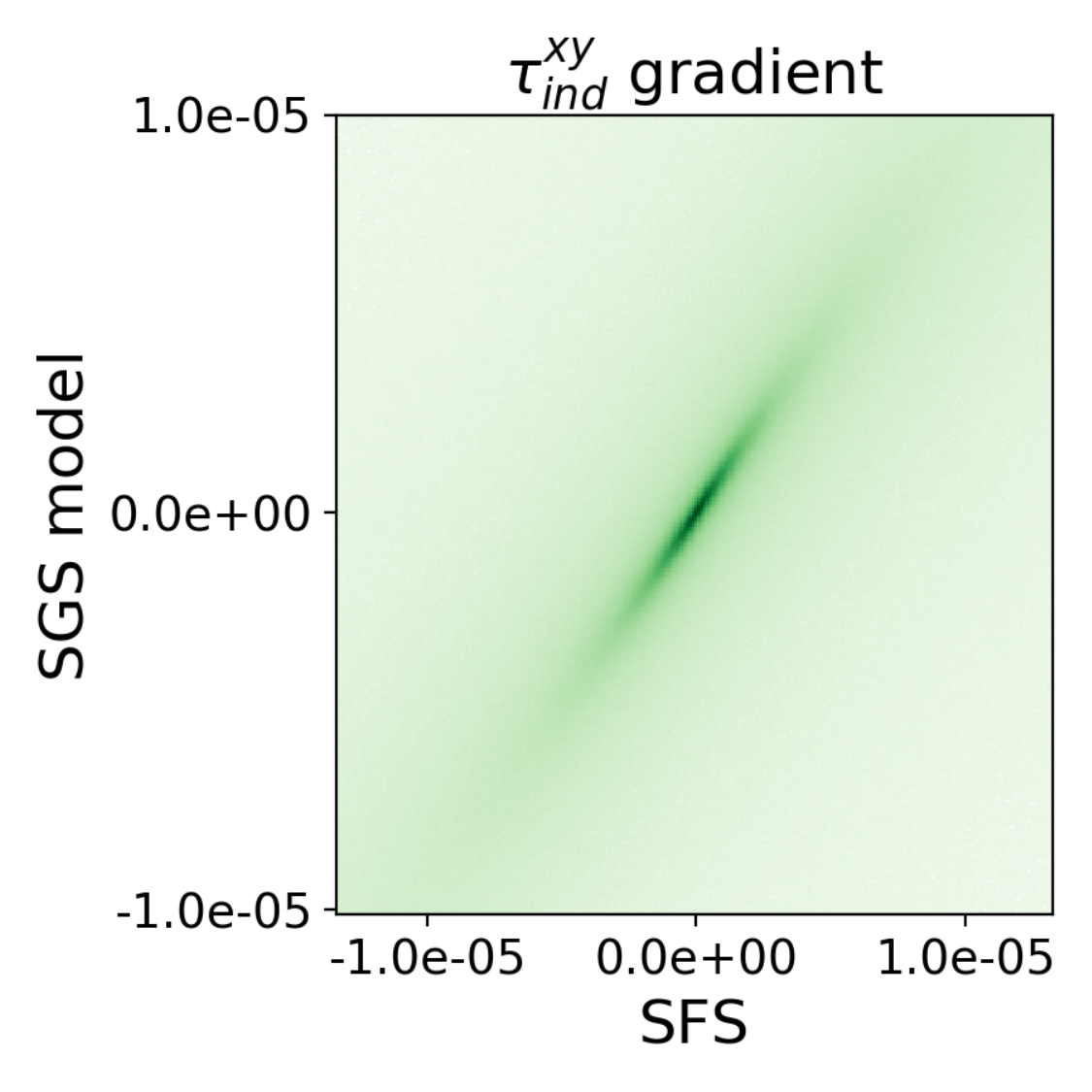}
	\includegraphics[width=0.31\linewidth]{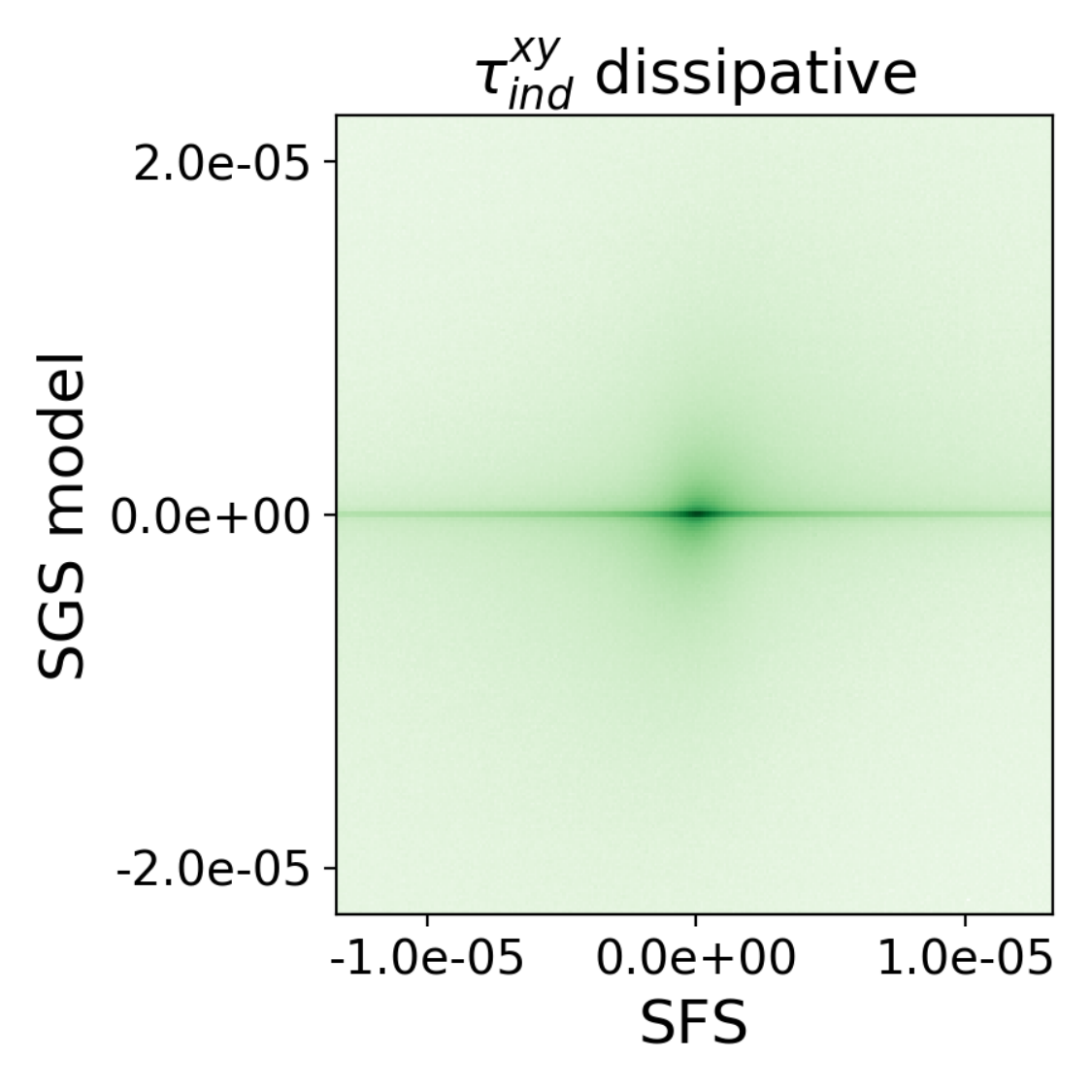}
	\includegraphics[width=0.31\linewidth]{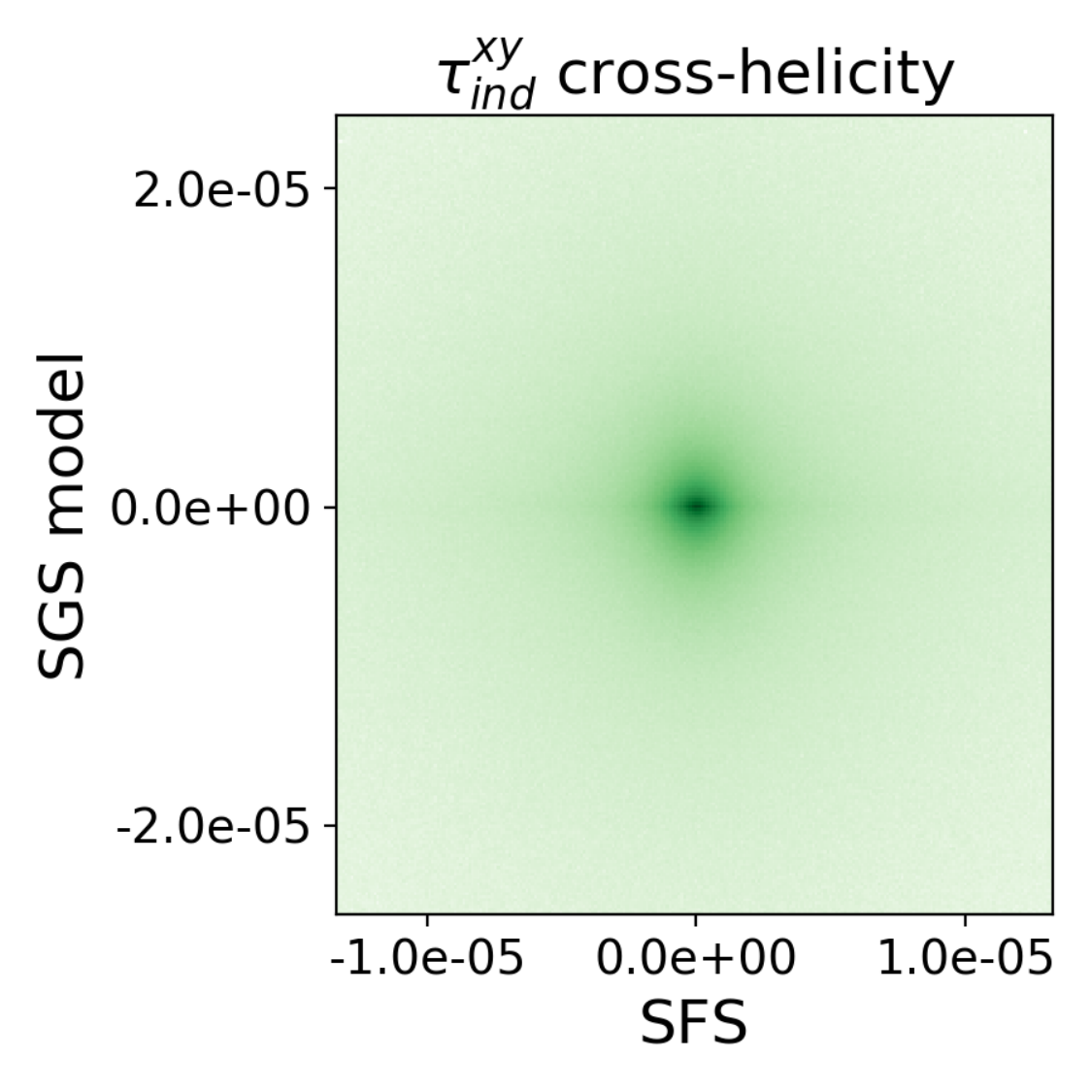}
	\caption{Correlation between the SFS terms $\overline{\tau}$ (x-axis), and the SGS terms $\tau$ (y-axis) for the model {\tt KH3D500} with $S_f=2$, at $t=10$. We show the correlations for $\tau_{\rm kin}^{zx}$ (left and middle top panels), $\tau_{\rm mom}^{zx}$ (top right), and $\tau_{\rm ind}^{xy}$ (bottom panels), for the gradient (left panels), Eddy-dissipative (middle), and cross-helicity models. For each axis, the range of values shown here is $[<\tau> \pm 0.25~\sigma]$ where $<\cdot>$ and $\sigma$ are the mean and standard deviation of the sample over the domain, for the considered snapshot. The colour scales as $f_{bin}^{0.3}$, where $f_{bin}$ is the relative frequency in each bin, in order to exaggerate the tails of the distributions. Very similar figures are obtained for other models, components and snapshots.}
	\label{fig:correlation_N500sf2} 
\end{figure*}

\begin{figure*}[ht] 
	\centering
	\includegraphics[width=0.24\linewidth]{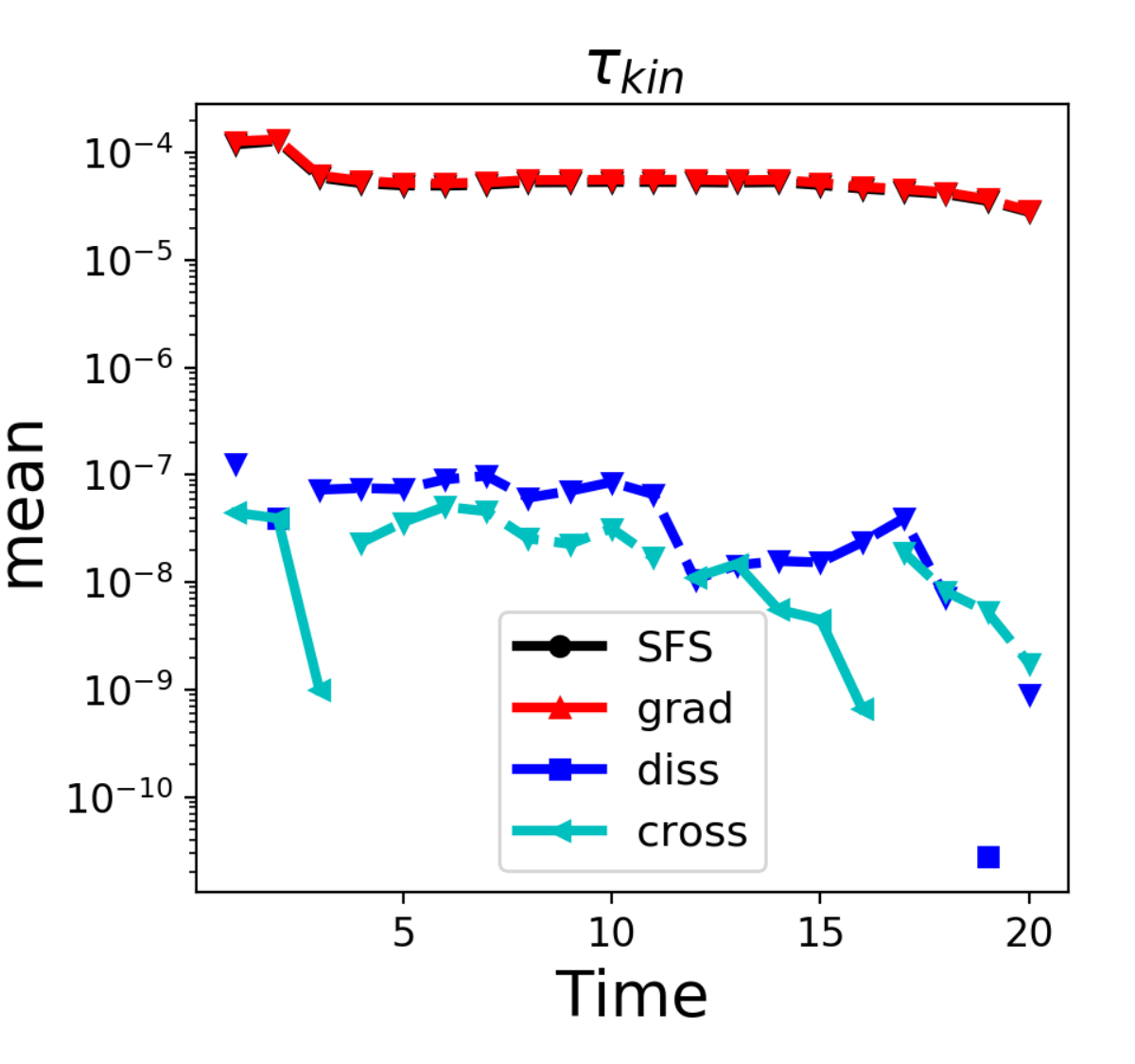}
	\includegraphics[width=0.24\linewidth]{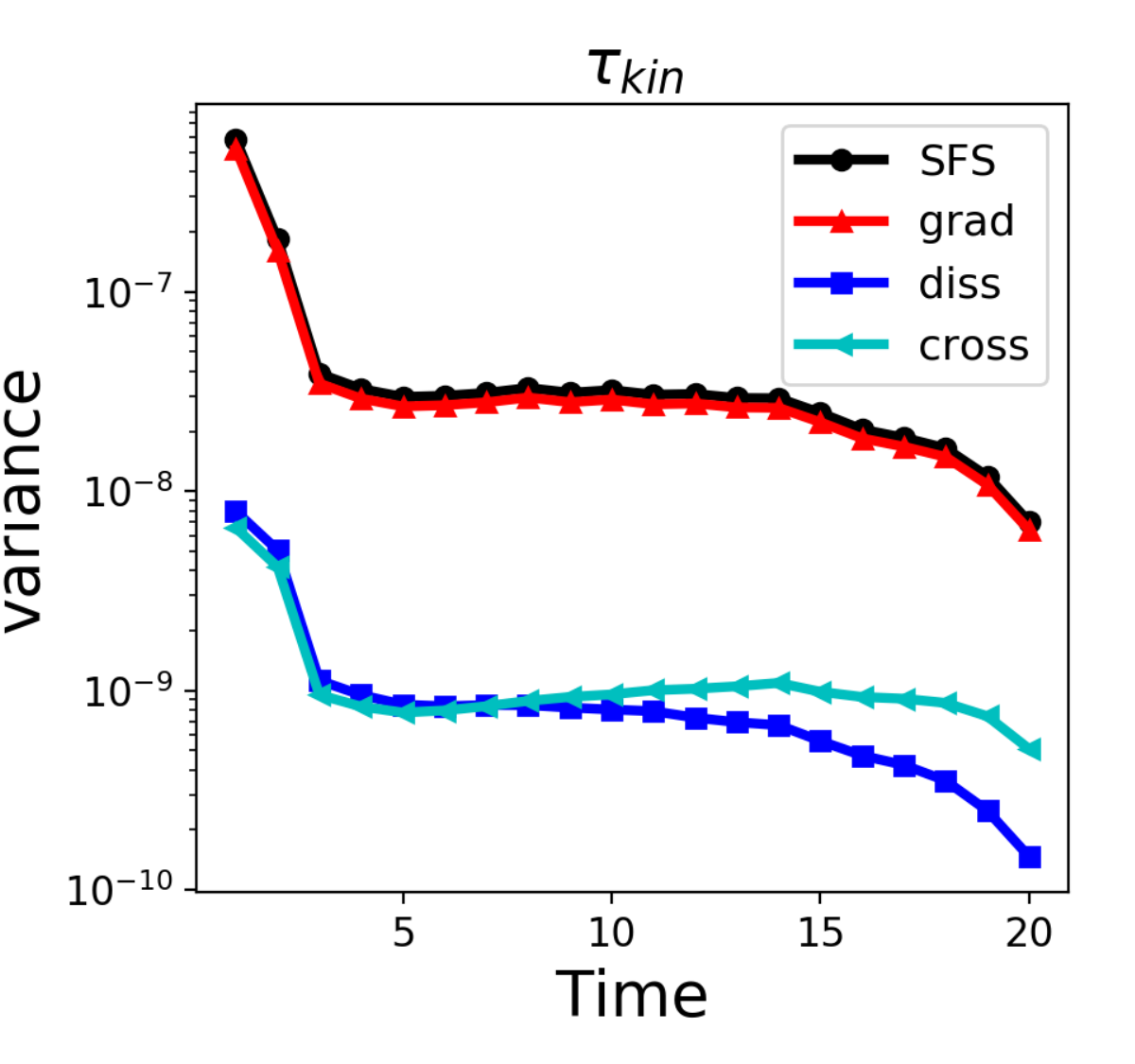}
	\includegraphics[width=0.24\linewidth]{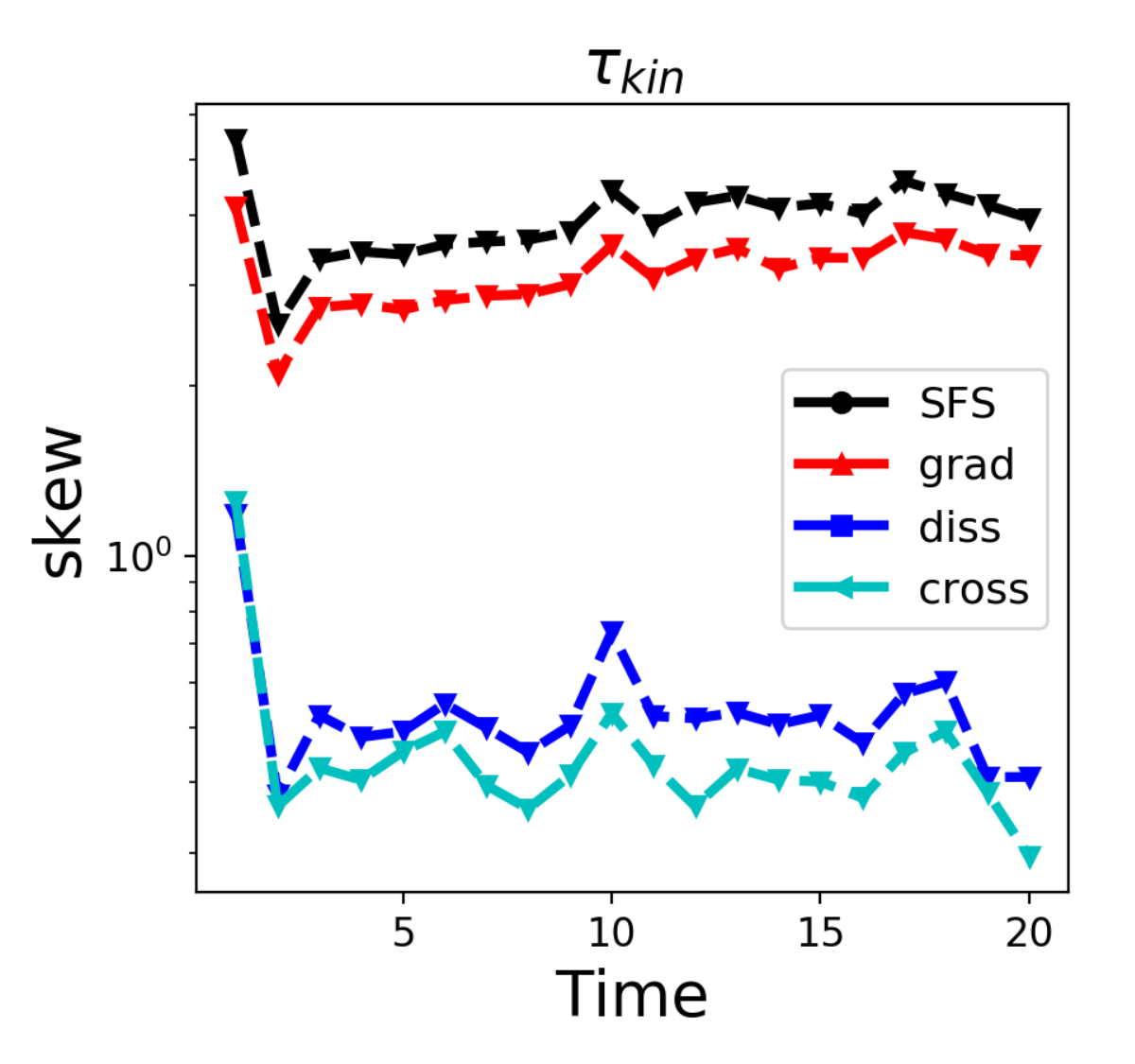}
	\includegraphics[width=0.24\linewidth]{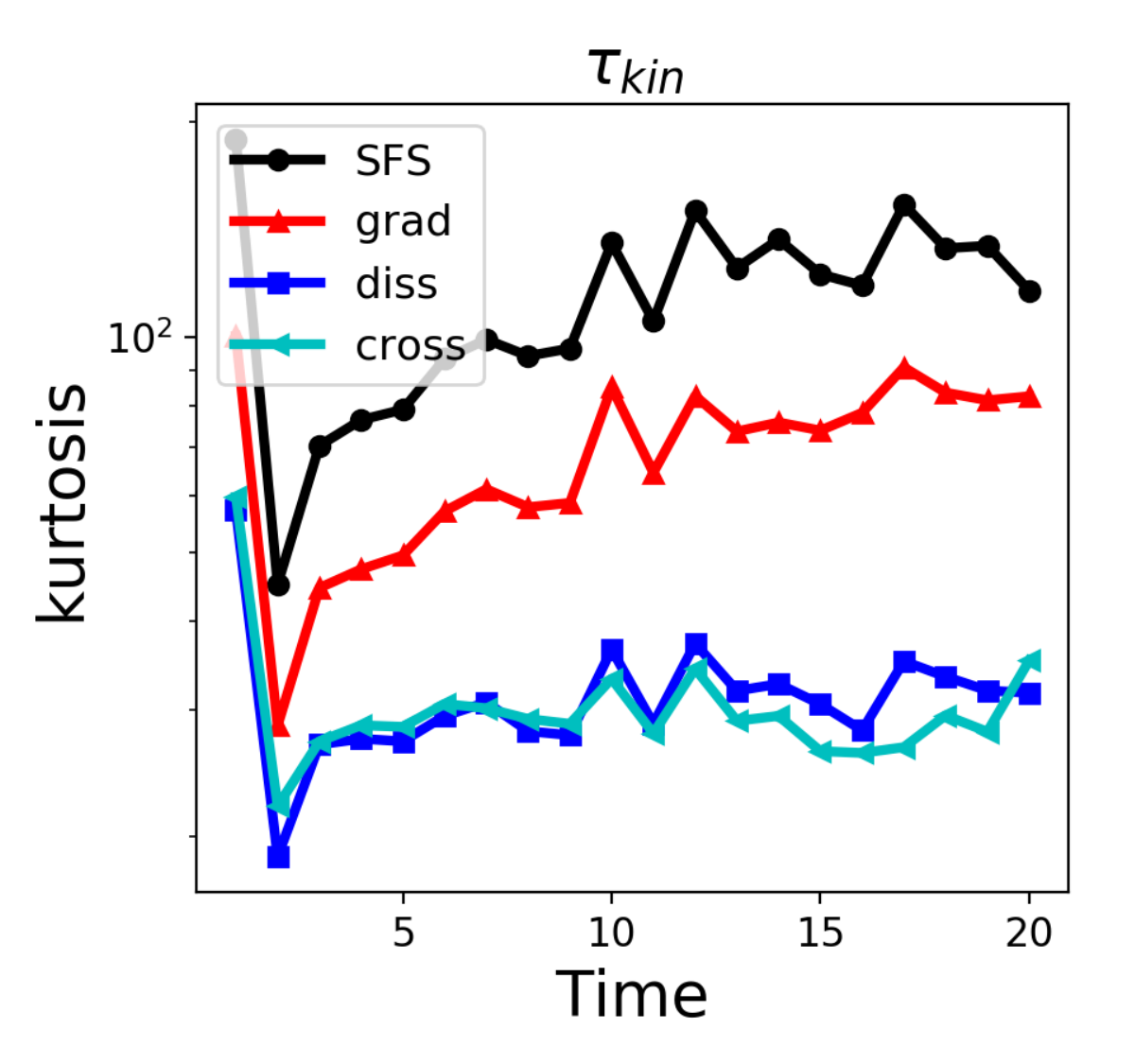}
	\caption{Comparison of the $\tau_{\rm kin}$ statistics: mean, variance, skew, and kurtosis (from left to right) for the SFS and different SGS models. We show the evolution for the simulation {\tt KH3D500}, with filter factor $S_f=2$; others are similars, with a progressive deviation between the gradient model and the SFS values for increasing $S_f$. In the mean and skew, breaking of the line means a change of sign, with the symbols of the legend indicating positive values, and $\bigtriangledown$ indicating negative values.}
	\label{fig:stats_kin} 
\end{figure*}

\begin{figure}[ht] 
	\centering
	\includegraphics[width=0.45\linewidth]{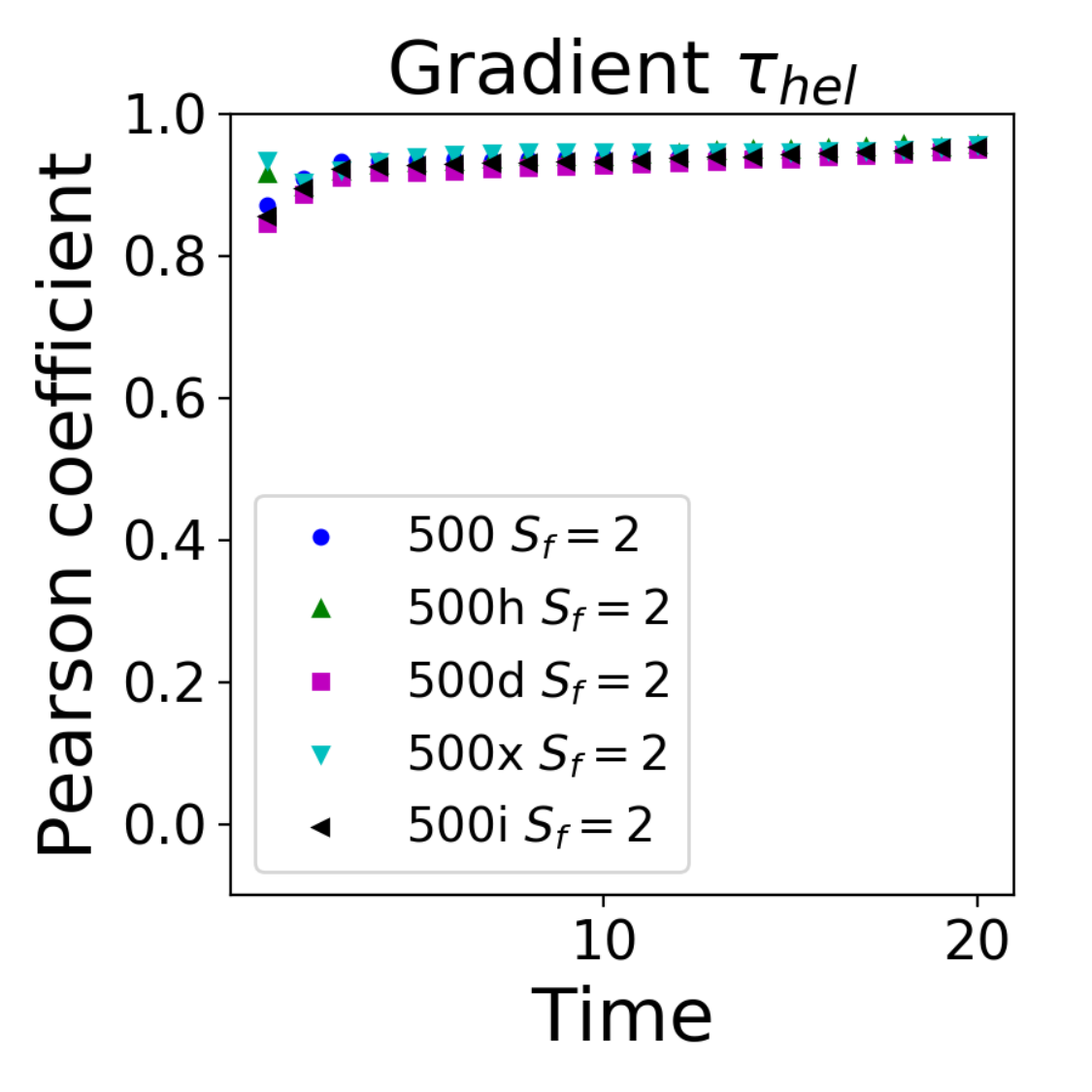}
	\includegraphics[width=0.45\linewidth]{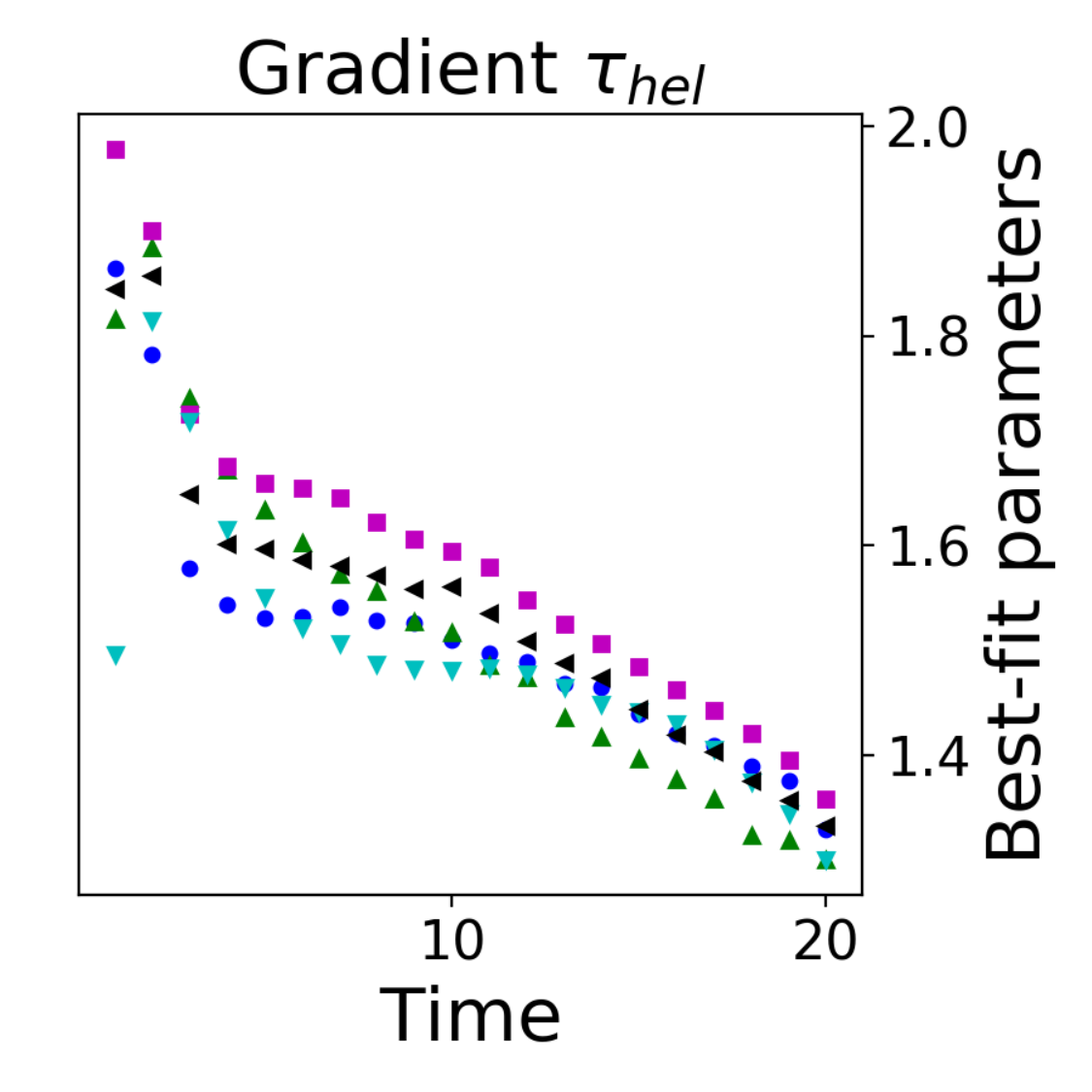}\\
	\includegraphics[width=0.45\linewidth]{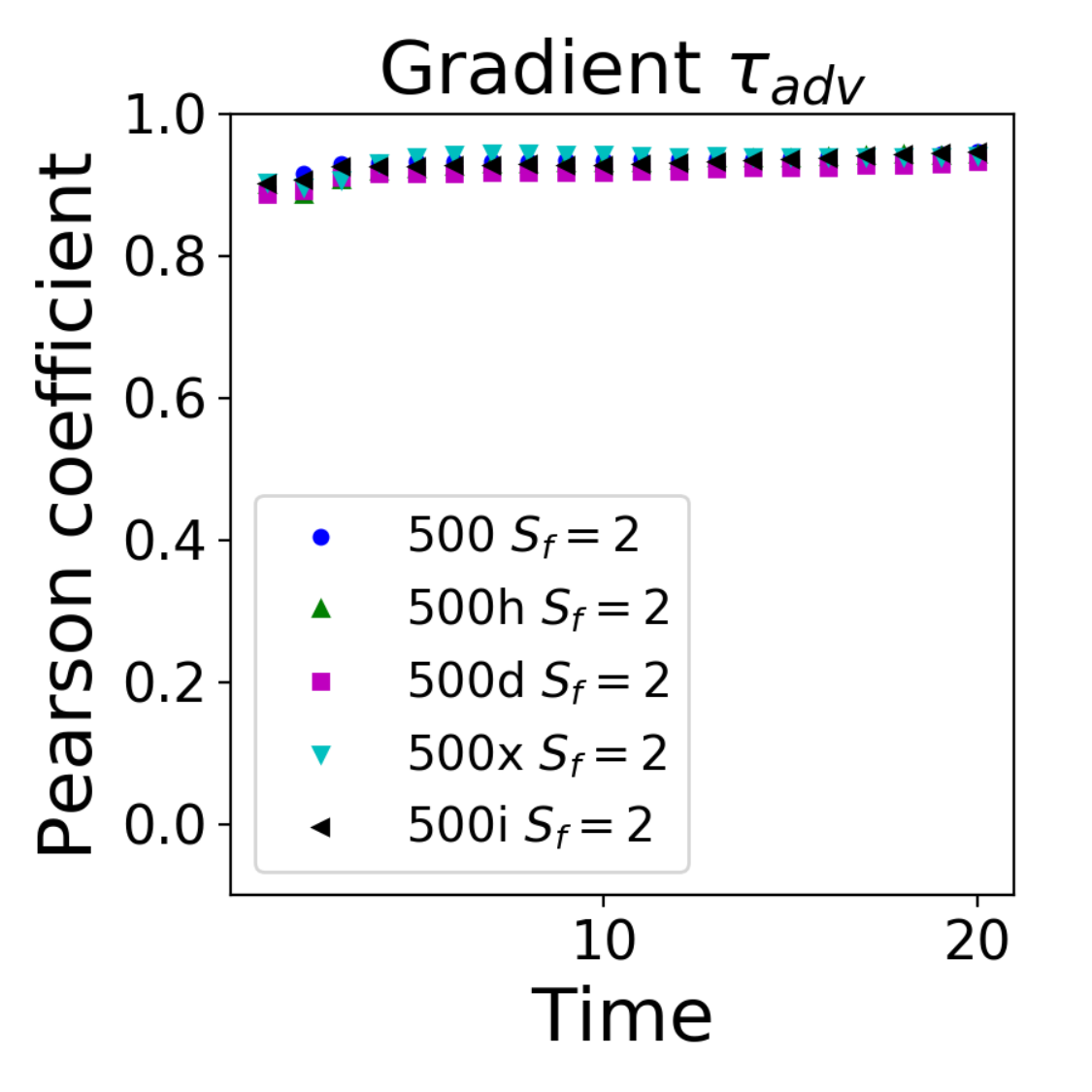}
	\includegraphics[width=0.45\linewidth]{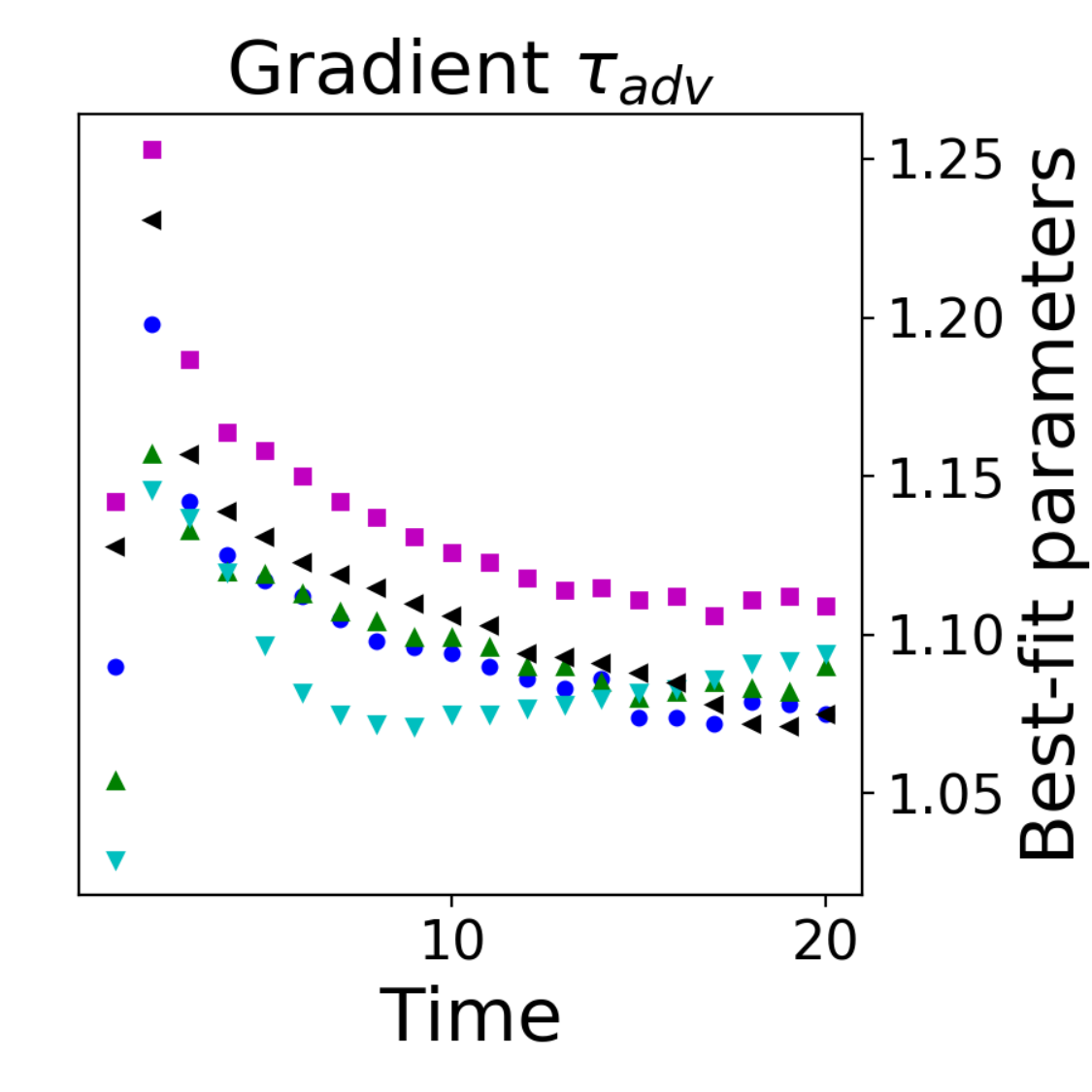}
	\caption{Evolution of the Pearson correlation coefficients ${\cal P}$ (left panels) and the best-fit parameters $C_{\rm best}$ (right panels) for $\tau_{\rm hel}$ and $\tau_{\rm adv}$, for all models with $500^3$ and $S_f=2$.}
	\label{fig:tau_hel_adv_N500sf2}
\end{figure}

\begin{figure*}[ht] 
	\centering
	\includegraphics[width=0.24\linewidth]{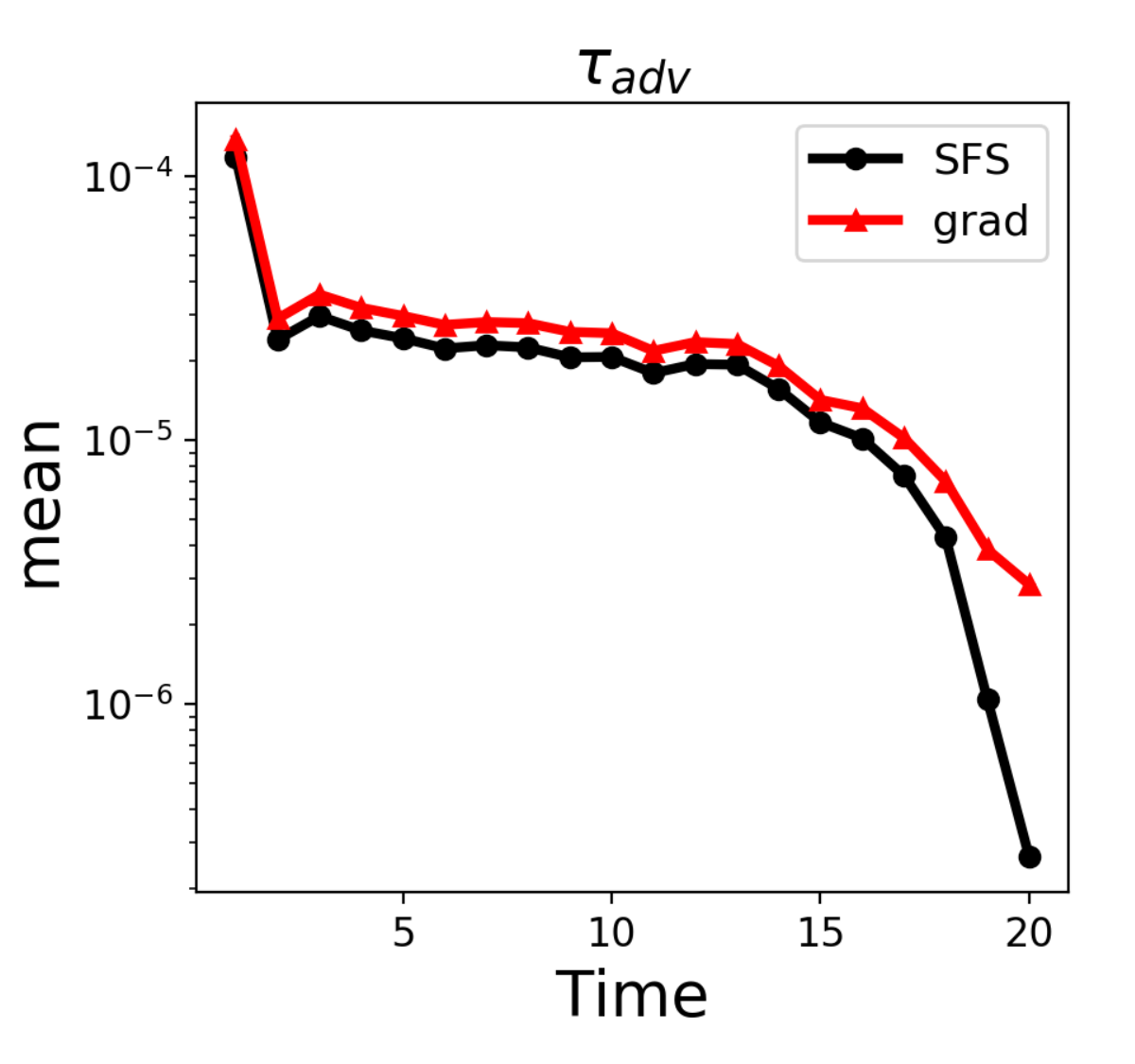}
	\includegraphics[width=0.24\linewidth]{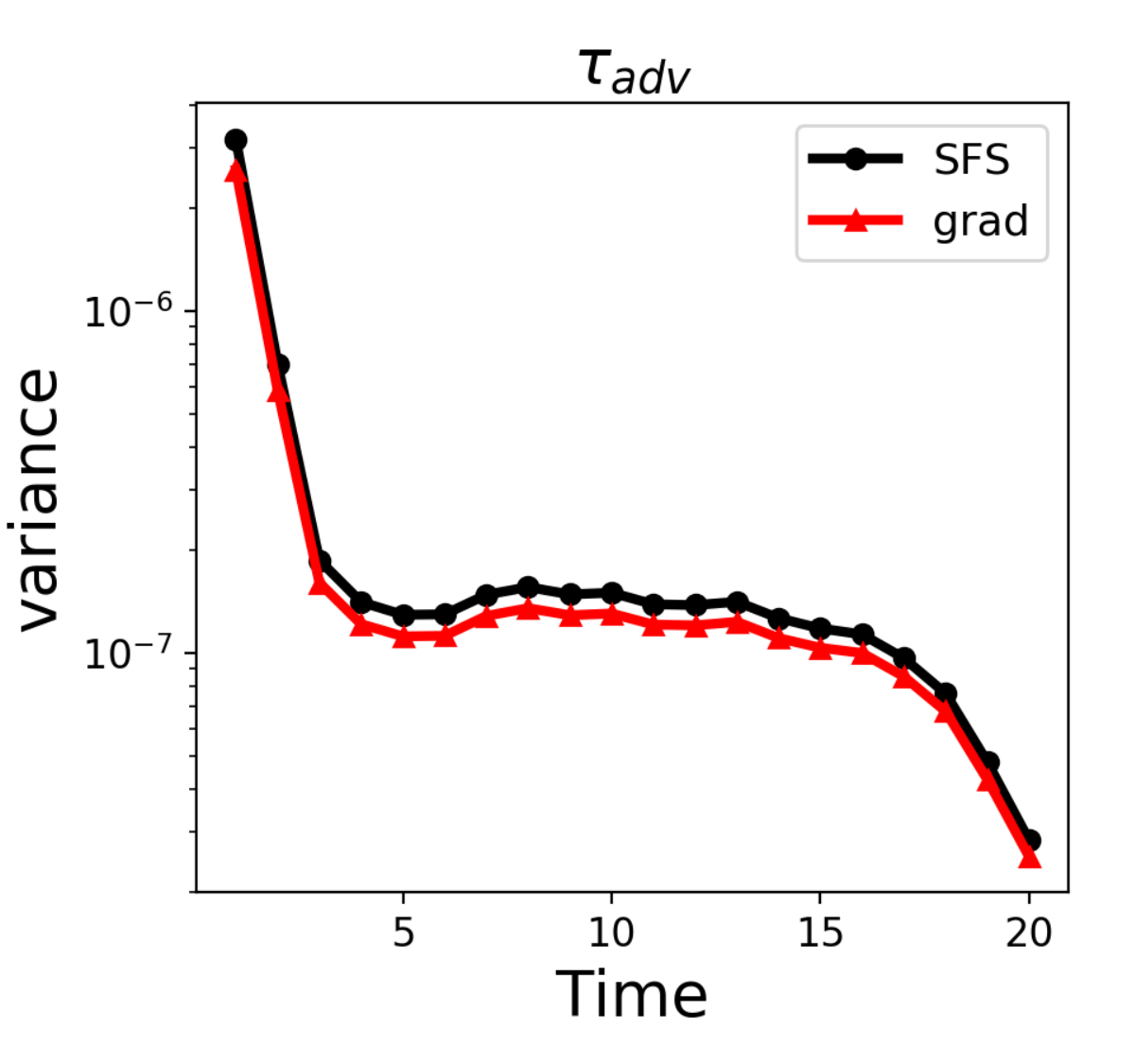}
	\includegraphics[width=0.24\linewidth]{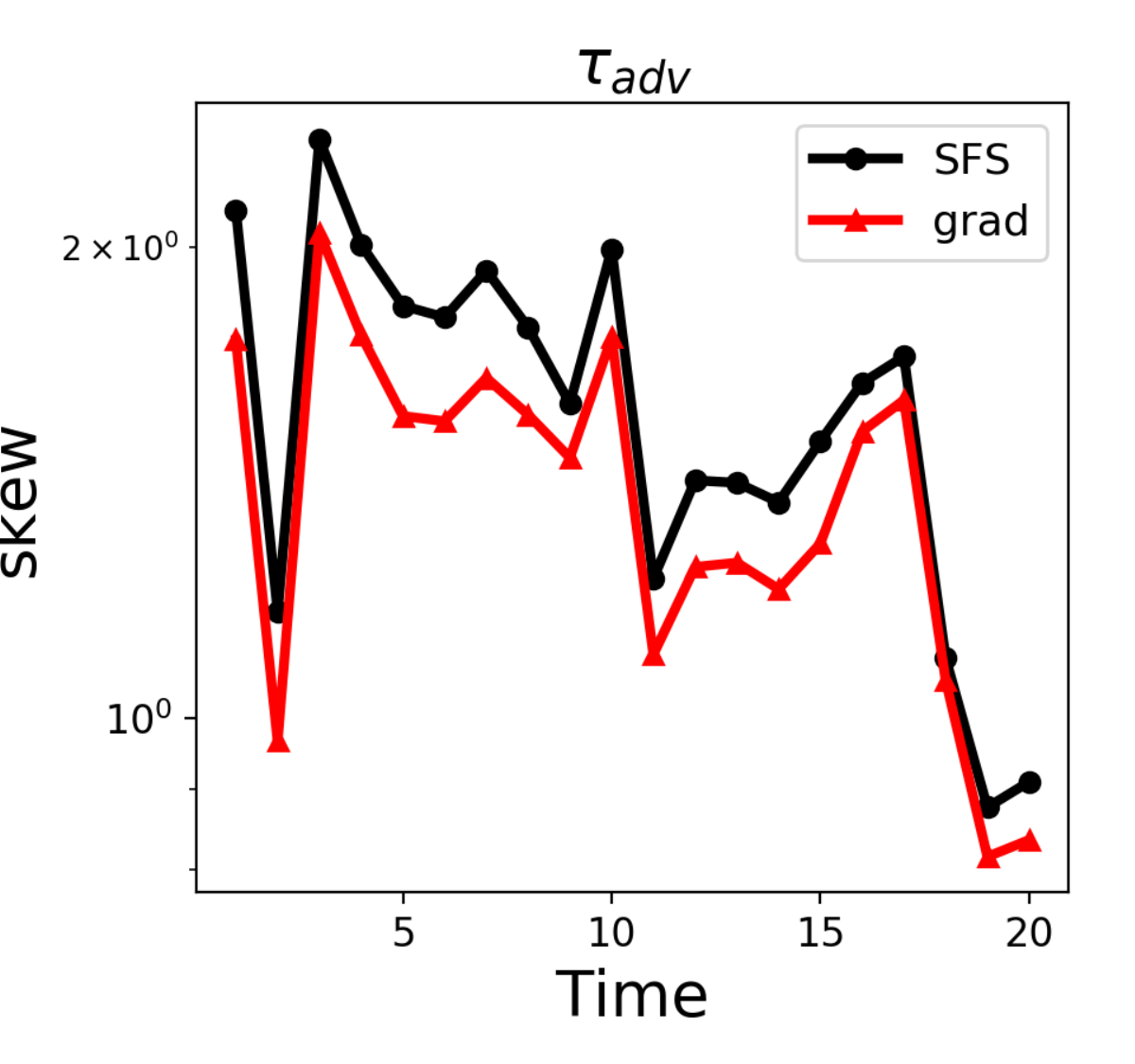}
	\includegraphics[width=0.24\linewidth]{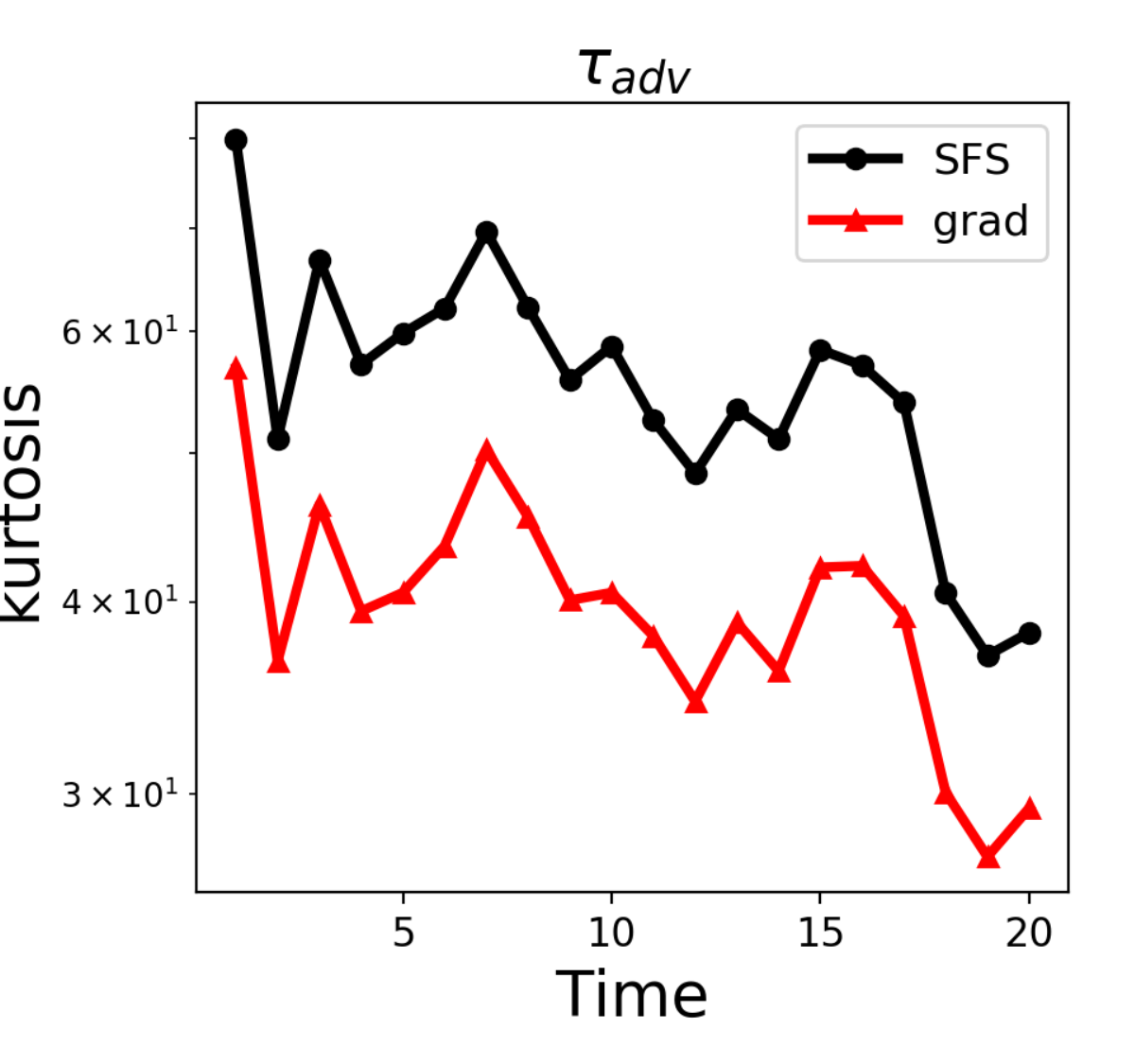}\\
	\includegraphics[width=0.24\linewidth]{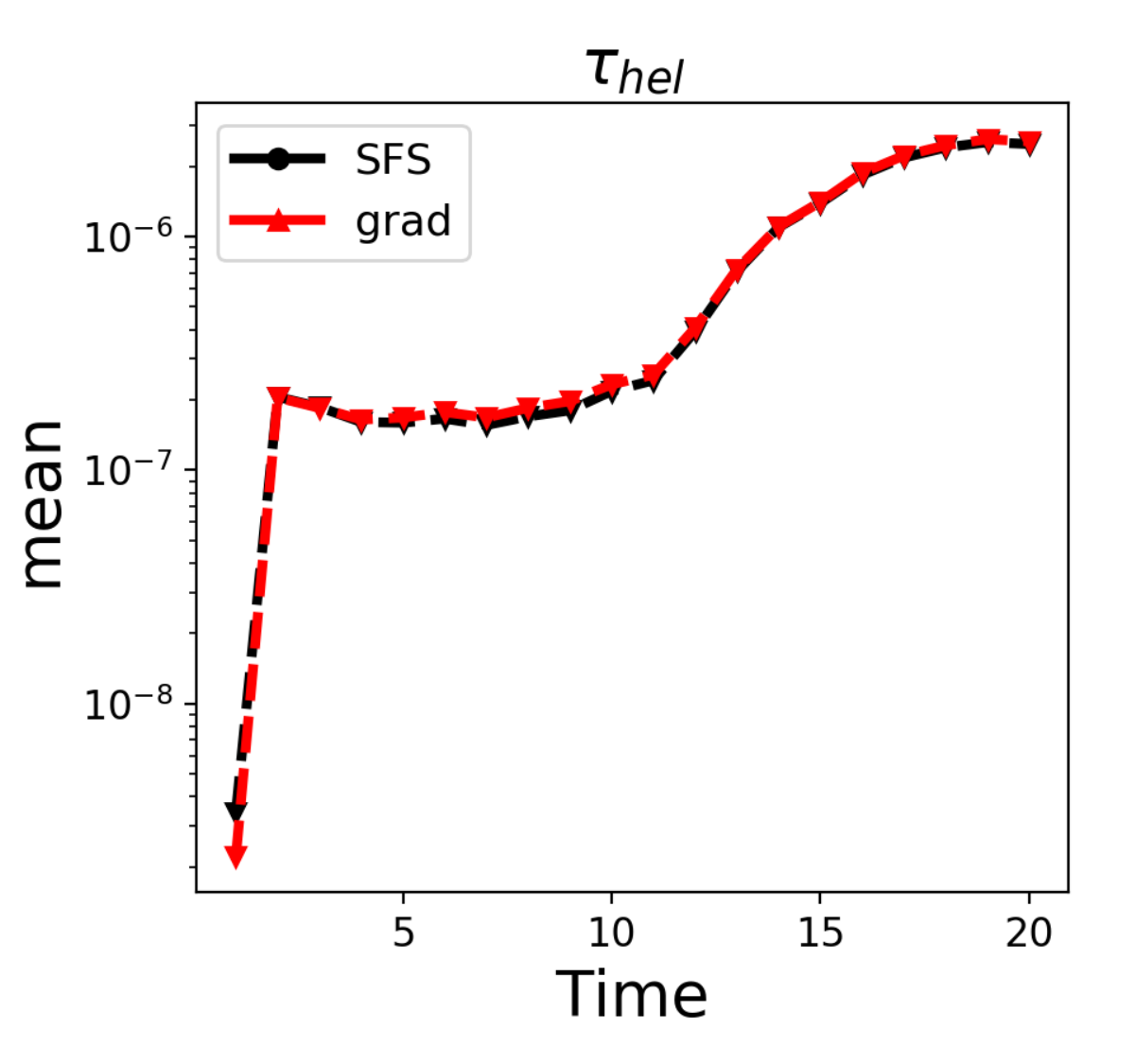}
	\includegraphics[width=0.24\linewidth]{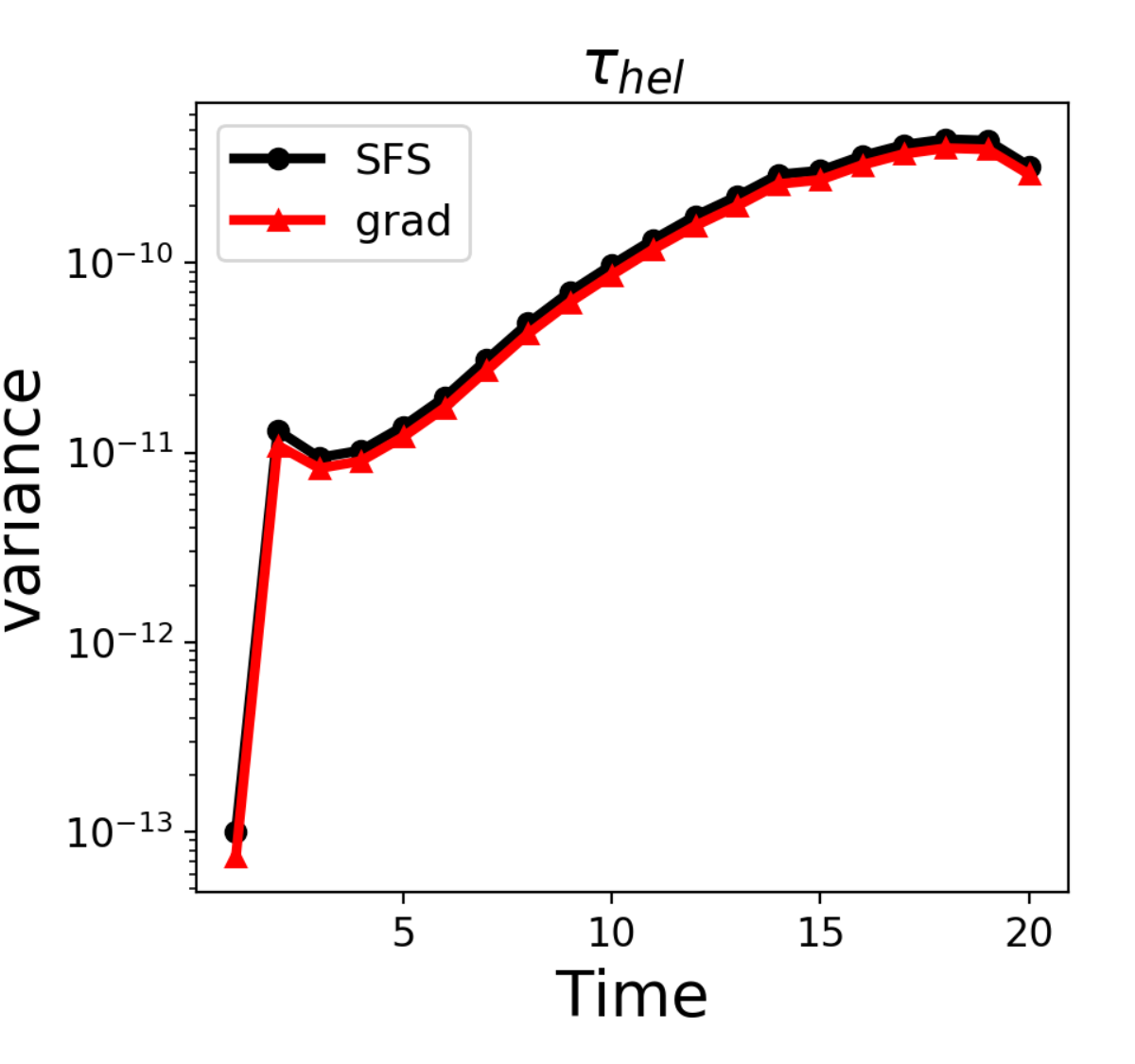}
	\includegraphics[width=0.24\linewidth]{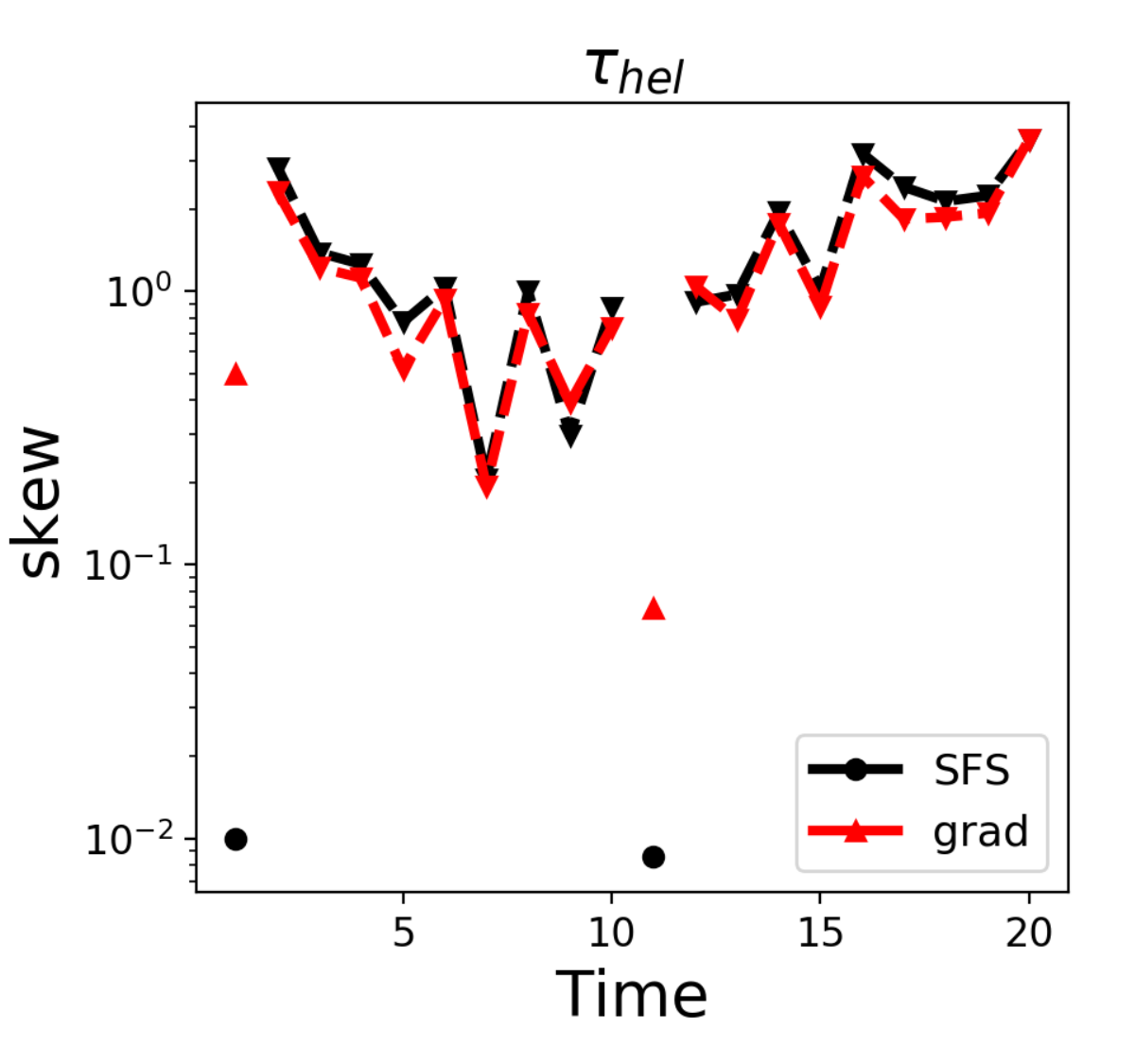}
	\includegraphics[width=0.24\linewidth]{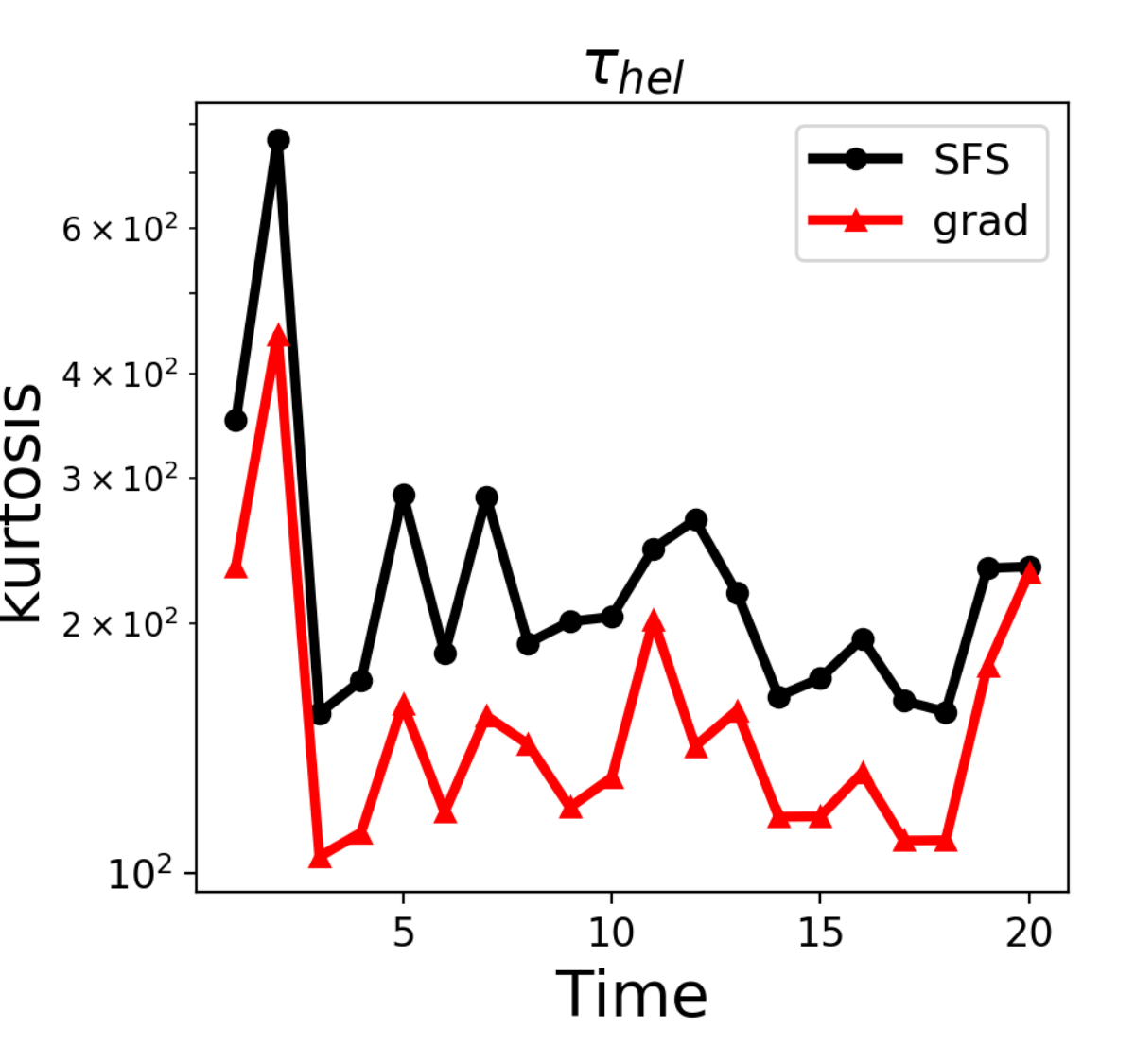}
	\caption{The same as Fig.~\ref{fig:stats_kin} but for the SFS-gradient comparison statistics of $\tau_{\rm adv}$ (top) and $\tau_{\rm hel}$ (bottom). }
	\label{fig:stats_adv} 
\end{figure*}

\begin{figure}[ht] 
	\centering
	\includegraphics[width=0.49\linewidth]{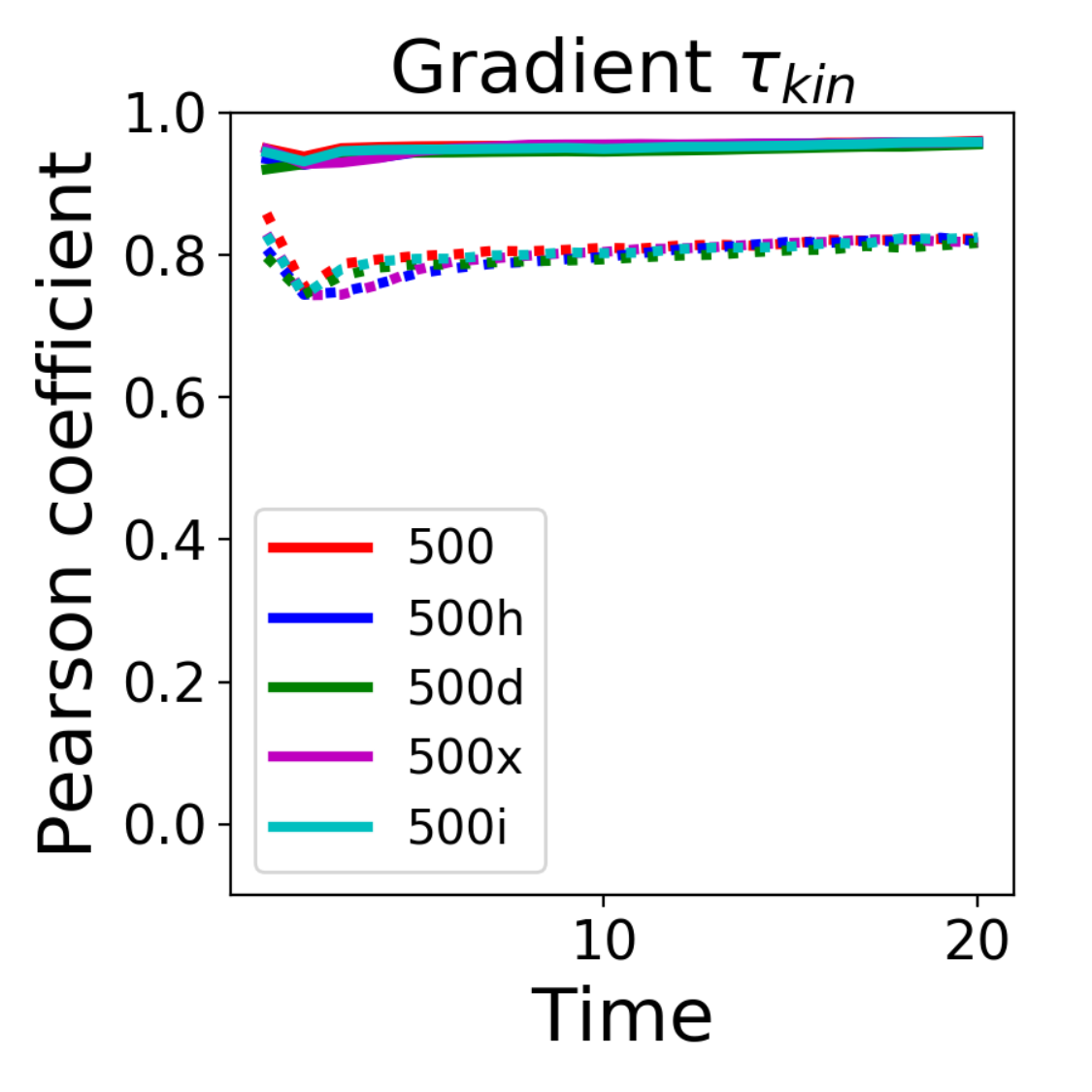}
	\includegraphics[width=0.49\linewidth]{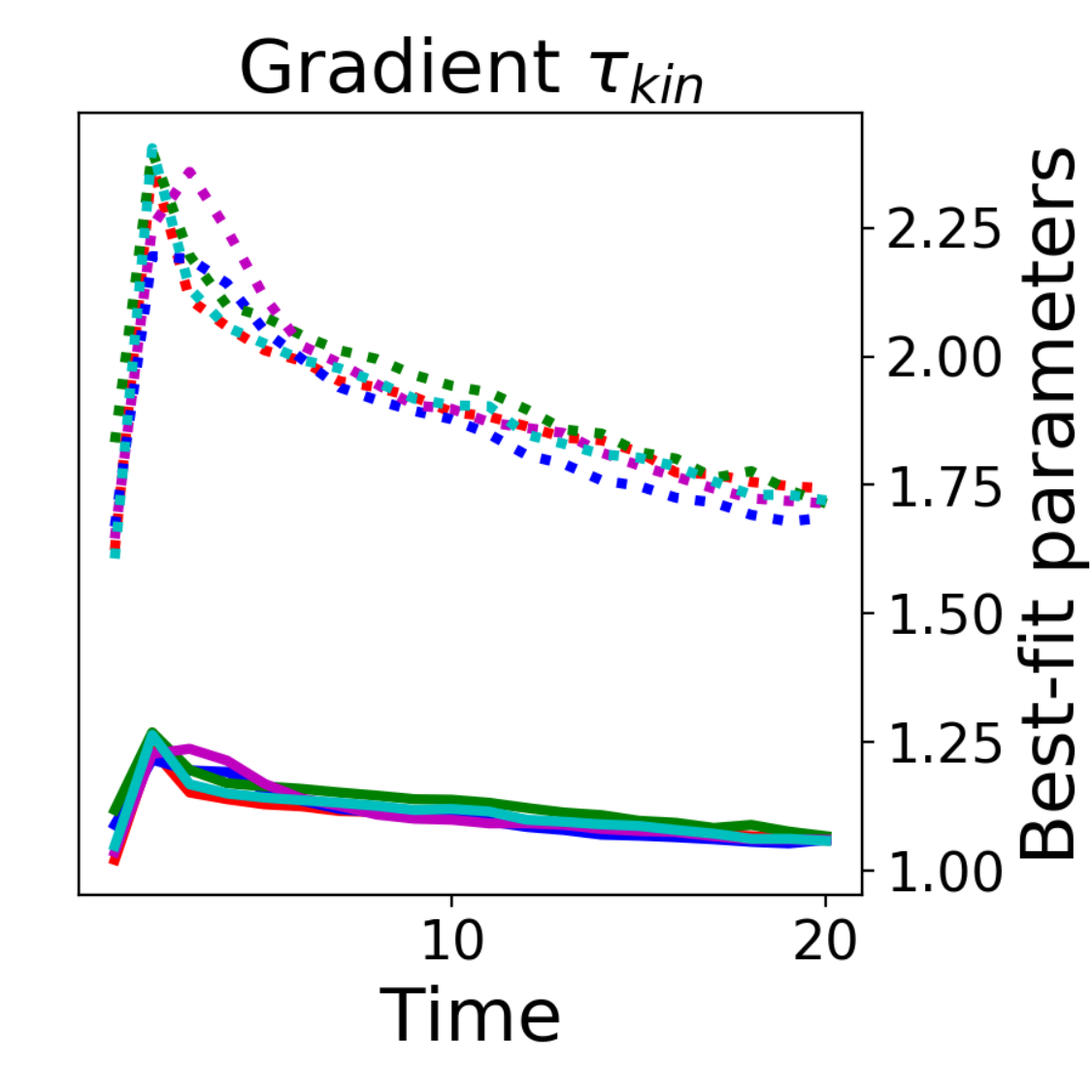}\\
	\includegraphics[width=0.49\linewidth]{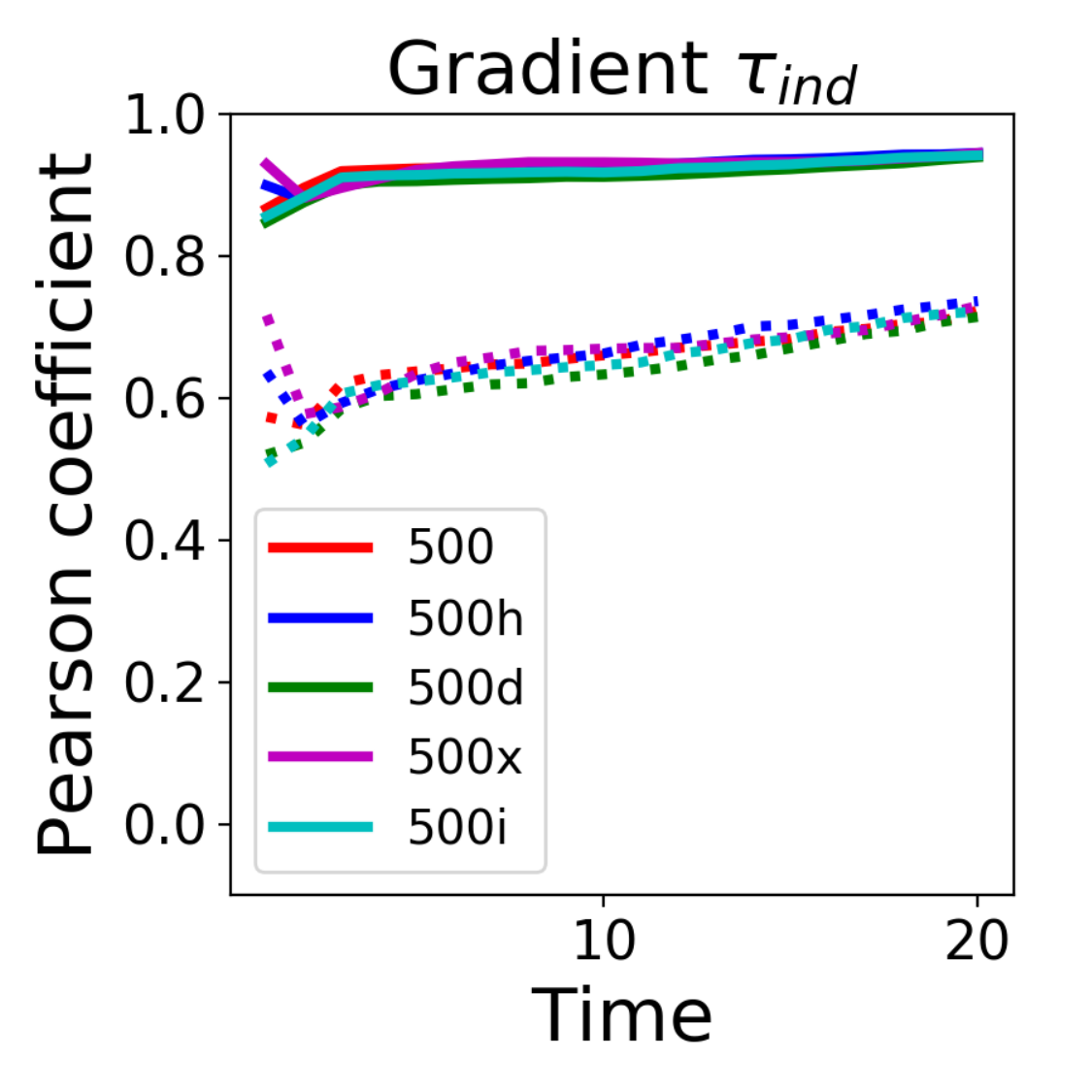}
	\includegraphics[width=0.49\linewidth]{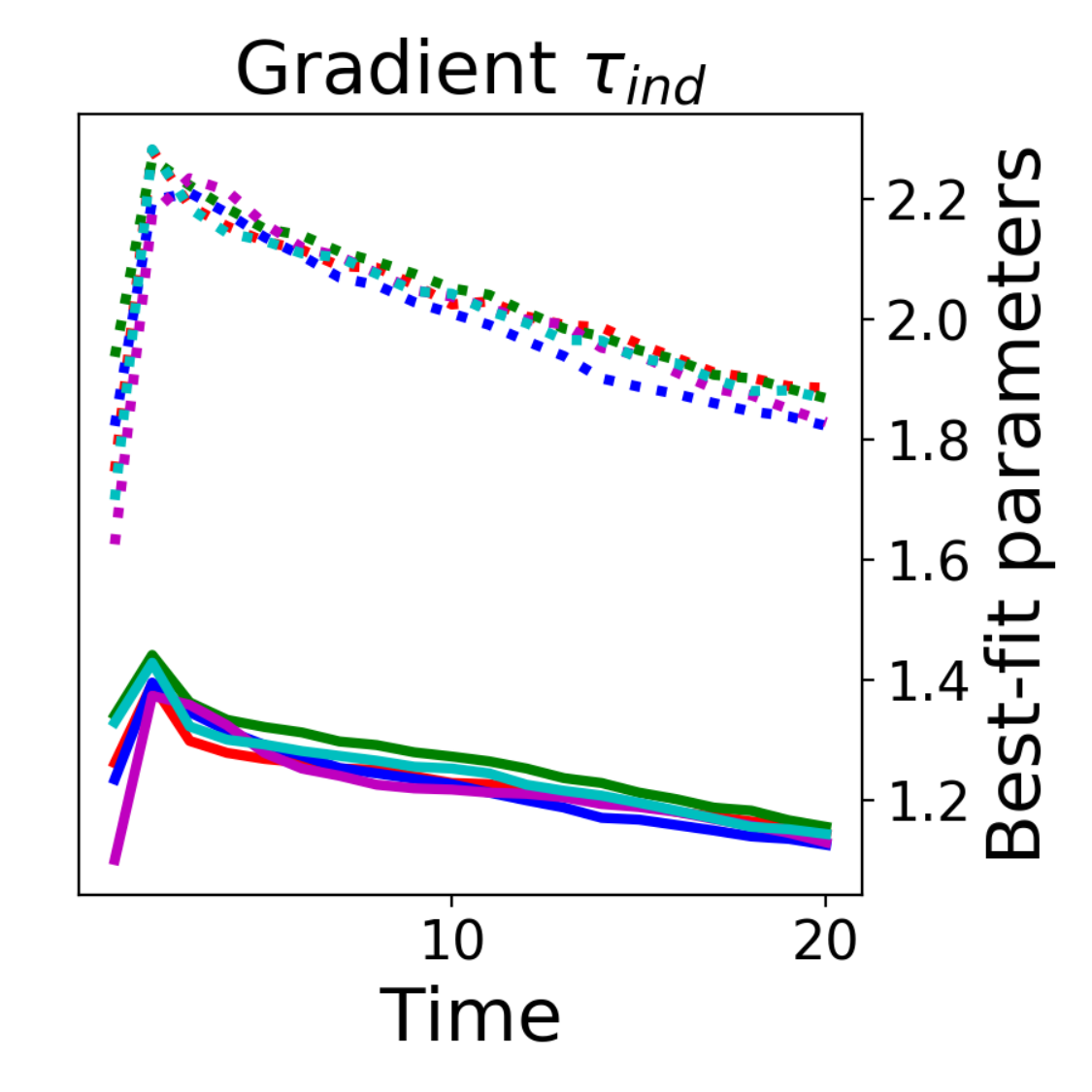}\\
	\includegraphics[width=0.49\linewidth]{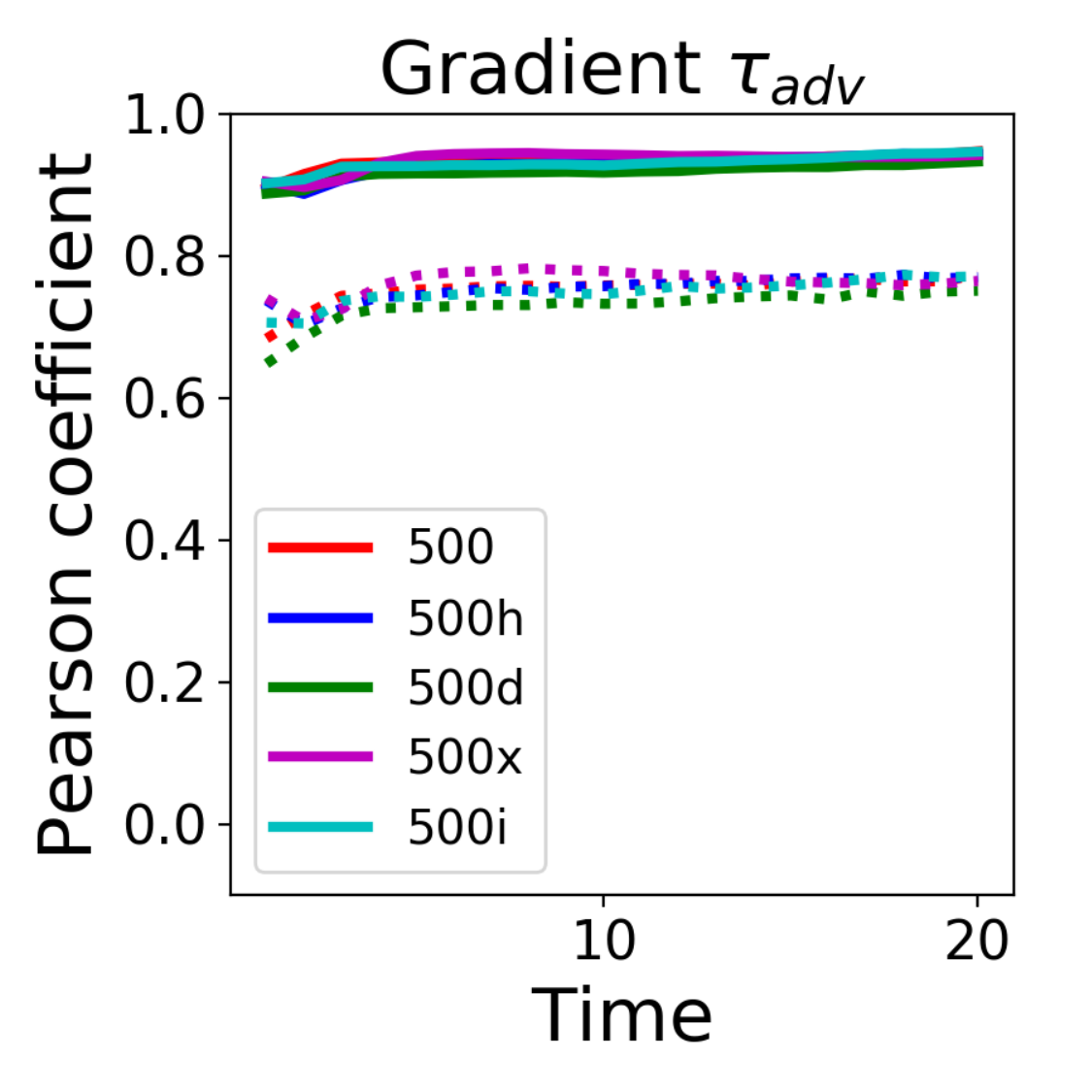}
	\includegraphics[width=0.49\linewidth]{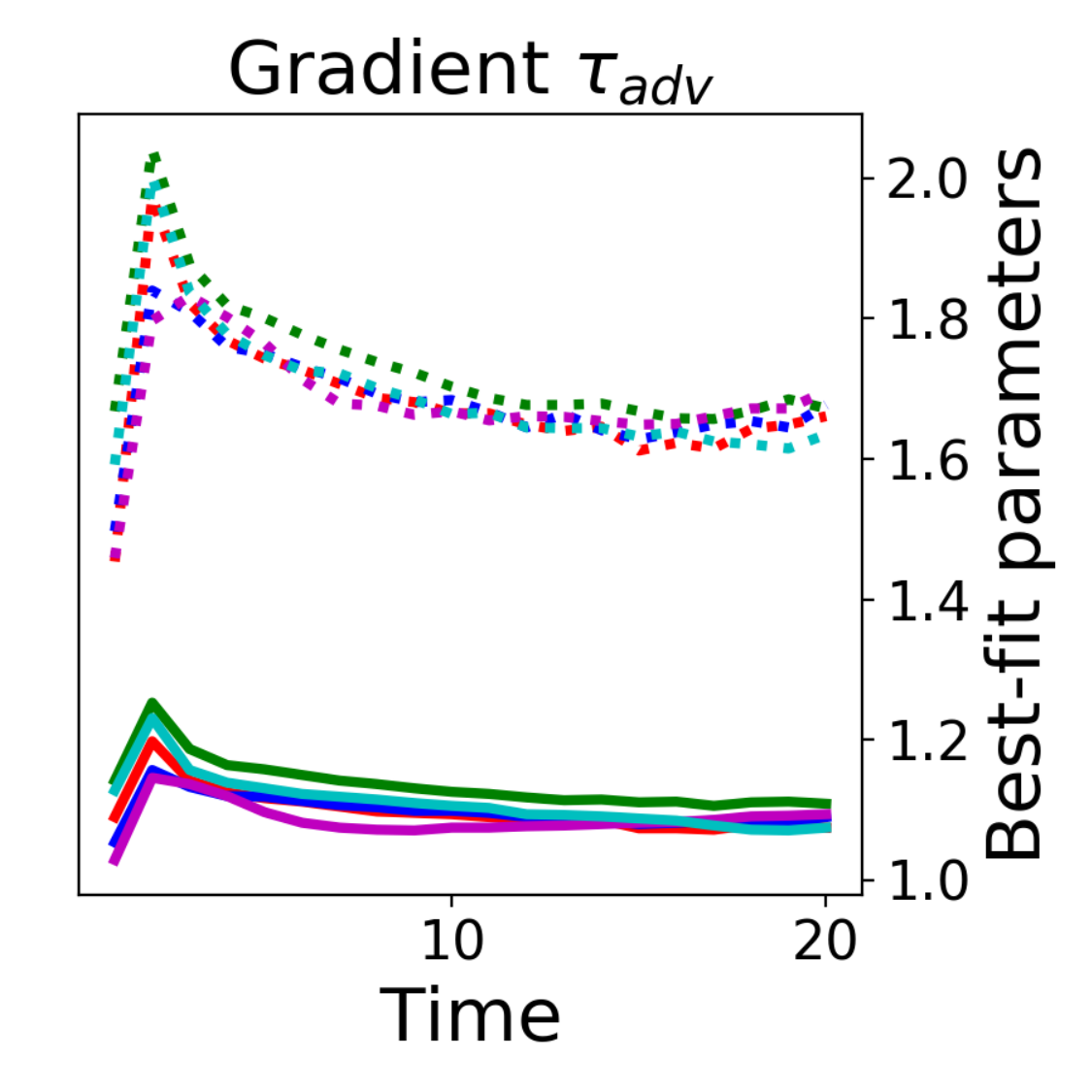}\\
	\includegraphics[width=0.49\linewidth]{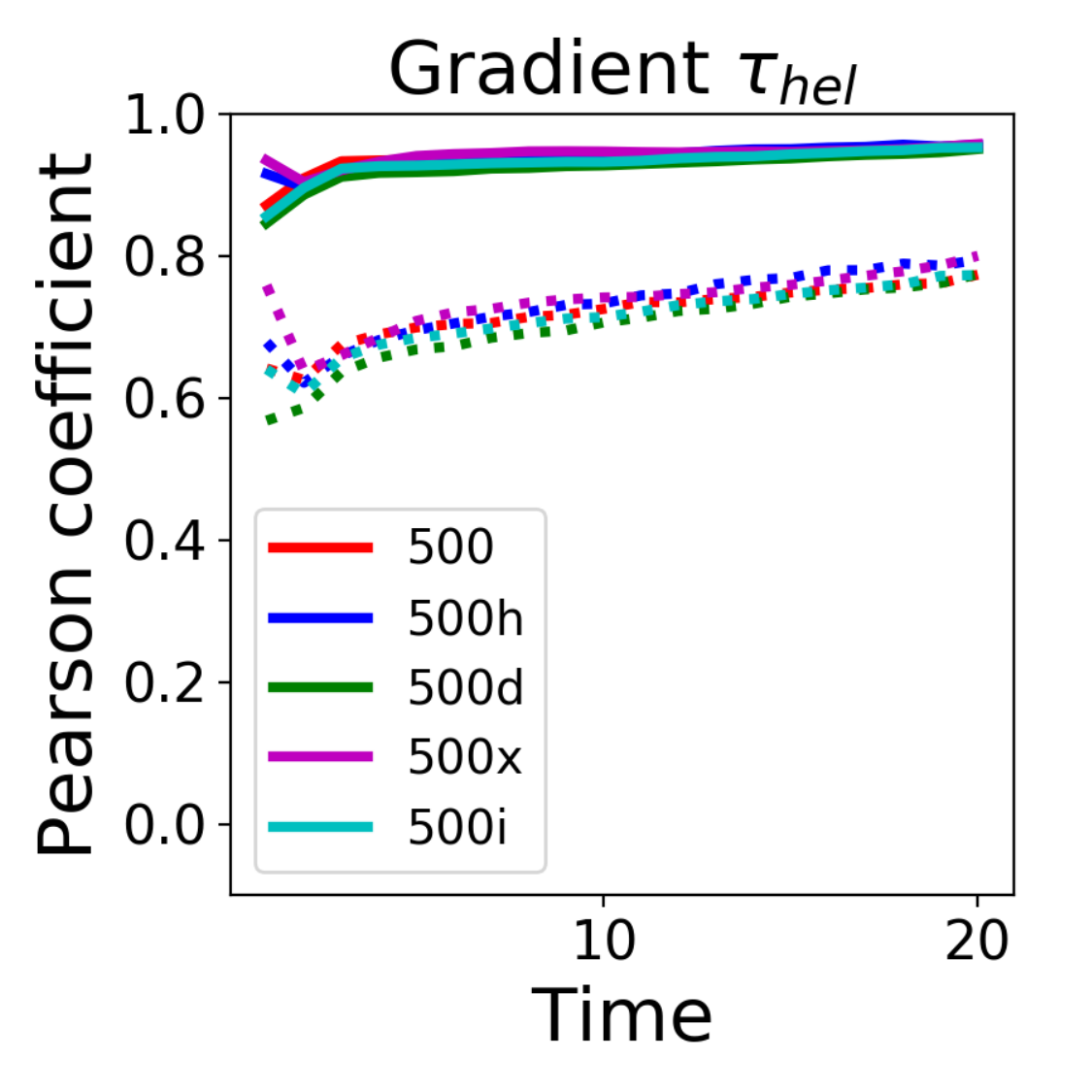}
	\includegraphics[width=0.49\linewidth]{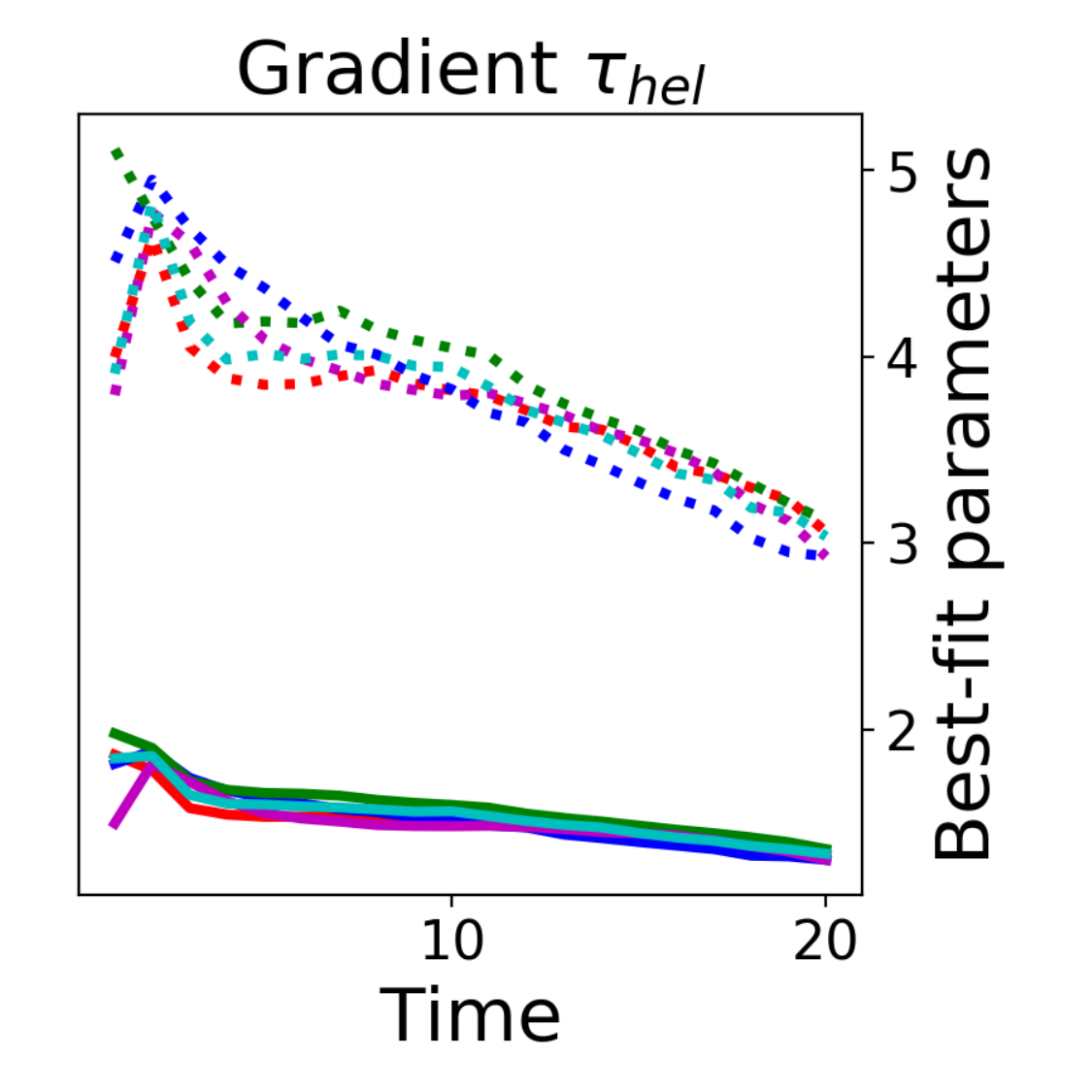}
	\caption{Evolution of the Pearson correlation coefficients ${\cal P}$ (left panels) and the best-fit parameters $C_{\rm best}$ (right panels) for the following gradient models (from top to bottom): $\tau_{\rm kin}$, $\tau_{\rm ind}$, $\tau_{\rm adv}$, $\tau_{\rm hel}$. We show all models with $N=500^3$, and different filter factors $S_f=2$ (solid lines) and $S_f=4$ (dots).}
	\label{fig:pearson_500} 
\end{figure}

\begin{figure}[ht] 
	\centering
	\includegraphics[width=0.49\linewidth]{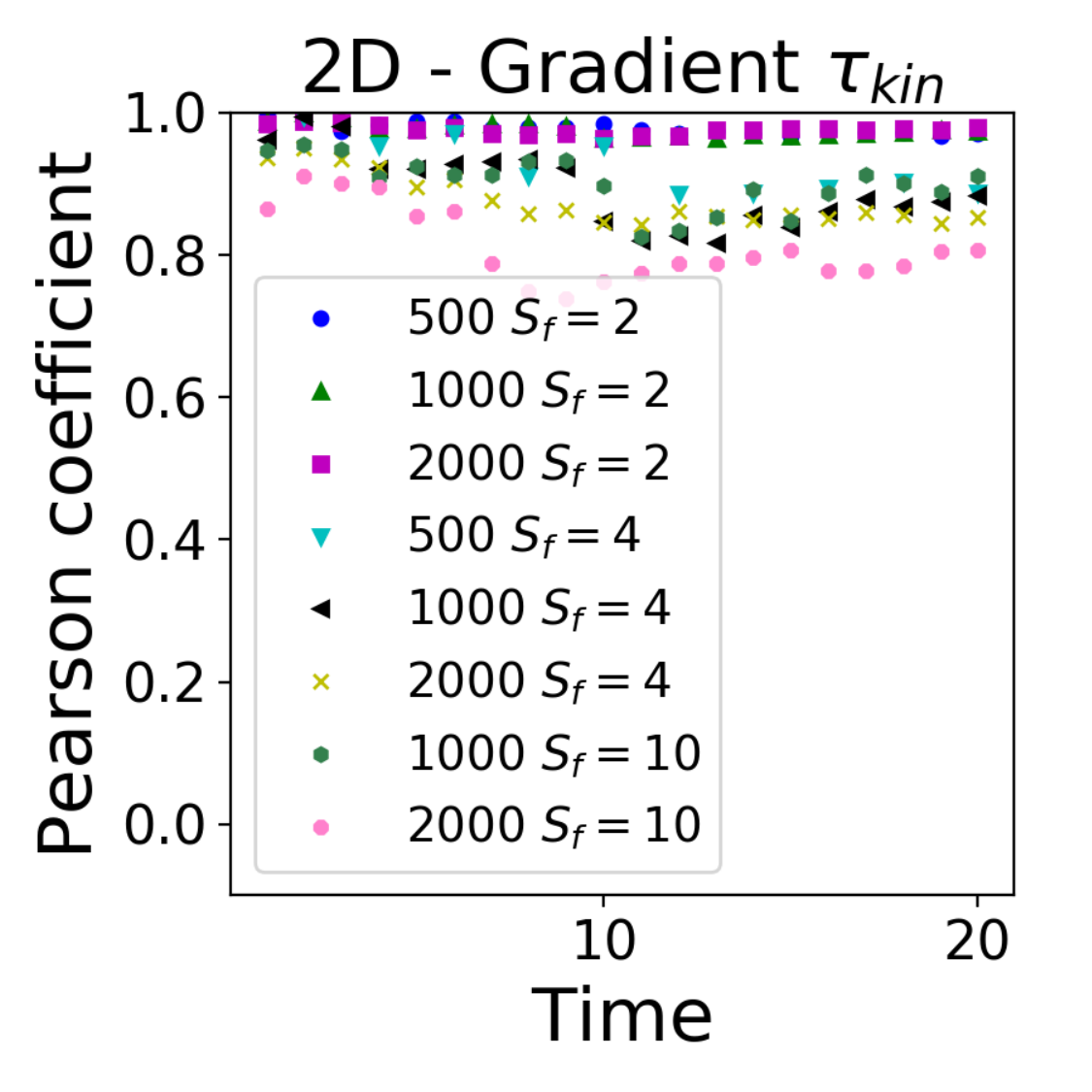}
	\includegraphics[width=0.49\linewidth]{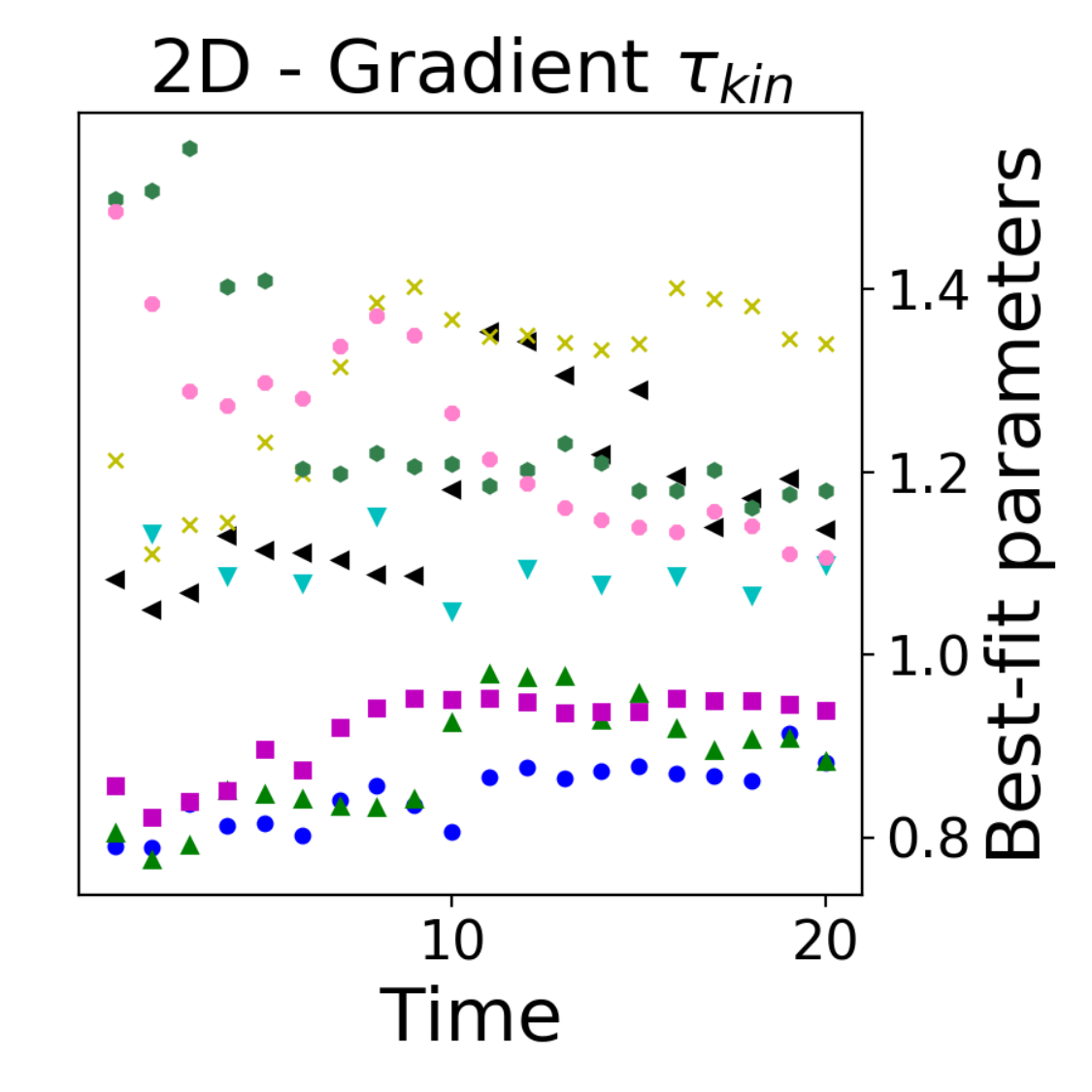}
	\caption{Evolution of the Pearson correlation coefficients ${\cal P}$ (left) and the best-fit parameters $C_{\rm best}$ (right) for the gradient $\tau_{\rm kin}$ for the 2D models with $N=500^2,1000^2,2000^2$, with filter factors $S_f=2,4,10$ (see Appendix~\ref{app:kh2d_periodic} for the description of the setup).}
	\label{fig:apriori_2D} 
\end{figure}

\begin{figure}[ht] 
	\centering
	\includegraphics[width=0.49\linewidth]{comp_SFall_tau_kin_bestfit_pearson_grad.pdf}
	\includegraphics[width=0.49\linewidth]{comp_SFall_tau_kin_bestfit_coef_grad.pdf}\\
	\includegraphics[width=0.49\linewidth]{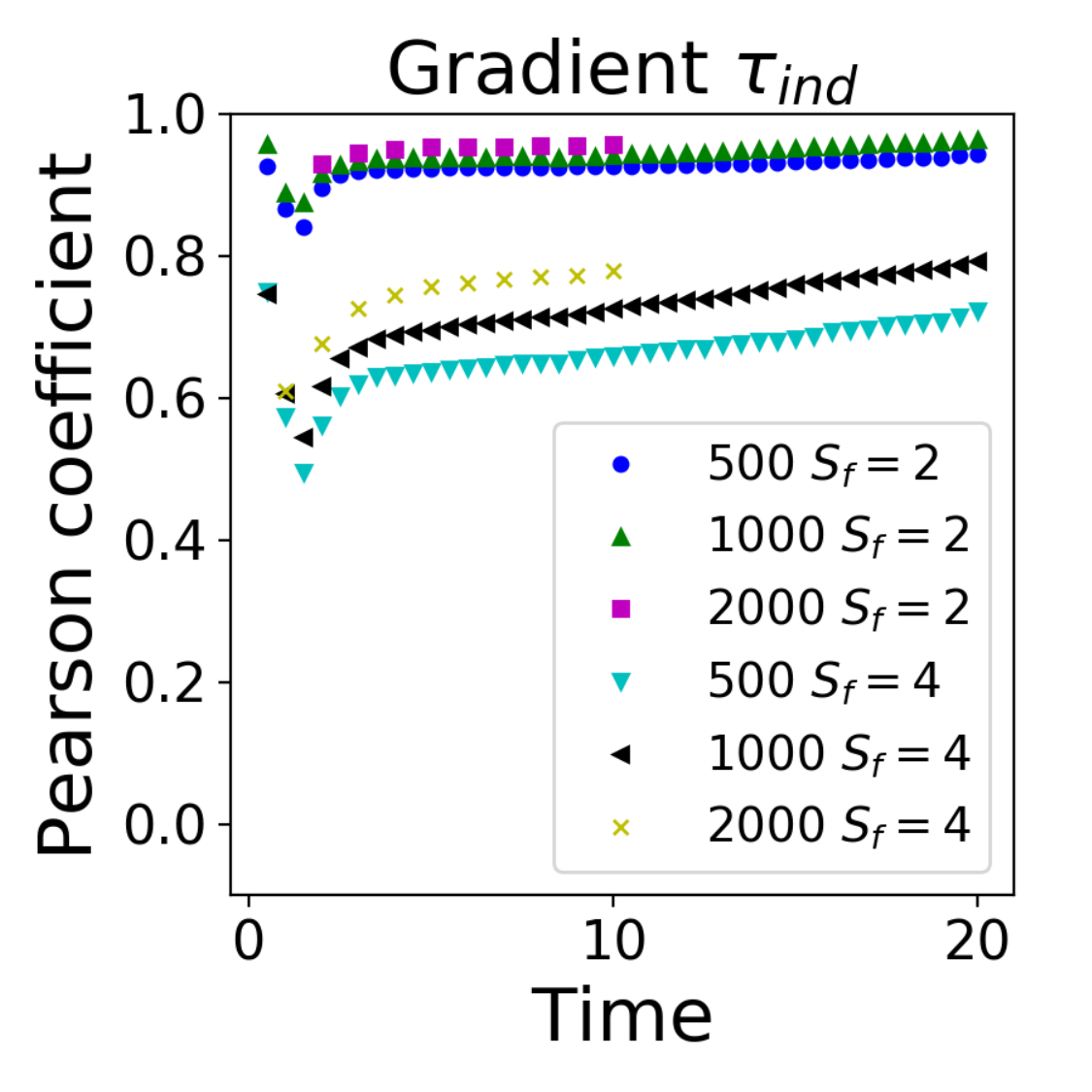}
	\includegraphics[width=0.49\linewidth]{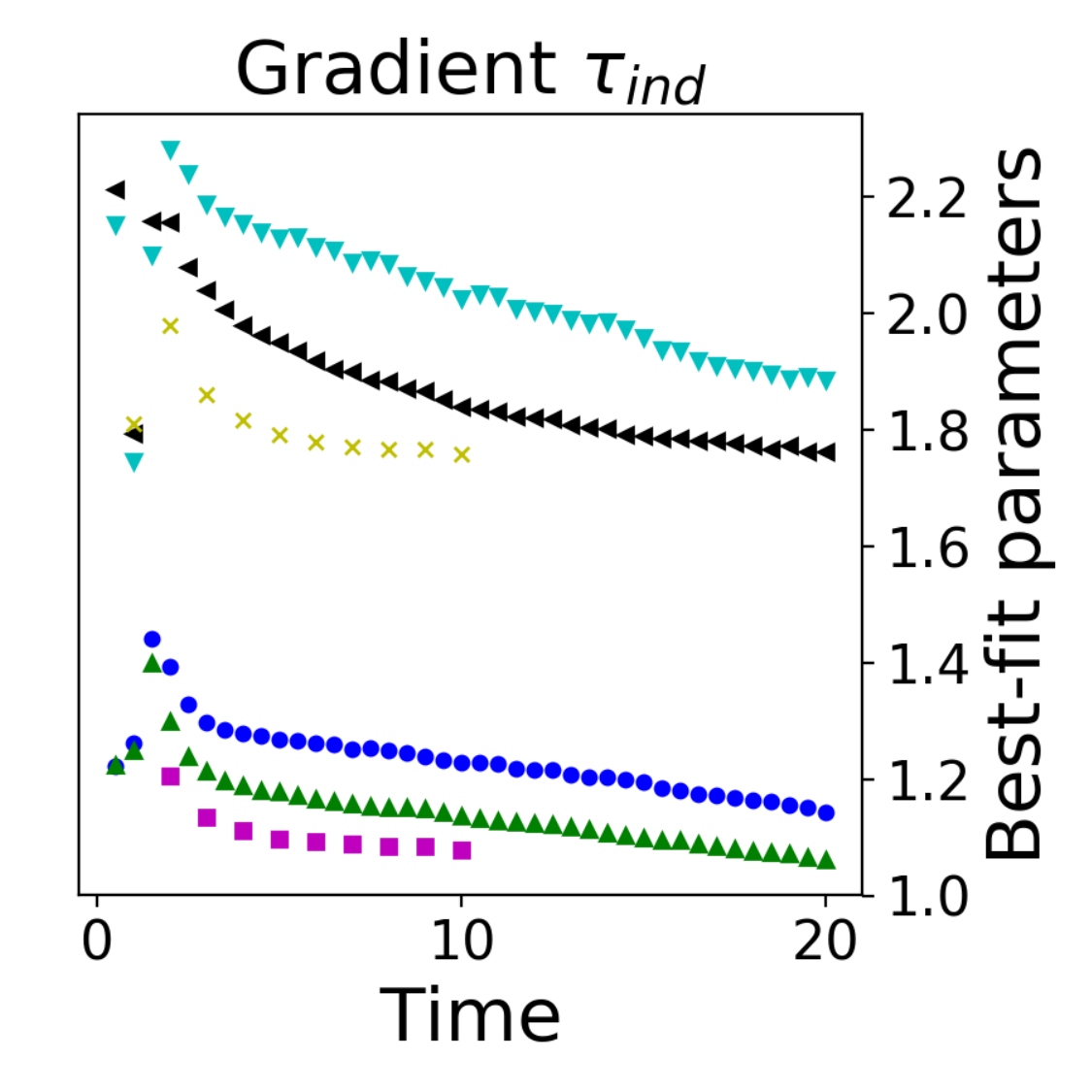}\\
	\includegraphics[width=0.49\linewidth]{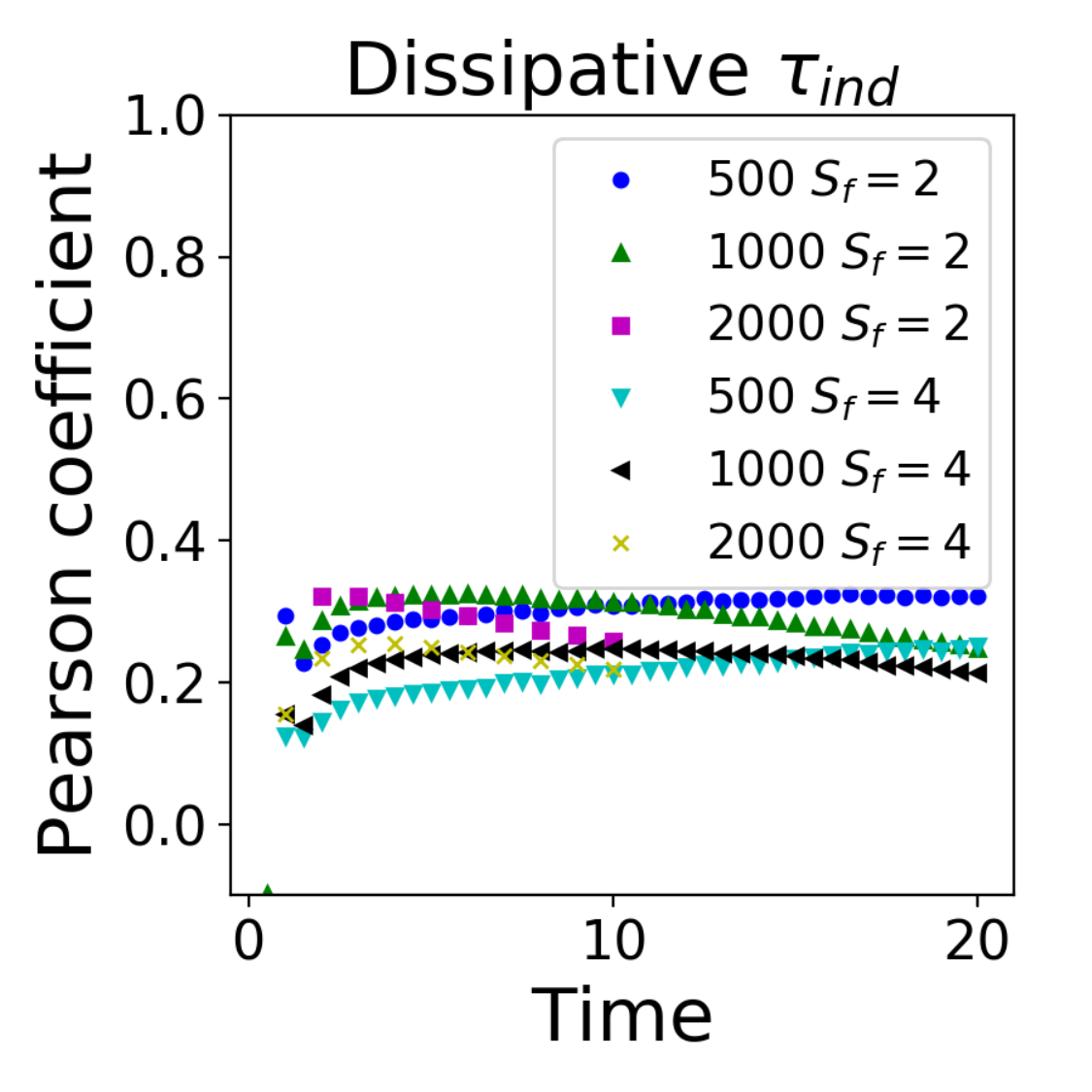}
	\includegraphics[width=0.49\linewidth]{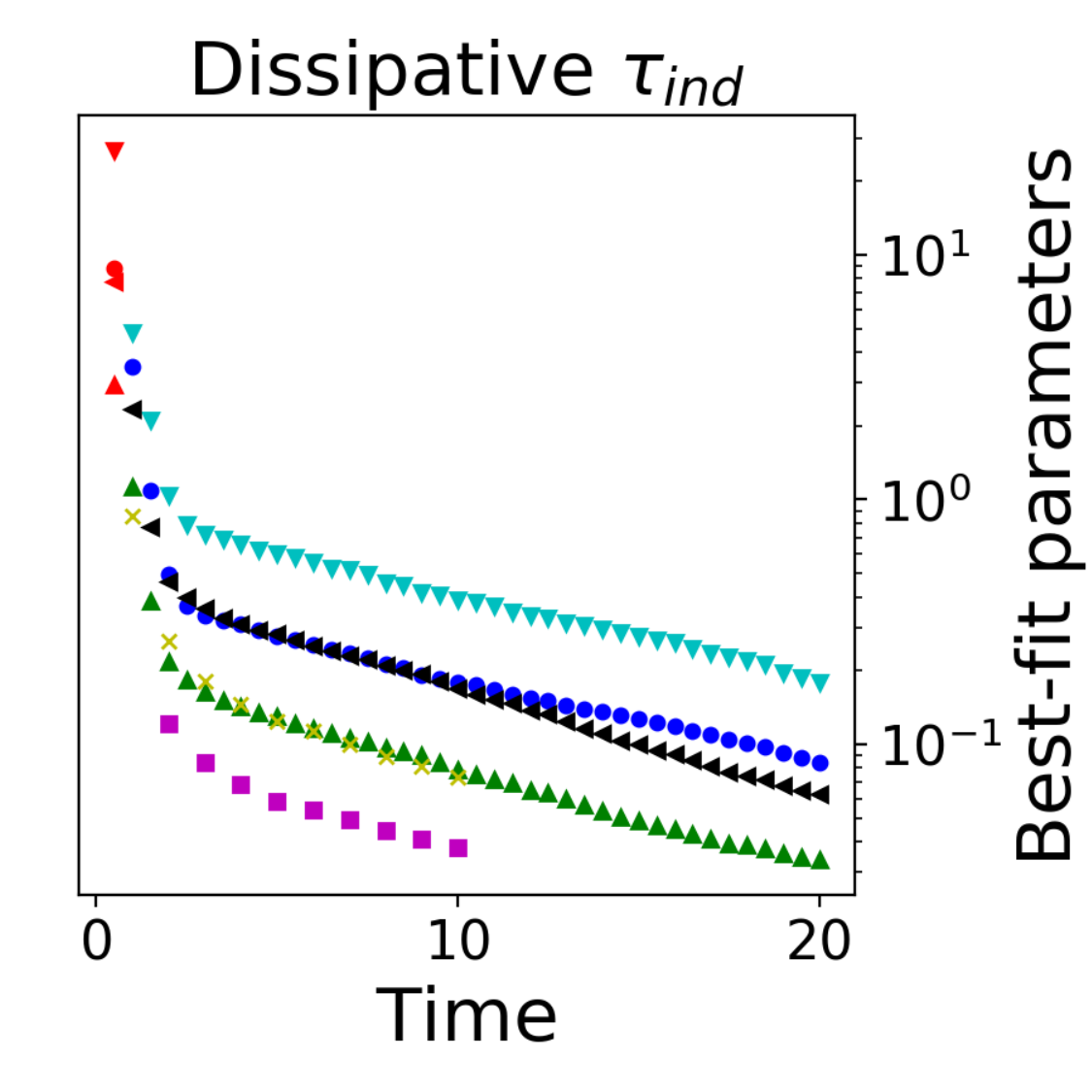}\\
	\caption{Evolution of the Pearson correlation coefficients (left) and the best-fit parameters (right) for the gradient $\tau_{\rm kin}$ (top) and $\tau_{\rm ind}$ (middle) and for the Eddy-dissipative  $\tau_{\rm ind}$ (bottom) tensors, for the models {\tt KH3D500, KH3D1000, KH3D2000}, and filter factors $S_f=2,4$.}
	\label{fig:pearson_sfall} 
\end{figure}

\begin{figure}[ht] 
	\centering
	\includegraphics[width=0.49\linewidth]{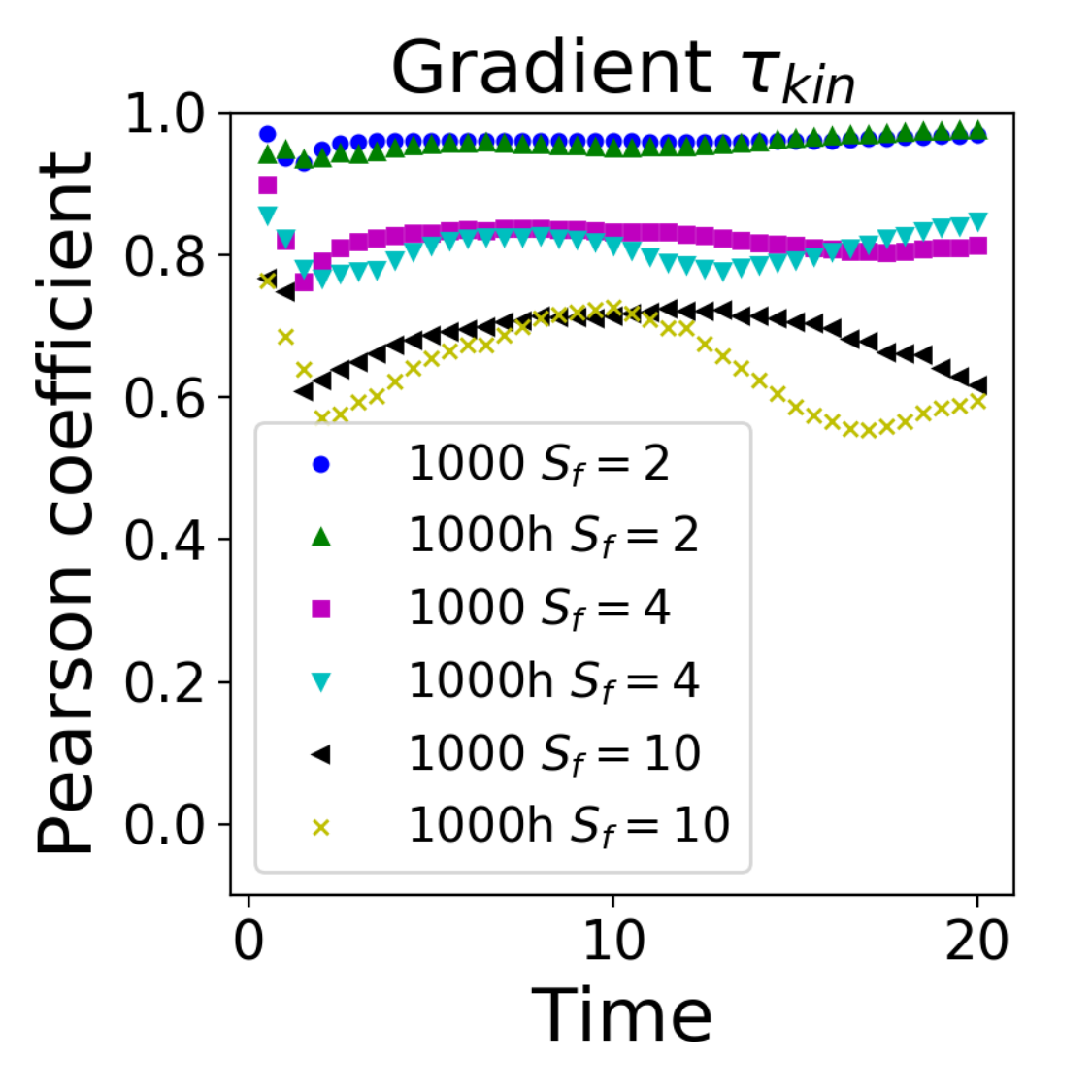}
	\includegraphics[width=0.49\linewidth]{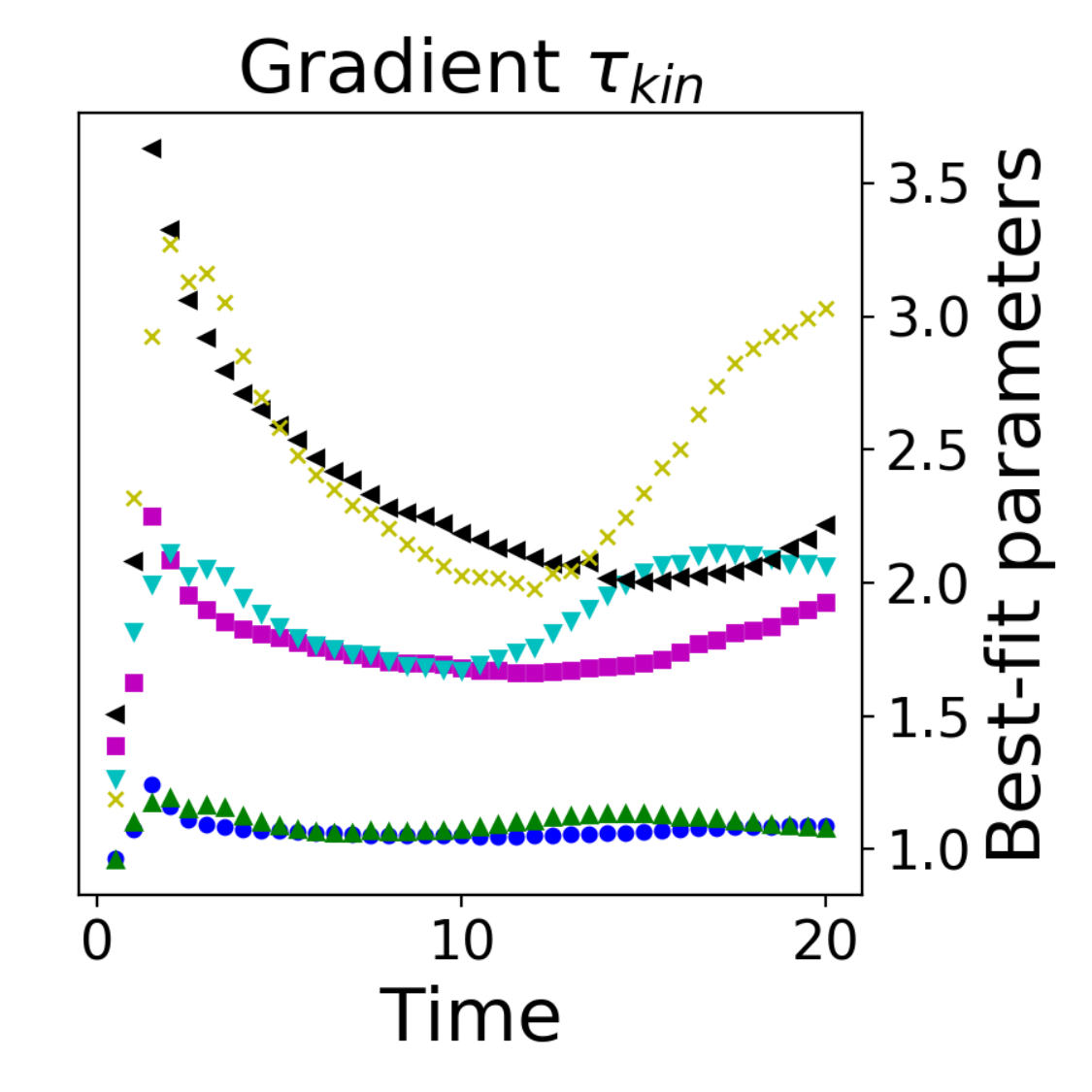}\\
	\includegraphics[width=0.49\linewidth]{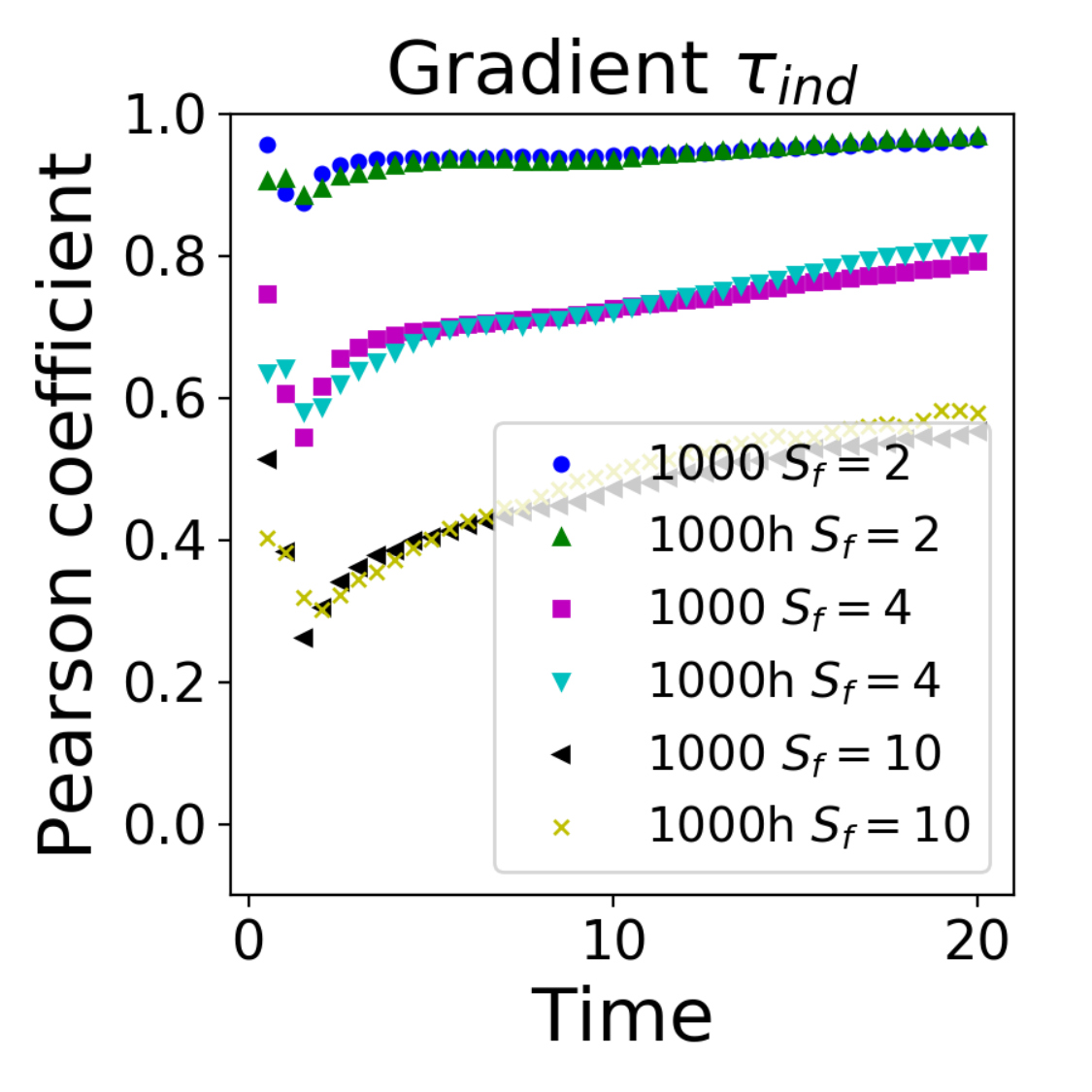}
	\includegraphics[width=0.49\linewidth]{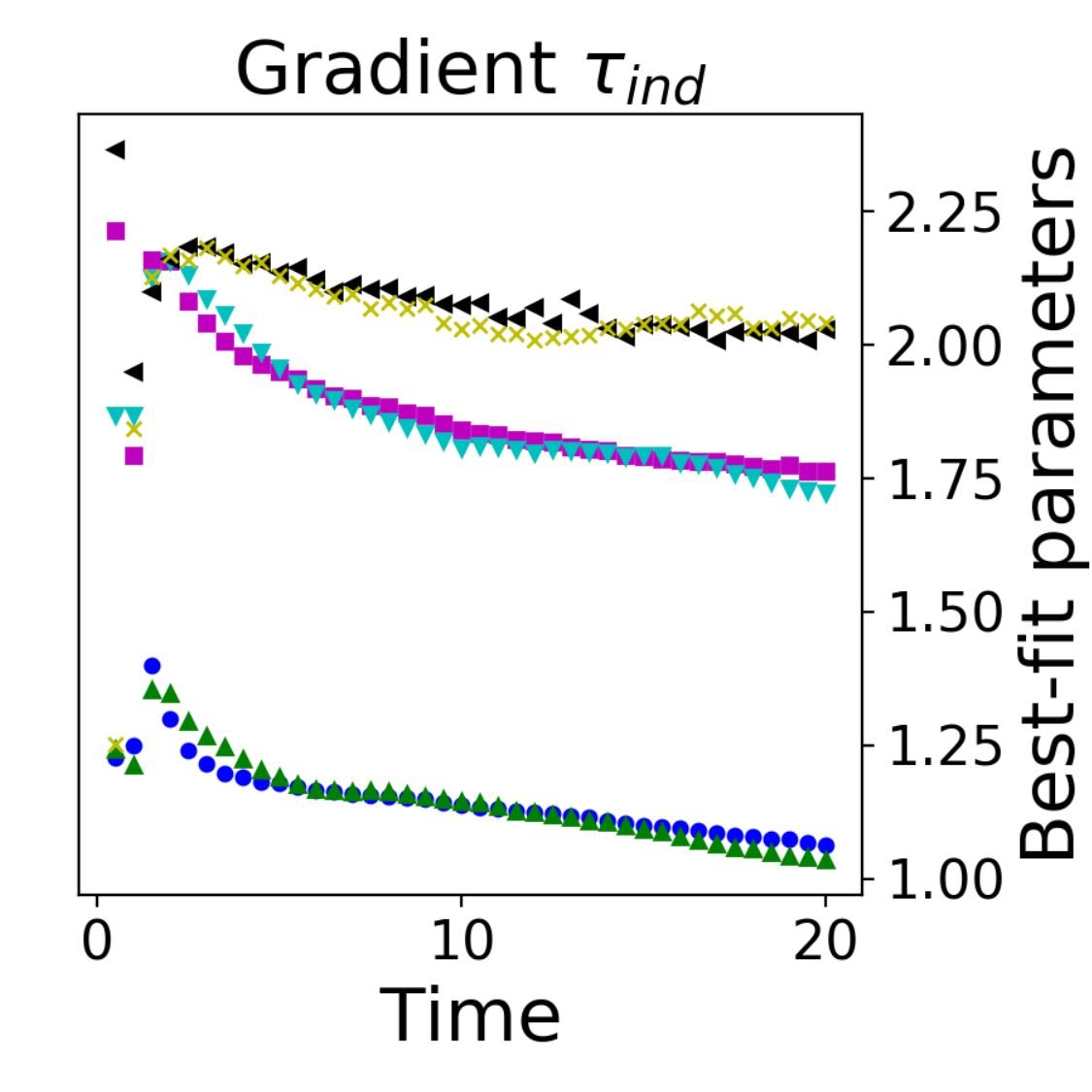}\\
	\includegraphics[width=0.49\linewidth]{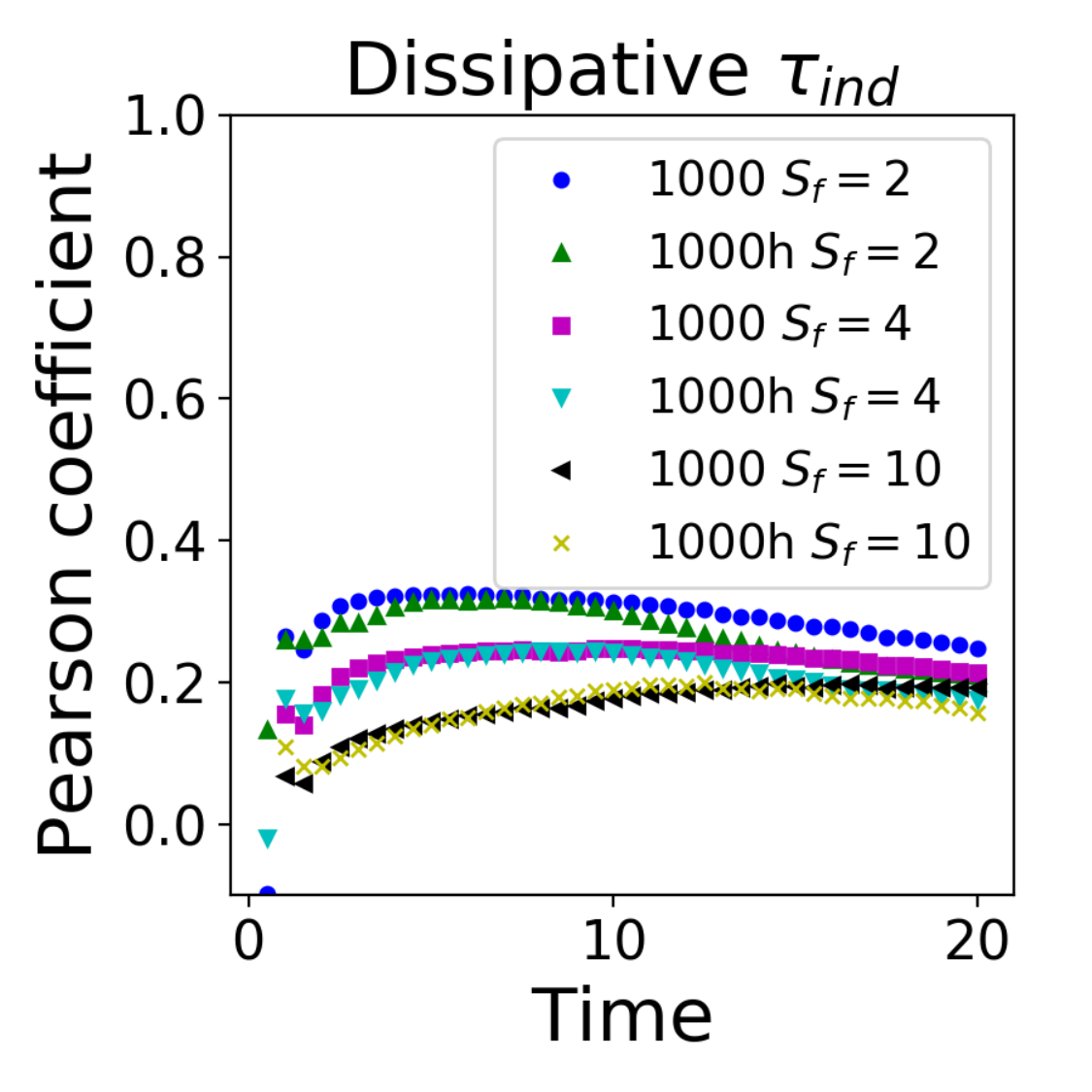}
	\includegraphics[width=0.49\linewidth]{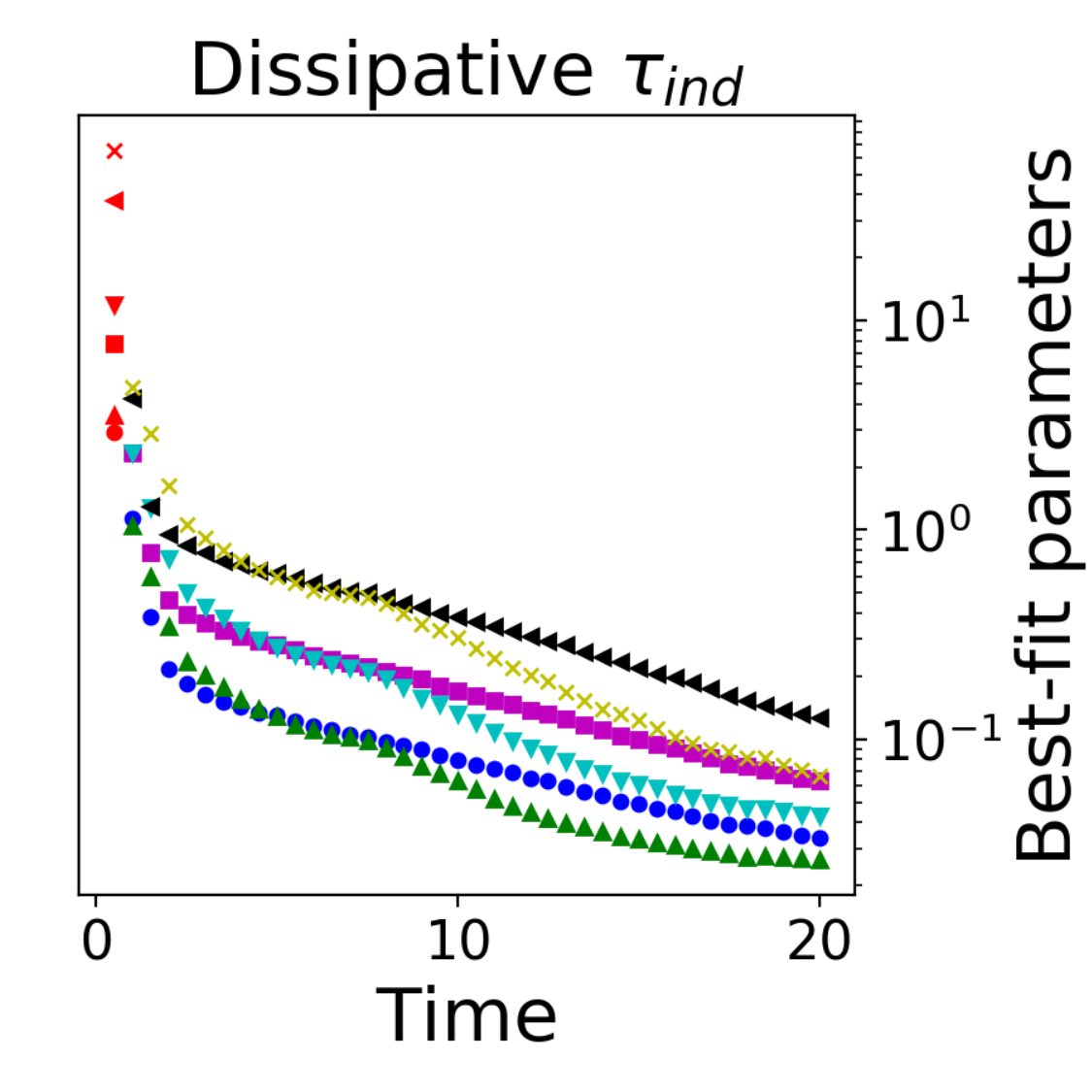}
	\caption{The same as Fig.~\ref{fig:pearson_sfall}, but for all the models with $N=1000^3$, and different filter factors $S_f=2,4,10$.}
	\label{fig:pearson_1000} 
\end{figure}

\begin{figure*}[ht] 
	\centering
	\includegraphics[width=0.32\linewidth]{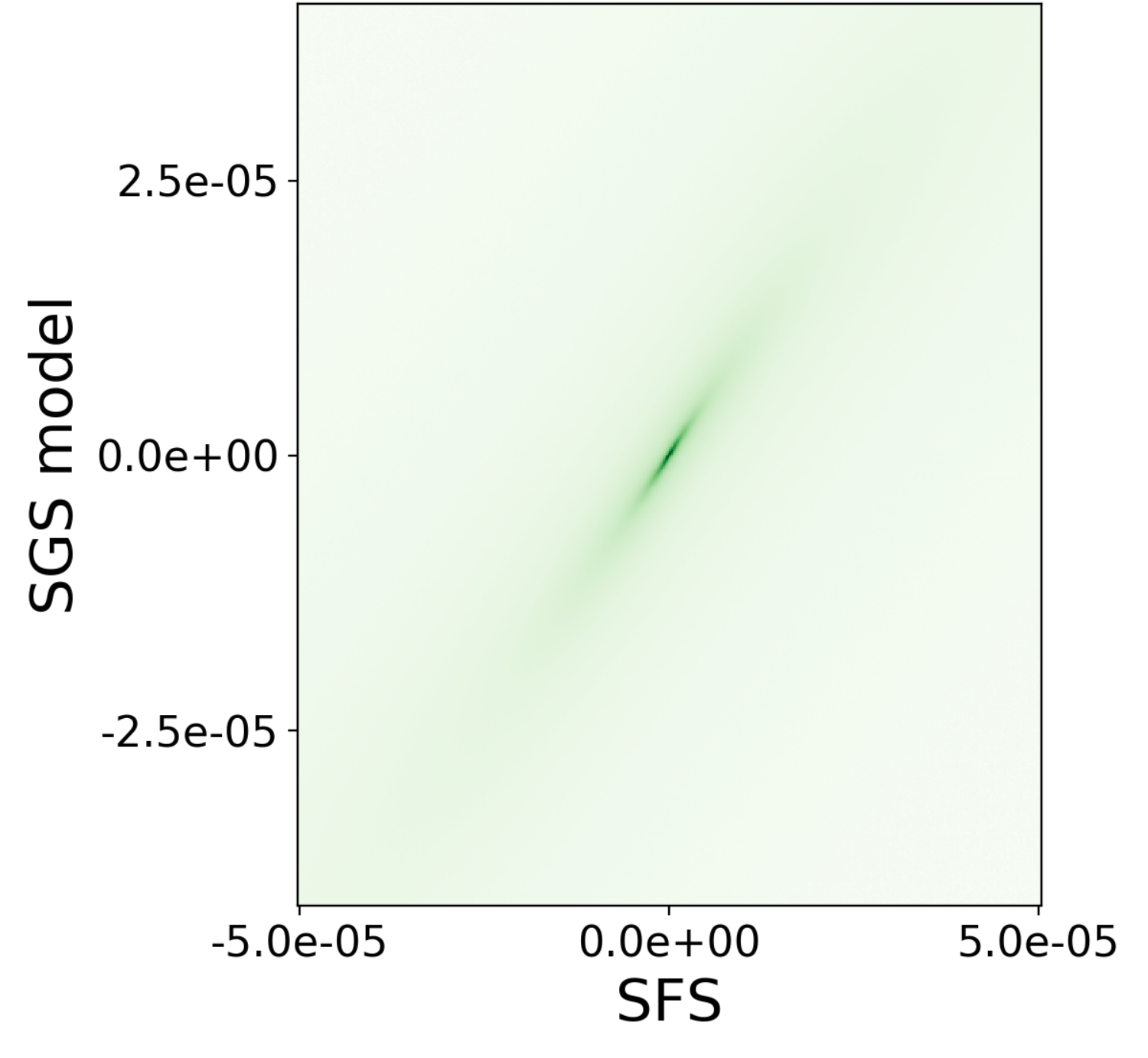}
	\includegraphics[width=0.32\linewidth]{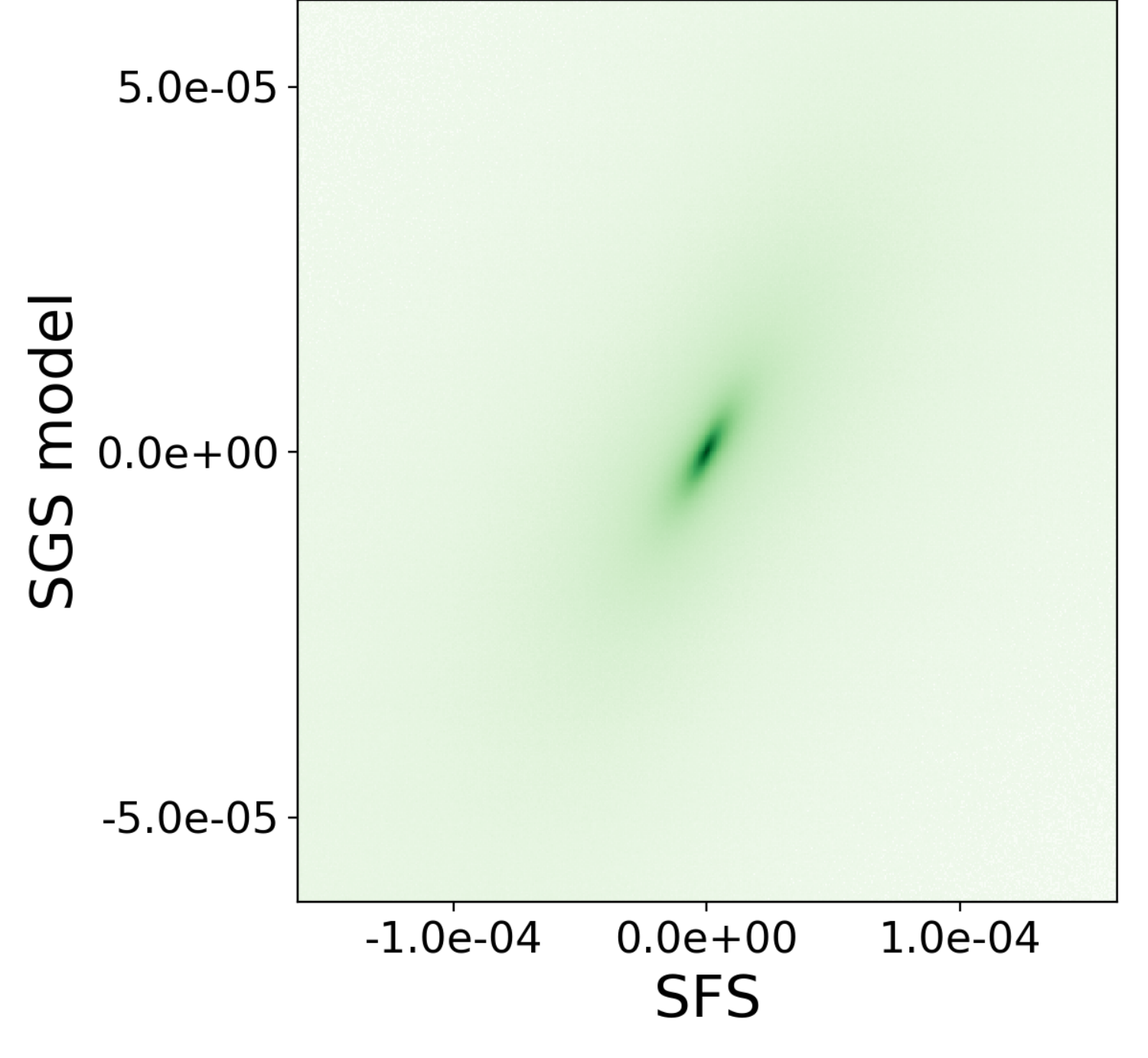}
	\includegraphics[width=0.32\linewidth]{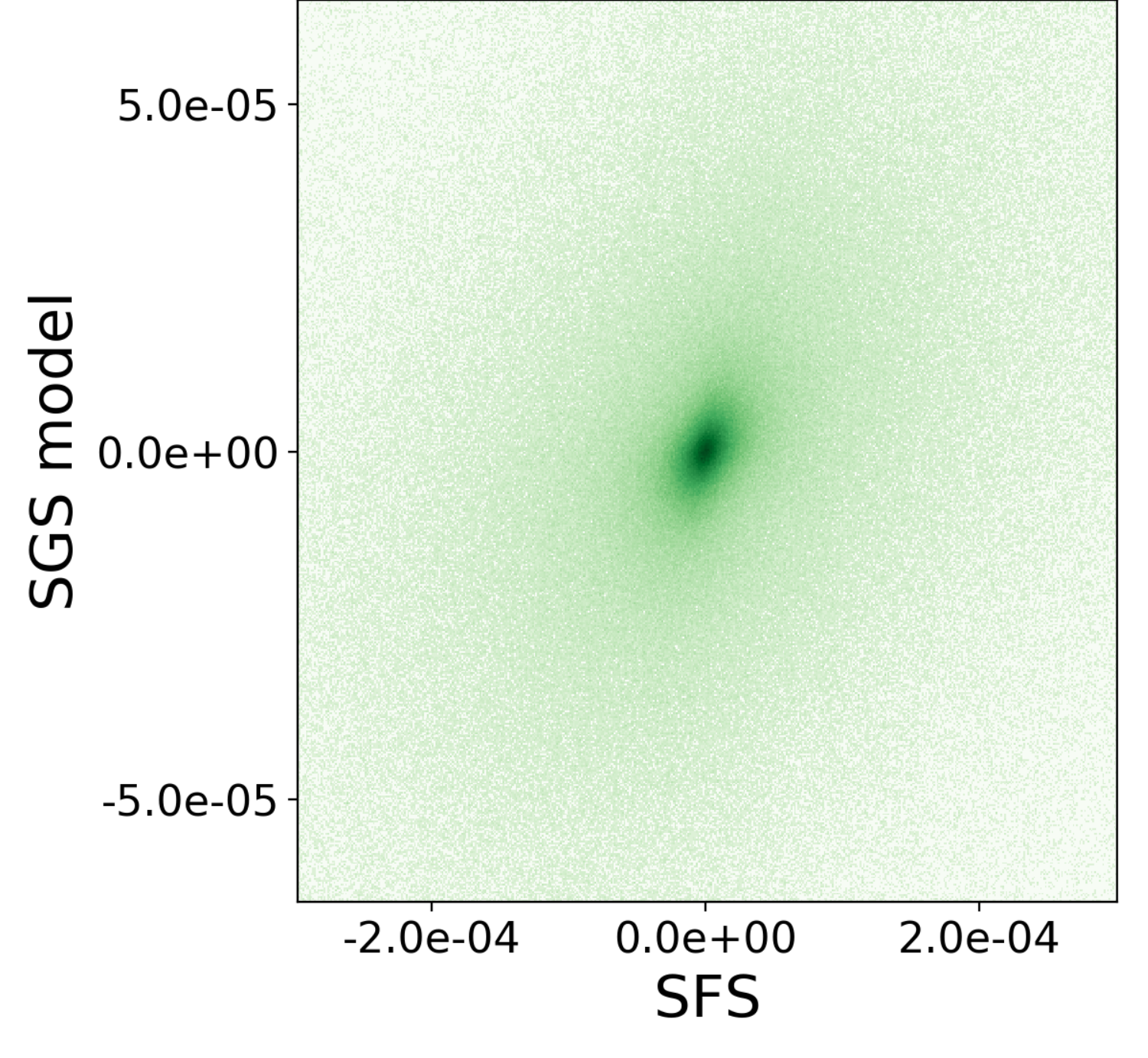}
	\caption{The same as Fig.~\ref{fig:correlation_N500sf2}, but for the $\tau_{\rm ind}^{xy}$ gradient model, for the simulation {\tt KH3D1000} at $t=4$, with filter factors $S_f=2,4,10$ (from left to right).}
	\label{fig:correlations_N1000SFall} 
\end{figure*}

\subsection{General behaviour and magnetic energy growth}\label{sec:general_behaviour}

In Fig.~\ref{fig:kh3d_evo_rho} we show the density distribution over the slice $z=0$, at $t=1,5,9$, for the highest resolution case, {\tt KH3D2000}. The following qualitative behaviour is the same in all cases. At the beginning, the instability develops as small-scale structures with modes given by the initial perturbation. The development starts at the shear layer, where the transfer of kinetic to magnetic energy (dynamo effect) is maintained fast up to $t \sim 2$ in all models (left panel). Then, a larger-scale mixing takes place (middle and right panel), during which the magnetic growth rate reduces. At $t\gtrsim 10-12$, the fluid approaches the isotropy and homogeneity.

In Fig.~\ref{fig:kh3d_evo_overall} we compare the evolution of the integrated kinetic, magnetic and internal energies for the different models. Notice that the initial kinetic energy varies by a factor of a few depending on the explored initial conditions (different colors in the figures), while the magnetic energy is the same in all cases (we always consider the same initial magnetic field). The initial value of the internal energy depends the initial pressure (i.e.,  on $\gamma$), so that there are large differences between the isothermal {\tt KH3D500i} and the other cases. For comparison purposed, we show, for each case, the evolution of the internal energy relatively to its initial value.

For all models, the KHI makes the magnetic energy grow very fast, by orders of magnitude, at the expense of the kinetic budget. Part of the kinetic energy is constantly dissipated into internal energy (right panel). 

At the end of our simulations the system does not reach yet a quasi-stationary state, and the dynamo mechanism is still acting, though at a slower pace. The magnetic energy has not saturated yet and will be, in the best case, limited by equipartition with kinetic energy, but our highest-resolution case shows is still far from it. Longer, possibly higher resolution runs would be needed to know when the magnetic energy saturates (and thus approaching a numerical convergence of the asymptotic solution), but this is not the aim of this paper.

A crucial thing to note is that we do not reach numerical convergence even with the high resolution employed. This is partially expected, since we employ no physical viscosity or resistivity (formally infinite fluid and magnetic Reynolds numbers), and we excite the entire spectrum of the instable modes (see \S~\ref{sec:initial_perturbation}). The more resolution we have, the higher is the wavenumber of the excited modes. Since the dynamo mechanism is more efficient at small scales, the net effect is a much faster rise of the magnetic field for higher resolution. In order to have numerical convergence, one should either (i) excite a single (or a fixed amount of) mode, which wavelength is larger than the resolution; (ii) employ a finite Reynolds number, in order to approach convergence when the grid size becomes comparable to the dissipation scales; or (iii) have a much higher resolution, which would allow an even faster rise of the magnetic field to its saturation value.

Note also that initially larger values of kinetic helicity (blue curves) provide similar growth rate, but a larger asymptotic values of the integrated magnetic energies. On the other hand, changing the initial density jump (green) or the equation of state (cyan) has an important effect on the rise of the internal energy, but not on the kinetic or magnetic energy evolution. This is compatible with the relatively small contribution of the pressure-related terms to the energy interchanges found in the literature~\cite{salvesen14}.

\subsection{Spectra}

Fig.~\ref{fig:kh3d_spectra} shows the kinetic (solid) and magnetic (dotted) radial-averaged spectra (see Appendix~\ref{app:spectra} for details about their practical calculation), for different models, at $t=5,10$ and 20 (first three panels).  They confirm the considerations done in the previous subsection, and add some insights related to the relevant scales and the dissipation of the numerical methods used.

As turbulence develops, the magnetic and kinetic spectral energy densities become comparable below a certain scale. The scale of equipartition evolves to larger values (i.e., smaller $k$) in time. At large scales, instead, the kinetic energy remains always dominant.

Spectra show an inertial range (between the integral scale and the scale at which the numerical dissipation becomes important). In this range, the kinetic spectrum shows the typical Kolmogorov slope\cite{kolmogorov41} (thin solid black line $\propto k^{-5/3}$). The magnetic spectrum, instead, peaks at small $k$, being possibly compatible with the Kazantsev slope\cite{kazantsev68} (dotted thin line $\propto k^{3/2}$). These results are largely independent on the specific perturbation or details of the initial data, as it is shown in Fig.~\ref{fig:kh3d_spectra_500} in the energy spectra at $t=10$ for different initial conditions.

The numerical dissipation of the finite-difference scheme is well visible as a change of slope (knee) in all the spectra, located at a scale a few times larger than the grid size. As the resolution increases, the knee moves to the right, but the overall shape is the same. We see the same knee in 2D, for all models and for other fifth-order methods, while the third-order WENO3 fails even to accurately reproduce the Kolmogorov slope at any range (see details in Appendix~\ref{app:2D_validation}). The dissipation visible in the spectra is quantitatively similar to what obtained in other finite-difference codes employing splitting schemes\cite{radice12,radice13}, and contrasts with the low-dissipative spectral schemes often used in turbulence box simulations.

Spectra give a further insight on the lack of numerical convergence already noted in the integrated quantities.
The more we refine the grid, the more fast-growing excited modes (high $k$) will be included, and the more effective is the kinetic-to-magnetic energy transfer. Since we have no viscosity in our equations, there is no limit to the value of $k$ at which the instability develops. For a given time, as the resolution increases, the magnetic spectra is roughly rescaled to larger values for all $k$, and the equipartition is reached at larger scales (smaller $k$). As mentioned in the previous subsection, this is due to the faster growth of the KHI at smaller scales, at which the dynamo mechanism is also more effective. This can be seen clearly in Fig.~\ref{fig:kh3d_img}, where we show the representative images for the $B_z$ component in the $z=0$ plane at $t=9$, for the same model with different resolutions, {\tt KH3D500, KH3D1000, KH3D2000}. It is clear how stronger small-scale structures are created in the best-resolved cases.

\section{A-priori results}\label{sec:apriori}

After the previous qualitatively description, let us focus on the a-priori fitting of SFS residuals with SGS models. We consider a simulation with a certain grid step $\Delta = L/N$, and its snapshot at a given time. We define the filter as a simple average over groups of $S_f^3$ cells, where we define $S_f$ as the filter factor and the filtered grid size, $\Delta_f = S_f\Delta$. The a-priori test consists in comparing each of the SFS residuals, $\tau$ (see \S~\ref{sec:mhd}) to a given SGS model, $\bar{\tau}$ (see \S~\ref{sec:sgs}). We evaluate the correlation between them (each consisting of $N_f^3$ values) by means of the Pearson correlation coefficient ${\cal P}$, and obtain the best-fit parameter $C_{\rm best}$ which enters in eq.~(\ref{eq:precoefficient}). Further details of this standard analysis are available in Appendix~\ref{sec:sfs_fit}.

Below, we will extensively compare the evolution in time of ${\cal P}$ and $C_{\rm best}$ (i.e., analyzing and fitting different snapshots), for different SGS models and/or for different simulations of Table~\ref{tab:models}.
Initially we calculate the values of ${\cal P}$ and $C_{\rm best}$ for each of the 21 independent components of the tensors  ($\tau_{\rm kin},\tau_{\rm mag},\tau_{\rm ind}, \tau_{\rm adv}, \tau_{\rm hel}$). However, after having checked that there are no significant differences among the different components, we will show only their average over all the components (for instance, the values for $\tau_{\rm ind}$ are the average among its three components).
	
Ideally, ${\cal P}$ should be as close to one as possible, and should not exhibit important variations in time: the model should be able to fit all snapshots. The values of $C_{\rm best}$ should be $\sim {\cal O}(1)$ for the gradient model. A limited variation over time supports the idea that it is possible to have an analytical model without a dynamical procedure to adjust the pre-coefficient. Besides the correlation properties, we will further compare the distributions of $\tau$ vs. $\bar{\tau}$ by looking at the 2D histograms and computing different high-order moments.

At the end of the section we have summarized the outcomes from these comparisons.

\subsection{Representative case: comparison between SGS models}

As a representative example, we consider the model {\tt KH3D1000}, filtered with $S_f=2$ (thus, averaging the $N=1000^3$ values to $N_f = 500^3$). In Fig.~\ref{fig:pearson_evo_1000sf2}, we show the evolution in time of ${\cal P}$, and $C_{\rm best}$ for the different SGS tensors.

As a first check, we compare in Fig.~\ref{fig:pearson_evo_1000sf2} the averaged values of ${\cal P}$ (top panels), and $C_{\rm best}$ (bottom panels) for the diagonal (first column) and off-diagonal (second column) components of $\tau_{\rm kin}$ for the gradient, Eddy-dissipative  and cross-helicity models. Differences between the values related to the diagonal and off-diagonal components are visible only at the very beginning, when the best-fit parameters for the kinetic SGS are positive for the diagonal terms (indicating dissipation), while the shear terms (off-diagonal) are negative. In this early stage of localized turbulence development (see left panel of Fig.~\ref{fig:kh3d_evo_overall}), we register the lowest values of $\cal P$ for all models. This is due to the high degree of anisotropy and discontinuity in the velocity and density field at the beginning. After this phase, correlation further improves and the best-fit parameters tend instead to be very similar, for all models and snapshots studied. Therefore, hereafter we focus only on the values of $\cal P$ and $C_{\rm best}$ averaged over all components of a given tensor.

Coming to the comparison between SGS models, we find that the gradient model performs better than the others, since the associated Pearson correlation coefficients are very close to one for all tensor components (${\cal P}\gtrsim 0.9$) at all times. Among the other models, the Eddy-dissipative and cross-helicity model for $\tau_{\rm kin}$ and $\tau_{\rm ind}$ have a lower, but still non-negligible Pearson correlation coefficient (${\cal P} \sim 0.1-0.3$). The same holds for the $\tau_{\rm ind}$ Alfv\'en model. On the other hand, the Eddy-dissipative model for $\tau_{\rm mag}$, and both the cross-helicity and vorticity models for $\tau_{\rm ind}$ show basically no correlation.

Furthermore, the best-fit parameters for the gradient model are fairly constant in time and $C_{\rm best} \sim {\cal O} (1)$ (see bottom panels of Fig.~\ref{fig:pearson_evo_1000sf2}). In comparison, {the \bf $C_{\rm best}$ values for the other SGS models} can change by orders of magnitude depending on the time considered (Eddy-dissipative model for $\tau_{\rm kin}$ and $\tau_{\rm ind}$, cross-helicity model for $\tau_{\rm mom}$ and Alfv\'en model for $\tau_{\rm ind}$).  The best-fit values $C_{\rm best}$ of the Eddy-dissipative and cross-helicity model are often below unity, in line with what found in literature. For models unable to fit SFS (very low ${\cal P}$), the best-fit parameters can even change sign (vorticity models, Eddy-dissipative model $\tau_{\rm mag}$, cross-helicity models for $\tau_{\rm ind}$), confirming their unreliability.

A closer look to this correlation can be seen, for a given time, by looking at the 2D distribution of the $N_f^3$ pairs of local values $\tau$-$\bar{\tau}$, for a given tensor component. In Fig.~\ref{fig:correlation_N500sf2}
we show a representative case, the model {\tt KH3D500} at $t=10$. These plots are very similar for all times, components and models, with only differences in the range of the value. We plot with colors the 2D SFS and SGS probability distribution for different terms, $\tau_{\rm kin}^{zx}$ (top panels) and $\tau_{\rm ind}^{xy}$ (bottom) calculated with gradient (left), Eddy-dissipative (center) and cross-helicity (right, for which we show $\tau_{\rm mom}^{zx}$ instead of $\tau_{\rm kin}^{zx}$, since the latter is not available) models. For these components, the distributions are quite symmetrical around zero, as expected for shear terms under isotropic conditions. A perfect correlation would be indicated by a single line with a finite slope corresponding to $1/C_{\rm best}$. Only the gradient model (left panels) is approaching such distribution, indicating its ability to systematically reproduce quite well the lost information on average. All the others (center and right) visually show a very poor correlation, as the values of ${\cal P}$ show.

Another way to compare the statistics is to look at the moments of the SFS and SGS distributions. In Fig.~\ref{fig:stats_kin} we compare the evolution of the mean, variance, skew and kurtosis of the SFS terms and the related different models $C_{\rm best}\tau_{\rm kin}$, for the simulation {\tt KH3D500}, with filter factor $S_f=2$. For the mean and the variance, the gradient model adheres to the SFS case, with a good performance also for the higher moments (within a factor $\lesssim 2$). On the other hand, the Eddy-dissipative and the cross-helicity models show smaller values for all quantities. This is consistent with the distribution comparisons in literature\cite{kessar16}. The same relative behaviour is valid for other tensors, resolutions, initial conditions and filter size.

\subsection{Performance of the new gradient terms}

The overall performance of the gradient model in the a-priori tests, as shown so far, are well known \citep{grete15,grete16,grete17,grete17b,grete17phd,kessar16}. Besides confirming those results for our compressible ideal MHD problem, we here show how the SFS tensors $\bar{\tau}_{\rm hel}$ and $\bar{\tau}_{\rm adv}$ can also be fit very well by the corresponding gradient models, with the values of ${\cal P}$ and $C_{\rm best}$ in line with the other tensors. This is shown in Fig.~\ref{fig:tau_hel_adv_N500sf2}, where we show them for the simulations with $500^3$ points, and $S_f=2$. Fig.~\ref{fig:stats_adv} compares the moments of their SFS and SGS distributions, for the {\tt KH3D500} case, $S_f=2$. It qualitatively behaves like the tensors shown above, with an important mismatch (still within a factor $<2$) only for the kurtosis.

Finally, we have studied how the compressibility terms $\propto \partial_i\rho/\rho$ affects the correlations, for this specific problem. The well-known kinetic, magnetic and induction tensors show a negligible change in the Pearson if we include the term or not. The same holds for $\tau_{\rm hel}$. On the other hand, if we neglect such correction, $\tau_{\rm adv}$ loses all the correlation with the SFS term $\bar{\tau}_{\rm adv}$, showing that the compressibility term is actually dominant. Notice however that the importance of each term in a SGS tensor generally depends on the problem. For instance, if we consider a case with much larger changes in density (which, again, is the standard case for astrophysical bodies), then the compressibility correction is fundamental and cannot be neglected.

Below, we explore the results for different initial conditions, resolutions and filter sizes.

\subsection{Dependence with initial conditions and dimensionality}

By comparing all the simulations with the same resolutions but different initial conditions, we notice that the obtained values of ${\cal P}$ and $C_{\rm best}$ are almost coincident with the previous case. An example is given in Fig.~\ref{fig:pearson_500}, where we show the results of the gradient models for $\tau_{\rm kin}$ and $\tau_{\rm ind}$ respectively, with filter factors $S_f=2$ and $S_f=4$. Differences with initial conditions are really minor, compared to the ones obtained by changing $S_f$, as we will see below. The same holds for the other SGS models, and other resolutions.

As a further check, we also changed the dimensionality of the problem. We consider a 2D case, with the periodic box setup described in Appendix~\ref{app:kh2d_periodic}, with $N=500^2,1000^2,2000^2$. We obtain very similar results regarding the Pearson correlation coefficients and the best-fit parameters, for all models and tensors. As a representative example, we show the results for $\tau_{kin}$, for the gradient model, in Fig.~\ref{fig:apriori_2D}. Actually, the best-fit parameters are even more constant than in the 3D case, and the Pearson correlation coefficient is slightly better. The degrading trend of the Pearson correlation coefficient for increasing filter factors $S_f$ (see below) is instead the same. The lack of dependence of the results on the dimensionality is remarkable, given the qualitative differences in the kinetic cascade in the two cases.

\subsection{Dependence with filter size and resolution}

Changing the resolution (i.e., $\Delta$) or the filter size $\Delta_f$ affects the scale range of the lost information. Therefore, the values of ${\cal P}$ and $C_{\rm best}$ can change. For all cases, the gradient model results to be the best one. However, there are some differences in the quantitative performance indicators.

First of all, in Fig.~\ref{fig:pearson_sfall} we show that, for a given $S_f$ (in this case 2 or 4), the dependency on the initial resolution is mild. If we maintain the same filter factor, the results are very similar. This reflects the fact that the relative amount of information lost, for instance, between $N=2000$ and $N_f=500$, is similar to the one lost between $N=500$ and $N_f=125$. On the other hand, by increasing the filter size, instead, the relative amount of "hidden" information increases. Thus, the fit gets worse for higher filter size, as shown with more detail in Fig.~\ref{fig:pearson_1000} for the model {\tt KH3D1000} and $S_f = 2,4,10$. Increasing the size of the filter, the values of ${\cal P}$ for the gradient model decrease from $\sim 0.95$ to $\sim 0.6-0.7$ for $\tau_{\rm kin}$ (top panels), and from $\gtrsim 0.9$ to $\sim 0.3-0.5$ for $\tau_{\rm ind}$ (middle). In Fig.~\ref{fig:correlations_N1000SFall} we show the 2D correlations of the gradient model $\tau_{\rm ind}^{xy}$ for the {\tt KH3D1000} simulation, comparing different size filters. It is evident how increasing the filter (i.e., the amount of information loss), the correlation degrades.

The values of $C_{\rm best}$ for the gradient model increase by up of a factor of a few as the filter factor increases, deviating from the theoretical unitary value. This can be understood taking into account that the higher-order ${\cal O}(\xi^2)$ terms will become more important for larger $S_f$. The same trends (degrading of $\cal P$ and mild increasing of the best-fit parameter, for increasing $S_f$) are seen for the Eddy-dissipative model (see $\tau_{\rm kin}$ in bottom panels of Fig.~\ref{fig:pearson_500}, and $\tau_{\rm ind}$ in Figs.~\ref{fig:pearson_sfall} and \ref{fig:pearson_1000}), and for the cross-helicity model for $\tau_{\rm kin}$. For the Alfv\'en  model, instead, the best-fit parameter scales as $\Delta_f$, as expected, since we defined it to be proportional only to $\Delta_f$, instead than $\Delta_f^2$. On the other hand, the vorticity model, the Eddy-dissipative model for $\tau_{\rm mag}$, and the cross-helicity model for $\tau_{\rm ind}$ do not show any apparent trend with filter size and resolution, probably because of the absence of a true minimum in the least square residuals function.

The study of the trends for varying filter factors allows to understand that even the best SGS models are able to fit the unseen tensors only until an effective improvement of resolution of a factor of a few.

\subsection{Summary of the a-priori results}

We summarize the findings obtained in this section as follows: \begin{itemize}
	\item The gradient model is able to fit very well (${\cal P}\sim 0.9$) the SFS residuals, for a limited filter size $S_f=2$.
	\item The best-value pre-coefficients show little variations in time, and $C_{\rm best} \sim 1$, as expected by theory.
	\item The gradient model is able to fit the new proposed terms, $\tau_{\rm hel}$ and $\tau_{\rm adv}$, besides the other traditional terms.
	\item The fitting performance of the gradient model degrades with the increasing size of the filter, but is independent on the original resolution: what matters is $S_f$. Reasonable correlations are seen also for $S_f=4,10$, indicating the potential ability to partially reproduce information hidden well deep inside the SGS (i.e., scales up to one order of magnitude smaller than the grid size).
	\item When several models are available for the same SGS term ($\tau_{\rm kin},\tau_{\rm mag},\tau_{\rm ind}$), the comparison of their performance show that the gradient model is always the best one for this particular problem, for all SGS terms.
	\item The out-performance of the gradient model is confirmed for all times considered and for all different initial conditions, resolutions and filter sizes.
	\item The comparison of the higher-order statistics (variance, skew, kurtosis) and the 2D-correlation plots confirm that the gradient model is able to fit the SFS residuals with no important systematic biases.
\end{itemize}

\section{A-posteriori test with LES}\label{sec:les}

While the a-priori test is informative about how well the functional form of a SGS tensor fits SFS residuals, the non-linear dynamics and their feedback over time can be tested only by means of direct implementation of such tensors in a LES. We implement the SGS gradient tensors, eqs.~(\ref{eq:taukin_grad})-(\ref{eq:tau_pres_grad_ideal}), as an additional part of the fluxes in the filtered MHD equations, as in eqs.~(\ref{eq:momentum_filtered}), (\ref{eq:ind_filtered}), (\ref{eq:energy_filtered}). Each SGS tensor includes the same free parameter $C$, eq.~(\ref{eq:precoefficient}). 

\begin{figure}[ht]
	\includegraphics[width=0.7\linewidth]{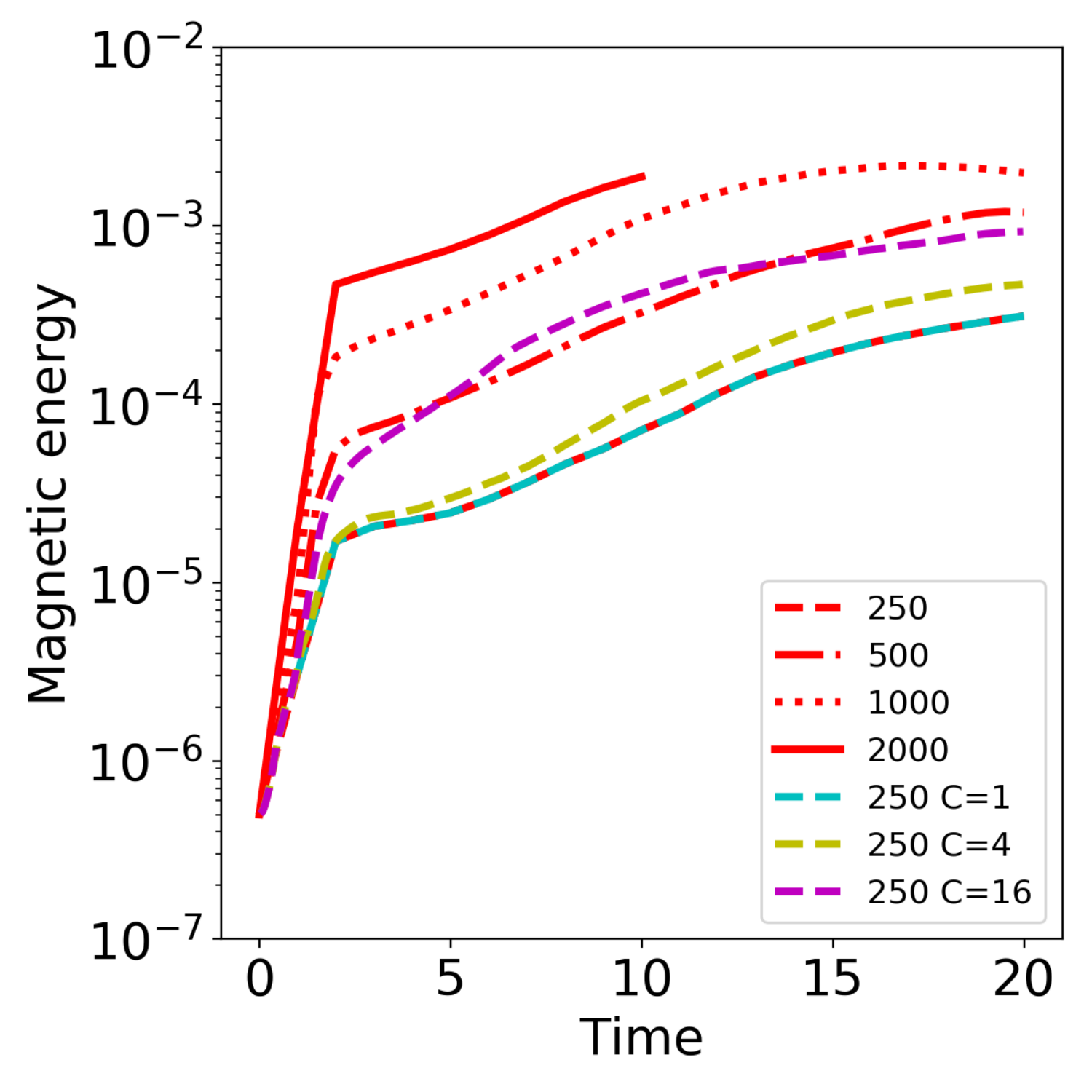}\\
	\includegraphics[width=0.7\linewidth]{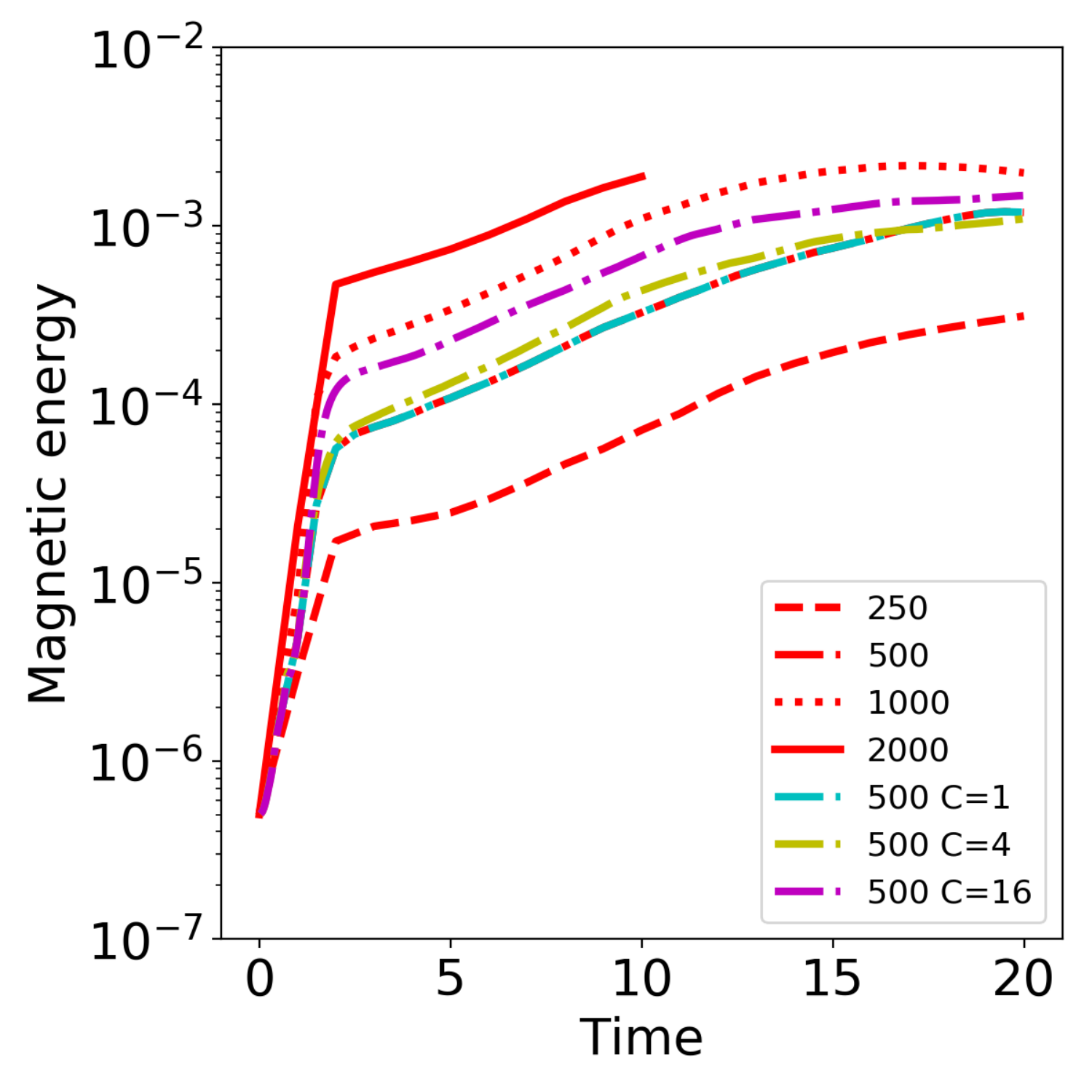}
	\caption{Evolution of the magnetic energy, for the models {\tt KH3D250}, {\tt KH3D500}, {\tt KH3D1000}, {\tt KH3D2000} without the SGS modelling, compared to the {\tt KH3D250} (top) or {\tt KH3D500} (bottom) with a SGS gradient model and $C=\{1,4,16\}$.}
	\label{fig:evolution_sgs}
\end{figure}

We consider the same 3D KHI setup and numerical methods as before, and use the fiducial models   {\tt KH3D250} and {\tt KH3D500} from table\S~\ref{tab:models}. Then we compare the results with the previous runs with no SGS included (\S~\ref{sec:general_behaviour}), at different resolutions. Fig.~\ref{fig:evolution_sgs} compares the magnetic energy evolution obtained without or with SGS models (setting $C=\{1,4,16\}$), for different resolutions. The first evident effect is that, by increasing the value of $C$, the magnetic energy attained increases as well.  Setting $C=1$ seems to have a negligible effect, which start to be visible for $C=4$. The magnetic energy value obtained by the gradient model for {\tt KH3D250} with $C=16$ is as high as the one obtained with {\tt KH3D500} without any SGS model. Similarly applying the SGS gradient model with $C=16$ to {\tt KH3D500} gives a growth of the magnetic field slightly smaller than {\tt KH3D1000}.

\begin{figure*}[ht]
	\centering
	\includegraphics[width=0.24\linewidth]{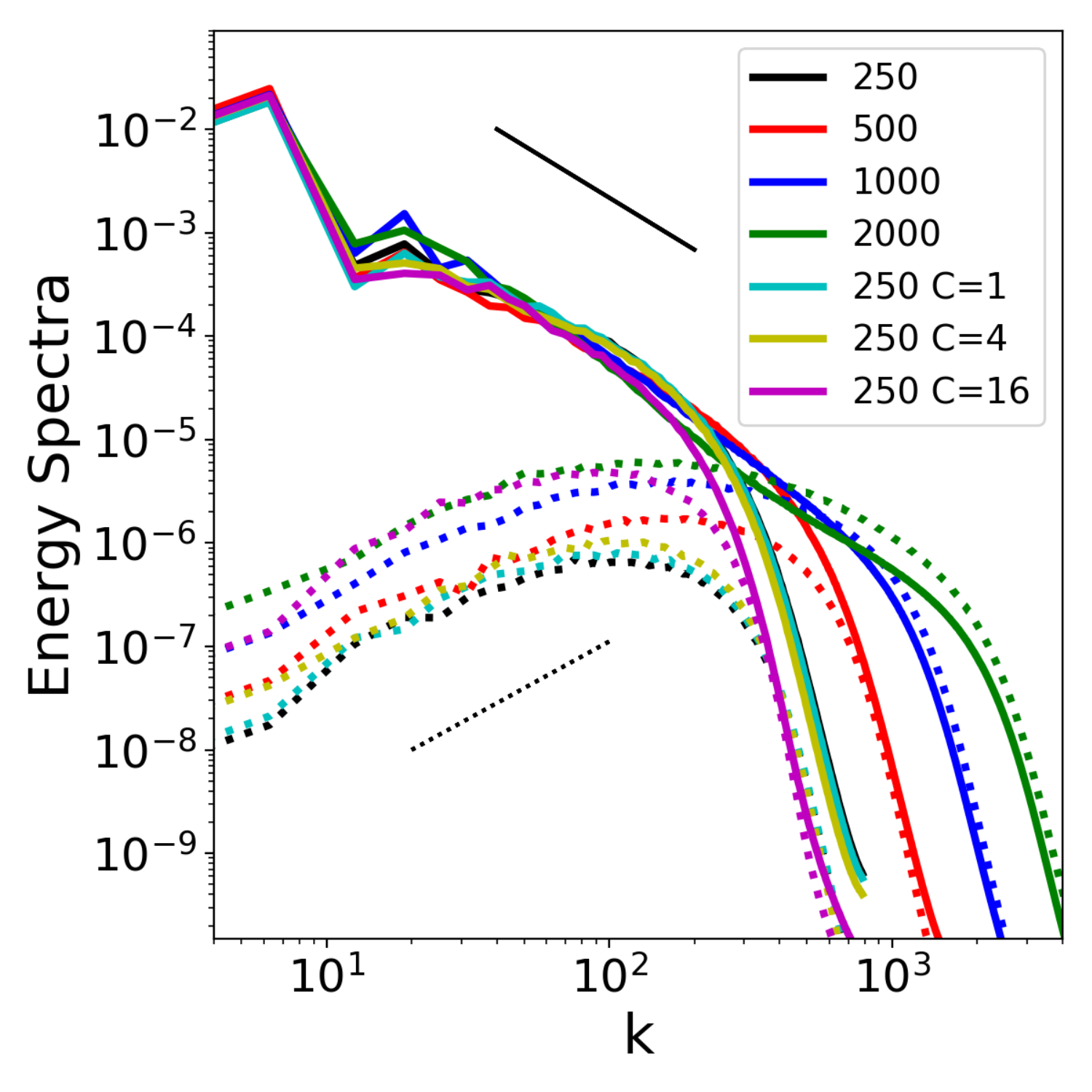}
	\includegraphics[width=0.24\linewidth]{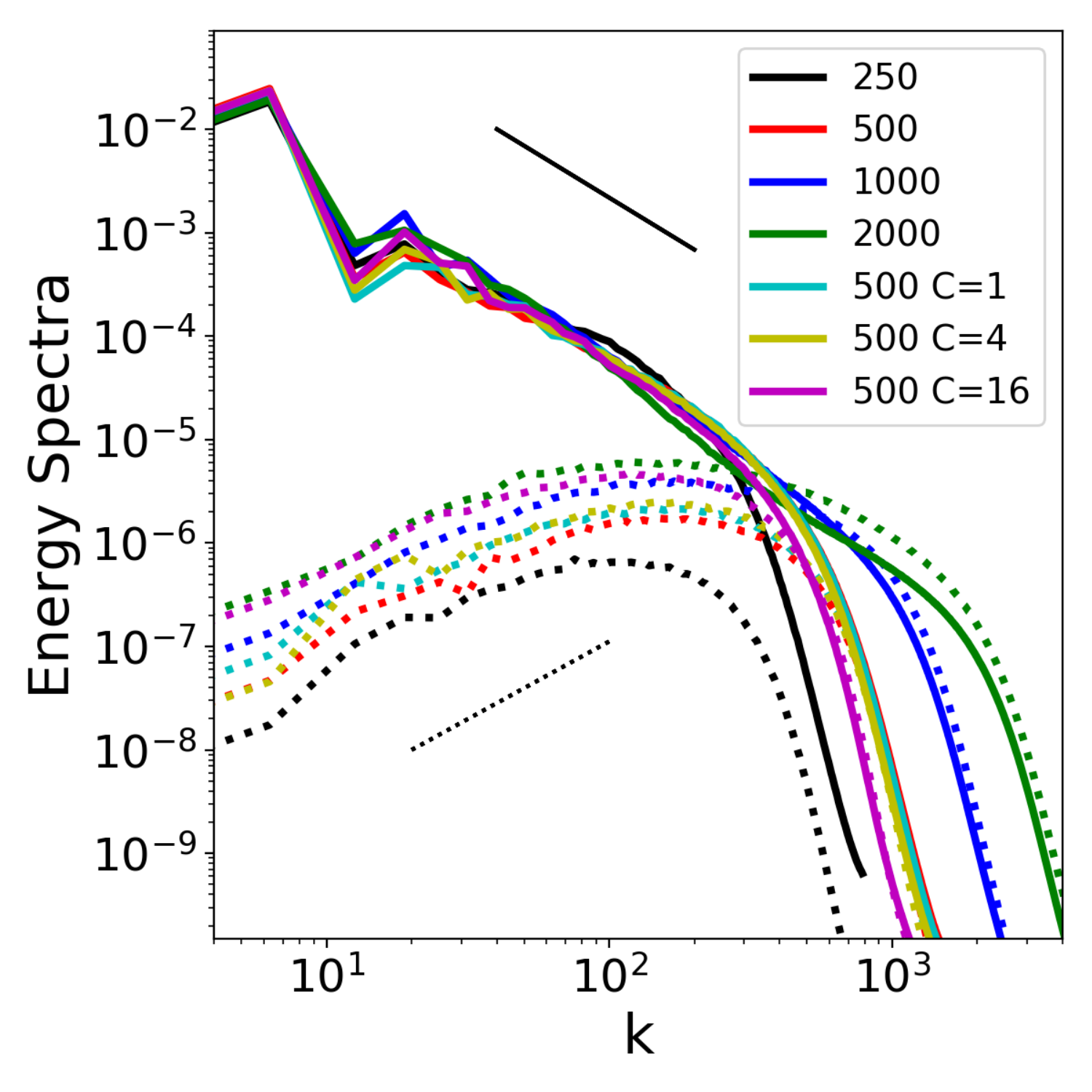}
	\includegraphics[width=0.24\linewidth]{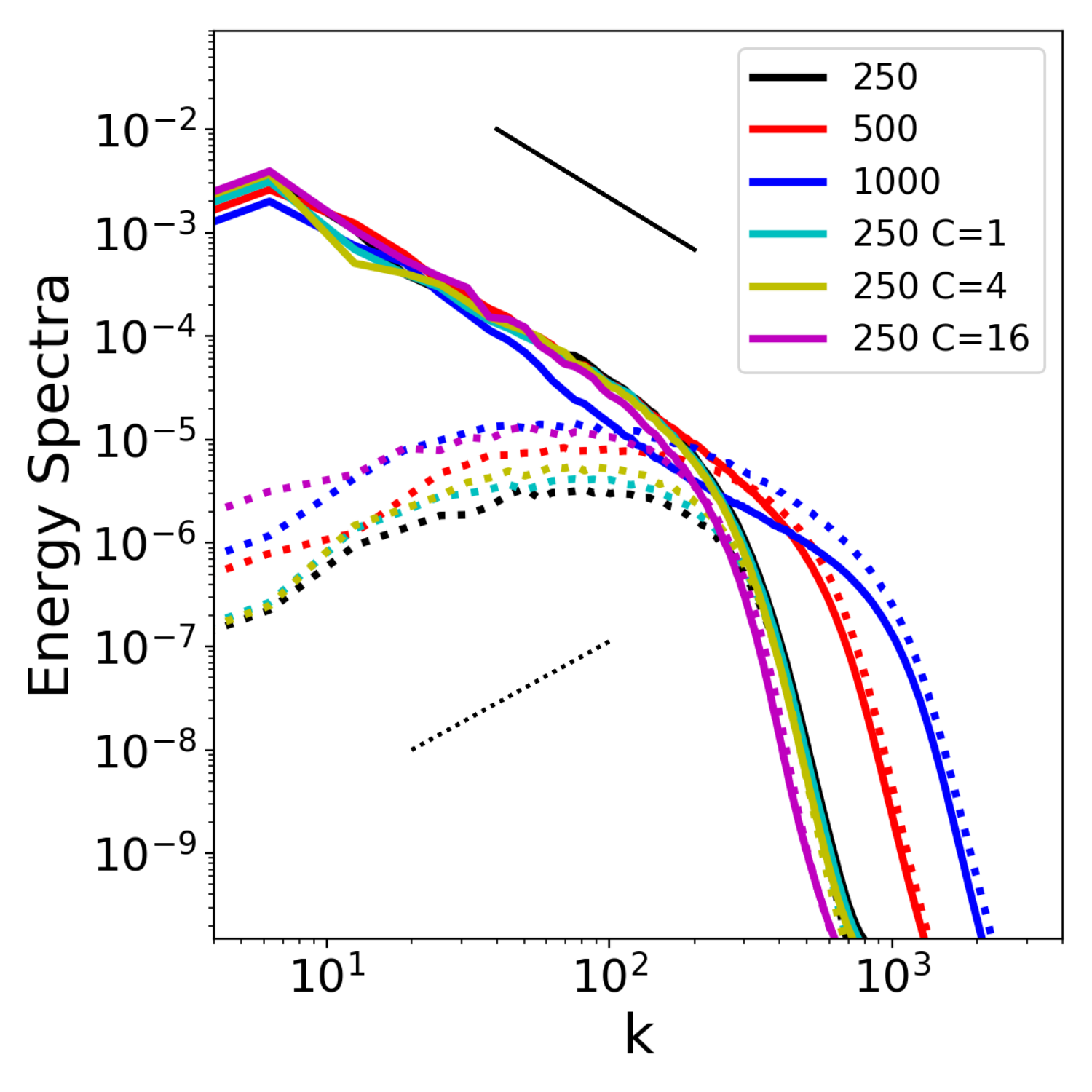}
	\includegraphics[width=0.24\linewidth]{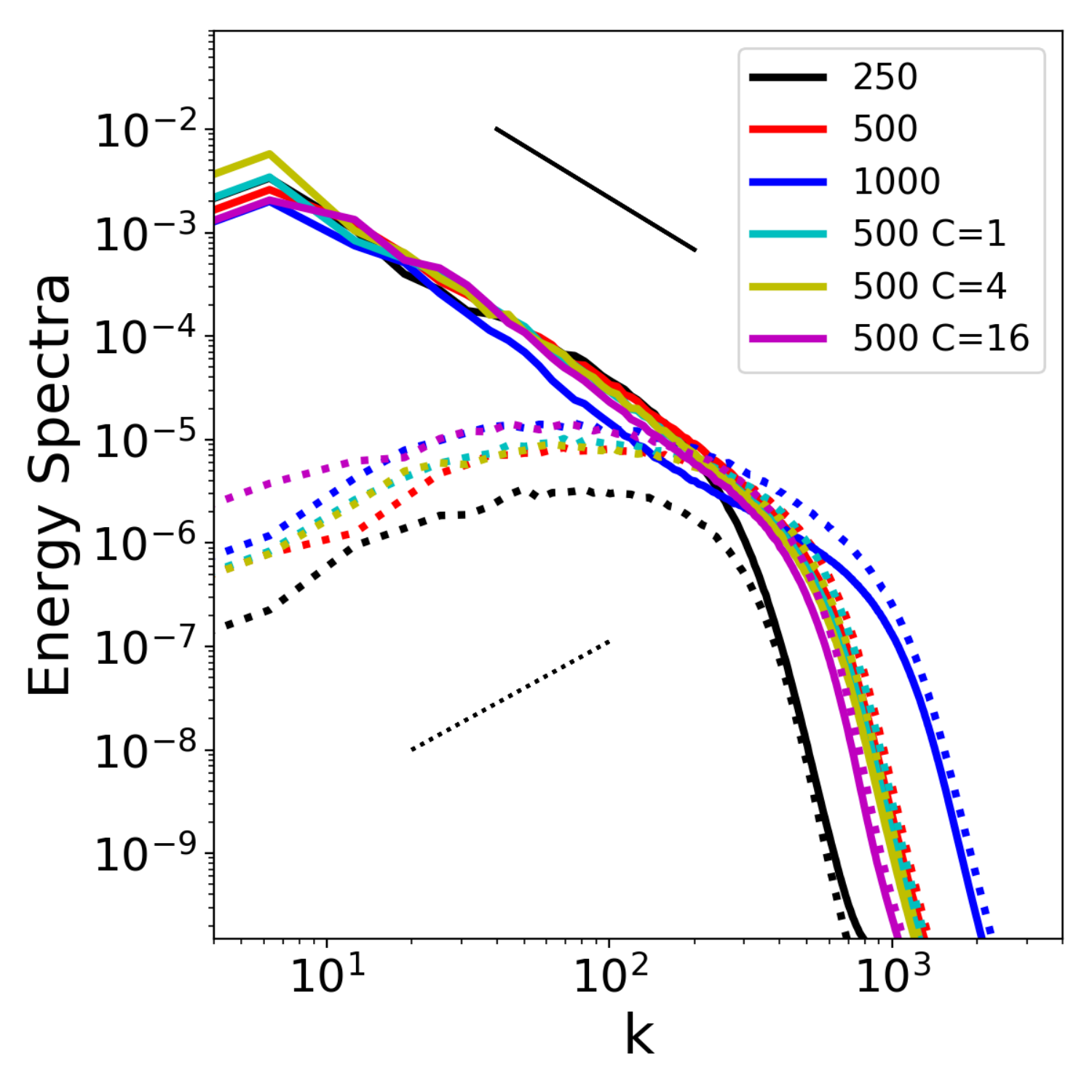}
	\caption{Comparison of KH3D spectra at $t=\{10,20\}$ (first and last two panels, respectively) for the models {\tt KH3D250}, {\tt KH3D500}, {\tt KH3D1000}, {\tt KH3D2000} (only at $t=10$) without the SGS modelling, compared to the {\tt KH3D250} (first, third) or {\tt KH3D250} (second, fourth) with a SGS gradient model and $C=\{1,4,16\}$.  The solid and dotted slopes indicate as references the Kolmogorov and Kazanstev spectral laws, $\propto k^{-5/3}$ and $\propto k^{3/2}$, respectively.}
	\label{fig:spectra_les}
\end{figure*}

More detailed information can be obtained by analyzing the spectra for the same cases, displayed in Fig.~\ref{fig:spectra_les} at $t=\{10,20\}$.
While the kinetic spectra is barely affected due to the dominance of the large scales, the magnetic spectra show some interesting features. At both times we can see that, including a SGS gradient model with $C=16$, we reach the same values of the spectra in the large scales (inertial range) as the highest-resolution cases without SGS models. Due to the intrinsic dissipation, the knee is still located at the same wavenumber, so that spectra unavoidably decays for high $k$. Since the magnetic spectra peak at small scales, a low resolution by definition cannot take into account an important part of the total magnetic energy. This is why the gain in total magnetic energy is limited, as it was shown in Fig.~\ref{fig:evolution_sgs}.

However, it is crucial to underline that the feedback on the large scales is properly captured by the SGS gradient model up to resolutions: the magnetic energy growth and spectra in the inertial range are comparable with what obtained using a resolution $\sim 4-8$ times higher.

Although these final results are encouraging, there is a mismatch between the a-priori values of $C_{\rm best}\sim 1-2$ and the ones that we need to implement in the LES (i.e., $C\gtrsim 10$) to have an important effect. A possible explanation might be that the Taylor expansion is not accurate enough and additional terms need to be included. In order to explore such possibility, we implemented the leading next-order terms in the $\tau_{\rm kin}$ and $\tau_{\rm ind}$ gradient models (following the already developed formalism\cite{vlaykov16}), setting them proportional to $C^2\xi^2$. We do not see any relevant change in the overall evolution (integrated energies and spectra), compared to the standard first-order gradient models considered above (again, with different values of $C=1,4,16$). This excludes the second-order terms are the reason of the mismatch.

The high values of $C$ can be instead understood considering the scheme's intrinsic dissipation. The gradient model is known generally to over predict back-scatter effects,\cite{leonard75} leading to an over-estimation of the transfer from SGS to resolved scales. This is favored especially for spectral methods, where the very low numerical dissipation shows accumulation of energy at the smallest resolved scales. This is why many works  try to improve the numerical stability considering a mixed model given by a superposition of both gradient and Eddy-dissipative models~\cite{muller02a,kessar16} . However, in our case the intrinsic dissipation of the scheme (visible from the steep spectral slope at high $k$) prevents this instability, unless we use high values of the parameter $C\gtrsim 20$.

This is probably the same reason why we need high values of $C$ to see some effects in the LES, much higher than the a-priori best-fit ones. As a matter of fact, the intrinsic dissipation partially damps the small-scales structures, i.e., the local gradients. Since the gradient model relies on them, we need a higher value of $C$ to compensate the damping. Note that this effect in the a-priori test is not visible, since we filter the simulation, reaching a scale $S_f \Delta$, where the intrinsic numerical dissipation is not that significant. Said in other words, if one looks at the spectra, the steep decay is  especially severe at the smallest scales $\sim \Delta$, so that applying the SGS model there (like in the a-posteriori LES tests) require a higher factor than if we apply it at $S_f\Delta$ (like in the a-priori test). We then expect that
the $C$-mismatch should decrease  by using a less dissipative numerical scheme, probably at the cost of a less stable solution.

\section{Conclusions}\label{sec:conclusions}

This work generalizes the SGS gradient model for the compressible ideal MHD with a generic equation of state $p(\rho,e)$, although we focus here on the specific case of the ideal gas. We have extended the formalism of filtering, including also the energy evolution equation.

We have considered a turbulent scenario with a very large Reynolds number (formally, infinite), which is common in astrophysics. In those cases, the finest available numerical resolutions are often very far from being able to capture all the relevant small scales and the integral ones. With this in mind, we have considered box simulations of magnetic KHI, with the initial excitation of perturbations which extend to the smallest possible wavelengths. If the modes are well below the best achievable resolution, the simulations do not show numerical convergence, in terms of total magnetic energy and its spectra. The absence of physical viscosity and/or resistivity in our runs, and the specific problem, imply not having a reference solution to compare with. This actually represents the main motivation for the use of SGS in astrophysical LES: to explore unaccessible scales at an affordable computational cost.

We have generated our computational code by using the platform {\it Simflowny}\cite{arbona13,arbona18}, endowed with HRSC high-order finite-difference schemes commonly used in numerical relativity applications, already described in depth in our previous work\cite{palenzuela18}. We stress that we use methods that may not be optimal for periodic box turbulence, but are adequate in astrophysical simulations with the development of strong shocks. As a matter of fact, many works in turbulence use spectral methods, which are particularly suitable for studies of smooth solutions (or with weak shocks) in periodic boxes due to their intrinsic low numerical dissipation. However, in a real problem, where the dominion is not periodical and the dynamics includes strong shocks, codes employing finite-difference or finite-volume schemes are preferred due to their robustness.
With this in mind, we have considered box simulations of magnetic KHI, with the initial excitation of perturbations which extend to the smallest possible wavelengths.

Note that the gradient model does not assume or mimic any physical behavior like dissipation. Instead, it extrapolates the non-linearity of the flux terms to the SGSs. Therefore, it can be applied in principle to any set of equations. This is a key issue for our specific problem, for which the efficiency of the dynamo effect (conversion of kinetic into magnetic energy) strongly depends on the smallest resolved scales, where the non-linear coupling between velocity and magnetic fields needs to be captured.

We have assessed in detail the gradient model, and compared it with a few more SGS models. Our results agree with what found in literature: the gradient model outperforms all the others in the a-priori tests (fitting SFS residuals), with best-fit parameters that do not deviate from the expected unity more than a factor of 2. We obtained our results for a remarkable variety of: dimensionality, initial conditions, resolution, time within a simulation and filter factor. The Pearson correlation coefficient is excellent (${\cal P} \gtrsim 0.9$) for a filter factor $S_f = 2$, and gradually degrades, being still decent (${\cal P}\gtrsim 0.5$) for a filter factor 10. The degradation of the performance is understandable: if the LES resolution is too coarse compared to the dynamically interesting scales, then the inclusion of the SGS model will not be able to represent all the missing scales.

Then we include the gradient model in the LES to perform a posteriori tests, less frequent in the literature, and compare it to the integrated energies and spectra coming from the runs with a higher resolution. The results can be summarized as follows: (i) we obtain a more effective dynamo mechanism, with a growth by up to a factor of a few of the magnetic energy, (ii) the gain of the magnetic energy comes from the large scales, where by adding the gradient with a large value of the parameter (i.e. $C=16$) we can reproduce a very similar magnetic spectrum to the one given by a much higher resolution. By definition, and due to the intrinsic dissipation, the magnetic energy stored at the small scales is not completely provided, but its feedback on the large scales (i.e., the overall large-scale dynamo) is effectively taken into account.

The use of such a large value of the parameter needed to reproduce the high-resolution results suffers a mismatch with the a-priori test. We think that this is due to the intrinsic dissipation of the finite-difference code, and less dissipative methods should need a smaller value of $C$. The intrinsic dissipation of our scheme also explains why our simulations are numerically stable for any value $\lesssim 20$, while spectral codes typically need to introduce an artificial viscosity (or use a mixed SGS model) for $C\sim 1$.

Although the precise value of $C$ to be used can be improved, here we stress how the implementation of the extended SGS gradient proposed manages to obtain our main aim for the considered problem: the growth to intermediate and large-scale magnetic energy, which cannot otherwise be captured. The ability of the gradient model to capture the SGS dynamics is limited to a factor of a few times the resolution used. Therefore, it can be seen as a computational-cost-effective (the extra cost is to add terms to the flux) way to capture the small-scale dynamics, equivalent to employing an effective higher resolution. This is especially important in order to improve the simulations, and possibly get a little bit closer to the numerical convergence, in the astrophysical turbulent scenarios mentioned above.

The main intrinsic limitation of the SGS gradient model arises from the mathematical assumption of having smooth fields: in the presence of strongly varying fields, the truncation to first-order in $\xi$ of the Taylor expansion becomes a poor approximation. 
Although this effect is not observed in our simulations, in other scenarios with strong shocks/discontinuities the gradient SGS model might loses part of its accuracy in describing the unresolved scales. A possible solution might be to include higher-order terms in the Taylor series. Nevertheless, we stress again that the LES combined with the gradient SGS model allows to include only part of the effects coming from the unseen dynamics, effectively digging information inside the numerical cell, but only to a certain effective depth (which is difficult to evaluate, since the performance of the gradient degrades gradually with the size filter).

This paper is the first step towards the generalization of SGS-gradient models in LES to the framework of General Relativity. An astrophysical case where it will be applied is the binary neutron star merger, which requires a dominion of thousands of km, and, at the same time, should ideally simulate the KHI happening in the shear layer forming during the collision, probably below the meter.

\begin{acknowledgments}
	We acknowledge support from the Spanish Ministry of Economy, Industry and Competitiveness grants AYA2016-80289-P and AYA2017-82089-ERC (AEI/FEDER, UE). CP also acknowledges support from the Spanish Ministry of Education and Science through a Ramon y Cajal grant. The computational time in the Barcelona Supercomputing Center's MareNostrum has been granted by the $17^{th}$ PRACE regular call (project Tier-0 GEEFBNSM, P.I. CP), and by the RES calls AECT-2018-1-0005 (P.I. DV), AECT-2019-1-0007 (P.I. DV). We thank B.~Mi\~nano for the technical support, S.~Liebling and F.~Carrasco for useful comments.
\end{acknowledgments}

\appendix

\section{Evaluation of spectra and fitting}

\subsection{Spectra}\label{app:spectra}

For a given field $f$ defined in a periodic box $[0,L]^3$, we use common {\tt python} functions to calculate its discrete fast Fourier transform $\hat{f}(\vec{k}) = \Sigma_{\vec{x}} f(\vec{x}) e^{-i \vec{k}\cdot\vec{x}}$, where the sum is performed over the $N^3$ spatial points $x_i\in [0:L)$ equally spaced in each direction, with $k_j = n~\Delta k$, where $\Delta k = \frac{2\pi}{L}$ and $n\in [0,N/2]$ is integer. We consider the radial coordinates of the Fourier space, describing it with $\Delta k$-wide radial bins also centered on $k_r=\{n~\Delta k\}$. Then, we calculate the spectra $\mathcal{E}(k)$ as averages $<\cdot>_{k_r}$ over the annular bins of the power density per unit of radial wavenumber in 3D:

\begin{eqnarray}
&& \mathcal{E}_k(k_r) = \frac{L^3 4\pi}{(2\pi)^3N^6} <k^2|\widehat{\sqrt{\rho}\vec{v}}|^2(\vec{k})>_{k_r} ~,\nonumber\\
&& \mathcal{E}_m(k_r) = \frac{L^3 4\pi }{(2\pi)^3N^6}<k^2|\hat{\vec{B}}|^2(\vec{k})>_{k_r} ~,\label{eq:spectra}
\end{eqnarray}
where $k^2 = k_x^2+k_y^2+k_z^2$. The normalization comes from the Parseval identity, so that $\int_0^{N/2} \mathcal{E}(k_r) dk_r = \int_V f^2(\vec{x}) dV$ where $f^2 = \rho v^2/2,B^2/2$ are the local energy densities in the real space. Note that this identity, expressed in terms of the integral in $k_r$, strictly holds if the fluid is isotropic, so that the average over the annular bin does not lose statistically relevant information (for instance, a strong dependence on power on the direction of $\vec{k}$). For further technical considerations about normalizations, caveats (e.g. the systematic noise introduced by the conversion to radial coordinates in the Fourier space) and possible corrective factors, we refer to a recent dedicated paper\cite{durran17}. Note that the calculation of spectra for simulations with a large number of points ($N\gtrsim 1000^3$) require a large computational memory. We use a parallelization by parallel slabs, with a dedicated {\tt python} package, recently provided\cite{mortensen16}. This calculation is different and more precise, compared with simplified methods used in other works, e.g. the spatial average over one dimension of the spectra obtained from the Fourier transform over the other two dimensions.\cite{beckwith11}

\subsection{A-priori fitting}\label{sec:sfs_fit}

For a simulation with a certain grid step $\Delta = L/N$, we consider a snapshot at a given time, and spatially filter all the evolved fields. The simplest recipe is to use a simple average groups of $S_f^3$ cells, where we define $S_f$ as the filter factor. This corresponds to apply a filter in the real space, with a box kernel of size
\begin{equation}
\Delta_f = S_f\Delta~, 
\end{equation}
obtaining filtered fields evaluated over $N_f^3=(N/S_f)^3$ points. When one considers a non-linear combination of conserved fields, the filtering process allows one to evaluate numerically the residuals contained between the scales represented by $N_f$ and $N$, which can be seen as an approximation of the formal definitions of $\overline{\tau}$, eqs.~(\ref{eq:tau_kin})-(\ref{eq:tau_pres}) and (\ref{eq:tau_adv})-(\ref{eq:tau_hel}). For instance, if a SFS tensor is formally defined as $\overline{\tau}(\vec{x}_f) = \overline{f}(\vec{x}_f)\overline{g}(\vec{x}_f) - \overline{fg}(\vec{x}_f)$, we can evaluate it at each of the $N_f^3$ positions of the filtered mesh $\{\vec{x}_f\}$, as 

\begin{equation}
\overline{\tau}(\vec{x}_f) = \frac{1}{S_f^{3}}\left[\Sigma_i f_i(\vec{x}_i) \Sigma_i g_i(\vec{x}_i) - \Sigma_i f_i(\vec{x}_i) g_i(\vec{x}_i)\right] ~, \label{eq:tau_example}
\end{equation}
where $i$ indicate each of the $S_f^3$ discrete positions considered inside the cell centered in $\vec{x}_f$.
Note that this estimation is not an exact evaluation of the loss information, since, by construction, it can only include the range of scales $[\Delta , S_f\Delta]$. The information for scales $<\Delta$ cannot be evaluated.

Once built each component of each SFS tensor, one can consider a given SGS model $\tau$. A measurement of the linear correlation between the numerical data and the different models can be estimated by the Pearson correlation coefficient, 
\begin{equation}
{\cal P} = {\rm Corr}\{\overline{\tau}^{ki}(\vec{x}),\tau^{ki}(\vec{x})\}~.
\end{equation}
Note that, due to the usually very large of degrees of freedom ($N_f^3$), the associated $p$-value is almost always very small, and will not be taken into account as a useful indicator.

While the Pearson correlation coefficient tests the functional form, one can also consider each SGS component with a parameter $C$ to be adjusted. Its best-fit value  can be calculated by the minimizing the L2-norm, $\Sigma\, [\overline{\tau}^{ki}(\vec{x}_f) - C\tau^{ki}(\vec{x}_f)]^{2}$, where the sum is performed over all the positions $\{\vec{x}_f\}$. The minimization gives simply:
\begin{equation}
C_{\rm best}^{ki} = \frac{\Sigma\,\overline{\tau}^{ki}(\vec{x}_f)\,\tau^{ki}(\vec{x}_f) }{\Sigma\,\tau^{ki}(\vec{x}_f)^2}~.
\end{equation}
This procedure can be repeated independently for each SGS component, for each tensor $\tau$.

\section{2D tests}\label{app:2D_validation}

\subsection{Open boundaries Kelvin Helmholtz test}\label{app:2D_tests}

As a validation test for the methods employed, we run a well-known 2D KHI test set-up\cite{obergaulinger10}, with open boundaries in the vertical direction. We set a rectangular domain in the $x-y$ plane, with side length $L_x=1$, $L_y=2$, centered in $(0,0)$, with a number of points $(N,2N)$. The initial conditions consist of $P=\rho=1$ everywhere, with an ideal equation of state with $\gamma=5/3$, with a $a_l$-thick shear layer along the $x$ direction, with the following initial conditions:

\begin{table}
	\begin{center}
		\caption{Table of 2D single layer models, with the theoretical and best-fit values ($1~\sigma$ error typically $\lesssim 2\%$) of the growth phase for the simulations run with WENO5Z.}
		\label{tab:khsl}
		\begin{tabular}{c@{$\quad$}c@{$\quad$}c@{$\quad$}c@{$\quad$}c@{$\quad$}c@{$\quad$}c@{$\quad$}c}
			\hline
			\hline
			model & N & $v_0$ & $a_l$ & $k_x$ & $B_0$ & $\alpha_{\rm an}$ & $\alpha_{\rm num}$ \\
			\hline
			{\tt grw1} & 50 & 0.645 & 0.05 & 1 & 0 & 1.73 & 1.53 \\
			{\tt grw2} & 100 & 0.645 & 0.05 & 1 & 0 & 1.73 & 1.70 \\
			{\tt grw3} & 200 & 0.645 & 0.05 & 1 & 0 & 1.73 & 1.75 \\
			{\tt grw4} & 400 & 0.645 & 0.05 & 1 & 0 & 1.73 & 1.75 \\
			\hline
			{\tt grw5} & 200 & 0.645 & 0.025 & 1 & 0 & 2.4 & 2.37 \\
			{\tt grw6} & 200 & 0.645 & 0.1 & 1 & 0 & 0.66 & 0.68 \\
			\hline
			{\tt grw7} & 200 & 0.3225 & 0.05 & 1 & 0 & 1.09 & 1.07 \\
			{\tt grw8} & 200 & 0.9215 & 0.05 & 1 & 0 & 1.77 & 1.79 \\
			\hline
			{\tt grw9} & 200 & 0.645$^\circ$ & 0.05 & 2 & 0 & 1.36 & 1.32 \\ 
			\hline
			{\tt grw10} & 200 & 0.645 & 0.05 & 1 & 0.129 & 1.69 & 1.70 \\
			{\tt grw11} & 200 & 0.645 & 0.05 & 1 & 0.258 & 1.56 & 1.54 \\
			\hline
			\hline
		\end{tabular}
		\begin{minipage}{.45\textwidth}
			$^\circ$ this value has previously incorrectly reported\cite{obergaulinger10} to be half of this, incompatible with the corresponding growth rate.
		\end{minipage}
	\end{center}
\end{table}

\begin{figure}[t] 
	\centering
	\includegraphics[width=0.49\linewidth]{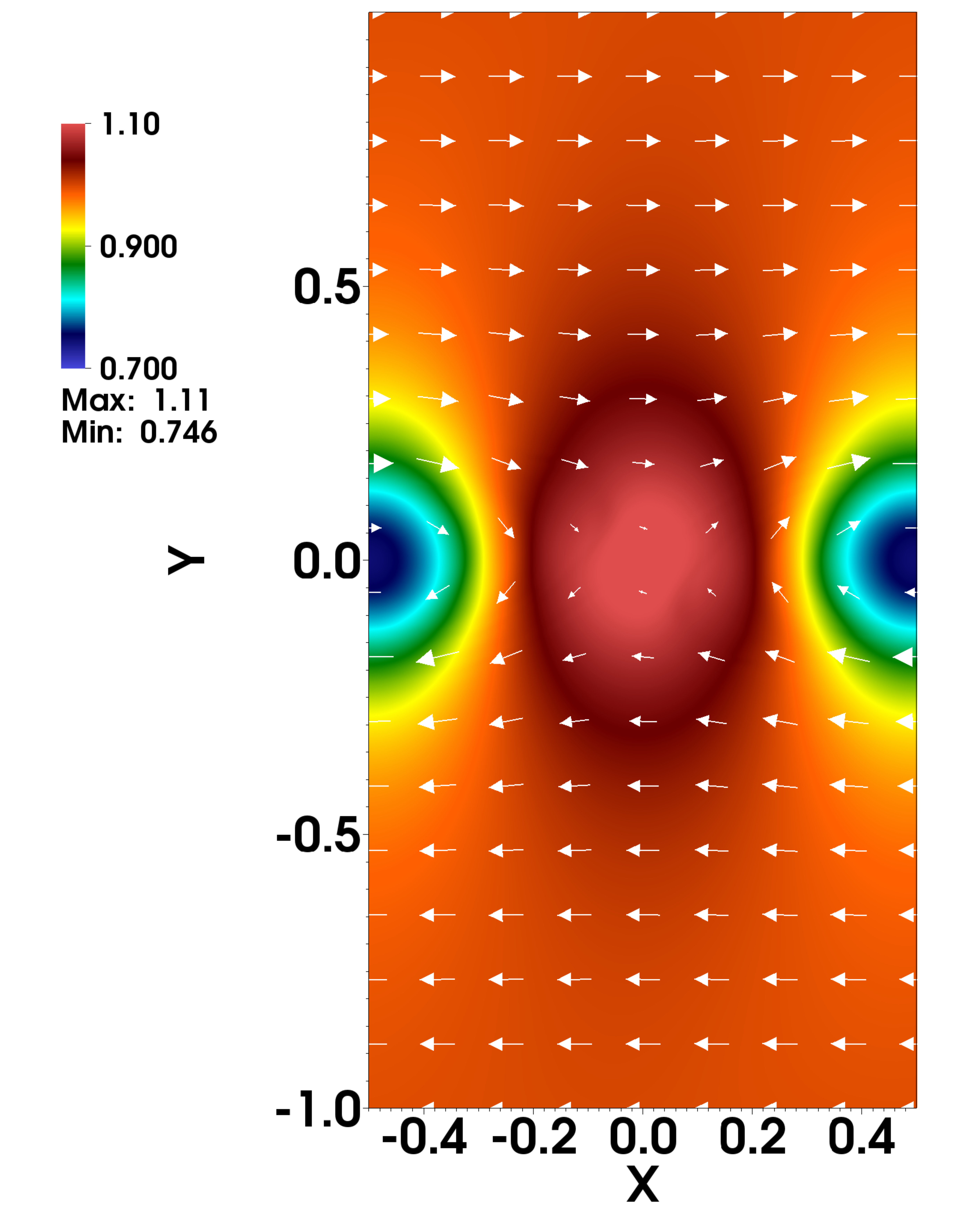}
	\includegraphics[width=0.49\linewidth]{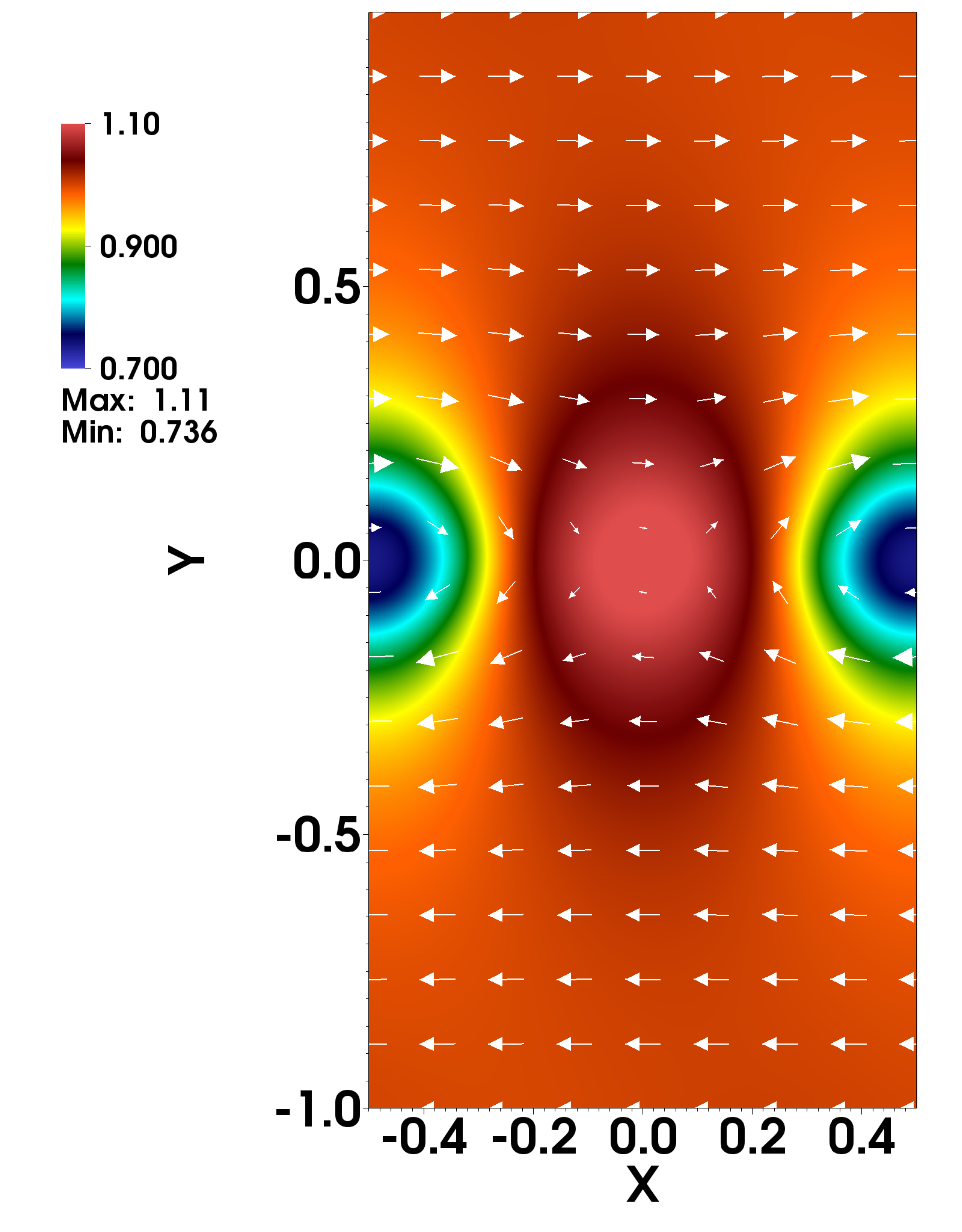}\\
	\includegraphics[width=0.49\linewidth]{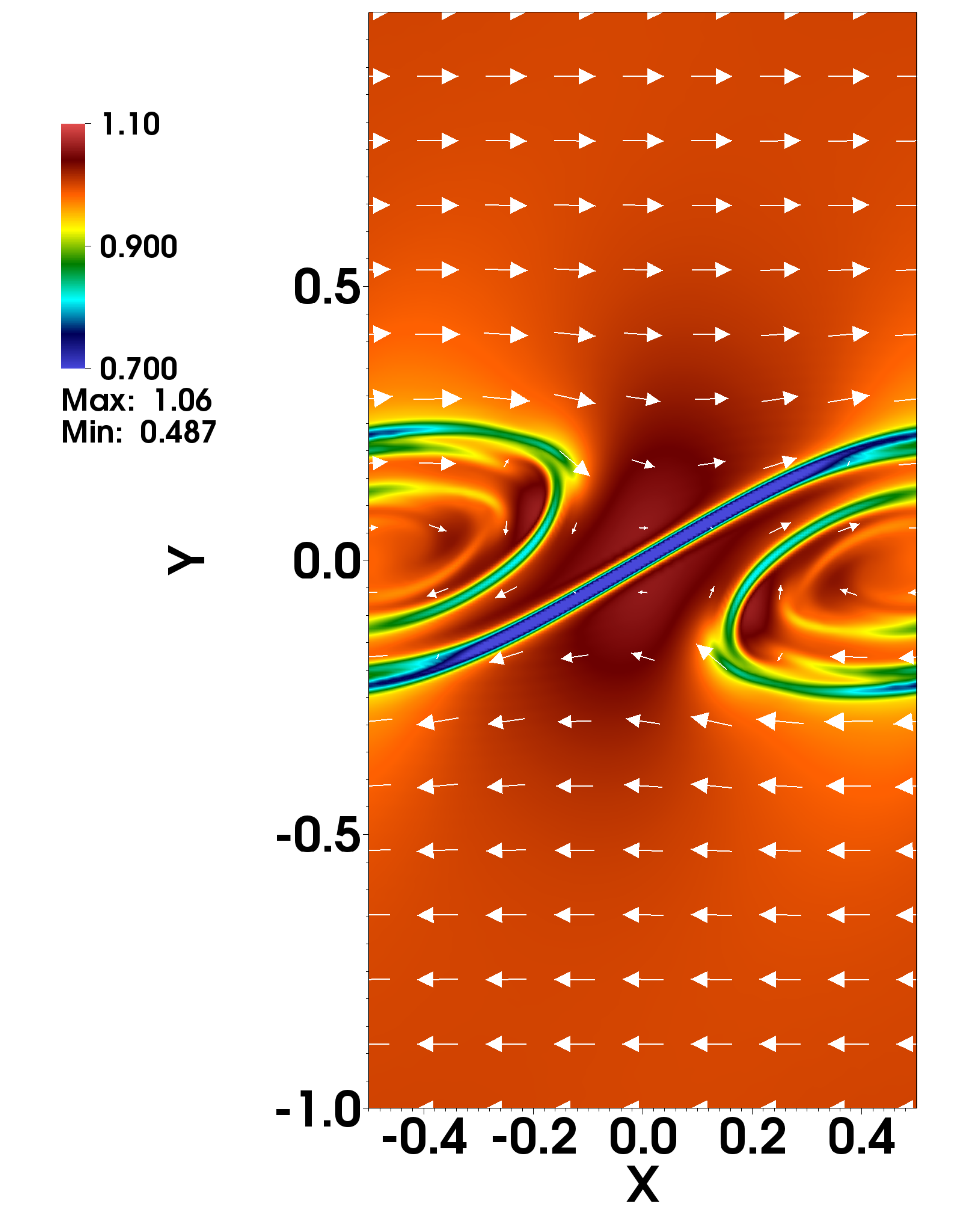}
	\includegraphics[width=0.49\linewidth]{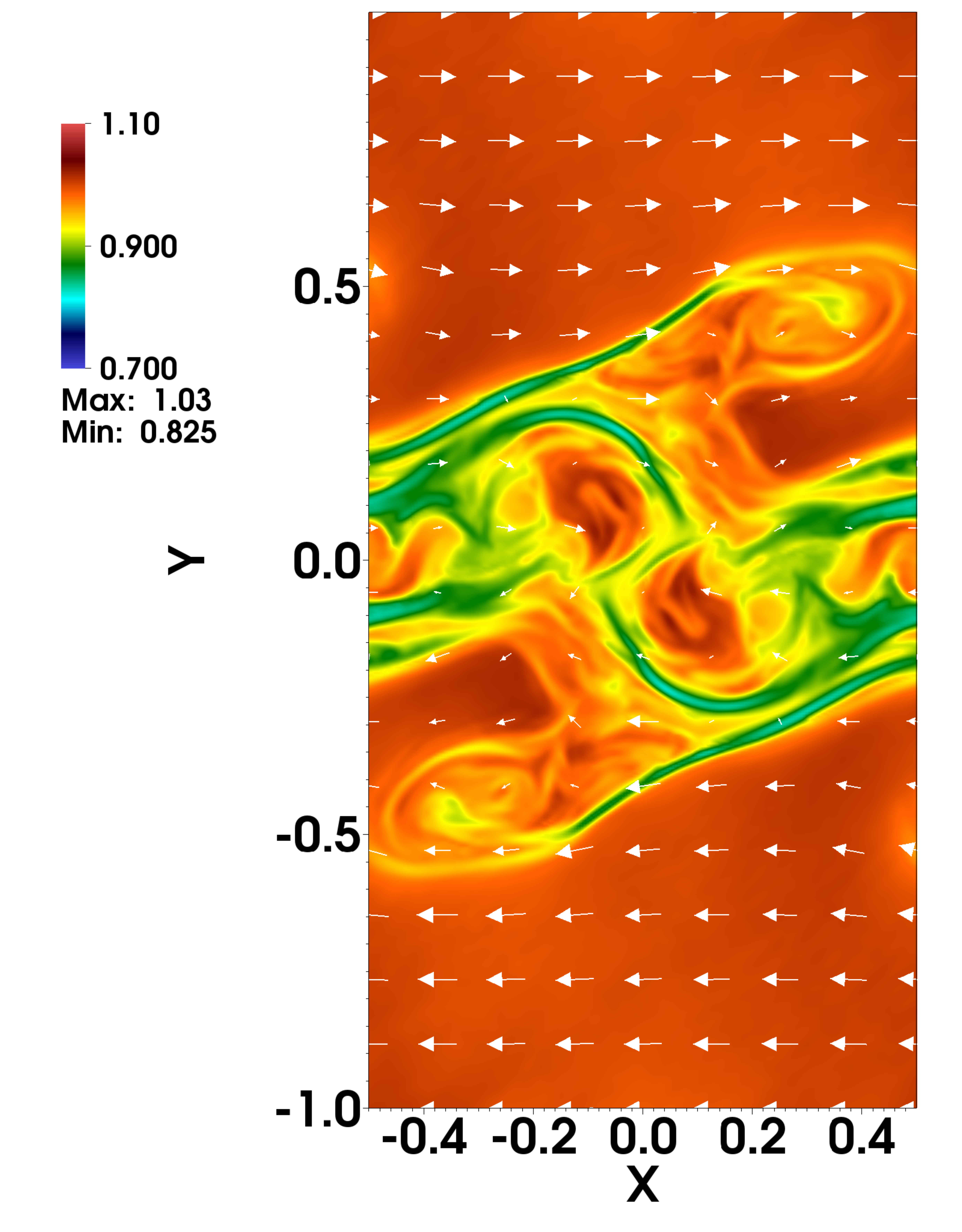}\\
	\includegraphics[width=0.49\linewidth]{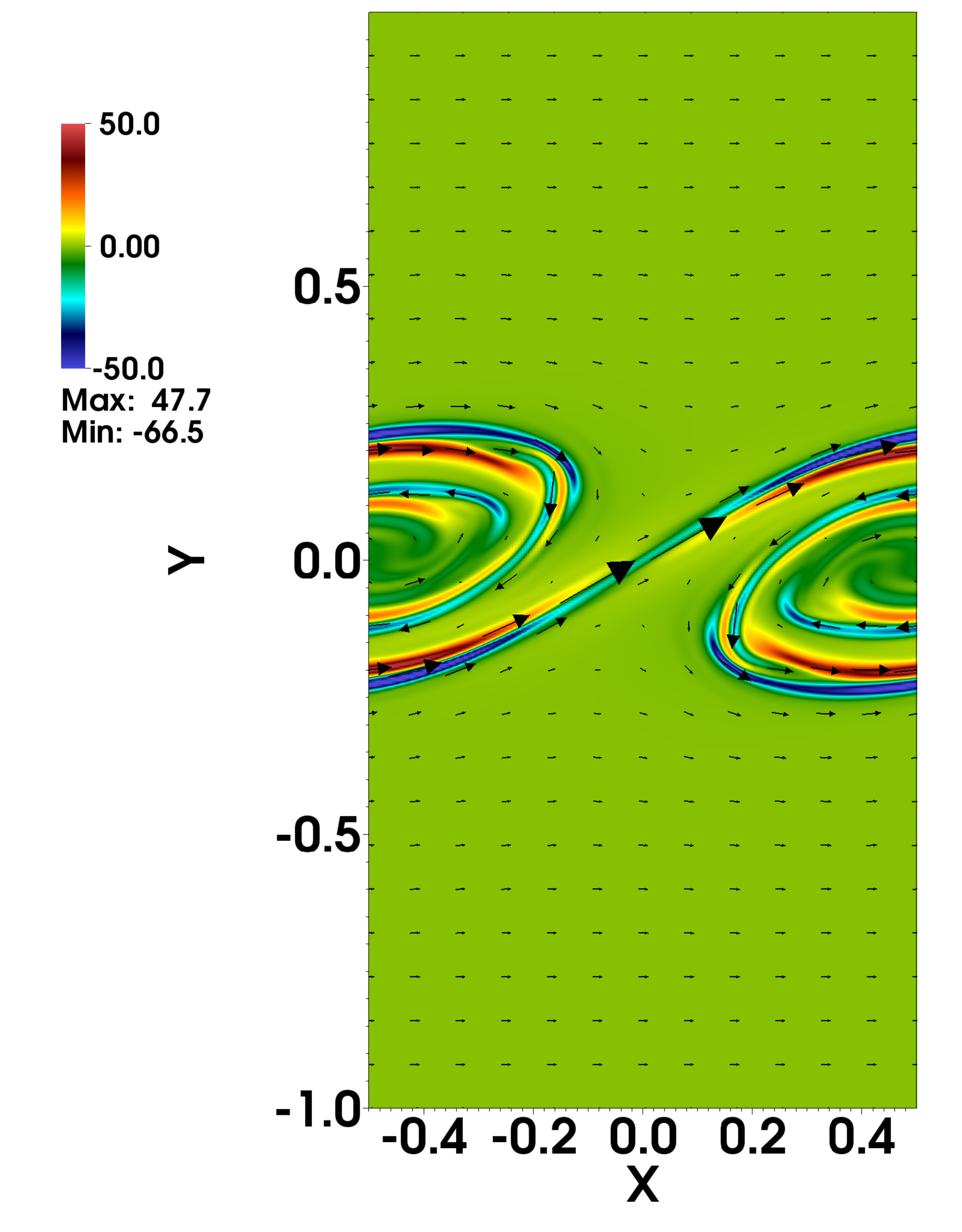}
	\includegraphics[width=0.49\linewidth]{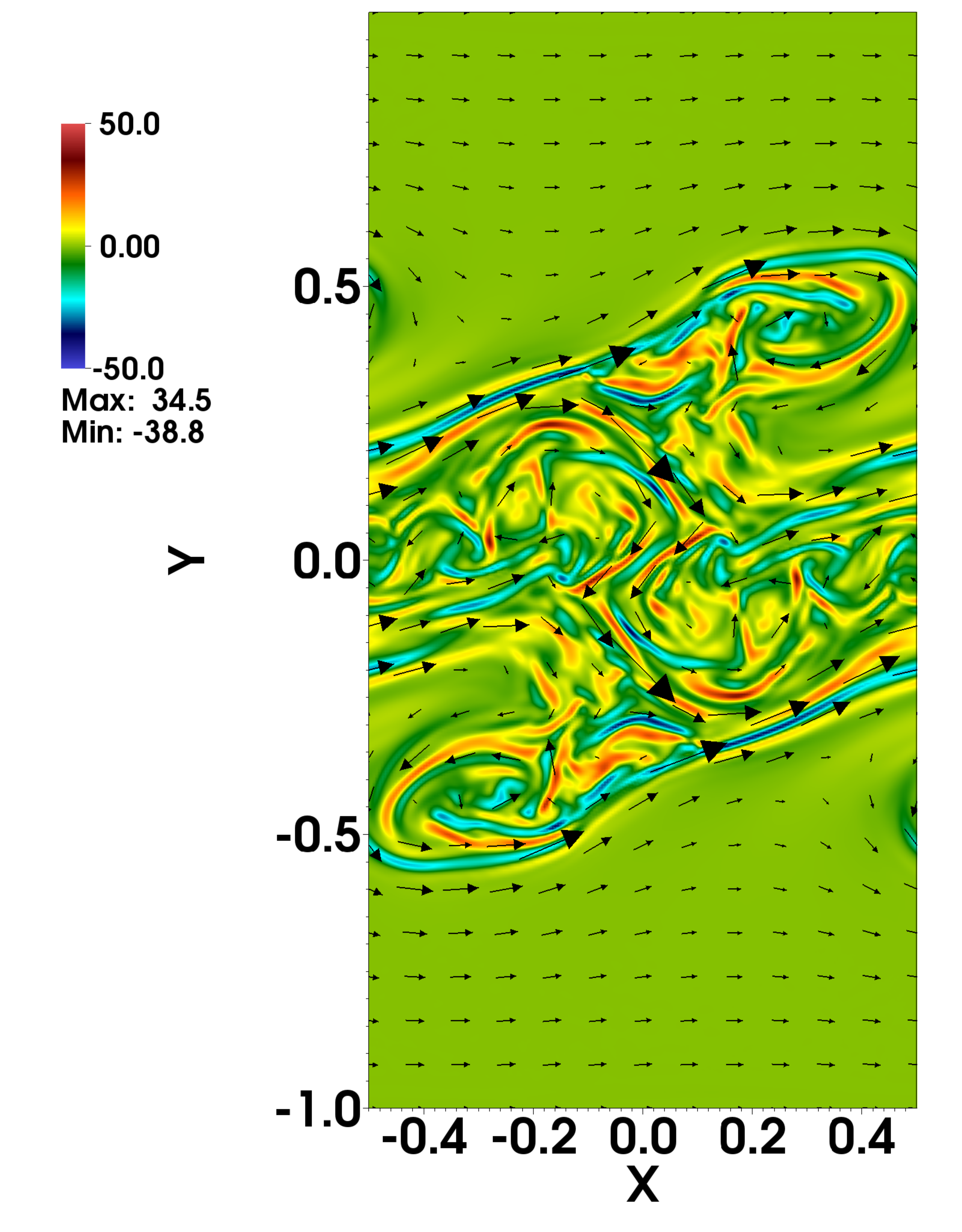}
	\caption{Development of KHI for the single layer problem, at $t=10$ (left panels) and $t=15$ (right panels). We show the density with colors and the velocity fields with white arrows for the model {\tt grw3} (top panels) and {\tt grw10} (middle panels), and the vorticity with colors and the magnetic fields with black arrows for the model {\tt grw10} (bottom panels).}
	\label{fig:khsl_grwplots} 
\end{figure}

\begin{figure}[ht] 
	\centering
	\includegraphics[width=0.75\linewidth]{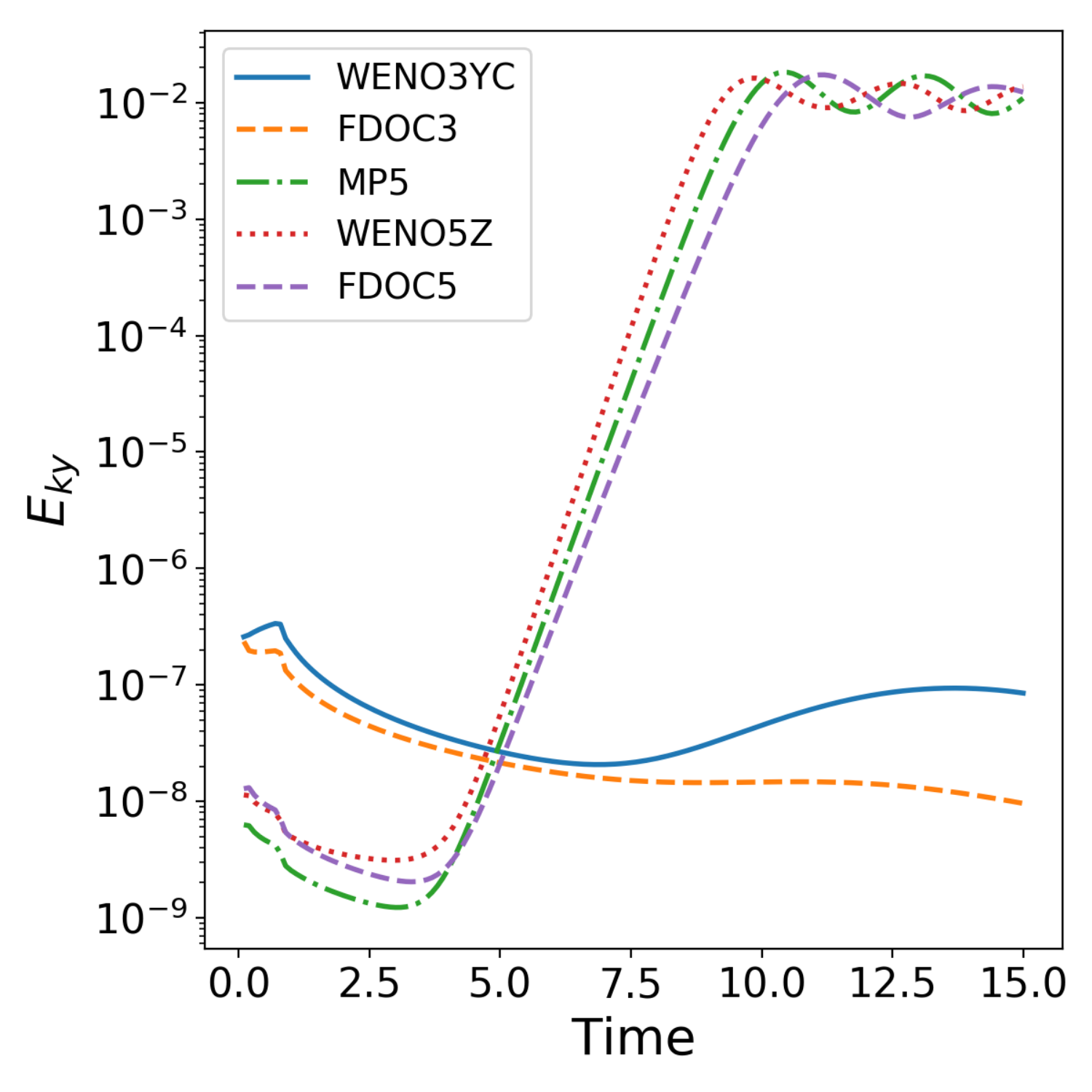}\\
	\includegraphics[width=0.75\linewidth]{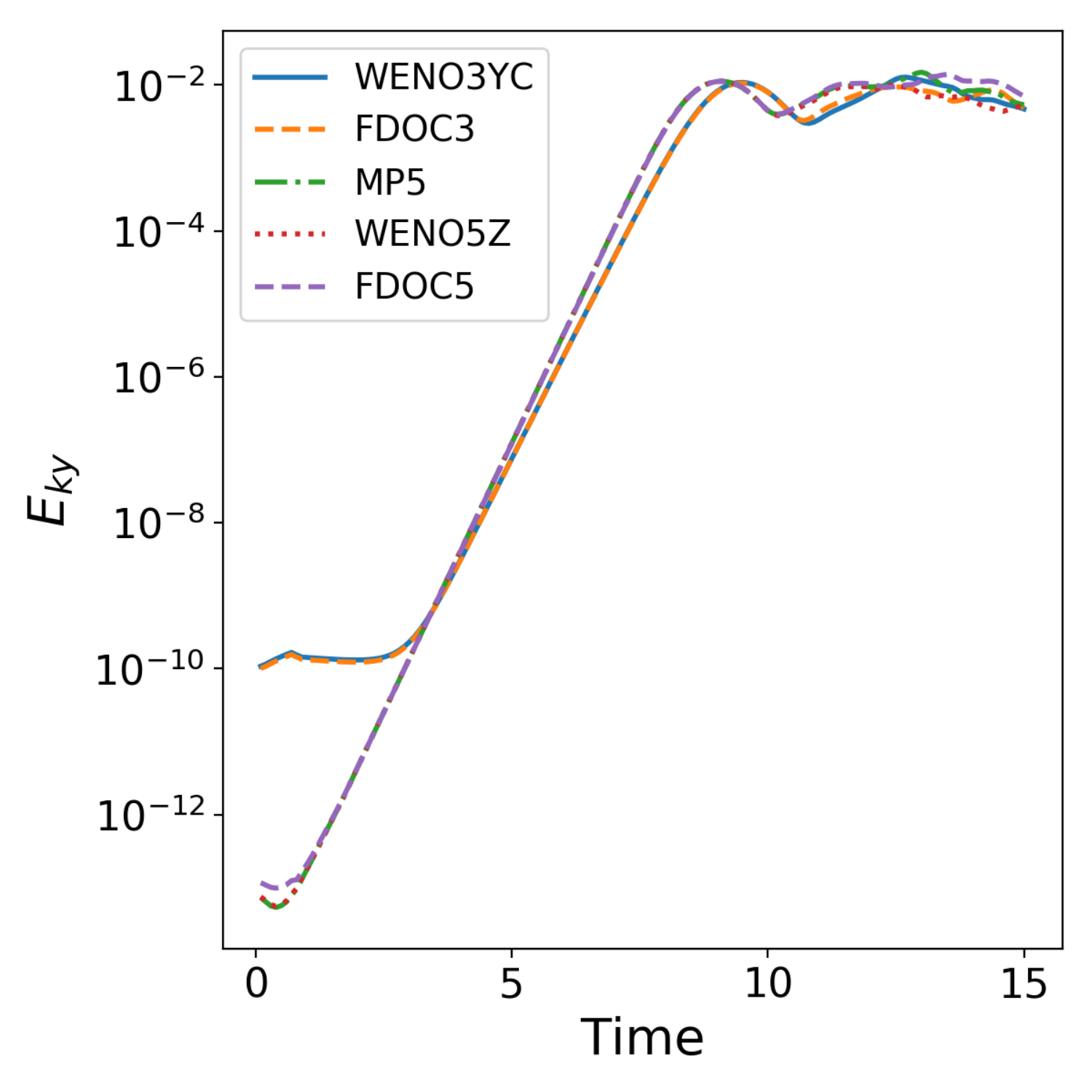}
	\caption{Comparison of the growth of $E_{ky}$ for the cases {\tt grw1} (top) and {\tt grw10} (bottom), with the five different reconstruction methods.}
	\label{fig:khsl_grw_methods} 
\end{figure}

\begin{eqnarray}
&& v_x = \frac{v_0}{2}\tanh{\frac{y}{a_l}}~, \label{eq:khslvxin} \\
&& v_y = \delta v_y \exp[-(y/4a_l)^2]\sin(2\pi k_x x)~, \\
&& B_x = B_0 ~, \\
&& B_y = 0~,
\end{eqnarray}
where $v_0$ is the shear velocity, and $\delta v_y\ll v_0$ represents the initial perturbation. We impose periodic boundaries in the $x$-direction, and open conditions (i.e., copying the values of the last cell to the ghost cell) in the $y$-direction (see a previous detailed study\cite{obergaulinger10} for the effects of changing the boundaries in the $y$-direction from open to reflecting, but this is not the focus of our work).

\begin{figure}[ht] 
	\centering
	\includegraphics[width=0.75\linewidth]{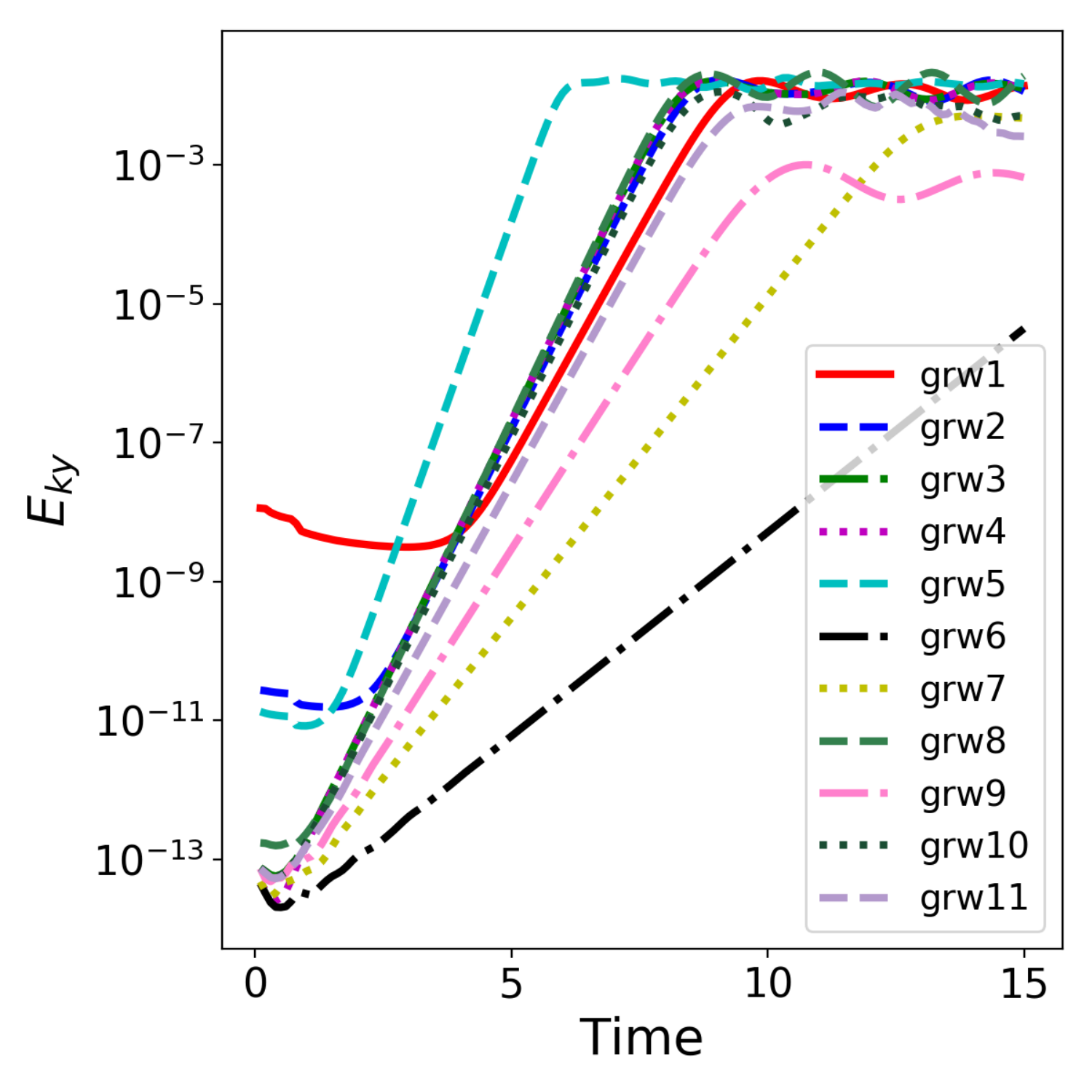}
	\caption{Comparison of the growth of $E_{ky}$ for the tested models with WENO5Z.}
	\label{fig:kh2d_methods} 
\end{figure}

\begin{figure}[ht] 
	\centering
	\includegraphics[width=0.75\linewidth]{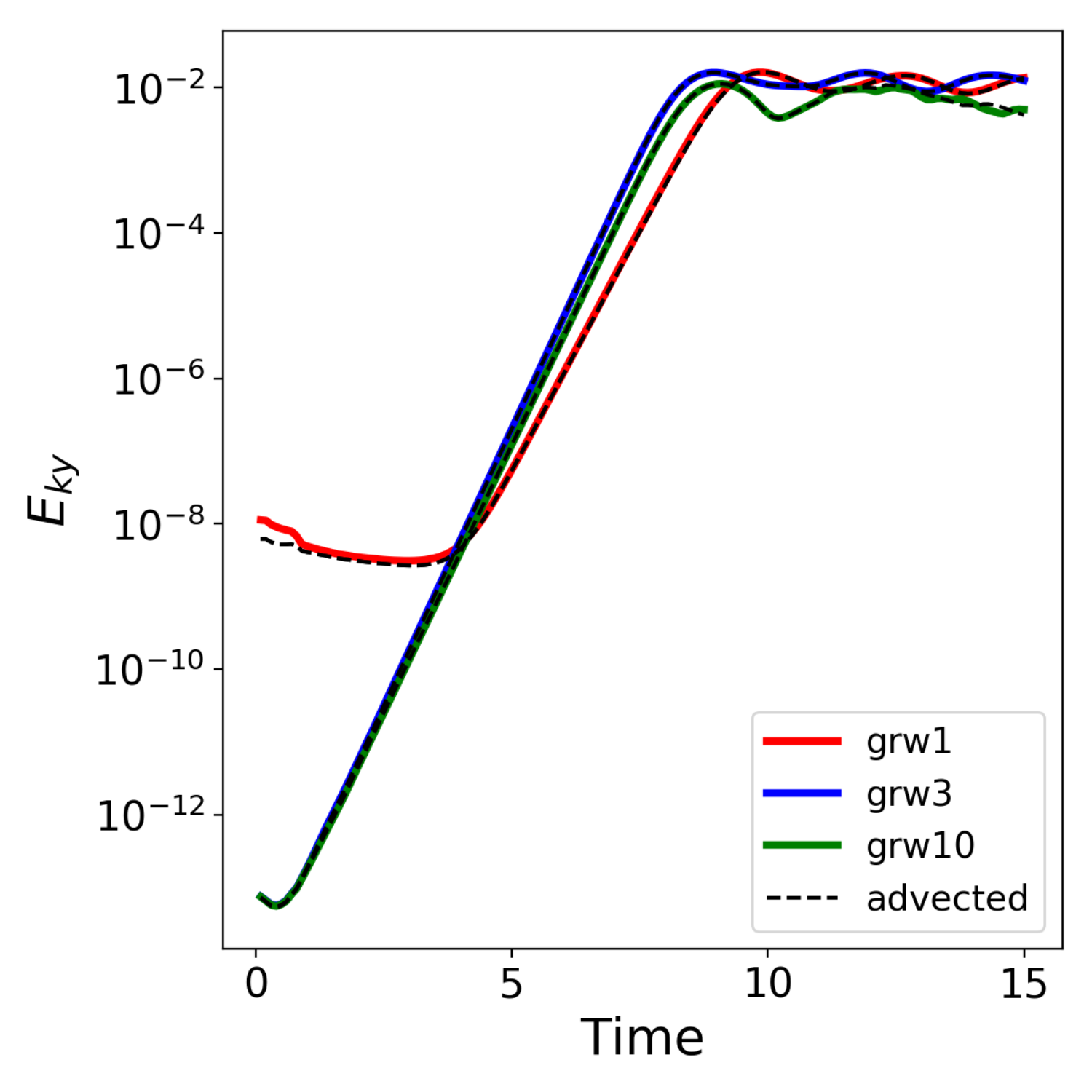}
	\caption{Comparison of the growth of $E_{ky}$ for simulations with WENO5Z, for the models {\tt grw1, grw3, grw10}, without (colors) or with (dashed thin lines) an initial additional uniform advection velocity $v_{x,\rm adv}=1$.}
	\label{fig:khsl_adv} 
\end{figure}

\begin{figure*}[ht] 
	\centering
	\includegraphics[width=0.24\linewidth]{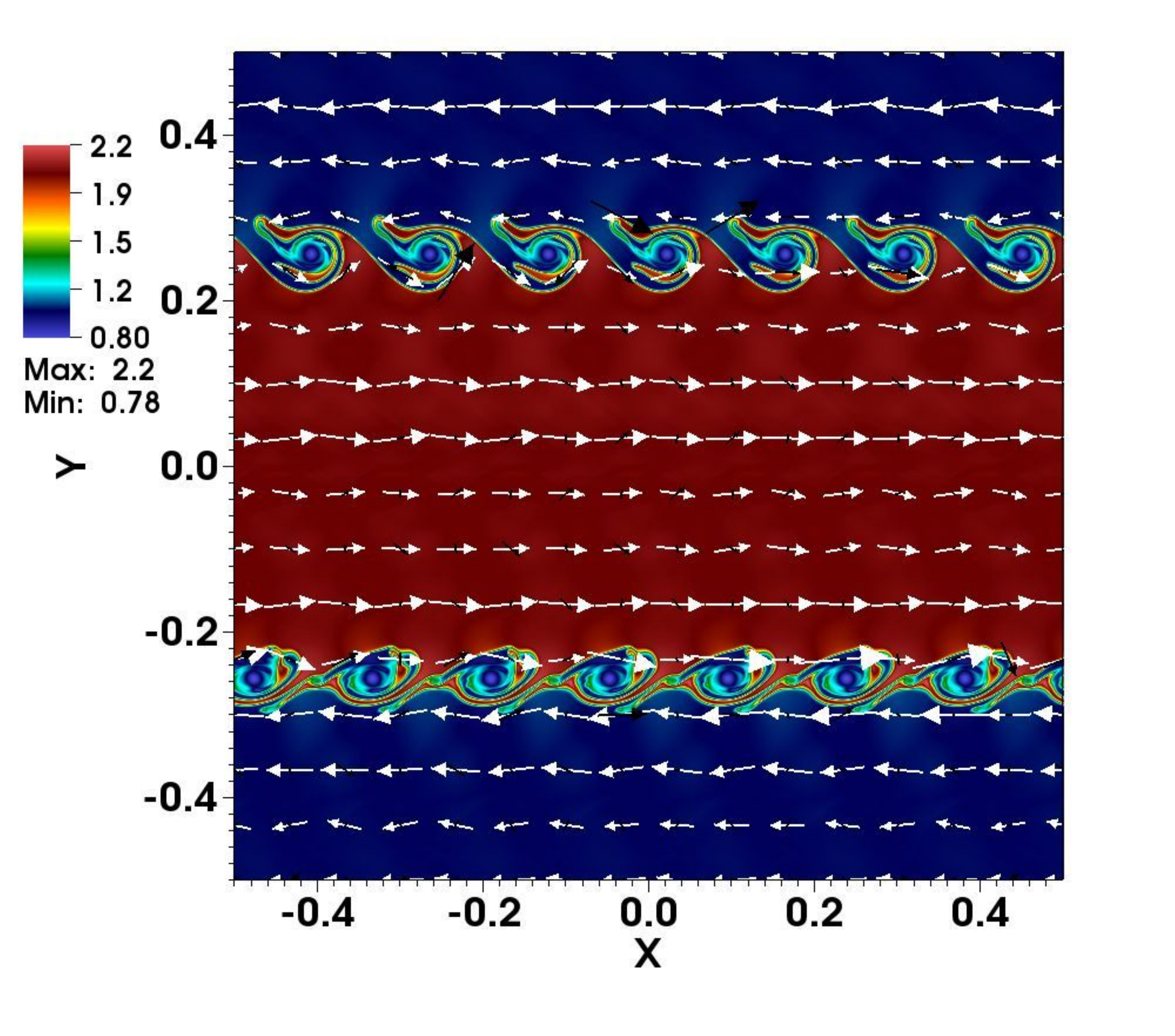}
	\includegraphics[width=0.24\linewidth]{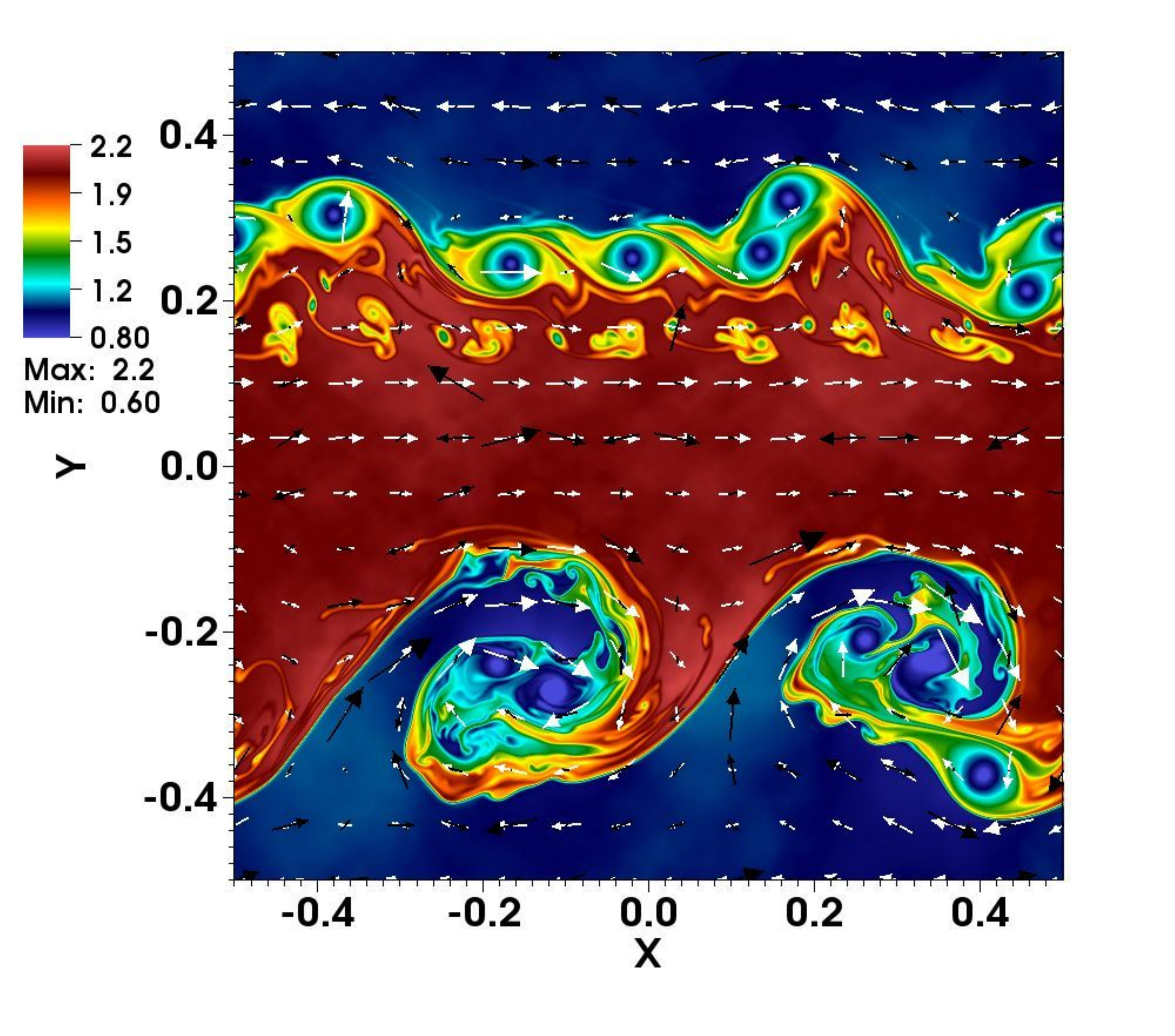}
	\includegraphics[width=0.24\linewidth]{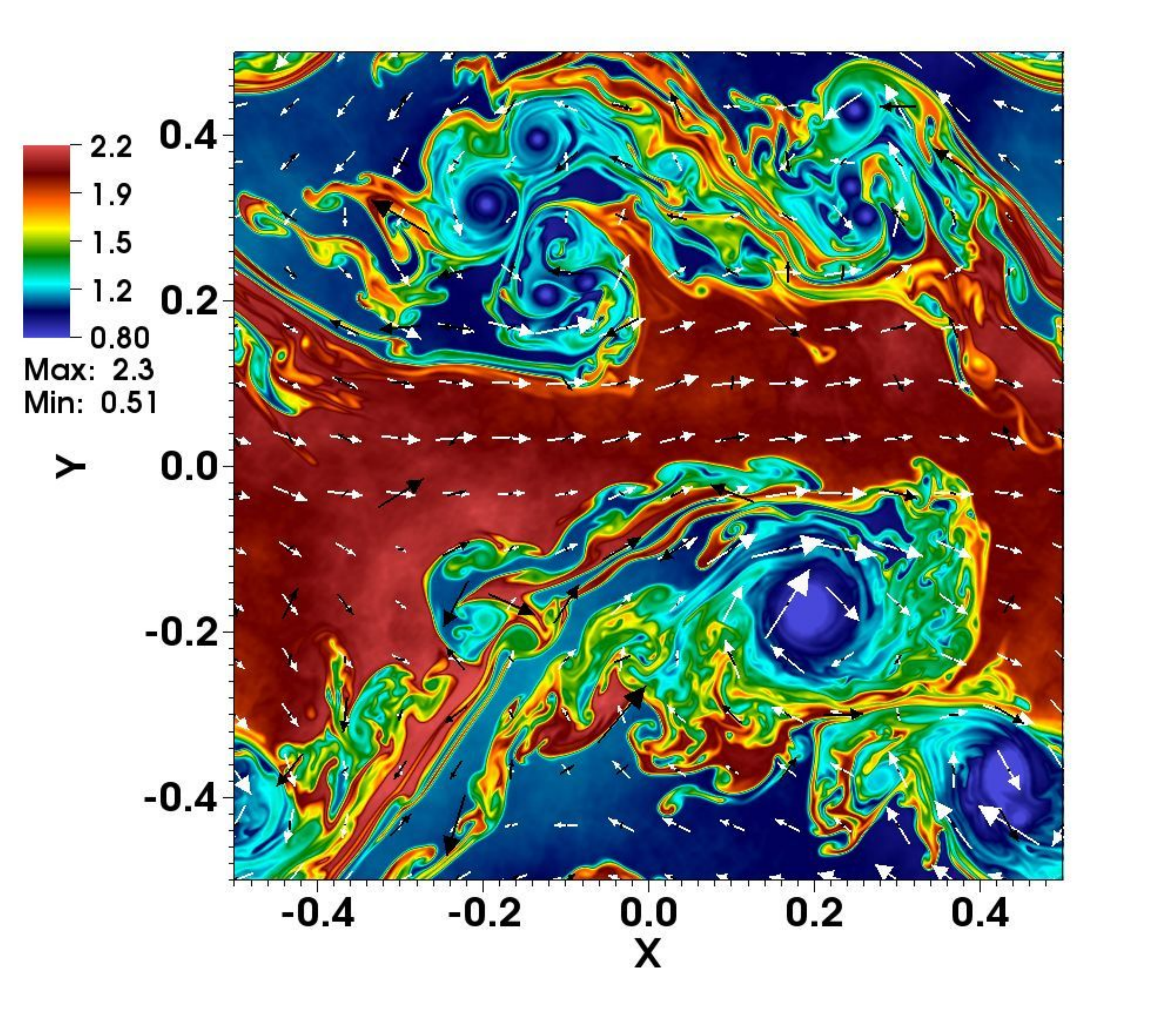}
	\includegraphics[width=0.24\linewidth]{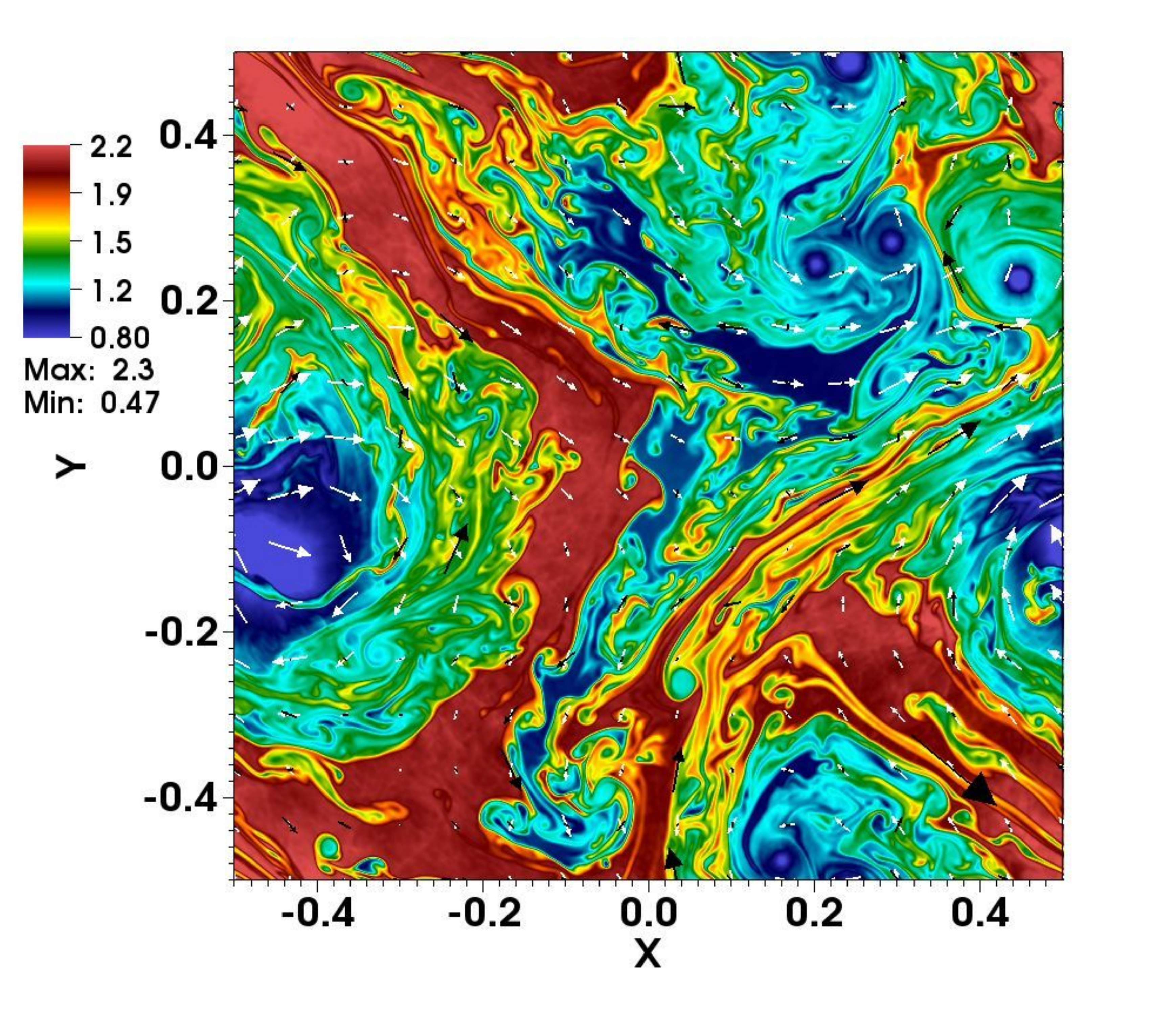}
	\caption{Image of the KHI simulation in 2D with $1000^2$ points at $t=\{0.5,3,5,7\}$ from left to right, respectively. Colors indicate the density, while white and black arrows represent the velocity and magnetic fields, respectively.} 
	\label{fig:kh2d_b_1000_evo} 
\end{figure*}

\begin{figure*}[ht] 
	\centering
	\includegraphics[width=0.33\linewidth]{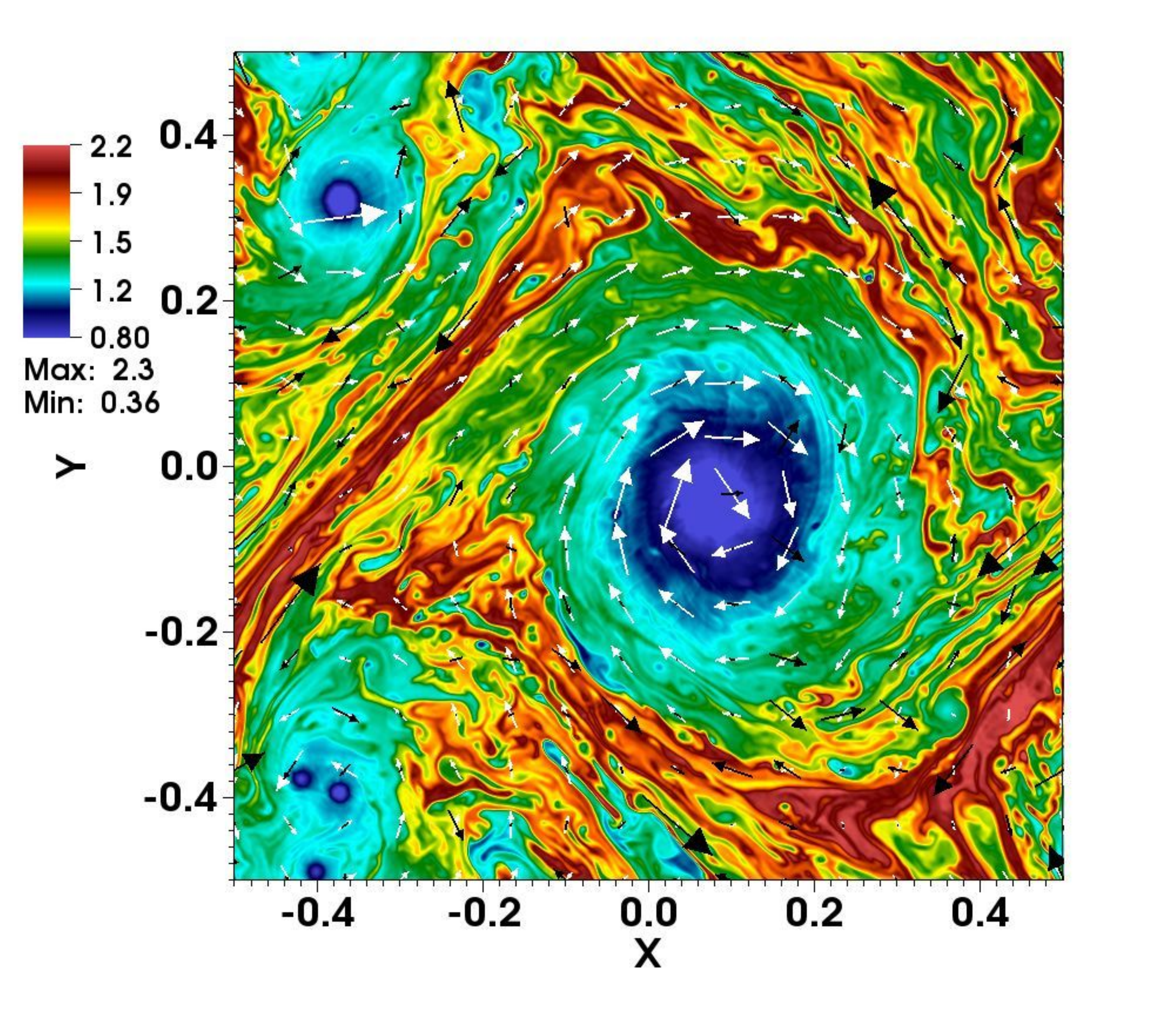}
	\includegraphics[width=0.33\linewidth]{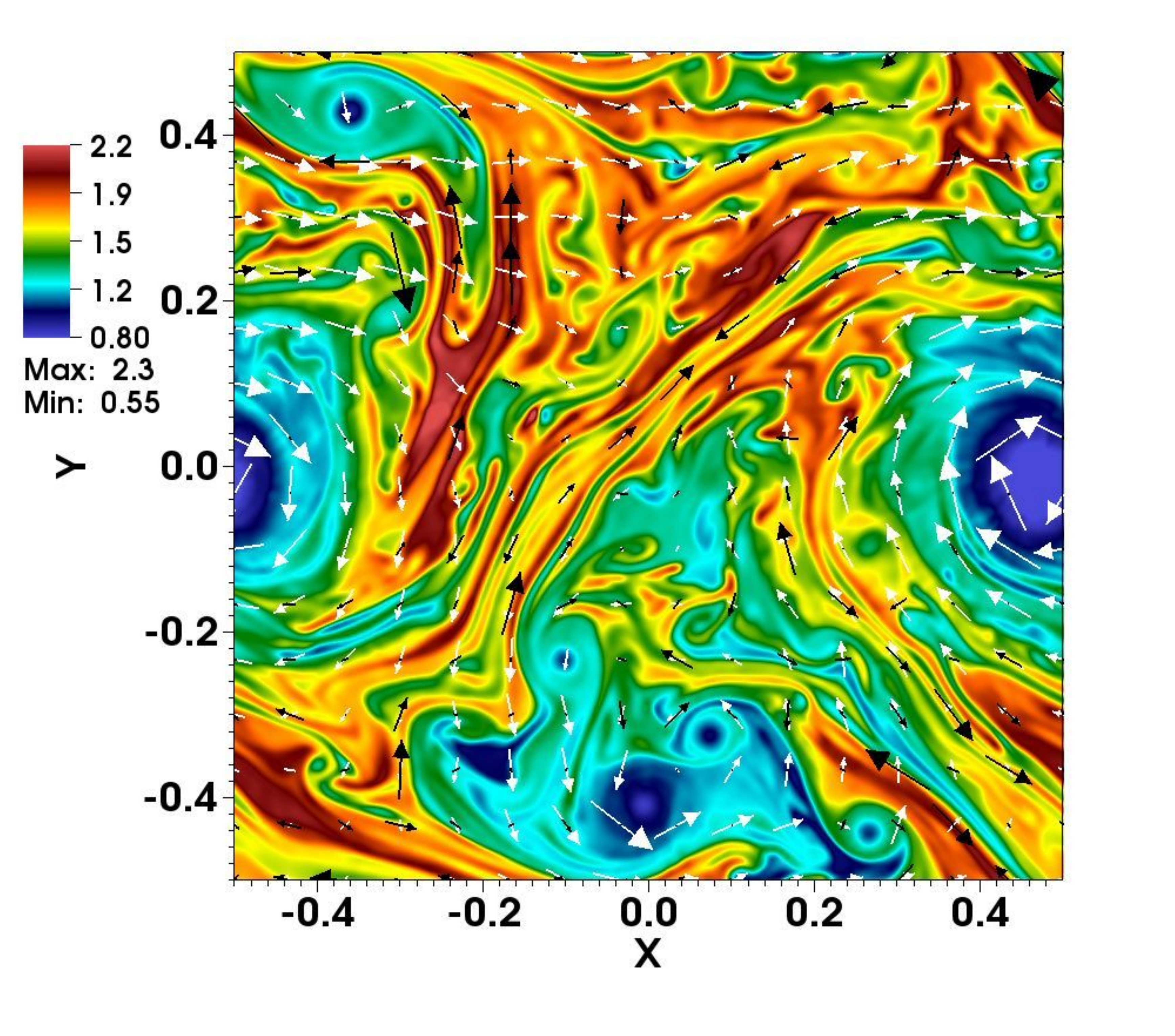}
	\includegraphics[width=0.33\linewidth]{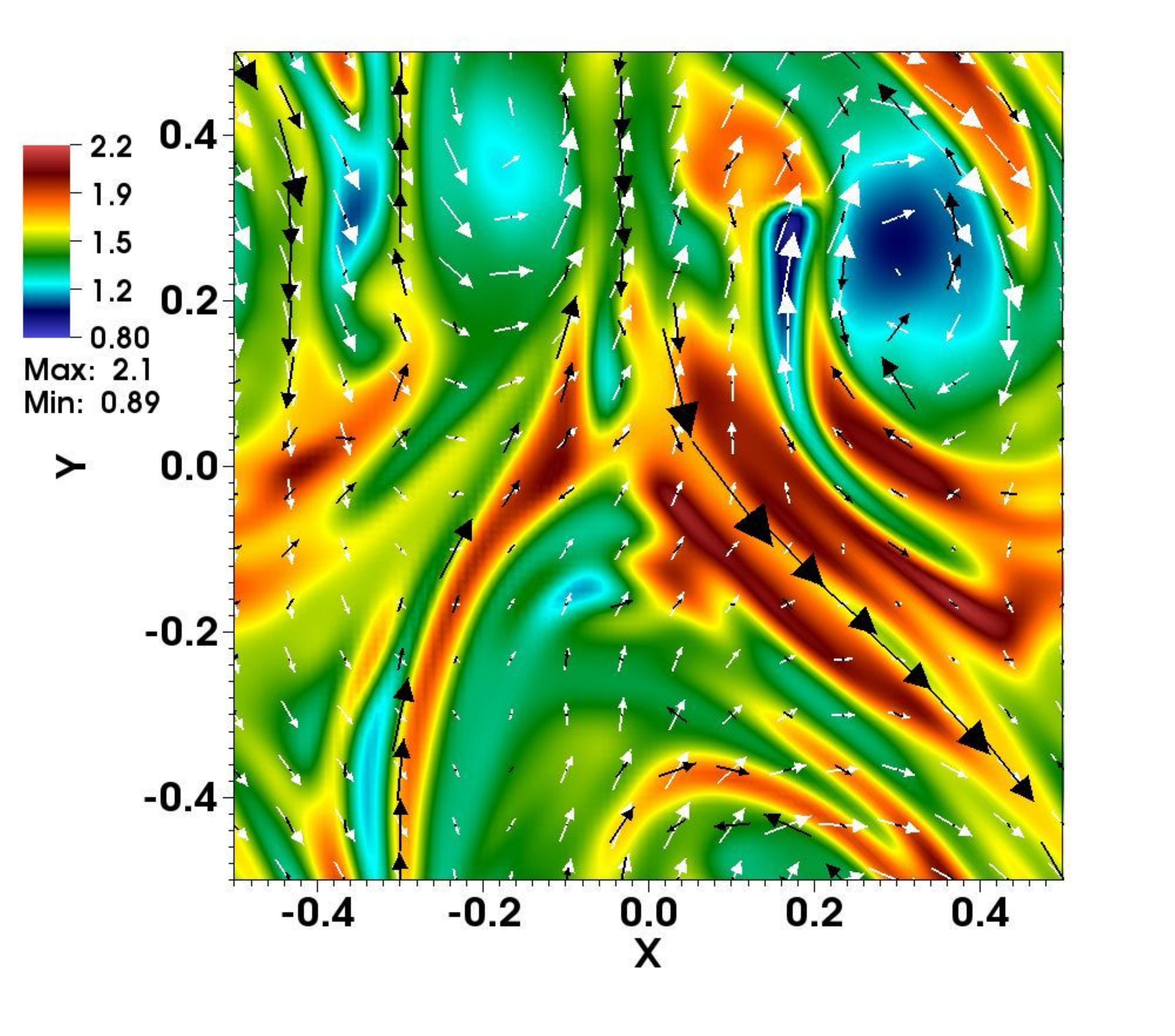}
	\caption{Image of the KHI simulation in 2D with $N^3=\{1000^2,400^2,100^2\}$ points from left to right, at time $t=10$. Colors indicate the density, while white and black arrows represent the velocity and magnetic fields, respectively.} 
	\label{fig:kh2d_b_img} 
\end{figure*}

We set $\delta v_y=10^{-6}$ and we considered different cases, as listed in Table~\ref{tab:khsl}, similar to Table A.1 in the reference paper.\cite{obergaulinger10} We use different reconstruction methods: WENO3YC, FDOC3, FDOC5, MP5, WENO5Z (for details about the parameters of these models and further tests, see our dedicated paper\cite{palenzuela18}). We follow the simulations well beyond the saturation time, with a time-step $\Delta t = 0.25 ~\Delta$, where $\Delta=L_x/N$ is the homogeneous resolution of the numerical grid.

As a first example, in Fig.~\ref{fig:khsl_grwplots} we show the plots of the density and velocity fields (white arrows), at $t=10$ and $t=15$ (left/right panels, respectively), for the reference model {\tt grw3} (top panels) and its magnetic version, {\tt grw10} (middle panels). For the latter, we also show in the bottom panel the plots of vorticity (with colors) and magnetic field (black arrows). After an initial phase, the instability is triggered and develops vortices, until a quasi-stationary configuration is reached.

A quantitative tracer of the degree of the turbulence developed at a given time during the development stage is the $y$-component of the kinetic energy integrated in the volume:

\begin{equation}\label{eq:ekiny}
E_{ky} = \int_V\frac{1}{2}\rho v_y^2~ {\rm d} V~, 
\end{equation}
After an initial time, $E_{ky}$ tends to grow exponentially until it saturates. We fit the exponential growth rate phase with a function $E_{fit} = E_0 e^{2\alpha_{\rm kh} t}$, and compare the best-fit values of $\alpha_{\rm kh}$ with the numerical values obtained by the numerical simulations available in literature\cite{obergaulinger10}, and the ones theoretically obtained by means of a theoretical stability analysis\cite{miura82}.

The growth rate and the saturation value of $E_{ky}$ are physically controlled by different values of $v_0$, $a_l$, $k_x$ and $B_0$, but not by the amplitude of the perturbation $\delta v_y$, which only influence the time at which the instability starts to develop. Furthermore, the numerical methods and resolution are crucial to determine the growth rate and the instability trigger time.

In Fig.~\ref{fig:khsl_grw_methods} we show the $E_{ky}$ growth for two of the eleven cases analyzed and listed in Table~\ref{tab:khsl}: {\tt grw1} (low resolution) and {\tt grw10} (magnetic case). Inspecting the first, low-resolution case, {\tt grw1}, we underline that the third-order methods, FDOC3 and WENO3YC, are not able to capture the instability in the low resolution case, while the fifth-order methods do, even though the growth rate is different. In particular, WENO5Z gives the higher value, slightly larger than MP5 and and FDOC5, and slightly smaller than the theoretical reported values\cite{obergaulinger10} (see last two columns of Table~\ref{tab:khsl}). Such differences disappear already with intermediate resolutions, $N=200$ (model {\tt grw3}, right panel). In this case, the third-order method still show a value of $\alpha_{\rm kh}$ slightly inferior than the theoretical value, which is instead perfectly matched by the simulations with the three fifth-order methods. The three of them are basically indistinguishable.

We then test all the cases of the benchmark paper\cite{obergaulinger10} with the method WENO5Z, which was the one with the best behavior. For all of them, we find an excellent agreement with the theoretical expectations\cite{miura82}, as shown in Table~\ref{tab:khsl}. One on side, the trends between the growth rate and the physical parameters $v_0$ (compare {\tt grw3, grw7, grw8}), $a_l$ ({\tt grw3, grw5, grw6}), $B_0$ ({\tt grw3, grw10, grw11}) and $k_x$ ({\tt grw3, grw9}) are correctly recovered. On the other hand, the differences among the models {\tt grw1, grw2, grw3} and {\tt grw4} show how the resolution allows one to better capture the instability. For this particular case, having $N \gtrsim 200$ seems to be enough to capture the instability. However, we underline that the importance of the numerical resolution is amplified if the initial perturbation mode ($k_x$) is larger: the smaller the scale, the more important the resolution and methods become.

As a final preliminary test with this setup, we examine the possible issue considered in literature\cite{hopkins15}: in presence of a uniform, Galilean boost in one direction (i.e., an additional advection velocity, constant everywhere), the development of the KHI can be numerically inhibited, due to the numerical dissipation which is enhanced by the value of the advecting velocity. Thus, we compare the same model ({\tt grw3} in this case), with and without an additional advection in the direction $x$ with a value comparable to the difference of velocities, $v_{x,\rm adv}=1$, such that, initially, eq.~(\ref{eq:khslvxin}) becomes $v_x = 1 + \frac{v_0}{2}\tanh{\frac{y}{a_l}}$. Fig.~\ref{fig:khsl_adv} shows the growth of the kinetic energy with and without the advection, with WENO5Z, for the problems {\tt grw1, grw3, grw10}. No important differences are noticed. We repeated the test with for the cases with lower-order methods (WENO3YC, FDOC3) and/or the low-resolution problems, and in any case we found an appreciable difference.

This first test shows that the numerical implementation is correct and able to reproduce the expected growth rate of a single-mode KHI, as long as at least one among the resolution and the accuracy order is high enough.

\subsection{Periodic 2D Kelvin Helmholtz tests}\label{app:kh2d_periodic}

We also tested the KHI in a periodic 2D box with a double mixing layer, similar to the 3D case presented in the main text. This has allowed us to compare the available results for 2D turbulent periodic boxes\citep{beckwith11,stone08,lecoanet16,mcnally12}.\footnote{See also Athena webpage:\\{\tt http://www.astro.princeton.edu/\~jstone/Athena/tests/}} We consider a square domain with ${x,y}$ both in the range $[-L/2,L/2]$ with $N$ points in each direction. The initial positions of the mixing layers are $y=\pm L/4$, as the initial set-up describes:

\begin{eqnarray}
&& \rho= \rho_0 \pm \delta \rho ~, \\
&& v_x = \pm 0.5 v_0 + \delta v_x ~, \label{eq:kh2dvxin} \\
&& v_y = \delta v_y ~, \\
&& p = p_0 ~,\\
&& B_x = B_0 ~,\\
&& B_y = 0~,
\end{eqnarray}
where the sign $\pm$ indicate the region $|y|>L/4$ and $|y|\leq L/4$, respectively. The perturbation in each $i$ direction is given either by an uniform random distribution within $[-\delta v_0,\delta v_0]$, or by $\delta v_i = \delta v_0 \cos(2\pi x n_i/L)$, where $n_i \in [0,N/2]$ are integers and they can be different in the two directions.

Hereafter we set $L=1$, $v_0=1$, $\rho=1$, $\delta \rho=0.5$, $p_0=2.5$, and we use an ideal equation of state with $\gamma=1.4$, with a sinusoidal perturbation with $\delta v_0 =0.01$, $n_x=4$ and $n_y=7$, unless specified otherwise. We impose periodic boundary conditions in all directions and we run with a time-step $\Delta t = 0.25~\Delta$, in order to be sure that the numerical errors are not dominated by the time discretization. We run up to $t=10$, a time long enough to reach the quasi-stationary state, and we run the same simulation with different resolutions, $N \in [100^2, 2000^2]$ with WENO5Z. 

We have run different values of the initial magnetic field. In general, the initially constant magnetic field partially inhibits the KHI. This feature is well known and is quantified by the Alfv\'en factor, defined as the ratio between hydro and Alfv\'en velocities, $A= v_0/\sqrt{B^2/\rho}$. When $A \lesssim 1$, the KHI is totally inhibited, as explained in detailed in previous works.\cite{obergaulinger10} In our case, we start from a value $A \sim 10^3$, thus the KHI takes place.

Fig.~\ref{fig:kh2d_b_1000_evo} shows the evolution for a $N=1000^2$ simulation with WENO5Z, including magnetic field with $B_0=0.001$. We show the density map (color scales) and the velocity and magnetic fields with arrows, at different times. 

The stages of the dynamics are similar to the 3D case: at the beginning, vortical structures are formed with a pattern given by the $k_x$ wave-numbers of the initial perturbation ($t\lesssim 1$ for the chosen values of $\delta v_y$). The symmetry of the patterns is broken when the structures at the two mixing layers start to interact ($t\gtrsim 1$). Then, the turbulence, initially concentrated around the mixing layer, tends to homogeneously spread over the entire dominion, thanks to the periodic boundary conditions. Meanwhile, the vortexes tend to merge (the well-known inverse kinetic cascade in 2D), and at late time ($t \gtrsim 10$ in this case) only two large vortexes of opposite vorticity sign survive. The turbulent dynamics reaches a saturated, fully developed state, dominated by these large vortexes, and, given the absence of any forcing, it decays, transforming kinetic energy into internal energy.

For our setup, when the system reaches a quasi-stationary state, the magnetic energy remains still significantly smaller then the kinetic energy, and is significantly stored in small scales. Kinetic and magnetic spectra show the same characteristics as the 3D case (main text), with the expected 2D Kolmorogov and Kazantsev slopes in the inertial range.

The timescales and the resolved structures mainly depend on the problem parameters ($B_0$, $\delta v_y$), and on the resolution, respectively. At this respect, Fig.~\ref{fig:kh2d_b_img} shows the plot of density with colors, and the velocity and magnetic fields at $t=10$, for decreasing resolution $N=1000^2,400^2,100^2$. It is evident how the the resolution affects the capability to capture the turbulent details.

We have also performed the a-priori fitting for different resolutions and filter sizes. We report a representative case in Fig.~\ref{fig:apriori_2D}, briefly mentioned in the main text. Overall, for all tensors and SGS modles results are really similar to the 3D ones.

We checked that the growth of $E_{ky}$, indicating the development of the turbulent state, is enhanced if high resolution or high-order methods are used, and MP5 and WENO5Z provide statistically compatible results, if the same resolution is used. The difference between two different numerical schemes are only stochastic: the particular position of the vortexes can be different, but the spectra and the integrated quantities are statistically indistinguishable. In the main text, all simulations are run with WENO5Z.

The same holds if we compare the fully developed turbulence obtained with different initial perturbations: the specific time of growth can change, but at saturation the runs are statistically equivalent. In this sense, we checked that the values of $\delta v_0$ or the specific form of the perturbation can affect the delay of the beginning of the exponential growth, but not the growth rate and the asymptotic value of $E_{ky}$. This holds always as long as the initial values of the perturbation is random, or it is sinusoidal but at least one among $n_x$ and $n_y$ of the initial perturbation has no common integer factors with $N$. When the three numbers have a common integer factor $C$, then the initial data are effectively periodic in the $x$-direction on a scale $L/C$. Therefore, in these cases, a mode with wavenumber $C$ and its higher harmonics are maintained during the entire simulation, since the $x$-scales larger than $L/C$ are not explored, thus preventing a proper spectral cascade (the spectrum is constituted by spike corresponding to the chosen modes and its harmonics) and the related full development of the turbulence. For instance, we numerically proved that, starting with $n_x=n_y=4$ or $n_x=n_y=8$, and $N=500$ (for which $C=4$), the simulations show a periodicity in the $x$-direction given by $L/4$, being 4 the common factor between $N$, $n_x$ and $n_y$. This is a further test for the correct numerical implementation, that shows no sign of numerical break of symmetry.

Summarizing, the 2D tests has allowed us to validate the capability of the code to simulate turbulent dynamics, and to identify WENO5Z as the best numerical method among the tested ones. That is what we use in the main text.

\bibliography{turbulence}
\newpage

\end{document}